\documentclass[a4paper,twoside,openright,12pt]{report}
\usepackage{jheppub}
\usepackage[nottoc,numbib]{tocbibind}
\usepackage{amsmath}
\usepackage{amssymb}
\usepackage{graphicx,color,slashed,mathtools,bm,bbm,float,placeins}
\usepackage{bbm} 
\usepackage{ifpdf}
\usepackage{setspace}
\usepackage{pdfpages}
\usepackage{emptypage}
\usepackage[bottom]{footmisc}
\newcommand{\be}{\begin{equation}}
\newcommand{\ee}{\end{equation}}
\newcommand{\bea}{\begin{equation}\begin{aligned}}
\newcommand{\eea}{\end{aligned}\end{equation}}
\newcommand{\ba}{\begin{eqnarray}}
\newcommand{\ea}{\end{eqnarray}}

\def\rmi{{\rm i}}
\def\rme{{\rm e}}
\def\ib{{\bar \imath}}
\def\jb{{\bar \jmath}}
\def\rmre{{\rm Re}}
\def\rmim{{\rm Im}}

\newcommand{\V}{\mathcal{V}}

\usepackage[parfill]{parskip} 
\usepackage{array} 
\usepackage{ifthen} 
\usepackage{setspace}

\def\sectionlineskip{\medskip} 
\def\sectionskip{\medskip} 

\title{Flux Compactifications, dS Vacua and the Swampland}
\author{Christoph Oliver Roupec MSc}
\abstract{One of the main goals of the field of string phenomenology is to describe the properties of our spacetime in low-energy effective models related to and consistent with string theory. It is known from observations that our universe undergoes accelerated expansion, which can be described in Einstein's theory of general relativity as a positive cosmological constant. Within string theory descriptions, this is realized by a scalar field with a positive value of its potential. Often times, this is modeled by having the scalar at a stable, positive minimum. One then obtains a de Sitter spacetime. Constructions of de Sitter spaces from string theory are not straight forward and face many criticisms. In this thesis we address several aspects related to this topic.\\
We investigate the description of general branes in supergravity and in particular describe the uplifting anti-$D3$-brane of the KKLT model, including all world-volume fields. Additionally, we translate the type IIB based KKLT model into type IIA and introduce a mechanism that allows for the rapid construction of many de Sitter solutions. Related to this we investigate models based on twisted $7$-tori from M-theory and their relation to type II supergravity. These setups can also yield de Sitter solutions using the same mass production mechanism.\\
One of the more recent and serious claims against the construction of de Sitter spaces in string theory, the de Sitter swampland conjecture, is also addressed here. We highlight the difficulty of obtaining a stable, positive vacuum energy in classical type IIA flux compactifications and their reliance on $O$-plane sources. Then, we turn our attention to the conjecture itself. Contrary to the initial claim, we find many unstable de Sitter points in numerical searches and go on to present our own refined conjecture that differs in several key aspects from the refined version of the original conjecture.\vspace{12pt}\\
This thesis is based on the publications \cite{Roupec:2018mbn,Banlaki:2018ayh,Andriot:2018mav,Cribiori:2019hod,Cribiori:2019bfx,Cribiori:2019drf,Cribiori:2019hrb,Cribiori:2020bgt} that I worked on during my doctoral studies.
}

\allowdisplaybreaks 

\begin{document}
\pagestyle{myplain}
\setlength{\parindent}{0pt}
\renewcommand{\arraystretch}{1.2} 


\pagenumbering{roman}
\setcounter{page}{0}

{\thispagestyle{empty}
\includegraphics[scale=0.08]{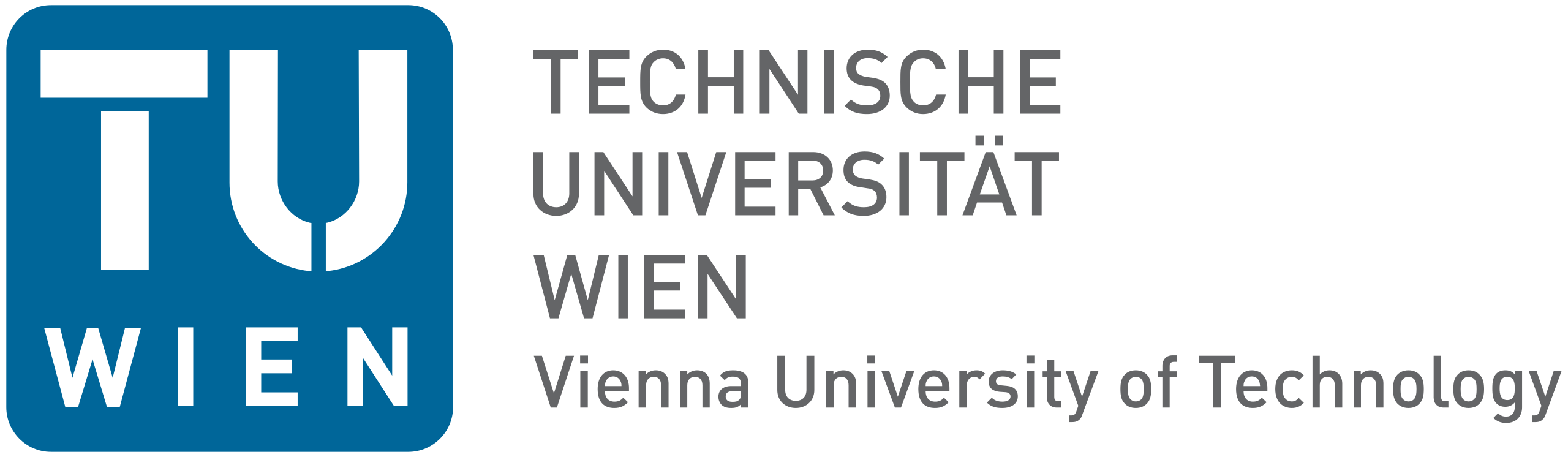}
\vspace{48pt}
\begin{center}
{\large
Dissertation\vspace{12pt}\\
{\Large\textbf{Flux Compactifications, dS Vacua and the Swampland}}\vspace{12pt}\\
ausgeführt zum Zwecke der Erlangung des akademischen Grades eines Doktors der Naturwissenschaften\vspace{12pt}\\
unter der Leitung von\vspace{12pt}\\
Privatdozent Timm Michael Wrase, PhD\\
E136\\
Institut für Theoretische Physik\vspace{12pt}\\
eingereicht an der Technischen Universität Wien\\
Fakultät für Physik\vspace{12pt}\\
von\vspace{12pt}\\
{\Large\textbf{Christoph Oliver Roupec}, MSc}\\
}
\end{center}
}

\newpage\thispagestyle{empty}\vspace*{\fill}
\null 
\vspace*{\fill}\newpage
\setcounter{page}{1}
\section*{Kurzfassung}
\addcontentsline{toc}{chapter}{Kurzfassung}
Eines der Ziele der String-Phänomenologie ist es, die grundlegenden Eigenschaften der Raumzeit des Universums bei niedrigen Energien in einer konsistenten, effektiven Theorie zu beschreiben. Von kosmologischen Observationen wissen wir, dass sich unser Universum beschleunigt ausdehnt. In Einsteins Relativitätstheorie kann dies durch eine positive kosmologische Konstante beschrieben werden. In der String-Theorie ist es möglich, diese durch ein skalares Feld zu beschreiben, welches in seinem Potential an einer Stelle mit positiver Energie sitzt. Oft wird ein stabiles Minimum des skalaren Potentials verwendet, um eine de Sitter-Raumzeit zu erhalten. Konstruktionen von de Sitter-Raumzeiten in der String-Theorie sind nicht trivial und werden oft kritisiert. In dieser Dissertation werden Themen behandelt, die mit Konstruktionen von de Sitter-Raumzeiten in effektiven Theorien, die von der String-Theorie kommen, zusammenhängen.\\
Ein wichtiger Bestandteil einer Klasse von Modellen sind (anti)-$Dp$-Branen, welche wir im Rahmen von Supergravitation beschreiben werden. Besonderes Augenmerk liegt auf der anti-$D3$-Brane im KKLT Modell, welche dazu dient, die Energie des Vakuums positiv zu machen. Wir liefern hierfür eine vollständige Beschreibung, inklusive aller Felder, die im Volumen der Brane vorkommen. Des Weiteren zeigen wir, wie das KKLT Modell, welches ursprünglich in Typ IIB-String-Theorie erfunden wurde, auch in Typ IIA mittels anti-$D6$-Branen funktioniert. Außerdem führen wir eine ``Massenproduktionsmethode" ein, welche es erlaubt, unkompliziert viele de Sitter-Vakua zu konstruieren. Aufbauend darauf untersuchen wir Modelle, die auf 7 Torus-Kompaktifizierungen von M-Theorie basieren. Wir stellen fest, dass auch solche Konstruktionen zu de Sitter-Raumzeiten führen können und zeigen, wie sie mit Typ II-Supergravitation zusammenhängen.\\
Eine neuere Entwicklung im Bereich der String-Phänomenologie sind die sogenannten Sumpfland-Vermutungen. Hier liegt unser Augenmerk auf der de Sitter-Vermutung, welche behauptet, dass konsistente Konstruktionen von de Sitter-Raumzeiten in der String-Theorie nicht möglich sind. Wir zeigen, dass es schwierig ist, de Sitter-Minima in klassischen Typ IIA-String-Kompaktifizierungen zu finden. Nachdem wir die ursprüngliche Behauptung durch explizite instabile Lösungen widerlegen, präsentieren wir unsere eigene, verbesserte de Sitter-Sumpfland-Vermutung, welche sich von der überarbeiteten Version der ersten Behauptung in einigen Aspekten unterscheidet.\vspace{12pt}\\
Die vorliegende Dissertation basiert auf den Publikationen \cite{Roupec:2018mbn,Banlaki:2018ayh,Andriot:2018mav,Cribiori:2019hod,Cribiori:2019bfx,Cribiori:2019drf,Cribiori:2019hrb,Cribiori:2020bgt}, an welchen ich während meines Doktoratsstudiums gearbeitet habe.

\newpage\thispagestyle{empty}\vspace*{\fill}
\null
\vspace*{\fill}\newpage

\section*{Abstract}
\addcontentsline{toc}{chapter}{Abstract}
One of the main goals of the field of string phenomenology is to describe the properties of our spacetime in low-energy effective models related to and consistent with string theory. It is known from observations that our universe undergoes accelerated expansion, which can be described in Einstein's theory of general relativity as a positive cosmological constant. Within string theory descriptions, this is realized by a scalar field with a positive value of its potential. Oftentimes, this is modeled by having the scalar at a stable, positive minimum. One then obtains a de Sitter spacetime. Constructions of de Sitter spaces from string theory are not straightforward and face many criticisms. In this thesis, we address several aspects related to this topic.\\
We investigate the description of general branes in supergravity and in particular describe the uplifting anti-$D3$-brane of the KKLT model, including all world-volume fields. Additionally, we translate the type IIB based KKLT model into type IIA and introduce a mechanism that allows for the rapid construction of many de Sitter solutions. Related to this, we investigate models based on twisted $7$-tori from M-theory and their relation to type II supergravity. These setups can also yield de Sitter solutions using the same mass production mechanism.\\
One of the more recent and serious claims against the construction of de Sitter spaces in string theory, the de Sitter swampland conjecture, is also addressed here. We highlight the difficulty of obtaining a stable, positive vacuum energy in classical type IIA flux compactifications and their reliance on $O$-plane sources. Then, we turn our attention to the conjecture itself. Contrary to the initial claim, we find many unstable de Sitter points in numerical searches and go on to present our own refined conjecture that differs in several key aspects from the refined version of the original conjecture.\vspace{12pt}\\
This thesis is based on the publications \cite{Roupec:2018mbn,Banlaki:2018ayh,Andriot:2018mav,Cribiori:2019hod,Cribiori:2019bfx,Cribiori:2019drf,Cribiori:2019hrb,Cribiori:2020bgt} that I worked on during my doctoral studies.

\newpage

\tableofcontents

\newpage
\newpage\thispagestyle{empty}\vspace*{\fill}
\null 
\vspace*{\fill}\newpage
\pagenumbering{arabic}
\chapter{For the Uninitiated}
The world we live in is a curious place, and we humans are trying to comprehend as much of it as possible. This lead to the eventual development of the different subjects of natural science. Physics represents the most fundamental of these fields, beating out mathematics due to formality. However, physics is still divided into many different categories ranging from applied, like condensed-matter physics, to theoretical and fundamental, sometimes bordering on the field of pure mathematics \footnote{The reader may feel free to consider physics as little more than ``applied mathematics''. If the reader is a mathematician they may find the mercy in their heart to forgive me for this statement.}. During the human pursuit to understand the laws of nature, physicists have pushed to ever more fundamental theories. This lead to the discovery of the theories of the very small with Quantum Mechanics, and the very large, described by General Relativity. Unfortunately, the combination of the two proves to be troublesome. Theories that attempt to quantize gravity are collected under the umbrella term of \emph{Quantum Gravity}. Perhaps the most promising theory to date is known as \emph{String Theory} and considers tiny, vibrating strings as the fundamental objects of the theory. String Theory does allow for a quantum theory that includes gravity but the connection to the physics at energy scales that we can probe in experiments is not clear. In particular, we need to be able to obtain both the Standard Model of Particle Physics and the spacetime properties of the universe from string theory. The present work is motivated by the latter.\\
In 1998, a curious property of the universe was detected. The space around us is expanding at an accelerated rate. This statement might seem mundane at first, but has important implications. For one, it implies that, as time  progresses, certain parts of the universe will become unobservable as not even light will be able to reach us. More importantly for this work, however, is that it leads to the following question: What causes this expansion? Due to the lack of any concrete ideas and the behavior of the expansion matching a positive and constant vacuum energy very precisely the name \emph{Dark Energy} has been coined. Geometrically, a universe with accelerated expansion can be described, on large scales, as a de Sitter spacetime, a geometry with positive and constant cosmological constant.\\
At this point the reader might ask themselves how this connects to string theory exactly. In string theory, unlike general relativity, we cannot add an arbitrary constant to our theory in order to obtain the desired spacetime geometry. In fact, the geometry is obtained in the process of building a model. One important feature of string theory is that it lives naturally in higher dimensions than the usual 3+1 that we observe in our daily lives and indeed also in all experiments to date. \emph{Superstring Theory} requires 10 dimensions in order to be consistent and anomaly-free. This theory, furthermore, has the property of \emph{supersymmetry}, which relates bosons with fermions. This is currently the most popular method in order to include fermions, such as electrons, in the theory. Supersymmetry, like extra dimensions, has also not been observed in experiments to this day.\\
This raises the question why theoretical physicists are still investigating string theory at all, if it describes a universe with a wrong number of spatial dimensions and includes a symmetry that is not observed in nature. The answer has to do with the notion of \emph{effective theories}. In short, full string theory is constructed to be valid up to energies close to the Planck scale. At significantly lower energies, however, not all details of string theory matter. Nature is then approximated sufficiently well\footnote{Read: Indistinguishable in experiments due to insufficient precision.} by an \emph{effective theory}. At the energies we currently probe, these theories would be the Standard Model and General Relativity. We thus expect that, in the process of going to lower energies, the extra dimensions \emph{become small} and supersymmetry \emph{breaks} at some point. Both of these things can be accomplished, in principle.\\
In order to go from 10 dimensions down to our familiar 4 we perform what is called a \emph{Compactification}. For this, we split spacetime in two parts, the 3+1 dimensions we observe, and a compact space with 6 dimensions. The choice of the internal geometry is crucial for the resulting effective theory. In the process of compactification, scalar fields, so-called moduli which describe the properties of this manifold, have to be considered. These moduli contain information about the size and shape of the space and are dynamical. During the compactification we need to \emph{stabilize} them such that their dynamics do not appear in the effective theory. While this is a difficult task in general, it also opens up possibilities. For example, if a modulus is stabilized at the minimum of a potential with a positive energy, this can act as a cosmological constant in the effective theory! Thus, a string compactification can solve the dark energy problem. The properties of the internal manifold also play a role in supersymmetry breaking, however, also other ingredients, such as \emph{D-Branes}, can play a role. In string theory, branes are extended objects on which strings can end. They play an important role in many models and, as already mentioned, can break supersymmetry, depending on their orientation.\\
At this point the reader might conclude that string theory allows for a realistic description of the nature we observe\footnote{The valued reader might also think that this is contrived beyond reason. Unfortunately, convincing them of the contrary would require more space than is available in this thesis.}. The devil, however, lies in the details. Currently, no model is known that reproduces even the basic properties of our spacetime and is without major criticism. There are numerous properties one needs to achieve that go beyond the scope of this simple overview. Due to this, it is necessary to tackle one problem at a time before one can hope to fit the pieces of the puzzle together. To this end, several different aspects of string compactifications will be described in this thesis in the hopes that they may contribute to the field as a whole and bring us closer to understanding the fundamental nature of the world we live in.

\chapter{Introduction}
The goals of applying string theory to our physical world are roughly split into two categories. One half is trying to reproduce the Standard Model of particle physics, one of the most successful physical theories in history. The other, is concerned with gravity and the large scale universe. From cosmological observations we know that the universe is expanding in an accelerated manner \cite{Planck:2018jri}. The two most popular attempts to describe this behavior are that either the spacetime of our universe is de Sitter or that it is undergoing quintessence. In the context of low-energy effective theories obtained from string theory, both situations are described by a scalar field in a potential. For de Sitter, the scalar rests at a (meta-) stable minimum with positive vacuum energy. In the case of quintessence, the scalar rolls slowly on a flat slope such that the positive vacuum energy changes only by an amount that is unobservable on timescales below the cosmological. In both scenarios, one needs to build low-energy effective models that are consistent with string theory. The framework of these models is supergravity. Unfortunately, supergravity does not sufficiently restrict models in order to ensure that they consistently lift to string theory. This is the concern of a field that has recently emerged under the name ``swampland program" \cite{Brennan:2017rbf,Palti:2019pca,vanBeest:2021lhn}. While this program has yielded several so-called swampland conjectures, here we are mostly interested in the de Sitter swampland conjecture \cite{Obied:2018sgi}, that effectively forbids dS solutions to be consistent with string theory. In this thesis, we will investigate several aspects related to the description of spacetime in string theory motivated models.\\
In chapter \ref{sec:antiD3}, based on \cite{Cribiori:2019hod,Cribiori:2020bgt}, we consider important ingredients that are used in one of the the most successful classes of constructions of de Sitter spaces from string theory, namely the anti-branes of the Kachru-Kallosh-Linde-Trivedi (KKLT) model \cite{Kachru:2003aw}. First, we review, in detail, non-linear supergravity and how it can be used in order to incorporate different contributions into the supergravity potentials. For this, constrained supermultiplets \cite{Rocek:1978nb,Lindstrom:1979kq,Samuel:1982uh,Komargodski:2009rz,DallAgata:2016syy,Ferrara:2016een} are used that allow us to write a situation that has broken supersymmetry in terms of a (seemingly) linear Lagrangian. In the second section of this chapter, we show how any $Dp$-brane can be written using this formalism \cite{Cribiori:2020bgt} and we investigate general, supersymmetry-breaking branes \cite{Sugimoto:1999tx,Antoniadis:1999xk,Angelantonj:1999jh,Aldazabal:1999jr,Angelantonj:1999ms,Dudas:2000nv,Pradisi:2001yv}. The focus will be on intersecting $D6$-branes and we clear up a long-standing misconception about the description of such setups in the $4d$ effective theory \cite{Villadoro:2006ia,Blumenhagen:2002wn,Kachru:1999vj,Cvetic:2001nr}. In particular, we show that the description of a supersymmetric brane intersecting another brane that breaks supersymmetry is always non-supersymmetric. Following this, we give the complete description of the anti-$D3$-brane in the KKLT setup \cite{Cribiori:2019hod}, including all world-volume fields \cite{GarciadelMoral:2017vnz}. This includes all bosonic and fermionic contributions and once again utilizes constrained multiplets.\\
In the next chapter \ref{sec:kkltconstr}, following \cite{Cribiori:2019bfx,Cribiori:2019drf,Cribiori:2019hrb}, we move on to construct de Sitter vacua of our own. After reviewing the basic KKL(MM)T model \cite{Kachru:2003aw,Kachru:2003sx}, we  port it from type IIB to IIA in section \ref{sec:IIAuplift} \cite{Cribiori:2019bfx}. In type IIA, we can only utilize anti-$D6$-branes in order to lift a stable anti-de Sitter solution to a positive vacuum energy. We work with the so-called STU-model in a particular setup with only $F_6$-flux. The superpotential incorporates non-perturbative corrections in all moduli directions that are vital to the success of the setup. We argue that these corrections can appear even for the volume modulus due to M-theory U-duality. The anti-brane is again introduced by the use of constrained superfields. We find explicit models where the masses of all moduli are positive and we are able to satisfy basic string theory consistency conditions such as a large internal volume and small coupling. In section \ref{sec:massprod} we expand upon a mechanism \cite{Cribiori:2019drf}, first discovered in \cite{Kallosh:2019zgd}, where, by going first to Minkowski space, it is possible to guarantee that the procedure yields a stable de Sitter solution after the uplift. This works as follows: One first constructs a Minkowski minimum with no flat directions. Then, a parametrically small downshift, included into the superpotential, will yield a stable AdS point. After this, one can perform an uplift either via an anti-$D3$-brane in type IIB or an anti-$D6$-brane in IIA. For this procedure, we utilize the Kallosh-Linde double exponent superpotential \cite{Kallosh:2004yh} for which we also proof that the resulting de Sitter point has positive masses if the Minkowski progenitor has no flat directions and the downshift is small. A further nice feature of the mass production procedure is that there is no need for an extreme fine-tuning of the uplift. Since we can vary the downshift, the uplift can be a lot larger than it would be if performed from Minkowski space directly. We also present several explicit examples, both in type IIA and IIB, based on different internal manifolds and show that the procedure works in each of them. In the second to last section \ref{sec:mtheory} of chapter \ref{sec:kkltconstr}, we consider more general, M-theory inspired models \cite{Cribiori:2019hrb}, based on compactifications on a generalized, twisted $7$-torus \cite{DallAgata:2005zlf,Duff:2010vy,Derendinger:2014wwa,Ferrara:2016fwe}. Interestingly, we find models that allow us to forego the racetrack potential by inclusion of tree-level flux terms. We investigate different kinds of models and their relation to type II supergravity compactifications. It is even possible to not include any non-perturbative contributions in certain directions, due to the included fluxes. A particular case relies on conjectured $S$-fluxes. Using these, we are able to build a model in type IIB that does not require any sort of non-perturbative contributions. We also present explicit, numerical solutions for all of these models and show that the mass production procedure works in each of them.\\
In chapter \ref{sec:swamplandconjectures}, we review the works \cite{Roupec:2018mbn,Banlaki:2018ayh,Andriot:2018mav} and turn our interest to the de Sitter swampland conjecture \cite{Obied:2018sgi} and related issues. We first briefly review the general idea of the swampland program \cite{Brennan:2017rbf,Palti:2019pca,vanBeest:2021lhn} before turning our attention to the de Sitter conjecture \cite{Obied:2018sgi,Ooguri:2018wrx} and its implications. After earlier considerations along similar lines \cite{Danielsson:2018ztv}, the original de Sitter conjecture \cite{Obied:2018sgi} proposed a bound on the ratio of the first derivative of the scalar potential to the value of the potential that effectively forbids any extremum with positive vacuum energy. In section \ref{sec:scaling}, we discuss scaling limits of a particularly simple compactification in type IIA \cite{Banlaki:2018ayh}, inspired by the AdS solutions of \cite{DeWolfe:2005uu}, where the $4$-flux is unconstrained. The goal is to find a limit where the solutions are trustworthy, with large internal volume and small string coupling. To that end, we review different ingredients that can appear in such compactifications and we also present an explicit example. Next, we show that the initial de Sitter swampland conjecture is too strict \cite{Roupec:2018mbn}, as it also forbids unstable extrema with positive vacuum energy. We highlight that such points have been found previously \cite{Caviezel:2008tf,Flauger:2008ad} and, in addition, present many new solutions that we obtained via a numerical search. Together with other issues of the original conjecture \cite{Denef:2018etk} this lead to a refinement in \cite{Ooguri:2018wrx}. Finally, in section \ref{sec:refconj} we present our own refinement of the de Sitter swampland conjecture \cite{Andriot:2018mav}. It is given as a single expression inequality on both the first and second derivative of the scalar potential which translates nicely into constraints on the slow-roll parameters of cosmology. We also show that our conjecture differs from the refined original one \cite{Ooguri:2018wrx} in some areas of parameter space. This has implications on the allowed cosmological models. In particular, our conjecture does allow for slow-roll inflation in certain cases and does restrict quintessence models.\vspace{12pt}\\
The present thesis is based on the works \cite{Roupec:2018mbn,Banlaki:2018ayh,Andriot:2018mav,Cribiori:2019hod,Cribiori:2019bfx,Cribiori:2019drf,Cribiori:2019hrb,Cribiori:2020bgt} published during the course of my doctoral studies at TU Wien under the supervision of Timm Wrase.

\chapter{D-Branes in 4d Supergravity}
\label{sec:antiD3}

\section{Non-linear Supersymmetry and Supergravity}
\label{sec:multiplets}
Before we can come to the two main topics of this chapter, non-supersymmetric branes, that we discussed in \cite{Cribiori:2020bgt}, and the complete action of the anti-$D3$-brane in the KKLT setup, as per our work \cite{Cribiori:2019hod}, we have to review some tools that we will be using that might not be as well known as necessary. These are non-linear supersymmetry, its description via constrained multiplets and the so-called new $D$-term. Since these methods play an important role throughout this thesis we dedicate this section to this purpose.
\subsection{Constrained Superfields in local Supersymmetry}
When supersymmetry is broken spontaneously, it is still realized in a non-linear manner on the multiplets in supergravity. This means that there is still an identification of fields, however, this identification now follows a non-linear transformation. Although originally conceived as a model to get massless neutrinos, we can use the Volkov-Akulov (VA) action \cite{Volkov:1972jx} as a simple example to highlight non-linear supersymmetry \cite{Kallosh:2016aep}:
\bea
S_{\text{VA}} &= -M^4 \int E^0 \wedge E^1 \wedge E^2 \wedge E^3\qquad \text{with}\,,\\
E^\mu &= d\sigma^\mu + \bar{\lambda} \gamma^\mu d\lambda\,.
\label{eq:VAorigin}
\eea
In this case, the $\lambda$ is the $4d$ Goldstino associated with the breaking of $\mathcal{N}=1$ supersymmetry. This action is still invariant under a non-linear transformation:
\be 
\delta_\epsilon = \epsilon +  \left(\bar{\lambda} \gamma^\mu \epsilon\right) \partial_\mu \lambda\,.
\ee
However, this can be re-written in a way that looks linear by the use of constrained superfields \cite{Rocek:1978nb,Lindstrom:1979kq,Samuel:1982uh,Komargodski:2009rz}. The principles of constrained superfields were investigated in these earlier works but only rather recently have they been re-discovered and applied in modern supergravity model building. For some examples see: \cite{Kallosh:2016aep,Vercnocke:2016fbt,GarciadelMoral:2017vnz,Kallosh:2018nrk,Kallosh:2019apq,Cribiori:2019hod,Cribiori:2019bfx,Kallosh:2019zgd,Cribiori:2020zoh,Cribiori:2020bgt}. Here, we can use a nilpotent chiral field in order to re-write the VA-action. A chiral superfield $X$ has an expansion in superspace that is given to be:
\be 
X = x + \sqrt{2} \psi \theta + F \theta^2\,.
\ee
Where $x$ is a scalar, $\psi$ a spinor and $F$ an auxiliary field. Note that $\psi \neq \lambda$ and that there is no simple identification between the $\psi$ in the expansion of $X$ and the fermion of the VA-action $\lambda$. Rather, they are related via a non-linear field redefinition \cite{Bergshoeff:2015jxa}. Finally, the superspace coordinates are $\theta$ and due to their Grassmannian nature this expansion is exact. Now, if we impose the nilpotency constraint:
\be 
X^2 = 0\,,
\ee
we find that the only remaining degree of freedom is the fermion $\psi$, as the scalar is fixed to be:
\be 
x = \frac{\psi^2}{2 F}\,,
\ee 
which requires $\langle F \rangle \neq 0$, for consistency. Using $X$, we then can write the Volkov-Akulov action \eqref{eq:VAorigin} as:
\be 
S_{\text{VA}} = \int d^4x \int d^2 \theta \int d^2 \bar{\theta} X \bar{X} + M^2 \left( \int d^4 x \int d^2\theta X + h.c.\,\right)\,.
\ee
After employing the nilpotency condition and performing the superspace integral, we can recover the original Volkov-Akulov action if we also perform the correct field redefinition \cite{Bergshoeff:2015jxa}, relating $\psi$ to $\lambda$.\\
Furthermore, if we decide to add the secondary constraint:
\be 
-\frac{1}{4} X \bar{D}^2 \bar{X} = M^2 X \qquad \Rightarrow \qquad F = M^2 + \text{fermions}\,
\label{eq:secconst}
\ee
to the above it is possible to write an equivalent action to the VA-action either as a pure $F$- or pure $D$-term:
\bea 
S_{\text{VA},F} &= \int d^4x \int d^2 \theta \int d^2 \bar{\theta} X \bar{X}\,,
\eea
\bea
S_{\text{VA},D} &=  M^2 \left( \int d^4 x \int d^2\theta X + h.c.\,\right)\,.
\eea
The equivalence of these two formulations is implied by the secondary constraint given in equation \eqref{eq:secconst}.\\
Importantly, the choice of constrained multiplet is not unique. Many different options are available \cite{Bandos:2016xyu,GarciadelMoral:2017vnz}. Not limiting ourselves to their applicability to the VA theory and repeating the nilpotent chiral field for completeness, we mention the following options:
\begin{itemize}
\item The nilpotent chiral field $X = x + \sqrt{2} \psi \theta + F \theta^2$ with $X^2=0$ gives:
\be 
X = \frac{\psi^2}{2F} + \sqrt{2}\theta \psi + \theta^2 F\,.
\ee
\item A chiral superfield $Y = y + \sqrt{2} \chi \theta + G^2 \theta^2$ satisfying $XY=0$ ($X^ 2 = 0$) yields: 
\be
Y = \frac{\psi \chi}{F} - \frac{\psi^2}{2F^2} G + \sqrt{2}\theta \chi + \theta^2 G\,.
\ee
\item One can remove the fermion from a chiral superfield $Z = z + \sqrt{2} \theta \omega + \theta^2 H$ by the condition $X \bar{D}_{\dot{\alpha}} \bar{Z}=0$ ($X^ 2=0$). Then both the fermion and the auxiliary field are determined:
\bea 
\omega &= \rmi \sigma^\mu \left(\frac{\bar{\psi}}{\bar{F}}\right) \partial_\mu z\,,\\
H &= -\partial_\mu \left(\frac{\bar{\psi}}{\bar{F}}\right) \bar{\sigma}^\nu \sigma^\mu \left(\frac{\bar{\psi}}{\bar{F}}\right) \partial_\nu z + \left(\frac{\bar{\psi}^2}{2\bar{F}^2}\right) \square z\,.
\eea
\item The constraint that removes the gaugino $\Lambda_\alpha$ from a real vector field $V$ is $X W_\alpha = 0$, where we used the field strength chiral superfield:
\be 
W_\alpha = -\frac{1}{4} \bar{D}^2 D_\alpha V = - \rmi \Lambda_\alpha + L^\beta_\alpha \theta_\beta + \sigma^\mu_{\alpha\dot{\alpha}} \partial_\mu \bar{\Lambda}^{\dot{\alpha}} \theta^2\,,
\ee
with:
\be 
L_\alpha^\beta = \delta^\beta_\alpha D - \frac{\rmi}{2} (\sigma^\mu \sigma^\nu)^\beta_\alpha F_{\mu\nu}\,.
\ee
The gaugino $\Lambda_\alpha$ can then be expressed as:
\bea
\Lambda_\alpha = & + \frac{\rmi}{\sqrt{2}}\frac{1}{F} L_\alpha^\beta \psi_\beta - \frac{\psi^2}{2F^2} \partial_\mu  \left( \frac{\bar{G}_{\bar{\dot{\alpha}}} \bar{L}^{\dot{\alpha}}_{\dot{\beta}}}{\sqrt{2} \bar{F}} \right) \bar{\sigma}^{\mu \dot{\beta}\gamma} \epsilon_{\gamma\alpha}\\
& - \frac{\rmi}{2} \frac{\psi^2}{F^2} \sigma^\mu_{\alpha \dot{\beta}} \bar{\sigma}^{\nu\dot{\beta}\gamma} \partial_\mu \left[ \frac{\bar{\psi}^2}{2 \bar{F}^2} \partial_\nu \left(\frac{L^\delta_\gamma \psi_\delta}{\sqrt{2} F}\right)\right]\\
& - \frac{1}{2} \frac{\psi^2}{F^2} \frac{\bar{\psi}^2}{\bar{F}^2} \left[\partial \left(\frac{\psi}{\sqrt{2}F}\right)\right]^2\partial_\mu \left(\frac{\bar{\psi}_{\dot{\alpha}} \bar{L}^{\dot{\alpha}}_{\dot{\beta}}}{\sqrt{2} \bar{F}}\right) \bar{\sigma}^{\mu\dot{\beta}\gamma} \epsilon_{\gamma \alpha}\,.
\eea
\end{itemize}
These expressions are valid for supersymmetric theories using the superspace formalism. How constrained multiplets can be used in the context of supergravity was discussed in \cite{DallAgata:2015zxp}. For a review about constrained superfields in SUSY and SUGRA see \cite{Cribiori:2018hxv}. We will review what is needed for the action of the anti-$D3$-brane in the KKLT setup in the following section.

\subsection{Supergravity and constrained Multiplets}
In this section, we focus on the explicit constrained multiplets we are going to use in order to describe the anti-$D3$-brane in the KKLT setup. This set is not unique but instead based on convenience and it is possible to find different descriptions. Here, we work in the conventions of \cite{Freedman:2012zz}, as opposed to the last section, where $4d$ fermions are described using four-component Majorana spinors. They are related to the spinors from before in the following way:
\be 
\Omega = \begin{pmatrix} \psi \\ \bar{\psi}\end{pmatrix}\,.
\ee
In order to bridge the gap between the notation of the previous section, where we dealt with constrained superfields in the superspace formulation, we quickly re-write the VA-model in the notation of this section. The chiral multiplet we consider here is called $X$ and consists of the scalar we call $X$ as well, a fermion $P_L \Omega$ and the auxiliary field $F$. We impose the same nilpotency condition $X^2=0$ as before, in order to remove the scalar, and find that it is given to be:
\be 
X = \frac{\bar{\Omega} P_L \Omega}{2F}\,.
\ee
Coming to the VA action, we finally see significant differences in the notation. In the notation of \cite{Freedman:2012zz}, we can write an invariant action for the chiral multiplet $X$ in the following way:
\bea 
S &= [X\bar{X}]_D + M^2 [X]_F\\
  &= \int d^4 x \left( - \bar{\Omega} P_L \slashed{\partial}\Omega + \frac{\bar{\Omega}P_L\Omega}{2F} \square \frac{\bar{\Omega}P_R\Omega}{2\bar{F}} + F \bar{F} + M^2 (F+\bar{F})\right)\,.
\label{eq:VAsusy}
\eea
Immediately, it is evident that the equation of motion for the auxiliary field changes as it now also appears as part of the composite expression replacing the supersymmetric partner of the Goldstino. Due to this fact, one has to be careful when going on-shell if supersymmetry is realized non-linearly.
\\The equation of motion for the auxiliary field is:
\bea 
F = &-M^2 - \frac{1}{4 M^6} \left(\bar{\Omega}P_R  \Omega\right) \square \left(\bar{\Omega}P_L\Omega \right) \\
&+ \frac{3}{16 M^{14}} \left(\bar{\Omega}P_R  \Omega\right)  \left(\bar{\Omega}P_L\Omega \right) \square \left(\bar{\Omega}P_R  \Omega\right) \square \left(\bar{\Omega}P_L\Omega \right)\,,
\eea
which yields the on-shell action that reproduces the Volkov-Akulov action \eqref{eq:VAorigin} up to a field redefinition \cite{Kuzenko:2011tj}:
\bea 
S = \int d^4 x \bigg( &- M^2 - \bar{\Omega}P_L \slashed{\partial} \Omega + \frac{1}{4M^4} \left(\bar{\Omega}P_L\Omega \right) \square \left(\bar{\Omega}P_R\Omega\right)\\
&- \frac{1}{16 M^{12}} \left(\bar{\Omega}P_R\Omega\right)\left(\bar{\Omega}P_L\Omega \right) \square \left(\bar{\Omega}P_R\Omega\right)\square \left(\bar{\Omega}P_L\Omega \right) \bigg)\,.
\eea

\subsection{Coupling to Gravity}
\label{sec:couplinggravity}
Since we are interested in a supergravity action, where our constrained multiplets couple to gravity, we will employ the superconformal approach of \cite{Freedman:2012zz}. The superconformal symmetry is very useful for computations but not realized in nature, hence we will need to break it in order to recover the typical Poincar\'{e} symmetry of spacetime. An important feature of the superconformal approach is that it is not necessary to do field redefinitions when going to the Einstein frame. The type of action that we employ here is of the following form:
\be
S = \left[-3 X^0 \bar{X}^0 \rme^{-\frac{K(X,\bar{X})}{3}}\right]_D + \left[(X^0)^3 W (X)\right]_F + \left[f_{AB}(X) \bar{\Lambda}^A P_L \Lambda^B \right]_F\,,
\label{eq:suconfact}
\ee
with a set of chiral multiplets $\{X^I\}$, $I = 0, \ldots , n$ and $\{\Lambda^A\}$, $A=1,\ldots , m$, a set of vector multiplets. $X^0$ plays the special role of the compensator field with Weyl weight $1$, as opposed to the other chiral multiplets with weight $0$. As is usual, $K$ is the Kähler potential, $W$ the superpotential and $f_{AB}$ the gauge kinetic function. In order to go from the superconformal theory to Poincar\'{e} supergravity the compensator is fixed to:
\be 
X^0 = \frac{1}{\kappa} \rme^{\frac{K}{6}}\,,
\ee
which introduces the Planck scale into the theory. In our conventions we will set $\kappa = 1$.\\
The supergravity equivalent to the global SUSY Volkov-Akulov model \eqref{eq:VAsusy} can be obtained by choosing Kähler- and superpotential:
\bea 
K &= X\bar{X}\,,\\
W &= W_0 + M^2 X\,,
\eea

with a nilpotent chiral multiplet $X$ that contains the Goldstino. Applying the superconformal action to this example with just one chiral multiplet we find:
\be 
S = \left[-3 X^0 \bar{X}^0 + X^0 \bar{X}^0 X\bar{X}\right]_D + \left[ (X^0)^3\, (W_0 + M^2 X) \right]_F\,.
\ee
In solving this model, one has to keep in mind to break the superconformal symmetry and use the constraint coming from the nilpotency before going on-shell. The obtained theory is the generalization of the supersymmetric Volkov-Akulov theory \eqref{eq:VAsusy} and is called pure de Sitter supergravity \cite{Farakos:2013ih,Dudas:2015eha,Bergshoeff:2015tra,Hasegawa:2015bza,Ferrara:2015gta} as it only has the graviton and its supersymmetric partner as degrees of freedom while the Goldstino is a pure gauge mode.
\subsection{Constrained Multiplets in Supergravity}
\label{sec:constmult}
In this section, we will review all constrained multiplets that will be needed for the description of the anti-$D3$-brane action in the KKLT setup and non-supersymmetric branes in general. It has been shown in \cite{DallAgata:2016syy} that there are ways to remove any component from a supersymmetry multiplet. Going beyond the scope of what we use here it is even possible to have off-shell linear supersymmetry by the use of Lagrange multipliers \cite{Ferrara:2016een}. 
\paragraph{For completeness, we start with the nilpotent chiral field $\mathbf{X}$ one final time.} It has the components $X = \{X,P_L \Omega, F\}$ and the constraint:
\be 
X^2 = 0\,,
\ee
gives the scalar in terms of the fermion in the following way:
\be 
X = \frac{\bar{\Omega}P_L \Omega}{2F}\,.
\label{eq:nilconst}
\ee
\paragraph{Constraining chiral multiplets $\mathbf{Y^i}$} with $Y^i = \{Y^i, P_L \Omega^i,F^i\}$ by the conditions:
\be 
X Y^i =0\,,
\ee
where here and in the following $X$ will be the nilpotent chiral field from above, gives the scalar of the multiplet as \cite{Brignole:1997pe,DallAgata:2015pdd}:
\be 
Y^i=\frac{\bar{\Omega}^i P_L\Omega}{F}-\frac{\bar{\Omega}P_L \Omega}{2F^2}F^i\,.
\ee
Since we once more have removed the scalar, the remaining degrees of freedom are fermionic in nature. These multiplets have been successfully used in order to describe the world-volume spinors of the anti-$D3$-brane \cite{Vercnocke:2016fbt,Kallosh:2016aep} which will be very useful for our work in the following.
\paragraph{The chiral multiplets $\mathbf{H^a}$} with $H^a =\{H^a,P_L \Omega^a,F^a\}$ will be constrained such that:
\be 
X\bar{H}^a = \text{chiral.}
\ee
This removes the fermion and the auxiliary field \cite{Komargodski:2009rz}\footnote{We remind the reader that we are using the supergravity conventions of \cite{Freedman:2012zz}.}:
\bea
P_L\Omega^a &= \frac{\slashed{\mathcal{D}} H^a}{\bar{F}} P_R \Omega\,,\\
F^a &= \mathcal{D}_\mu \left(\frac{\bar{\Omega}}{\bar{F}}\right) \gamma^\nu \gamma^\mu \left(\frac{P_R\Omega}{\bar{F}} \right)\mathcal{D}_\nu H^a + \frac{\bar{\Omega}P_R\Omega}{2\bar{F}^2} \square H^a\,.
\eea
With this we have removed all fermionic degrees of freedom from the multiplets $H^a$. Note, that a superpotential $W(H)$ will not lead to masses for the scalars in $H^a$ but instead to fermionic Goldstino interactions. Scalar masses can be introduced into the theory via the Kähler potential if desired.
\paragraph{Finally, we will use a chiral field strength multiplet $\mathbf{P_L \Lambda_\alpha}$} with components $\{P_L \Lambda_\alpha,(P_L\chi)_{\beta\alpha},F^\Lambda_\alpha\}$. We further specify the components:
\bea 
(P_L\chi)_{\beta\alpha} &= \sqrt{2}\left[-\frac{1}{4} (P_L \gamma^{ab} C)_{\beta \alpha}\hat{F}_{ab} + \frac{\rmi}{2} D(P_LC)_{\beta\alpha}\right]\,,\\
\hat{F}_{ab} &= e^\mu_a e^\nu_b (2 \partial_{[\mu} A_{\nu]} + \bar{\psi}_{[\mu}\gamma_{\nu]}\lambda)\,,\\
F_\alpha^\Lambda &= (\slashed{\mathcal{D}}P_R \Lambda)_\alpha\,.
\label{eq:fieldstrencomp}
\eea
Here, $\hat{F}$ is the covariant vector field strength and $D$ a real auxiliary field. The matrix $C_{\alpha\beta}$ is used to raise and lower spinor indices and satisfies $C^T = -C$. $P_L \Lambda_\alpha$ is analogous to the superfield strength $W_\alpha $ in the superspace approach. The constraint we employ removes the gaugino and is:
\be 
X P_L \Lambda_\alpha = 0\,.
\label{eq:fermconst}
\ee
This will set the lowest component of the field to:
\bea
P_L \Lambda_\alpha = &+\frac{1}{F} (\bar{\Omega}P_L\chi)_\alpha - \frac{X}{F} \slashed{\mathcal{D}}_\alpha^{\,\beta} \left(\frac{(\bar{\Omega} P_R\chi)_\beta}{\bar{F}}\right)\\
&+\frac{X}{F}\slashed{\mathcal{D}}_\alpha^{\,\beta} \left[ \frac{\bar{X}}{\bar{F}} \slashed{\mathcal{D}}_\beta^{\,\gamma} \left(\frac{(\bar{\Omega}P_L\chi)_\gamma}{F}\right)\right]\\
&-\frac{X\bar{X}}{F^2\bar{F}^2} (\slashed{\mathcal{D}}\slashed{\mathcal{D}}X)_\alpha^{\,\beta} (\gamma^\mu)_\beta^{\,\delta}(\mathcal{D}_\mu\bar{\Omega}P_R\chi)_\delta\,.
\eea
Here, $(\bar{\Omega}P_L\chi)_\alpha = \Omega^\beta(P_L)_\beta^{\,\gamma} \chi_{\gamma\alpha}$ and we have to remember that the scalar in $X$ is given as $X = (\bar{\Omega} P_L \Omega)/(2F)$. With this, $P_L \Lambda_\alpha$ describes an abelian gauge vector and nothing else.
\subsection{The new D-Term}
\label{sec:newD}
The final ingredient we will need is the so called new Fayet-Iliopoulos (FI) $D$-term \cite{Cribiori:2017laj,Cribiori:2018dlc,Cribiori:2018hxv,Kuzenko:2018jlz,Antoniadis:2018cpq}. This term will play an important role in constructing the action for the anti-$D3$-brane. It will help to resolve an issue about the gauge kinetic function which will be discussed in section \ref{sec:D3barKKLT}. We are only going to focus on what we need in the following and thus start by considering the chiral field strength multiplet $P_L \Lambda_\alpha$ with components $\{P_L\Lambda_\alpha,(P_L\chi)_{\beta\alpha},F_\alpha^\Lambda\}$ and weight $3/2$. The gaugino is $P_L \Lambda_\alpha$, $F_{ab}$ is the $U(1)$ covariant field strength and $D$ a real auxiliary field, see \eqref{eq:fieldstrencomp} and the paragraph below that equation for more details. The standard Fayet-Iliopoulos $D$-term \cite{Fayet:1974jb,Freedman:1976uk}:
\be 
S_{\text{FI}}=-\xi \int d^4x \sqrt{-g_4}D\,,
\label{eq:FIDterm}
\ee
breaks supersymmetry in supergravity spontaneously, however, it requires a gauging of the $R$-symmetry, which prevents, for example, a gravitino mass in the Lagrangian. The new FI-term does not require gauged $R$-symmetry while keeping all the nice properties we desire. Utilizing a chiral compensator field $X^0$ and a real multiplet $D$, that has components $\{D,\slashed{\mathcal{D}}\Lambda,0,\mathcal{D}^b \hat{F}_{ab},-\slashed{D}\slashed{D}\Lambda,-\square D\}$\footnote{For conventions please refer to \cite{Freedman:2012zz}.}, we write the new FI $D$-term as:
\bea 
S_{\text{FI, new}} &= - \xi \left[ \frac{\omega^2\bar{\omega}^2}{\Sigma(\bar{\omega^2})\bar{\Sigma}(\omega^2)}X^0\bar{X}^0D\right]_D\\
&= -\xi \int d^4x \sqrt{-g_4}X^0\bar{X}^0 D + \ldots\,,
\eea
where the dots represent fermionic terms. Furthermore, we used the multiplets:
\bea 
\omega^2 &= \frac{\bar{\Lambda}P_L \Lambda}{\left(X^0\bar{X}^0\right)^2}\,,\\
\bar{\omega}^2 &= \frac{\bar{\Lambda}P_R\Lambda}{\left(X^0\bar{X}^0\right)^2}\,,
\label{eq:Dnew}
\eea
with components:
\be 
\bar{\Lambda}P_L\Lambda = \{\bar{\Lambda}P_L\Lambda\,,\sqrt{2}P_L\left(\rmi D-\frac{1}{2}\slashed{\hat{F}}\right)\Lambda\,,2\bar{\Lambda}P_L\slashed{\mathcal{D}}\Lambda+\hat{F}^-\cdot\hat{F}^--D\}\,,
\ee
and generalizations of the chiral projectors on superspace to the superconformal formalism in $\Sigma$ and $\bar{\Sigma}$ \cite{Ferrara:2016een}. These projectors act on multiplets with weights $(\text{Weyl, chiral})$ $ = (w,\pm(w-2))$ and produce new multiplets with weights:
\bea 
\Sigma&:\,(w,+w-2)\; \to \;(w+1,w+1)\,,\\
\bar{\Sigma}&:\,(w,-w+2)\; \;\to (w+1,w-1)\,,
\eea
that are chiral or anti-chiral. The action of the projectors on $\omega^2$ is:
\be 
\bar{\Sigma}(\omega^2)=\left(X^0\bar{X}^0\right)^{-2} \left(\frac{1}{2}F_{\mu\nu}F^{\mu\nu}+\frac{\rmi}{4}\frac{\epsilon^{\mu\nu\rho\sigma}}{\sqrt{-g_4}}F_{\mu\nu}F_{\rho\sigma}-D^2 + \ldots\right)\,,
\ee
where we have once again neglected fermionic contributions.\\
The new $D$-term \eqref{eq:Dnew} has only $D$ as its pure bosonic sector. All further terms are required by the superconformal symmetry that our formalism employs. After going to Poincar\'{e} supergravity by fixing the compensator $X^0 = \kappa^{-1} \rme^{K/6}$, as discussed in section \ref{sec:couplinggravity}, the new $D$-term yields:
\be 
S_{\text{FI,new}} = -\xi \int d^4x \sqrt{-g_4} \rme^{\frac{K}{3}}D + \ldots\,,
\label{eq:Dnewres}
\ee
with $\kappa = 1$. Like the original FI $D$-term, this expression will break supersymmetry spontaneously when added to the standard supergravity action due to the fact that the auxiliary field $D$ acquires a non-zero vacuum expectation value. The fermionic contributions in the $\ldots$ can be removed by setting the Goldstino equal to zero via a unitary gauge choice. However, there remains one issue. Comparing the result of the new $D$-term \eqref{eq:Dnewres} to the original FI $D$-term \eqref{eq:FIDterm}, a contribution of the Kähler potential appears. This spoils the Kähler invariance of the theory. Luckily, this can be remedied by performing a substitution \cite{Antoniadis:2018oeh}:
\be 
X^0 \bar{X}^0 \,\to\,X^0\bar{X}^0 \rme^{-\frac{K}{3}}\,.
\ee
With this replacement, the new $D$-term \eqref{eq:Dnew} can be viewed as containing two real multiplets with weights $(-2,0)$ and $(4,0)$, respectively:
\bea 
R_1 &=\frac{\omega^2\bar{\omega}^2}{\Sigma(\bar{\omega}^2)\bar{\Sigma}(\omega^2)}\,,\\
R_2 &= X^0\bar{X}^0 \rme^{-\frac{K}{3}}D\,.
\eea
The expansion in components begins with the lowest component of $R_2$ while $R_1$ provides higher-order fermionic terms that are required to complete the superconformal embedding. This method is readily generalized in order to complete any arbitrary real multiplet $R_2$ with weights $(4,0)$. We will use this procedure in order to complete the action of the anti-$D3$-brane in the KKLT scenario.

\section{Non-supersymmetric Branes}
\label{sec:nonsusybranessec}
Supersymmetry has not been observed in any experiments or natural phenomena thus far. If we want to investigate whether or not string theory can describe the universe we live in, we have to consider effective low-energy theories that come from string theory and that break supersymmetry during the compactification process spontaneously. When supersymmetry is spontaneously broken, a non-linear transformation of the fields can still be realized and in that case, we are able to use the tools of non-linear supersymmetry or supergravity and constrained multiplets to describe it. One way to achieve spontaneous supersymmetry breaking is by including objects in the theory that achieve this. In particular, here we are interested in $Dp$-brane supersymmetry breaking \cite{Sugimoto:1999tx,Antoniadis:1999xk,Angelantonj:1999jh,Aldazabal:1999jr,Angelantonj:1999ms,Dudas:2000nv,Pradisi:2001yv}, for a review see \cite{Mourad:2017rrl}. Other ingredients, like for example $Op$-planes, can also break supersymmetry in a similar way. Importantly, if supersymmetry is broken at or near the string scale, it is not possible to restore it below that scale.\\
Here, we focus on the description of non-supersymmetric $Dp$-branes in flux compactifications on $SU(3)\times SU(3)$ structure manifolds \cite{Lust:2008zd}. These theories compactify $10d$ superstring theory to $4d$, $\mathcal{N}=1$ supergravity. A space-filling $Dp$-brane can then break the remaining supersymmetry, not broken by the background, spontaneously. Our main goal here is to find the correct description of the supersymmetry breaking object in terms of the resulting supergravity. It is important to stress that the correct $10d$ results are known \cite{Villadoro:2006ia,Blumenhagen:2002wn,Kachru:1999vj,Cvetic:2001nr} but they have often been wrongly interpreted in the language of the $4d$ supergravity as a $D$-term. This section is based on \cite{Cribiori:2020bgt} where we rectified this issue and gave the correct description of any SUSY breaking $Dp$-brane utilizing non-linear supergravity and constrained multiplets.

\subsection[$Dp$-Branes in flat Space]{$\mathbf{Dp}$-Branes in flat Space}
\label{sec:flatspacebranes}
Let us review $Dp$-branes in type II supergravity on flat space $\mathbb{R}^{9,1}$ \cite{Polchinski:1998rr,Cederwall:1996pv,Aganagic:1996pe,Cederwall:1996ri,Bergshoeff:1996tu,Aganagic:1996nn}, for some basic facts. On this background we have maximal $\mathcal{N}=8$ supersymmetry with $32$ supercharges, conveniently packaged into two Majorana-Weyl spinors $\varepsilon_1$ and $\varepsilon_2$. Since we will not specify whether we consider type IIA or IIB supergravity, the chirality of the spinors is irrelevant. Nevertheless, we mention that in IIA the spinors have the same chirality while in IIB they have opposite chirality. If we add a $Dp$-brane in the directions $x^0,\,x^1,\,\ldots,\,x^p$ it will break half of the supersymmetry spontaneously \cite{Polchinski:1998rr}. This divides the supercharges into two groups. The ones acting linearly on the $Dp$-brane world-volume fields satisfy:
\be 
\varepsilon_1 = \Gamma_{01\ldots p} \varepsilon_2 =: \Gamma_{Dp}\varepsilon_2\,,
\ee
with $\Gamma_{01\ldots p} = \Gamma_0 \Gamma_1 \ldots \Gamma_p$ where we used the gamma matrices of $10$ dimensional flat space. On the other hand, the non-linearly realized supersymmetries have a sign flip in their behavior:
\be 
\varepsilon_1 = -\Gamma_{Dp}\varepsilon_2\,.
\ee
Meanwhile, an anti-$Dp$-brane behaves exactly opposite to a $Dp$-brane:
\be
\begin{matrix}
\text{linear:} & \qquad &\varepsilon_1 = - \Gamma_{Dp} \varepsilon_2 =: \Gamma_{\overline{Dp}} \varepsilon_2\,,\\
\text{non-linear:} & &\varepsilon_1 = \, \Gamma_{Dp} \varepsilon_2 = -\Gamma_{\overline{Dp}} \varepsilon_2\,.
\end{matrix}
\ee
In general, we can put a $Dp$-brane at an arbitrary angle, where it extends along the directions $x^0,\,x^1,\,\ldots ,\, x^{p-1}$ while it intersects the axis $x^p$ with an angle $\varphi$ in the plane $(x^p,x^{p+1})$. In this case the unbroken supercharges satisfy:
\be 
\varepsilon_1 = \left(\cos (\varphi) \Gamma_{01\ldots p} + \sin (\varphi) \Gamma_{01\ldots(p-1)(p+1)} \right)\varepsilon_2 =: \Gamma_{Dp}(\varphi) \varepsilon_2\,,
\ee
and once again, the broken charges have a minus instead of a plus in their relation. Of course, in flat space, this distinction between branes, anti-branes and branes at some angle $\varphi$ is rather forced, as we simply could change our coordinate system in order to transform any of the above cases into a simple $Dp$-brane and thus they are physically equivalent. If the background is more involved, like, for example,  $SU(3)\times SU(3)$ structure manifolds, this will no longer be the case.\\
Nevertheless, we mention that:
\bea
\Gamma_{Dp} (0) = \Gamma_{Dp}\,,\\
\Gamma_{Dp} (\pi) = \Gamma_{\overline{Dp}}\,.
\eea
Another way to get a physical distinction between different brane orientations is to add a reference object, like an $Op$-plane. Let us assume we have an $Op$-plane along the directions $x^0,\,x^1,\,\ldots ,\, x^{p-1}$ which projects out half of the supersymmetry. We are left with $16$ linearly realized supercharges:
\be 
\varepsilon_1 = \Gamma_{01\ldots p} \varepsilon_2 =: \Gamma_{Op}\varepsilon_2\,.
\ee
If we add a $Dp$-brane without angle to this setup, nothing will change for the supersymmetry. On the other hand, adding an anti-$Dp$-brane will break the remaining $16$ supercharges and supersymmetry will be broken completely. Naturally, the case of a $Dp$-brane at an angle $\varphi$ is the most interesting one. There, the $16$ charges preserved by the background\footnote{In this sense, we include the $Op$-plane in the background.} will be realized as a combination of linear and non-linear transformations on the brane world-volume fields. This combination will be necessarily non-linear in total, which will be one of the main conclusions of this section.\\
Another way to break supersymmetry via branes is by considering a world-volume flux $\mathcal{F}=B+F$, with $B$ the pullback of the Kalb-Ramond field and $F=dA$ a typical field strength on the brane. As discussed in \cite{Johnson:2003glb}, such a situation for a $D(p+1)$-brane is $T$-dual to two $Dp$-branes intersection at an angle $\varphi$ such that:
\be 
\mathcal{F} = \tan (\varphi) dx^p \wedge dx^{p+1}\,.
\ee
Due to this duality, we expect that branes with world-volume flux generically break some of the supersymmetry that would be preserved in the absence of flux. In particular, the linearly preserved supersymmetries now satisfy \cite{Koerber:2010bx}:
\be 
\varepsilon_1 = - \frac{1}{\sqrt{g+\mathcal{F}}} \sum_{2n+l=p+1} \frac{1}{n!l!2^n} \epsilon^{a_1\dots a_{2n}b_1 \ldots b_l} \mathcal{F}_{a_1 a_2 } \ldots \mathcal{F}_{a_{2n-1}a_{2n}} \Gamma_{b_1 \ldots b_l} \varepsilon_2 =: \Gamma_{Dp}^{\mathcal{F}}\varepsilon_2\,,
\ee
where here our convention is $\epsilon^{01\dots p} = -1$ and the pullback of the gamma matrices onto the brane world-volume is $\Gamma_a = \partial_a y^M \Gamma_M$.\\
The examples we have discussed thus far in flat space can be generalized to other spaces. Then, the internal manifold can already break some of the supersymmetry and we have to consider the Gauß law for all ingredients. This generally leads to reduced supersymmetry, some examples are:
\be 
\begin{matrix}
     T^6 &\quad&\Rightarrow&\quad& \mathcal{N}=8\,,\\
     T^2 \times K3 &\quad&\Rightarrow&\quad& \mathcal{N}=4\,,\\
     CY_3 &\quad&\Rightarrow&\quad& \mathcal{N}=2\,.
\end{matrix}
\ee
Onto these backgrounds we still can add $Op$-planes, removing some supersymmetry, before we add our $Dp$-branes. 

\subsection{Linear + non-linear = non-linear}
In this subsection we show that the description of a combination of linear and non-linear supersymmetry in the end necessarily has to be non-linear. If we have a background that has preserved supercharges $\{Q^A\}$, $A =1,\,\ldots,\,N$ and then add a supersymmetry breaking source, for example a $Dp$-brane, only a subset of the supersymmetry charges, that can be empty, will still be realized linearly: $\{Q^a\}$, $a=1,\,\ldots,\,n<N$. We then want to investigate the behavior of the other remaining charges $\{Q^i\}$, $i=n+1,\,\ldots,\,N$.\\
For this, let us consider a simplified situation where only one set of scalars $\phi^A$ is related to fermions $\lambda^A$ via supersymmetry. The transformation can be written as:
\be 
\delta \lambda^A = S^A \varepsilon + \slashed{\partial}\phi^A \varepsilon + \ldots\,,
\ee
with complex constants $S^A$ that are not all equal to zero and with additional terms in the $\ldots$ that are not relevant for our discussion here. Splitting the constants as $S^A = \{S^a,\,S^i\}$, with $S^a = 0$ related to the unbroken supersymmetries, we break supersymmetry spontaneously in the directions that are related to the fermionic shifts $S^i\neq0$. We call a supersymmetry transformation non-linear if the shift, related to $\{S^A\}$ is present:
\bea 
\delta_s \lambda^a = 0\,,\\
\delta_s \lambda^i = S^i \varepsilon\,.
\eea
One may wonder if it is possible to remove the shift by some sort of allowed transformation. For example, focusing on the set $\{\lambda^i\}$ that has non-linear supersymmetry, we define:
\be 
v=K_{i\jb}\lambda^iS^\jb =: \lambda^T g \bar{S}\,,
\ee
with the Kähler metric $K_{i\jb}$. The action of the shift on the new field $v$ is given to be:
\be 
\delta_s v = K_{i\jb} S^i S^\jb \varepsilon = \left( S^T K S \right) \varepsilon\,,
\ee
in matrix notation where, in this case, $K$ is still the Kähler metric. With this we can redefine the remaining fields as:
\be 
\tilde{\lambda}^i = \lambda^i - \frac{S^i}{S^T K \bar{S}}\,,
\ee
with:
\be 
\tilde{\lambda}^i K_{i\jb} S^\jb = 0\,.
\ee
The $\tilde{\lambda}^i$ now have the property that they do not have a shift part in their supersymmetry transformations:
\be 
\delta_s \tilde{\lambda}^i = 0\,.
\ee
With this we have transformed all degrees of freedom initially present in the $\{\lambda^i\}$ into the $\tilde{\lambda}^i$ and $v$. Importantly, for this to be consistent, we require that $ \left( S^T K S \right) = \| S \|^2 \neq0$. This means that the best thing we can achieve is to transform the non-linear behavior into one direction. We conclude that supersymmetry, realized as a combination of linear and non-linear transformations, will in total always require a non-linear description. An important caveat is that the parameter, giving the SUSY breaking scale, has to contain a constant piece that cannot be removed via field redefinitions. Such cases are readily available, for example, non-supersymmetric $Dp$-branes wrapping tori in the internal manifold. In such a scenario the SUSY breaking parameter is related to the wrapping numbers and cannot be removed.\\
Compactifications that include SUSY breaking sources thus necessarily require a non-linear realization of the transformations on the world-volume fields. This can, for example, lead to $4d$ theories with all supersymmetries realized non-linearly which are called de Sitter supergravity \cite{Antoniadis:2014oya,Dudas:2015eha,Bergshoeff:2015tra,Hasegawa:2015bza,Kallosh:2015sea,Schillo:2015ssx}.

\subsection{Supersymmetry Breaking and Unitarity}
When any symmetry is broken at some scale $F$ it is expected that it will, likewise, be restored when going to, or above, that scale. If we naively follow our effective theory to this scale we anticipate that new fields or other degrees of freedom appear in order to complete the symmetry. This manifests itself in unitarity violations of the effective theory. Examples of this happening include the Volkov-Akulov theory \cite{Volkov:1972jx,Volkov:1973ix} or when coupling a very heavy scalar in a chiral multiplet to supergravity \cite{Casalbuoni:1988sx}. We will study this problem in the case of some examples in this subsection.\\
Before we begin let us mention one notable exception to the above statement. In inflation the Hubble scale $H$ changes the SUSY breaking scale to be:
\be 
F \sim M_P^2 \sqrt{m^2_{(3/2)} + H^2 } \,,
\ee
with the gravitino mass $m_{(3/2)}$ and $M_P$ the Planck scale \cite{Kallosh:2000ve,DallAgata:2014qsj,Ferrara:2015tyn,Carrasco:2015iij,Ferrara:2016een}. In the case of a vanishing scalar potential $V=0$ the gravitino mass $m_{(3/2)}$ would give the SUSY breaking scale. When considering non-linear supergravity models of inflation one would need to keep this change in mind. For simplicity, we will only consider situations where we can safely neglect the Hubble scale.

\paragraph{In the KKLT scenario} we have a $4d$ AdS vacuum with $\mathcal{N}=1$ supersymmetry. The vacuum gets lifted to de Sitter by the inclusion of an anti-$D3$-brane at the bottom of a warped throat which breaks SUSY spontaneously at the same time \cite{Kachru:2003aw,Kachru:2003sx}. Here, the breaking scale is given by the uplift energy in turn corresponding to the warped down string scale. Likewise, the whole tower of open string states is warped down as well. Hence, when going to the SUSY breaking scale, one expects these states to become relevant and invalidate the initial effective theory.\\
For the KKLT scenario this is not the whole story. When considering the throat to be of Klebanov-Strassler (KS) type \cite{Klebanov:2000hb}, there is a non-trivial 3-cycle at its bottom. This introduces a warped down KK-scale that can be below the string scale, meaning that massive states enter at this scale instead. When anti-$D3$-branes at the bottom of the KS-throat polarize into an $NS5$-brane \cite{Kachru:2002gs}, it can wrap a meta-stable cycle and linear supersymmetry can be restored by the KK modes \cite{Aalsma:2018pll}.
\paragraph{Two $\mathbf{Dp}$-branes that enclose a small angle $\mathbf{\varphi \ll 1}$} have a supersymmetry breaking scale given by $\varphi/\sqrt{\alpha^\prime}$, with the string tension $1/(2\pi \alpha^\prime)$. Letting $\varphi$ become ever smaller will give us an arbitrarily small SUSY breaking scale. Still, it is not possible to describe such a situation using standard supergravity, as we have already shown. The way supersymmetry is restored when going to the breaking scale is by a set of open string states that stretch between the two branes \cite{Berkooz:1996km,Anastasopoulos:2011hj,Anastasopoulos:2016cmg}. Remember, this means that our or initial effective theory breaks down and no longer correctly describes the physics. It is important to find a new theory that replaces it. In the case of two branes at very small angle this can be done by guessing. It is natural to expect an $SU(2)$ gauge theory for two branes on top of each other. If the angle is non-zero a scalar will get an expectation value and the $SU(2)$ theory breaks to $U(1)\times U(1)$. Interestingly, in this example, one finds an infinite tower of states even though we have a particularly small SUSY breaking scale. It, however, is possible to find examples that lack such an explicit tower.
\paragraph{Two branes intersecting perpendicular on a toroidal orbifold} preserve supersymmetry partially. Again, we can rotate one brane slightly away from the special angle in order to break supersymmetry completely. In this case the SUSY breaking scale can be below both the KK and string scale and thus no obvious heavy states are present that can restore the symmetry at the breaking scale. The solution is somewhat different in this case. While supersymmetry is still realized non-linearly, the theory is nevertheless almost invariant under linear supersymmetry and the world-volume fields are in almost regular multiplets. The contributions from the different fields then cancel in divergent cross sections and hence no new massive states are required.\\ 
In two of the examples above we found that supersymmetry gets restored by a tower of states coming in at the symmetry breaking scale, as opposed to a single additional state. This means that at this scale the initial low-energy effective theory breaks down and needs to be replaced. This seems to be a generic feature for theories that break supersymmetry by the inclusion of some source, that is not specifically aligned in order to avoid this conclusion (see our last example).\\
This, fairly strong claim, requires some remarks. First, the significant changes do not need to affect all sectors of the theory. For example, in the KKLT model the SUSY breaking source sits at the bottom of the warped throat. It follows that effects that appear in the bulk will not be changed significantly due to the warping. Secondly, it might be that sources that (almost) preserve supersymmetry are favored by nature, as they minimize energy. This means that seemingly contrived setups could actually be more natural than expected. Finally, if supersymmetry is broken by background effects, like fluxes, and sources are supersymmetric we would not expect to encounter heavy states near the breaking scale.

\subsection[Non-supersymmetric Branes in $4d$, $\mathcal{N}=1$ Theories]{Non-supersymmetric Branes in $\mathbf{4d}$, $\mathbf{\mathcal{N}=1}$ theories}
We now move on to describe the inclusion of supersymmetry breaking branes in $4d$, $\mathcal{N}=1$ supergravity formalisms. $Dp$-branes are used in a variety of string models including applications to cosmology \cite{Grana:2005jc,Douglas:2006es} and for constructing standard model like theories \cite{Blumenhagen:2006ci,Blumenhagen:2005mu}. For convenience we will neglect the model dependent world-volume fields on the branes as their inclusion is technically difficult. Including all of these world-volume fields has been done in certain cases, see for some examples \cite{Vercnocke:2016fbt,Kallosh:2016aep,Kachru:2003aw,Aalsma:2017ulu,GarciadelMoral:2017vnz,Cribiori:2019hod}. Furthermore, here we will restrict to compactifications that initially preserve $\mathcal{N}=2$ supersymmetry and then add an orientifold projection that removes one supersymmetry and leaves us with $\mathcal{N}=1$ SUSY. The internal manifolds will thus be either CY3-folds in the fluxless case or general $SU(3)$ structure manifolds. 
\paragraph{Let us mention a peculiarity of the $\mathbf{D3}$-brane} before we come to the description of general $Dp$-branes in supergravity. In order to not violate Lorentz symmetry branes need to be spacetime filling. For the case of $D3$-branes this means that they cannot be at arbitrary angles. For a type IIB compactification on a CY3-fold we have $\mathcal{N}=2$ supersymmetry of the background. Introducing an orientifold projection we find:
\be
\varepsilon_1 = \Gamma_{0123}\varepsilon_2 =: \Gamma_{O3} \varepsilon_2\,,
\ee
Putting, on the other hand, a $D3$-brane along some angle $\varphi$ in the $(x^3,x^4)$ plane gives:
\be 
\varepsilon_1 = \left( \cos (\varphi) \Gamma_{0123} + \sin (\varphi) \Gamma_{0124} \right) \varepsilon_2\,,
\ee
meaning we would break Lorentz invariance in the $4d$ directions since for $\sin (\varphi)\neq0$ the above is proportional to $\Gamma_{0124}$. One concludes that the only allowed angles for a $D3$-brane with $4$ extended directions are $\{0,\pi\}$, corresponding to what is usually called a brane or an anti-brane. Note that, according to the discussion in section \ref{sec:flatspacebranes} this also excludes $D3$-branes with non-zero flux $\mathcal{F}$, even for the allowed values of $\varphi$.
\paragraph{For $\mathbf{Dp}$-branes with $\mathbf{p>3}$} we find that they still need to extend in the non-compact space and, internally, wrap a $p-3$ cycle $\Sigma$. The action consists of a DBI part as well as a CS part and is given to be:
\be 
S_{Dp} = \underbrace{T_{Dp} \int_{M^{3,1}\times \Sigma} d^4x \,d^{p-3}y \rme^{-\phi_\Sigma}\sqrt{\det (-g_\Sigma + \mathcal{F}|_\Sigma)}}_{DBI} \;\underbrace{- T_{Dp} \int_{M^{3,1}\times \Sigma} C|_\Sigma \wedge \rme^{\mathcal{F}|_\Sigma}}_{CS}\,,
\label{eq:Dpbraneact}
\ee
with the brane tension $T_{Dp}$, the dilaton $\phi$, the metric $g$ and the sum over all RR-fields $C$. Furthermore, $|_\Sigma$ denotes the pullback of the quantities onto the brane world-volume. In the following all quantities will be understood as pulled back automatically. Considering branes on either CY3 or $SU(3)$ structure manifolds we note that these types of spaces do not have non-trivial $1$- or $5$- cycles, which rules out $D4$- and $D8$-branes. The manifolds are described by a Kähler $(1,1)$-form $J$ and a holomorphic $(3,0)$-form $\Omega$. We note that this is only strictly true in the Calabi-Yau (CY) case whereas in the case of $SU(3)$ structure manifolds one has a real $2$-form $J$ and a complex $3$-form $\Omega$ as such manifolds are not Kähler, see \cite{Grana:2005jc} for a detailed description. With these words of caution out of the way we will continue to slightly abuse these terms and treat both cases in the same way. In \cite{Blumenhagen:2006ci} it was show that one can write the DBI part of a supersymmetric $Dp$-brane action using the characteristic forms of the internal manifold:
\bea
S_{DBI,5} &= T_{D5} \int_{M^{3,1}\times \Sigma_2} d^4x \, \rme^{-\phi} \,J\,,\\
S_{DBI,6} &= T_{D6} \int_{M^{3,1}\times \Sigma_3} d^4x \, \rme^{-\phi} \, \rmre(\Omega)\,,\\
S_{DBI,7} &= T_{D7} \int_{M^{3,1}\times \Sigma_4} d^4x \, \rme^{-\phi} \,\frac{1}{2}\,\left(J\wedge J-B\wedge B \right)\,,\\
S_{DBI,9} &= T_{D9} \int_{M^{3,1}\times \Sigma_6} d^4x \, \rme^{-\phi}\, \left(\frac{1}{6} \, J \wedge J \wedge J - \frac{1}{2}\,J \wedge B \wedge B \right)\,,\label{eq:susypbranes}
\eea
where $B$ is the Kalb-Ramond 2 form and we work in string frame with zero gauge flux $F$. These equations are called calibration conditions and readily generalize to non-vanishing gauge flux $F$ \cite{Martucci:2005ht,Martucci:2006ij,Martucci:2011dn}. For anti-$Dp$-branes similar calibration conditions hold where the main difference to the equations \eqref{eq:Dpbraneact} and \eqref{eq:susypbranes} is a minus sign for the CS-part of the action \cite{Kallosh:2018nrk}.
\paragraph{The description of general non-supersymmetric branes} will require the tools of non-linear supergravity and we will show in the following how it is possible to correctly describe such branes in the low-energy effective SUGRA action. This generalizes the result of \cite{Kallosh:2018nrk}, where the same was done for anti-$Dp$-branes. One main ingredient for this procedure is the nilpotent chiral field $X$ that we already discussed in section \ref{sec:multiplets}. When supersymmetry is broken spontaneously, the transformation is still realized in a non-linear way but fields no longer sit in regular multiplets. Nevertheless, constrained multiplets can be used in order to describe fields in such a case. The nilpotent chiral field $X$ is able to correctly describe the Goldstino related to spontaneous supersymmetry breaking by the brane. This will be an important topic in the next section \ref{sec:D3barKKLT} where we discuss one particular setup in great detail. See the same section as well for a detailed discussion how the brane provides the Goldstino.\\
To begin with, we consider the theory before the addition of the supersymmetry breaking brane. It is described by a Kähler potential $K_{b4}$ and a superpotential $W_{b4}$\footnote{Read $b4$ as ``before''.}. Additionally a gauge kinetic function and potential $D$-terms could be present in the theory but we will ignore these possibilities for now. The potentials are obtained by performing a string theory compactification and they depend on chiral multiplets $\Phi^a$. The goal now is to add a general brane to this setup and compute the backreaction onto the fields $\Phi^a$. This backreaction is given by the scalar potential $V_{Dp}(\Phi^a,\bar{\Phi}^a)$ that newly appears due to the inclusion of the brane. In the language of supergravity, this can be achieved by a modification of the Kähler- and superpotential:
\bea
K &= K_{b4} + \rme^{K_{b4}} \frac{X\bar{X}}{V_{Dp}}\,,\\
W &= W_{b4} + X\,.
\label{eq:newsugra}
\eea
Calculating the scalar potential using these potentials with the usual formula yields:
\be 
V = \rme^K \left(K^{I\bar{J}} D_I W \overline{D_J W} - 3 |W|^2 \right)\Big|_{X=0}  = V_{b4} + V_{Dp}\,,
\ee
which is exactly what one expects. Note that the indices $I$ and $J$ run over all fields, including the nilpotent field $X$. In the end, one has to set that field equal to zero as it only contains fermionic degrees of freedom. This means that:
\bea 
K^{X\bar{X}}|_{X=0} &= \rme^{-K_{b4}} V_{Dp}\,,\\
D_X W|_{X=0} &= 1\,.
\eea
It turns out that \eqref{eq:newsugra} actually gives the correct result, even for backgrounds that do not preserve supersymmetry. Furthermore the description is not necessarily limited to $Dp$-branes. Indeed, any SUSY breaking ingredient that can be described using a nilpotent chiral field can be incorporated in this way. For this reason we occasionally replace $V_{Dp}$ by $V_{\text{new}}$ in the following, when we discussing more general results. One thing that is not quite obvious is that $V_{\text{new}}$ is a real function of the fields $\Phi^a$, which it needs to be since we include it in the Kähler potential $K$. Nevertheless, this should be always the case since $V_{\text{new}}$ has to be a function of the closed string degrees of freedom that get packaged into the fields $\Phi^a$. One more thing to note is that in more general cases the Goldstino might not be completely given by the SUSY breaking ingredient. If the background fields also break supersymmetry the Goldstino will become a linear combination of the fermion contained in the nilpotent chiral field $X$ and the fermions in the $\Phi^a$. This can, for example, happen if the backreaction of the SUSY breaking source is significant and influences the background fields.\\
The main task in applying this process to new examples is finding the correct form of $V_{\text{new}}$. This might seem simple, given the expression in \eqref{eq:newsugra}, but can become rather involved for some cases. In particular, when one wants to include all world-volume fields in a given setup. Thus far, this has only been completely done for the anti-$D3$-brane in the KKLT setup, see \cite{Cribiori:2019hod} or the following section \ref{sec:D3barKKLTmain}. The full result reduces exactly to our proposal here if one sets all world-volume fields to zero. The reason that the simple prescription here works is that it seems to be an universal feature, at least for $Dp$-branes, that the Goldstino can be packaged into a nilpotent chiral field $X$ while the features of the background, the brane dimension and other things get put into a scalar function $V_{\text{new}}$. We will give some backup to this claim by investigating some examples in the following where we are able to validate this claim by recovering the correct results. Let us mention that thus far, whenever this method was carefully carried out, it did give the correct answer.

\subsection{Examples of genuine non-supersymmetric Branes}
\paragraph{As a first example we consider the anti-$\mathbf{D3}$-brane,} which is of particular interest due to its inclusion as an ingredient in the KKLT model \cite{Kachru:2003aw,Kachru:2003sx}. Depending on the setup, the brane is either placed in the bulk of the internal manifold or at the bottom of a warped throat. This gives different expressions for the scalar potential:
\bea 
V^{\text{bulk}}_{\overline{D3}} &= \frac{\mu^4}{\left(-\rmi(T-\bar{T})\right)^3}\qquad \text{for the brane in the bulk,}\\
V^{\text{throat}}_{\overline{D3}} &= \frac{\mu^4}{\left(-\rmi(T-\bar{T})\right)^2}\qquad \text{for the brane at the bottom of a warped throat.}
\eea
Here $T$ is the single Kähler modulus of the model and $\mu$ a parameter related to the number of branes and their tension. We will discuss the KKLT scenario in much more detail in the following sections of this thesis. Here, we focus on the description using \eqref{eq:newsugra}. 
\\\newpage The Kähler potentials for the two different cases are:
\bea 
K^{\text{bulk}} &= -3 \log \left[ -\rmi(T-\bar{T})\right] + \frac{X \bar{X}}{\mu^4}\,,\\
K^{\text{throat}} &= -3 \log \left[-\rmi (T-\bar{T})\right] + \frac{X\bar{X}}{\mu^4\left(-\rmi (T-\bar{T})\right)} = - 3 \log \left[ \left(-\rmi (T-\bar{T})\right) - \frac{X\bar{X}}{3\mu^4} \right]\,,
\eea
where $X^2=0$ is required to perform the last step in the second line.\\
The validity of the above description was verified in \cite{Kallosh:2014wsa,Bergshoeff:2015jxa} and this formalism was further used in \cite{GarciadelMoral:2017vnz,Cribiori:2019hod} to incorporate all world-volume fields of the anti-D3-brane in the KKLT setup, see also the following section \ref{sec:D3barKKLTmain}.

\paragraph{Our next example will be intersecting $\mathbf{D6}$-branes} which are also the main motivation for our considerations here. Intersecting $Dp$-branes can be used as ingredients in flux compactifications and for building Standard Model like scenarios \cite{Blumenhagen:2006ci,Blumenhagen:2005mu}. We focus on situations where the background initially preserves $\mathcal{N}=1$ supersymmetry and the branes are responsible for the full breaking of SUSY. Such constructions were initially viewed as being describable as a $D$-term in standard supergravity. In the following we will review the arguments that were used for this conclusion and then we will give the correct description using our method. Let us, however, clarify that the $10d$ picture was always correct. The description was obtained using the correct DBI and world sheet action in $10d$ and the conclusions drawn from this are, to the best of our knowledge, correct. However, the $4d$ interpretation of the behavior as a $D$-term is the wrong point of view. It is necessary to use non-linear supersymmetry for that purpose.\\
Let us review $D6$-branes wrapping some 3-cycle $\Sigma_3$ in a type IIA compactification on $T^6/(\mathbb{Z}_2 \times \mathbb{Z}_2)$ as our example \cite{Villadoro:2006ia}. For this we define, in our conventions, the 3-form:
\be 
\Omega_\Sigma = \int_{\Sigma_3} \rme^{-\phi} \Omega\,,
\ee
and we find the scalar potential originating from the DBI action, as:
\be 
V_{\text{DBI}} = T_{D6} \, \rme^{-2\phi} \, vol_6 \sqrt{\rmre (\Omega_\Sigma)^2 + \rmim (\Omega_\Sigma)^2}\,,
\label{eq:D6DBIact}
\ee
with $vol_6$ the volume of the internal space. In the case that the brane is supersymmetric one finds that $\rmim (\Omega_\Sigma) = 0$ and the DBI part of the scalar potential reduces to $V_{\text{DBI}} =T_{D6} \rme^{-2\phi} vol_6 \rmre (\Omega_\Sigma)$, as $\rmre (\Omega) > 0$ for SUSY branes. This actually follows from the action of the $D6$-brane given in equation \eqref{eq:susypbranes} when going to $4d$ Einstein frame. Changing to Einstein frame in four dimensions can be done by writing:
\be 
\int d^{10}x \sqrt{-g^{\text{string}}_{10}} \rme^{-2\phi} R_{10} = \int d^4x \sqrt{-g_4^{\text{string}}} vol_6 R_4 + \ldots\,,
\ee
and then letting the metric $g^{\text{string}} \to \rme^{4 \phi} vol_6^{-2} g^{\text{Einstein}}$. In particular, this means that the actions in \eqref{eq:susypbranes} only pick up a factor of $\rme^{4 \phi} vol_6^{-2}$ when changing frames. In \cite{Villadoro:2006ia} the authors followed \cite{Blumenhagen:2002wn} and re-wrote the scalar potential in the following way:
\bea
V_{\text{DBI}} &= V_F + V_D\,,\\
V_F &= T_{D6} \rme^{4 \phi} vol_6^{-2} \rmre (\Omega)_\Sigma\,,\\
V_D &= T_{D6} \rme^{4 \phi} vol_6^{-2} \left( \sqrt{\rmre(\Omega_\Sigma)^2 + \rmim(\Omega_\Sigma)^2} - \rmre (\Omega_\Sigma)\right)\,.
\eea
For supersymmetric branes this reduces correctly since then $V_D = 0$ which has to be the case since the $D$-term cannot uplift the vacuum energy without breaking supersymmetry. One can then attempt to re-write everything in terms of a gauge kinetic function and a $D$-term:
\bea 
T_{D6} &= \rme^{-2\phi}vol_6 \mathcal{P}\,,\\
\rmre(f) &= T_{D6} \sqrt{\rmre(\Omega)^2+\rmim(\Omega)^2} = T_{D6} \rmre(\Omega) + \mathcal{O} \left(\frac{\rmim(\Omega_\Sigma)}{\rmre(\Omega_\Sigma)}\right)\,,
\eea
where the first line describes the moment map $\mathcal{P}$.\\
We now neglect the higher-order terms in the above expressions and use everything to write:
\bea 
V_D &= T_{D6} \, \rme^{4 \phi} \, vol_6^{-2} \frac{\;\rmim(\Omega_\Sigma)^2}{\rmre(\Omega_\Sigma)} \frac{1}{1+\sqrt{1+\left(\frac{\rmim(\Omega_\Sigma)}{\rmre(\Omega_\Sigma)}\right)^2}}\,\\
&= \frac{1}{2\rmre(f)} \, \mathcal{P}^2 \, \frac{2}{1+\sqrt{1+\left(\frac{\rmim(\Omega_\Sigma)}{\rmre(\Omega_\Sigma)}\right)^2}}\,.
\eea
One could argue that this is compatible with standard $\mathcal{N}=1$ supergravity for as long as $|\rmim(\Omega_\Sigma)|/|\rmre(\Omega_\Sigma)| \ll 1$ \cite{Villadoro:2006ia}. Strictly speaking, however, this is only true iff $\rmim(\Omega_\Sigma) = 0$, which is, for example, the case for supersymmetric branes. Then, everything can be written using a standard $D$-term and standard supergravity is applicable. If the brane is not supersymmetric, higher-order terms need to appear to mend this discrepancy. These are exactly provided by non-linear supergravity. There is one more relevant example we would like to mention where standard supergravity formulas can correctly describe the situation. When anti-$D6$-branes wrap SUSY cycles that have the opposite orientation we also have $\rmim(\Omega_\Sigma)=0$ and everything works out. In \cite{Kallosh:2018nrk} the correct description of anti-branes using non-linear supergravity and constrained fields was given. In the following we will generalize their result to arbitrary branes.

\subsection{Non-linear SUGRA Description for any Brane}
\paragraph{Let us begin by describing arbitrary $\mathbf{D6}$-branes} by utilizing the method in equation \eqref{eq:newsugra}. This is easy because all we have to do is use the DBI contribution to the scalar potential and find the modified Kähler and superpotential:
\bea 
K &= K_{b4} + \rme^{K_{b4}} \frac{X\bar{X}}{T_{D6} \, \rme^{-2\phi} \, vol_6 \sqrt{\rmre (\Omega_\Sigma))^2 + \rmim (\Omega_\Sigma)^2}}\,,\\
W &= W_{b4} + X\,,
\eea
where we again used the nilpotent chiral field $X$. As we mentioned before, the brane provides the Goldstino via the only degree of freedom of $X$ and we neglect other, model specific, world-volume fields. This readily generalizes to multiple $D6$-branes by adding a sum over the individual contributions in the above formulae. The Goldstino, contained in $X$, is a linear combination of the fermions coming from each brane individually. Including world-volume fields in this description, in particular in the case where the background contributes to the supersymmetry breaking due to the backreaction of the brane, would significantly complicate the description and is model dependent.\\
In every case one still needs to make sure that the scalar potential contribution, $V_{\text{new}}$, only depends on the closed string degrees of freedom and is a real function. To be more specific $V_{\text{new}}$ has to be a real function of the complex structure moduli $\mathcal{Z}^N$. Let us give an argument that this holds in our simple example here before we move on to general branes. For this we note that the complex structure moduli in type IIA can be written as:
\be 
\mathcal{Z}^N = \int_{\Sigma^N} \left(\frac{1}{2} C_3 + \rmi \rme^{-\phi} \sqrt{vol_6}\,\rmre(\Omega)\right)\,,
\ee
with the $\Sigma^N$ a basis of the orientifold odd 3-homology. Using this we conclude that:
\be 
\rmim(\mathcal{Z}^N) = \int_{\Sigma^N} \rme^{-\phi} \sqrt{vol_6} \, \rmre(\Omega)\,,
\ee
and since the dilaton and the volume are always given by the closed string moduli $\rmre(\Omega_\Sigma)$ is a real function of the complex structure moduli. For the imaginary part of $\Omega_\Sigma$ that also gives contributions to the scalar potential we have to do a little bit more work as it does not appear in the above expression for the $\mathcal{Z}^N$. The resolution lies in the orientifold involution of the $O6$-plane that is part of the example we consider. It maps:
\be 
\sigma:\quad \Omega \to \bar{\Omega}\,,
\ee
which in turn allows us to write $\rmim(\Omega_\Sigma)$ as a real function in terms of $\rmre(\Omega_\Sigma)$, for details see \cite{Grimm:2004ua}. All of this leads us to conclude that $V_{\text{new}}$ in our example here is indeed a real function of the closed string degrees of freedom and the formalism works as intended.

\paragraph{We now generalize further to give the description for arbitrary $\mathbf{Dp}$-branes} utilizing the same tools as above. For the $Dp$-brane actions we will take the results found in \cite{Martucci:2005ht,Martucci:2006ij}. As our Ansatz for the metric we use:
\be 
ds^2 = \rme^{2\mathcal{A}(y)} g_{\mu \nu} dx^\mu dx^\nu + g_{mn} dy^m dy^n \,,
\ee
with the warp factor $\mathcal{A}(y)$ that depends only on the internal coordinates. Furthermore, we will assume that the internal space has $SU(3)\times SU(3)$ structure and use the pure spinors $\hat{\Psi}_1$ and $\hat{\Psi}_2$ to describe the manifold \cite{Martucci:2005ht}. Importantly, the definition of those spinors depends on whether we consider type IIA or type IIB theory. We re-iterate that the branes fill the 4 extended dimensions of spacetime and they wrap $(p-3)$-cycles $\Sigma$ in the internal manifold. Using a subscript $\Sigma$ to denote the pullback of a quantity onto the worldvolume of the brane and $(p-3)$ for quantities where we are only interested in the $(p-3)$-form part of the expression we define:
\bea 
\mathcal{W}_m \, d\sigma^1 \wedge \ldots \wedge d\sigma^{p-3} &= \frac{(-1)^p}{2} \left[\rme^{3\mathcal{A}-\phi} \left( i_m + g_{mk}\,dy^k \wedge \right) \hat{\Psi}_2 \right]_\Sigma \wedge \rme^{\mathcal{F}}\Big|_{p-3}\,,\\
\mathcal{D} \, d\sigma^1 \wedge \ldots \wedge d\sigma^{p-3} &= \left[\rme^{4\mathcal{A}-\phi} \rmim(\hat{\Psi}_1)\right]_\Sigma \wedge \rme^{\mathcal{F}}\Big|_{p-3}\,,\\
\Theta \, d\sigma^1 \wedge \ldots \wedge d\sigma^{p-3} &= \left[\rme^{4\mathcal{A}-\phi} \rmre(\hat{\Psi}_1)\right]_\Sigma \wedge \rme^{\mathcal{F}}\Big|_{p-3}\,.
\eea 
Utilizing these quantities we re-write the DBI action of the $Dp$-brane as:
\bea
S_{\text{DBI},\,p} &= T_{Dp} \, \int_{M^{3,1}} d^4x \int_\sigma d^{p-3}y \, \rme^{-\phi|_\Sigma} \sqrt{\det(-g|_\Sigma + \mathcal{F}|_\Sigma)}\\
&= T_{Dp} \, \int_{M^{3,1}} d^4x \int_\sigma d^{p-3}\sigma \, \sqrt{\Theta^2 + \rme^{4\mathcal{A}} \mathcal{D}^2 + 2 \rme^{2\mathcal{A}} g^{mn} \mathcal{W}_m \mathcal{W}_n}\,,
\eea
which reduces to the previous $D6$- brane solution in equation \eqref{eq:D6DBIact} for $p=6$, $\mathcal{W}_n = 0$ and when going to Einstein frame in $4$ dimensions.\\
In general, $\mathcal{W}_n$ will always be zero for us since it is the derivative of a holomorphic superpotential that includes the model dependent contributions from the open string degrees of freedom. Since we choose to neglect world-volume fields, except the Goldstino, we thus set $\mathcal{W}_n = 0$. Now we still need to argue that the remainder, $\Theta^2 + \rme^{4\mathcal{A}} \mathcal{D}^2$, is a real function of the closed string moduli. The general argument is that the moduli are a combination of the RR axions and the imaginary parts of the NSNS sector fields which we will call $\mathcal{Z}$ again for convenience. Similar to the $D6$ example above, $\rmim(\mathcal{Z})$ can be written directly as a function of $\rmre(\hat{\Psi}_1)$ and $\rmre(\mathcal{Z})$ can be related back to $\rmim(\mathcal{Z})$ as discussed in \cite{Grana:2005jc}. Hence, we write generally:
\bea 
S_{\text{DBI},\,p} &= T_{Dp} \, \int_{M^{3,1}} d^4x \int_\sigma d^{p-3}\sigma \, \sqrt{\Theta^2 + \rme^{4\mathcal{A}} \mathcal{D}^2 + 2 \rme^{2\mathcal{A}} g^{mn} \mathcal{W}_m \mathcal{W}_n}\,\\
&= T_{Dp} \, \int_{M^{3,1}} d^4x \int_\sigma d^{p-3}\sigma \, \sqrt{ f(\rmim(\mathcal{Z}^a))+g(\rmim(\mathcal{Z}^a))}\,.
\eea
Here, $f(\rmim(\mathcal{Z}^a))$ and $g(\rmim(\mathcal{Z}^a))$ are real functions and we recover a supersymmetric $Dp$-brane for $g(\rmim(\mathcal{Z}^a)) = 0$. Since the behaviors of different brane setups get incorporated into real functions depending on the closed string moduli, this procedure is applicable in very general setups.
\subsection{Non-supersymmetric Branes - Interim Summary}
In this section we have discussed the correct description of branes in $4d$, $\mathcal{N}=1$ supergravity, independent of their orientation and if they preserve supersymmetry. We have focused on type II compactifications where the background initially preserves $\mathcal{N}=2$ SUSY and added an orientifold projections that projects out one of these supersymmetries. Previously, while the $10d$ description and the results drawn from them have been correct, the re-packaging of  the brane behavior into the language of $4d$ SUGRA has been done incorrectly. Here we showed a general way to include any brane using non-linear supersymmetry and constrained superfields. Our method, given in equation \eqref{eq:newsugra}, allows us to find the correct description. It is important to check that the scalar potential, coming from the supersymmetry breaking source, $V_{\text{new}}$ has to be a real function of the closed string degrees of freedom. We have shown how this can be argued for general branes. When neglecting the world-volume fields on the branes and backreactions on the background, as we did here, the brane will be the sole provider of the Goldstino. It is the only degree of freedom of the nilpotent field $X$ in our description. Including the world-volume fields of the brane is model dependent and in general rather cumbersome. Also, it then can happen that the Goldstino is a linear combination of the fermion in $X$ and the other world-volume fermions in such cases. Likewise, it is possible that the brane backreacts onto the background which in turn leads to supersymmetry breaking by the background. If this is the case, the Goldstino will be a composite expression as well.\\
The result we obtained here is rather general and, in fact, should be applicable for other SUSY breaking sources, other than branes. Examples include $KK$-monopoles or $NS5$-branes. Other interesting applications for the procedure given here are specific models where one can study the situation in more detail and include model specific contributions, such as world-volume fields. We will investigate an explicit example in the anti-$D3$-brane in the KKLT setup in the next section.

\section[The Anti-$D3$-Brane in the KKLT Scenario]{The Anti-$\mathbf{D3}$-Brane in the KKLT Scenario}
\label{sec:D3barKKLTmain}
The anti-$D3$-brane in the KKL(MM)T scenario \cite{Kachru:2003aw,Kachru:2003sx} provides the uplifting contribution to the the scalar potential that raises the minimum from anti-de Sitter space to de Sitter. The description of the anti-brane in terms of $4d$, $\mathcal{N}=1$ supergravity has been studied extensively in the last couple of years. Here we extend this description by including all world-volume fields in a complete effective action for the KKLT scenario. As will be shown, we find that the action breaks supersymmetry spontaneously, which might be expected as the anti-$D3$-branes in the KKLT model can be viewed as excited states in a supersymmetric description \cite{Kachru:2002gs}.\\
The connection of the description of the anti-$D3$-brane to the uplifting contributions included in the KKLT scalar potential has only been fully understood fairly recently \cite{Ferrara:2014kva,Kallosh:2014wsa,Bergshoeff:2015jxa,Kallosh:2015nia,Garcia-Etxebarria:2015lif}. Nevertheless, a full description of the action has not been found before our contribution. The action consists of a bosonic part with three complex world-volume scalars and an $U(1)$ gauge field and a fermionic contribution that contains four 4 dimensional fermions. In the context of flux compactifications the fermionic part is only know up to quadratic terms \cite{Grana:2002tu,Grana:2003ek,Marolf:2003ye,Tripathy:2005hv,Martucci:2005rb,Bergshoeff:2013pia}. Naively, one would expect that the focus of the investigation would be on the bosonic part, as people typically dislike dealing with spinors. The usual argument would be that the fermionic contributions then are given by supersymmetry. However, the bosonic contributions on the brane can be projected  out using an orientifold and thus, the focus has largely been on the fermionic part of the action \cite{Kallosh:2014wsa,Bergshoeff:2015jxa,Kallosh:2015nia,Garcia-Etxebarria:2015lif,Dasgupta:2016prs,GarciadelMoral:2017vnz}. The combination of the fermionic action with the bosonic uplifting contribution can be merged into a Volkov-Akulov type action \cite{Volkov:1973ix} which can be written using constrained superfields in $4d$, $\mathcal{N}=1$ supergravity. In \cite{GarciadelMoral:2017vnz} the action for all GKP background fields and the four world-volume fermions has been derived. In \cite{Cribiori:2019hod} we extended the description to include the remaining contributions, namely the world-volume scalars and the $U(1)$ gauge field. This will give a complete description of the uplifting anti-$D3$-brane in the KKLT background in terms of an effective $4d$, $\mathcal{N}=1$ supergravity action.

\subsection[Action of the Anti-$D3$-Brane in KKLT]{Action of the Anti-$\mathbf{D3}$-Brane in KKLT}
\label{sec:D3barKKLT}
Before we can find the complete effective description of the anti-$D3$-brane we will review the action in the GKP \cite{Giddings:2001yu} and the KKLT \cite{Kachru:2003aw,Kachru:2003sx} background. The fields that will appear in our description are:
\begin{itemize}
\item the axio-dilaton $\tau = C_0 + \rmi \rme^{-\phi}$,
\item a single Kähler modulus $T$,
\item complex structure moduli $U^A$.
\end{itemize}
The bosonic and fermionic parts of the $\overline{D3}$-brane action are reviewed in the following.
\paragraph{The bosonic action} consists of a DBI and a Chern-Simons part that combine to give the complete expression. In $4d$ Einstein frame these can be written as: 
\bea 
S^{\text{DBI}} &= - \int d^4x \sqrt{-\text{det}\left( P \left[ g_{\mu \nu} + \rme^{-\phi/2}B_{\mu\nu}\right] + \rme^{-\phi/2} F_{\mu\nu}\right)}\,,\\
S^{\text{CS}} &= -\int P\left[(C_0 + C_2 + C_4)\wedge \rme^{B_2}\right] \wedge \rme^F\,.
\label{eq:DBICSact}
\eea 
In our conventions here the string length is $l_s = 2 \pi \sqrt{\alpha^\prime} = 1$, $B_2$ is the NSNS Kalb-Ramond field and $F_{\mu\nu}$ denotes the field strength of the $U(1)$ gauge field on the brane\footnote{Note that we have re-scaled the $U(1)$ field strength by $2\pi$ compared to other typical conventions. Likewise, the action has been re-scaled by $1/(2\pi)$ in order to remove the brane tension $T_3 = (2\pi)^{-3} (\alpha^\prime)^{-2} = 2\pi$.}. $P$ is the pullback operation onto the brane world-volume.\\
The GKP background \cite{Giddings:2001yu} is warped which makes the identification of the Kähler modulus $\rmim(T)$ rather tedious \cite{Frey:2008xw}. In the case of a single Kähler modulus a fixed overall scaling for all terms in the action exists. We will use the Ansatz \cite{deAlwis:2016cty}:
\be 
ds^2 = \rme^{-6u(x)} \left(1+\frac{\rme^{-4\mathcal{A}(z)}}{\rme^{4u(x)}}\right)^{-\frac{1}{2}} g_{\mu\nu} dx^\mu dx^\nu + \rme^{2u(x)} \left(1+\frac{\rme^{-4\mathcal{A}(z)}}{\rme^{4u(x)}}\right)^{\frac{1}{6}} g_{a\bar{b}} dz^a dz^{\bar{b}}\,,
\ee
to identify $\rmim(T)$ later on. In this Ansatz the non-compact dimensions are labeled $\mu,\,\nu =0,1,2,3$ while the internal directions are $a,\,\bar{b}=1,2,3$. The six-dimensional compact volume is given by $\rme^{6u} = vol_6$ and $\mathcal{A}(z)$ is the warp factor, depending only on the internal directions. The purpose of this metric is to interpolate between the bulk region and the throat. Note that this Ansatz does not solve the mixed parts of the Einstein equations, where the non-compact and internal coordinates mix. This will, however, not have any effect on our conclusions. Since we want to investigate the anti-$D3$-brane action at the bottom of a strongly warped throat we are interested in the regime where $\rme^{-4\mathcal{A}}\gg \rme^{4u}$. In this limit the metric becomes:
\be 
ds^2 = \rme^{-4 u(x) + 2\mathcal{A}(z)} g_{\mu\nu} dx^\mu dx^\nu + \rme^{\frac{4}{3} u(x) - \frac{2}{3} \mathcal{A}(z)} g_{a\bar{b}} dz^a dz^{\bar{b}}\,.
\ee
Utilizing this we proceed our investigation of the DBI action \cite{McGuirk:2012sb} in \eqref{eq:DBICSact}: 
\bea
S^{\text{DBI}} = - \int d^4x \sqrt{-g_4} \bigg(& \rme^{4 \mathcal{A}(H,\bar{H}) - 8 u(x)} + \frac{1}{2}  \rme^{\frac{4}{3} \mathcal{A}(H,\bar{H}) - \frac{8}{3} u(x)} g_{a\bar{b}}(H,\bar{H}) \partial_\mu H^a \partial^\mu \bar{H}^{\bar{b}} \\&+ \frac{1}{4} \rme^{-\phi(H,\bar{H})} F_{\mu\nu} F^{\mu\nu} + \ldots \bigg)\,.
\label{eq:DBIact}
\eea
Here, the $H^a$ are the  world-volume scalars that give the position of the brane, entering the action via the pullback and the ellipses denote suppressed, higher-order terms. In the following we will consider the brane to be sitting at some point in the strongly warped throat and then the $H^a$ will denote small fluctuations around that initial position.\\
The DBI-part of the action yields all kinetic terms for the scalar fields. In $4d$, $\mathcal{N} = 1$ supergravity this this can be described using the Kähler potential \cite{DeWolfe:2002nn}:
\be 
K = -3 \log [-\rmi (T-\bar{T}) + k(H,\bar{H})]\,,
\label{eq:barD3Kpot}
\ee
with $k(H,\bar{H})$ the part of the Kähler potential that incorporates the metric of the compact manifold:
\be 
\partial_{H^a} \partial_{\bar{H}^{\bar{b}}} k(H,\bar{H}) \approx \frac{1}{6} \rme^{\frac{4}{3}(\mathcal{A}+u)} g_{a\bar{b}}\,.
\ee
Subleading terms are again neglected \cite{Baumann:2007ah}. Importantly, the no-scale structure of the Kähler potential is not violated by this contribution. The volume of the internal manifold depends on the open and closed string degrees of freedom and gets modified by $k(H,\bar{H})$:
\be 
vol_6 = \rme^{6u} = \left(-\rmi (T-\bar{T}) + k(H,\bar{H})\right)^{\frac{3}{2}}\,.
\ee
The Chern-Simons part of the action simplifies in the GKP background \cite{Giddings:2001yu} to:
\be
S^{\text{CS}} = -\int \left(\frac{1}{2} C_0 (H,\bar{H}) F\wedge F + C_4 (H,\bar{H}) \right)\,,
\ee
where one has  to keep in mind that the contributions from $B_2$ and $C_2$ are projected out by the orientifold in this setup. We then expand the quantities around the position of the brane where we use, as mentioned before, $H^a$ as the distance from the brane sitting initially at $(H^a,\bar{H}^a) = (0,0)$ in the warped throat. To leading order we get:
\bea 
S^{\text{CS}} &= -\int \left(\frac{1}{2} C_0 (H,\bar{H}) F\wedge F + C_4 (H,\bar{H}) \right)\\
&=-\int \left( \frac{1}{2} C_0(0,0)F\wedge F + C_4(0,0)+ \ldots\right)\,.
\eea 
We then use that $C_0(0,0) = \rmre(\tau) $ and $C_4(0,0) = \alpha(z,\bar{z}) \sqrt{-g_4} dx^0 \wedge dx^1 \wedge dx^2 \wedge dx^3$ and find for the Chern-Simons action:
\be 
S^{\text{CS}}=-\int d^4x \sqrt{-g_4} \left(-\frac{1}{8 \sqrt{-g_4}}\rmre(\tau) \epsilon^{\mu \nu \rho \sigma} F_{\mu \nu} F_{\rho \sigma} + \alpha(H,\bar{H}) + \ldots \right)\,.
\label{eq:CSact}
\ee
In order to merge the two parts of the bosonic action we first combine the first term in the DBI action \eqref{eq:DBIact} with the second term in the CS action \eqref{eq:CSact} to:
\be 
\Phi_\pm := \rme^{4\mathcal{A}(H,\bar{H})-8u} \pm \alpha(H,\bar{H})\,,
\ee
where the sign difference is for branes (minus) and anti-branes (plus). For the D3-brane the equations of motion in the GKP background enforce  $\rme^{4\mathcal{A}(H,\bar{H})-8u} = \alpha(H,\bar{H})$ and thus $\Phi_-=0$ but for the anti-brane the DBI and CS contributions add up and we find the combined action for the anti-$D3$-brane to be:
\bea 
S^{\overline{D3}}_{\text{bosonic}}= -\int d^4x \sqrt{-g_4} \bigg(&2 \rme^{4\mathcal{A}(H,\bar{H})-8u} + \frac{1}{2} \rme^{\frac{4}{3} \mathcal{A}(H,\bar{H})-\frac{8}{3} u(x)} g_{a\bar{b}} \partial_\mu H^a \partial^\mu \bar{H}^{\bar{b}} \\&+ \frac{1}{4} \rmim(\tau)F_{\mu\nu} F^{\mu \nu} - \frac{1}{8 \sqrt{-g_4}} \epsilon^{\mu \nu \rho \sigma} F_{\mu \nu} F_{\rho \sigma} + \ldots \bigg)\,.
\label{eq:simpbosact}
\eea 
It is evident that, combining the DBI with the CS action, we get a typical Maxwell term for the $U(1)$ gauge field.  For a D3-brane in our setup we would have preserved $\mathcal{N}=1$ supersymmetry and one would use a gauge kinetic function $f(\tau)=-\rmi \tau$ for the supergravity description. At this point it is important to note that the gauge kinetic function has to be holomorphic and may only depend on the axio-dilaton. The coupling at the brane for the $U(1)$ gauge interaction is given by $\rmre(f(\tau)) = \rmim(\tau) = \rme^{-\phi}$. On the other hand, $\rmim(f(\tau)) = - \rmre(\tau) = -C_0$, controlling the theta term of the Maxwell part of the action. If we now want to consider our anti-$D3$-brane, the Chern-Simons action comes with a sign difference and thus it seems that, in order to match equation \eqref{eq:CSact}, we would need an anti-holomorphic gauge kinetic function, satisfying $f(\bar{\tau})= \rmi \bar{\tau}$. Supersymmetry does not allow for this since $\tau$ is part of an unconstrained chiral multiplet. The resolution of this issue is one of the main points of this chapter and will be discussed later on in \ref{sec:sugraact}. The important physical conclusion will be that the anti-$D3$-brane preserves non-linear $\mathcal{N}=1$ supersymmetry.\\
Moving on, we can identify a scalar potential:
\be 
V^{\overline{D3}}(H,\bar{H}) = \Phi_+ = 2 \rme^{4\mathcal{A}(H,\bar{H})-8u}\,,
\label{eq:barD3scalpotorig}
\ee
that gives the attraction of the brane towards the bottom of the warped throat, as the warp factor $\mathcal{A}$ is minimized there. As usual, we expand the scalar potential in terms of the $H^a$ and $\bar{H}^{\bar{b}}$ around the point of attraction:
\bea 
V^{\overline{D3}}(H,\bar{H}) = 2 \rme^{4 \mathcal{A}_0-8u_0} \big( 1 &+ 2 H^a H^b \partial_{H^a} \partial_{H^b} \mathcal{A}_0 + 4 H^a \bar{H}^{\bar{b}} \partial_{H^a} \partial_{\bar{H}^{\bar{b}}} \mathcal{A}_0  \\&+ 2 \bar{H}^{\bar{a}} \bar{H}^{\bar{b}} \partial_{\bar{H}^{\bar{a}}} \partial_{\bar{H}^{\bar{b}}} \mathcal{A}_0 + \ldots \big)\,.
\eea
The quantities with index ``$0$" are evaluated at $H=0$ and we once more neglect higher-order contributions, suppressed by the string scale \cite{McGuirk:2012sb}. From the first term in this expression we get the typical uplifting term from an anti-$D3$-brane in a warped throat which scales like $1/vol_6^{4/3}$ \cite{Kachru:2003sx}.\\
Before we continue, a remark about the KKLT scenario is in order. The volume modulus in a GKP background is flat and together with the scalar potential of the anti-$D3$-brane this would lead to an instability for the Kähler modulus. This gets remedied via the inclusion of non-perturbative corrections, usually from gaugino condensation on D7-branes in the bulk. The argument then is that, since the non-perturbative effect happens in the bulk and we consider the anti-$D3$-brane at the bottom of a highly warped throat, these contributions can safely be neglected. Issues around these non-perturbative corrections have been discussed in the literature to a great extend \cite{Moritz:2017xto,Moritz:2018sui,Kallosh:2018wme,Moritz:2018ani,Kallosh:2018psh,Gautason:2018gln,Hamada:2018qef,Kallosh:2019axr,Kallosh:2019oxv,Hamada:2019ack,Carta:2019rhx,Gautason:2019jwq,Kachru:2019dvo} and, as of now, it seems reasonable to proceed along, following the argument given here.

\paragraph{The fermionic part of the anti-$\mathbf{D3}$-brane action} has to contain a Goldstino, responsible for the spontaneous breaking of supersymmetry by the anti-brane. The goal of this part of the thesis is  to offer a complete description of the anti-$D3$-brane in the KKLT setup in the language of $4d$, $\mathcal{N}=1$ supergravity. For this, the Goldstino will be described using a constrained superfield, to be precise, it will be contained in a nilpotent chiral field. \\
Before we come to the fermionic action of the anti-$D3$-brane we need to figure out how the anti-$D3$-brane action contains the Goldstino. General $Dp$-brane actions in flux backgrounds are know only up to quadratic orders in fermions \cite{Grana:2002tu,Grana:2003ek,Marolf:2003ye,Tripathy:2005hv,Martucci:2005rb,Bergshoeff:2013pia}. Discussions focused on the anti-$D3$-brane in this context can be found in \cite{Kallosh:2014wsa,Bergshoeff:2015jxa,GarciadelMoral:2017vnz,McGuirk:2012sb}. \\
There are four world-volume fermions on the anti-$D3$-brane, part of the $SU(3)$ holonomy group of the internal manifold. One singlet $\lambda$ and three $\chi^i$, forming a triplet. The behavior of these fermions is determined by the imaginary  self dual (ISD) part:
\be 
G_3^{\text{ISD}} = \frac{1}{2} \left( G_3 - \rmi \star_6 G_3 \right)\,,
\ee
of the background 3-flux \cite{Bergshoeff:2015jxa,McGuirk:2012sb}:
\be 
G_3 = F_3 - \rmi \rme^{-\phi} H_3 = dC_2 - \left(C_0 + \rmi \rme^{-\phi} \right) H_3\,.
\ee
The flux gives rise to the following contributions:
\begin{itemize}
\item The $(0,3)$-flux gives a mass to the singlet $\lambda$.
\item The interactions between the singlet $\lambda$ and the triplet $\chi^i$ are determined by non-primitive $(2,1)$-flux.
\item The primitive $(2,1)$-flux gives the masses for the triplet $\chi^i$.
\end{itemize}
In a supersymmetric GKP background the anti-$D3$-brane is the sole source of supersymmetry breaking and then $\lambda$ will be identified as the Goldstino. It does not mix with the fermion triplet $\chi^i$ and does not receive a mass \cite{Bergshoeff:2015jxa}.\\
If the background already has broken supersymmetry there is a Gukov-Vafa-Witten (GVW) superpotential \cite{Gukov:1999ya}:
\be 
W_{\text{GVW}} = \int G_3 \wedge \Omega\,,
\label{eq:GVWpot}
\ee
that is non-zero. Here, the characteristic 3-form $\Omega$ of the Calabi-Yau gives rise to an F-term for the modulus $T$:
\be 
D_T W_{\text{GVW}} = K_T W_{\text{GVW}} \neq 0\,,
\ee
where the Kähler-covariant derivative with respect to $T$ acting on $W$ is: $D_T W = \partial_T W + (\partial_T K ) W = K_T W$. It follows that $G_3$ has to contain a non-vanishing $(0,3)$ contribution which leads to spontaneous supersymmetry breaking due to the background itself. In this scenario a closed string fermion will be identified as the Goldstino. Adding an anti-$D3$-brane to such a background, $\lambda$ would no longer be the Goldstino and, in fact, will get a mass from the $(0,3)$ part of the 3-flux $G_3$.\\
Our interest here is the KKLT scenario where, in addition to a non-vanishing GVW superpotential \eqref{eq:GVWpot}, we have non-perturbative corrections also depending on the Kähler modulus $T$ of form\footnote{See also section \ref{sec:kklt}.}:
\be 
W_{\text{np}} = A \rme^{\rmi a T}\,.
\ee
The parameter $A$ can, in general, depend on the moduli but, under certain conditions, we can neglect this as those terms are suppressed. The non-perturbative corrections allow us to find a supersymmetric solution to:
\be 
D_T \left( W_{\text{GVW}} + W_{\text{np}} \right)=0\,,
\ee
which will give a supersymmetric AdS vacuum with $\partial_T W_{\text{np}} = - K_T (W_{\text{GVW}} + W_{\text{np}})$. If we add one or more anti-$D3$-branes to this setup we can lift the AdS minimum to dS. However, then the anti-brane is the source of supersymmetry breaking and needs to come with a massless Goldstino. On the other hand, we saw that the background has non-zero $(0,3)$ flux which means the fermion singlet $\lambda$ obtains a non-vanishing mass term. Furthermore, all $\chi^i$ also obtain masses and thus can't act as the Goldstino either \cite{Bergshoeff:2015jxa}. The solution to this peculiar problem is that the $(0,3)$ part of the $G_3$ flux localizes on the source of the non-perturbative contributions \cite{Baumann:2010sx,Dymarsky:2010mf}. Namely, the stack of $D7$-branes that is localized in the bulk of the internal manifold, as opposed to the location of the anti-$D3$-brane at the bottom of a warped throat. If the flux is pulled back onto the anti-brane the contribution vanishes due to the warping. Then, the fermion singlet $\lambda$ does not obtain a mass and can act as the Goldstino. Note however that, once we go beyond the probe limit, this will not be exactly true anymore. The backreaction of the uplifting anti-$D3$-brane onto the geometry will shift the Kähler modulus $T$ away from the position of the supersymmetric minimum and then $D_TW\neq0$. Thus, the Goldstino becomes a combination of the singlet $\lambda$ and the fermionic partner of $T$.\\
With this discussion out of the way we go on to investigate the couplings of the four fermions provided by the anti-$D3$-brane to the moduli $\tau$, $T$ and $U^A$ from the closed string sector. The fermionic action of the anti-$D3$-brane \cite{Bergshoeff:2015jxa,GarciadelMoral:2017vnz,Grana:2002tu,McGuirk:2012sb} to quadratic order in the world-sheet fermions is given to be \cite{Bergshoeff:2015jxa,Martucci:2005rb,Bergshoeff:2005yp}:
\bea 
S_{\text{fermionic}}^{\overline{D3}} = 2 \int d^4x & \sqrt{-g_4} \bigg[ \rme^{4\mathcal{A} - 8u} \bar{\theta}\Gamma^\mu \left( \nabla_\mu - \frac{1}{4} \rme^\phi F_\mu \tilde{\Gamma}_{0123}\right)\theta\\
&+ \frac{1}{8\cdot 4!} \bar{\theta} \bigg(\rme^{\frac{16}{3} \mathcal{A} - \frac{32}{3} u} \Gamma^{\mu mnpq}F_{\mu mnpq} - 2 \rme^{\frac{8}{3} \mathcal{A} - \frac{16}{3} u} \Gamma^{\mu\nu\rho mn} F_{\mu\nu\rho mn}\bigg) \tilde{\Gamma}_{0123}\theta\\
&- \frac{\rmi}{24} \rme^{6\mathcal{A} - 12u} \rme^{\frac{1}{2} \phi} \left( G_{mnp}^{\text{ISD}} - \bar{G}_{mnp}^{\text{ISD}}\right) \bar{\theta} \Gamma^{mnp} \theta \bigg]\,,
\label{eq:fermact}
\eea
where we are in Einstein frame, as per usual. Greek letters label the non-compact directions $0,1,2,3$ and Latin letters the internal directions $4,5,6,7,8,9$. $\theta$ is the Majorana-Weyl spinor of type IIB theory with 16 components and we will decompose it into $4d$ Weyl spinors \cite{Bergshoeff:2015jxa,Grana:2002tu}. The gamma matrix $\Gamma$ has unwarped but curved indices while $\tilde{\Gamma}$ has flat indices. In supergravity language \cite{Freedman:2012zz} the fermions are an $SU(3)$ singlet $P_L\lambda$ and three fermions $P_L\chi^i$, forming a triplet. In \cite{Bergshoeff:2015jxa} the kinetic and the $G$-flux part of \eqref{eq:fermact} were considered in a fixed background and we will use these results here. There it was found that all other terms vanish. Here, we consider a dynamical background where these terms will give important contributions \cite{Cribiori:2019hod}.\\
First off, we will consider the spin connection term where we adopt the convention to put a tilde on flat quantities and indices. It is given to be:
\be
\bar{\theta} \Gamma^\mu \nabla_\mu \theta = \bar{\theta} \Gamma^\mu \left(\partial_\mu + \frac{1}{4} \omega_\mu^{\,\tilde{a}\tilde{b}} \tilde{\Gamma}_{\tilde{a}\tilde{b}} + \frac{1}{4} \omega_\mu^{\,i\ib} \tilde{\Gamma}_{i\ib}\right) \theta\,,
\ee
where the indices run over $\tilde{a},\tilde{b} = 0,1,2,3$ and $i,\ib=1,2,3$. Terms with mixed indices in the spin connection, like $\omega_\mu^{\,\tilde{a}i}$, vanish and thus are omitted. In the above expression the first two terms form the usual covariant derivative while the last term introduces an interaction of the fermions and the complex structure moduli $U^A$. In order to write the spin connection in terms of the Vielbein and the moduli we note that our metric is block diagonal, consisting of the $4d$ block of the non-compact directions and a $6d$ block of the internal manifold. This immediately tells us that the Vielbein is also block diagonal. Then, the internal part of the Vielbein satisfies:
\be 
e^a_i g_{a\bar{b}} e^{\bar{b}}_{\ib} = \delta_{i\ib}\,.
\ee
Since the geometry of the compact Calabi-Yau (CY) manifold is described by the Kähler modulus $T$ and the complex structure moduli $U^A$ this means that the ``internal" Vielbein $e^a_i$ is a function of those moduli. Furthermore, the holomorphic 3-form $\Omega_{abc}=\epsilon_{ijk} e^i_a e^j_b e^k_c$ of the CY manifold does not depend on the $\bar{U}^A$ and thus we know that $e^i_a$ and its inverse can only be a function of $T$, $\bar{T}$ and $U^A$. Accordingly, we write:
\be 
\partial_\mu e^a_i = (\partial_T e^a_i)\partial_\mu T + (\partial_{\bar{T}} e^a_i) \partial_\mu \bar{T} + (\partial_{U^A} e^a_i) \partial_\mu U^A\,.
\ee
The Kähler modulus gives the overall volume of the CY manifold and since we only have one such modulus the volume has to depend on some power of $(T-\bar{T})$, to leading order:
\be 
\partial_T e^a_i = -\partial_{\bar{T}} e^a_i \propto e^a_i\,.
\ee
With this the internal part of the spin connection simplifies to:
\be 
\omega_\mu^{i\ib} = e^{\bar{a}i}\partial_\mu e^\ib_{\bar{a}} - e^{a\ib} \partial_\mu e^i_a = e^{\bar{a}i}(\partial_{\bar{U}^A} e^\ib_{\bar{a}} ) \partial_\mu \bar{U}^A - e^{a\ib} (\partial_{U^A} e^i_a)\partial_\mu U^A\,.
\label{eq:spincon}
\ee
This expression can be further simplified due to the fact that we only have a single Kähler modulus and thus only one $(1,1)$-form. Using the spin connection we can define a 2-form. If it is in cohomology it has to be proportional to the Kähler form $J$ of the Calabi-Yau manifold. In flat indices this means it is proportional to $\delta_{i\ib}\,$:
\be 
\omega_{\mu i\ib}\, e^i_a e^\ib_{\bar{b}} \propto \delta_{i\ib} e^i_a e^\ib_{\bar{b}}\,.
\ee
With this we can use $\omega_\mu^{i\ib} \tilde{\Gamma}_{i\ib} = \omega_\mu^{i\ib} \delta_{i\ib} \delta^{j\jb} \tilde{\Gamma}_{j\jb}/3$ to write the following fermionic contribution that has not been considered in earlier works:
\bea
\bar{\theta} \Gamma^\mu \omega_\mu^{i\ib} \tilde{\Gamma}_{i\ib} \theta  
&= \frac{1}{3}\omega_\mu^{i\ib} \delta_{i\ib} \delta^{j\jb} \tilde{\Gamma}_{j\jb} \\
&= \frac{1}{3} \omega_\mu^{k\bar{k}} \delta_{k\bar{k}} \left( 3 \bar{\lambda} P_R \gamma^\mu \lambda - \delta_{i\jb} \bar{\chi}^\jb P_R \gamma^\mu \chi^i\right)\\
&= \frac{1}{3} \delta_{i\ib} \left( e^{\bar{a}i} (\partial_{\bar{U}^A} e^\ib_{\bar{a}} ) \partial_\mu \bar{U}^A - e^{a\ib} (\partial_{U^A} e^i_a) \partial_\mu U^A\right)\left( 3 \bar{\lambda} P_R \gamma^\mu \lambda - \delta_{j\jb} \bar{\chi}^\jb P_R \gamma^\mu \chi^j\right)\,.
\eea
The next new contribution comes from the $F_\mu$ term in \eqref{eq:fermact}. Keeping in mind that $F_\mu = \partial_\mu C_0 = \partial_\mu \rmre(\tau)$ it is straight forward to compute:
\bea
\rme^\phi F_\mu \bar{\theta} \Gamma^\mu \tilde{\Gamma}_{0123} \theta &= \frac{(\partial_\mu \rmre(\tau)}{\rmim(\tau)} \bar{\theta}\Gamma^\mu \tilde{\Gamma}_{0123} \theta\\
&=-\rmi \frac{(\partial_\mu \rmre(\tau)}{\rmim(\tau)} \left( \bar{\lambda} P_R \gamma^\mu \lambda + \delta_{i\jb} \bar{\chi}^\jb P_R \gamma^\mu \chi^i\right)\,.
\eea
There are two more new contribution in \eqref{eq:fermact} that we need to investigate, both of them related to $C_4$. The first comes from the derivative coupling to the axion $C_4$. For this we need to know that a Calabi-Yau manifold with a single Kähler modulus has one $(2,2)$-form $Y_{2,2}$. The normalization of this form is such that it evaluates to $1$ when integrated over the 4-cycle $\Sigma_4$. The volume or Kähler modulus $T$ is made up of this 2-form and $J\wedge J$, with $J$ the characteristic 2-form of the Calabi-Yau. Since $T$ is described in $4d$ supergravity via a chiral multiplet it furthermore has to be holomorphic. Using $Y_{2,2}$ to build a basis we find that the Kähler modulus is given as:
\bea
T :&= \int_{\Sigma_4} \left( C_4 - \frac{\rmi}{2} J\wedge J\right)\\
&=\int_{\Sigma_4} c_4(x^\mu) Y_{2,2} + \rmi \rmim(T) \int_{\Sigma_4} Y_{2,2}\,,
\eea
where we also read of that $c_4(x^\mu)\, Y_{2,2} = -c_4(x^\mu)\, [2 \rmim(T)]^{-1} J\wedge J$. We can now use that $\rmi e_i^u e_\ib^{\bar{u}} J_{u\bar{u}} = \delta_{i\ib}$, with curved indices $u$ and $\bar{u}$ to find the first contribution from $C_4$ to the fermionic action of the anti-$D3$-brane:
\bea 
\frac{1}{4!} \rme^{\frac{4}{3}\mathcal{A} - \frac{8}{3} u} \bar{\theta} \Gamma^{\mu n p q r} F_{\mu n p q r} \theta 
&= \frac{\rmre(T)}{2\rmim(T)} \bar{\theta} \delta_{i\ib}\delta_{j\jb}\Gamma^\mu \tilde{\Gamma}^{i\ib j\jb}\tilde{\Gamma}_{0123} \theta\\
&= -\rmi \frac{\partial_\mu \rmre(T)}{\rmim(T)} \left( 3\bar{\lambda} P_R \gamma^\mu \lambda - \delta_{i\jb} \bar{\chi}^\jb P_R\gamma^\mu \chi^i \right)\,.
\eea
The remaining term coming from $C_4$ can be related to the above one due to the self-duality of $dC_4$ in $10d$. Using this property one finds that:
\be 
F_{\mu n p q r} \Gamma^{\mu n p q r} = -2 \rme^{\frac{8}{3} \mathcal{A} + \frac{16}{3} u} F_{\mu\nu \rho mn} \Gamma^{\mu \nu \rho mn} \tilde{\Gamma}_\star\,.
\ee
Finally, we can combine all parts from above in order to find the complete fermionic part of the anti-$D3$-brane action up to total derivatives:
\bea 
S_{\text{fermionic}}^{\overline{D3}} = 2 \int d^4x \sqrt{-g_4} \rme^{4\mathcal{A}-8u} 
\bigg[& \bar{\lambda} P_R \gamma^\mu \nabla_\mu \lambda + \delta_{i\jb} \bar{\chi}^\jb P_R \gamma^\mu \nabla_\mu \chi^i\\
+& \frac{\rmi}{4} \frac{\left(\partial_\mu \rmre(\tau)\right)}{\rmim(\tau)} \left( \bar{\lambda} P_R \gamma^\mu \lambda + \delta_{i\jb} \bar{\chi}^\jb P_R \gamma^\mu \chi^i\right)\\
-& \frac{\rmi}{4} \frac{\left(\partial_\mu \rmre(T)\right)}{\rmim(T)} \left(3 \bar{\lambda} P_R \gamma^\mu \lambda - \delta_{i\jb} \bar{\chi}^\jb P_R \gamma^\mu \chi^i\right)\\
+& \frac{1}{12} \omega_\mu^{k\bar{k}} \delta_{k\bar{k}} \left(3 \bar{\lambda} P_R \gamma^\mu \lambda - \delta_{i\jb} \bar{\chi}^\jb P_R \gamma^\mu \chi^i\right)\\
+&  \frac{1}{2} m \bar{\lambda} P_L \lambda + m_i \bar{\lambda} P_L \chi^i + \frac{1}{2} m_{ij} \bar{\chi^i}P_L \chi^j + \text{c.c.}\bigg]\,,
\label{eq:simpfermact}
\eea
where we wrote everything in terms of $4d$ spinors.
\\As discussed, the masses depend on the the imaginary self dual 3-flux $G_3^{\text{ISD}}$. They are given as:
\bea 
m =&\, \rmi \,\frac{\sqrt{2}}{12}  \rme^{2\mathcal{A}-4 u}\, \rme^{\frac{\phi}{2}} \bar{\Omega}^{abc}\bar{G}_{abc}^{\text{ISD}}\,,\\
m_i =& - \frac{\sqrt{2}}{4} \rme^{2\mathcal{A}-4 u}\, \rme^{\frac{\phi}{2}} e_i^a \bar{G}_{ab\bar{c}}^{\text{ISD}} J^{b\bar{c}}\,,\\
m_{ij} =& \,\rmi \,\frac{\sqrt{2}}{8} \rme^{2\mathcal{A}-4 u}\, \rme^{\frac{\phi}{2}} \left(e_i^c e_j^d + e_j^c e_i^d \right) \Omega_{abc}g^{a\bar{a}}g^{b\bar{b}} \bar{G}_{d\bar{a}\bar{b}}\,.
\label{eq:fermmass}
\eea
Since we consider the KKLT setup with gaugino condensation on a stack of $D7$-branes in the bulk of the internal manifold, the pullback of the $G_3$-flux onto the anti-brane does vanish. Furthermore, in this background the $(2,1)$ part of $G_3^{\text{ISD}}$ is primitive and thus $G_{ab\bar{c}}^{\text{ISD}} \, J^{b\bar{c}} = 0$ and thus we find that, for our particular setup, two of the mass terms from above vanish:
\be
m = \, 0\qquad \text{and}\qquad
m_i = \, 0\,.
\ee
As already discussed, this means that $\lambda$ remains massless and is the Goldstino related to supersymmetry breaking. The fermion triplet $\chi^i$ meanwhile does acquire non-zero masses. Hence, we have a unique Goldstino arising from the source of supersymmetry breaking, namely the fermion singlet $\lambda$.

\paragraph{The complete anti-$\mathbf{D3}$-action for the KKLT model} from the perspective of string theory is the sum of the bosonic part \eqref{eq:simpbosact} and the fermionic contribution \eqref{eq:simpfermact}, with the masses $m$ and $m_i$ set to zero:
\bea 
S^{\overline{D3}} 
=& S^{\overline{D3}}_{\text{bosonic}} + S^{\overline{D3}}_{\text{fermionic}}\\
=& - \int d^4x \sqrt{-g_4} \bigg[ 2 \rme^{4\mathcal{A}-8u} + \frac{1}{2} \rme^{\frac{4}{3} \mathcal{A}-\frac{8}{3}u}\partial_\mu H^a \partial^\mu \bar{H}^{\bar{b}}\\
&\qquad\qquad\qquad\;\;\; + \frac{\rmim(\tau)}{4} F_{\mu\nu}F^{\mu\nu} - \frac{\rmre(\tau)}{8} \frac{\epsilon^{\mu\nu\rho\sigma}}{\sqrt{-g_4}} F_{\mu\nu}F_{\rho\sigma}\bigg]\\
& + 2 \int d^4x \sqrt{-g_4} \rme^{4\mathcal{A}-8u} \bigg[ \bar{\lambda} P_R \gamma^\mu \nabla_\mu \lambda + \delta_{i\jb} \bar{\chi}^\jb P_R \gamma^\mu \nabla_\mu \chi^i\\
&\qquad\qquad\qquad\qquad\qquad + \frac{\rmi}{4} \frac{\left(\partial_\mu \rmre(\tau)\right)}{\rmim(\tau)} \left( \bar{\lambda} P_R \gamma^\mu \lambda + \delta_{i\jb} \bar{\chi}^\jb P_R \gamma^\mu \chi^i\right)\\
&\qquad\qquad\qquad\qquad\qquad - \frac{\rmi}{4} \frac{\left(\partial_\mu \rmre(T)\right)}{\rmim(T)} \left(3 \bar{\lambda} P_R \gamma^\mu \lambda - \delta_{i\jb} \bar{\chi}^\jb P_R \gamma^\mu \chi^i\right)\\
&\qquad\qquad\qquad\qquad\qquad + \frac{1}{12} \omega_\mu^{k\bar{k}} \delta_{k\bar{k}} \left(3 \bar{\lambda} P_R \gamma^\mu \lambda - \delta_{i\jb} \bar{\chi}^\jb P_R \gamma^\mu \chi^i\right)\\
&\qquad\qquad\qquad\qquad\qquad + \frac{1}{2} m_{ij} \bar{\chi}^i P_L \chi^j + \frac{1}{2} \bar{m}_{\ib\jb} \bar{\chi}^\ib P_R \chi^\jb\bigg]\,.
\label{eq:simpbraneact}
\eea

\subsection[The $D3$-Brane Action and Supergravity]{The $\mathbf{D3}$-Brane Action and Supergravity}
\label{sec:susyD3}
Even though our goal is to describe the anti-$D3$-brane action in the KKLT background it will be useful to first have a look at the supersymmetric action of the normal $D3$-brane. The results of this section will proof useful  for the investigation of the anti-$D3$-brane case. In particular, we will match the action with the standard $\mathcal{N}=1$ supergravity action for a single vector multiplet and three chiral multiplets because for the $D3$-brane equation \eqref{eq:simpbosact} reduces exactly to that. First, the ``uplift" term of the bosonic action vanishes for the supersymmetric $D3$-brane as do the fermionic mass terms. Taking into account the sign flip in the RR-fields only one vector multiplet, containing $\lambda$ and $A_\mu$, and three chiral multiplets with $H^a$ and $\chi^a := e^a_i \chi^i$ remain. The $D3$-brane action is then given to be:
\bea 
S^{D3}
=& S^{D3}_{\text{bosonic}} + S^{D3}_{\text{fermionic}}\\
=& - \int d^4x \sqrt{-g_4} \bigg[ \frac{1}{2} \rme^{\frac{4}{3} \mathcal{A}-\frac{8}{3}u}\partial_\mu H^a \partial^\mu \bar{H}^{\bar{b}}\\
&\qquad\qquad\qquad\;\;\; + \frac{\rmim(\tau)}{4} F_{\mu\nu}F^{\mu\nu} - \frac{\rmre(\tau)}{8} \frac{\epsilon^{\mu\nu\rho\sigma}}{\sqrt{-g_4}} F_{\mu\nu}F_{\rho\sigma}\bigg]\\
& + 2 \int d^4x \sqrt{-g_4} \rme^{4\mathcal{A}-8u} \bigg[ \bar{\lambda} P_R \gamma^\mu \nabla_\mu \lambda + \delta_{i\jb} \bar{\chi}^\jb P_R \gamma^\mu \nabla_\mu \chi^i\\
&\qquad\qquad\qquad\qquad\qquad - \frac{\rmi}{4} \frac{\left(\partial_\mu \rmre(\tau)\right)}{\rmim(\tau)} \left( \bar{\lambda} P_R \gamma^\mu \lambda + \delta_{i\jb} \bar{\chi}^\jb P_R \gamma^\mu \chi^i\right)\\
&\qquad\qquad\qquad\qquad\qquad + \frac{\rmi}{4} \frac{\left(\partial_\mu \rmre(T)\right)}{\rmim(T)} \left(3 \bar{\lambda} P_R \gamma^\mu \lambda - \delta_{i\jb} \bar{\chi}^\jb P_R \gamma^\mu \chi^i\right)\\
&\qquad\qquad\qquad\qquad\qquad + \frac{1}{12} \omega_\mu^{k\bar{k}} \delta_{k\bar{k}} \left(3 \bar{\lambda} P_R \gamma^\mu \lambda - \delta_{i\jb} \bar{\chi}^\jb P_R \gamma^\mu \chi^i\right)\bigg]\,.
\label{eq:d3act}
\eea
\paragraph{First we will investigate the derivative coupling of the fermions to $\mathbf{\tau}$.} The chiral multiplets containing the fermions do not carry any $U(1)$ charge but couple to all scalars due to the Kähler covariant derivative. The relevant part of the supergravity action is \cite{Freedman:2012zz}:
\bea
\mathcal{L}^{\text{SUGRA}} \supset &-\delta_{i\jb}\bar{\chi}^\jb P_R \gamma^\mu \left(\partial_\mu - \frac{1}{4}[\partial_\mu \tau\partial_\tau K - \partial_\mu \bar{\tau}\partial_{\bar{\tau}} K] \right) \chi^i\\
&-\frac{1}{2} \delta_{i\jb} \bar{\chi}^\jb P_R \gamma^\mu \Gamma^i_{k\tau}\partial_\mu \tau \chi^k - \frac{1}{2} \delta_{i\jb} \bar{\chi}^i P_L \gamma^\mu \Gamma^\jb_{\bar{k}\bar{\tau}} \partial_\mu \bar{\tau} \chi^{\bar{k}}\,.
\eea 
If we choose the Kähler potential for $\tau$ to be:
\be 
K^{(\tau)} = - \log [-\rmi (\tau -\bar{\tau})]\,,
\ee
we obtain exactly the coupling proportional to $\partial_\mu \rmre(\tau)$ in \eqref{eq:d3act} from the square brackets above. Furthermore, the prefactor also is correct which tells us that the mixed index Christoffel symbols $\Gamma^i_{j\tau}$ have to be zero.
\paragraph{For the couplings to $\mathbf{\partial_\mu \rmre(T)}$} we have to investigate the very similar looking part of the Lagrangian:
\bea
\mathcal{L}^{\text{SUGRA}} \supset & - \delta_{i\jb} \bar{\chi}^\jb P_R \gamma^\mu \left( \partial_\mu -\frac{1}{4} [\partial_\mu T \partial_T K - \partial_\mu \bar{T} \partial_{\bar{T}} K]\right) \chi^i\\
&- \frac{1}{2} \delta_{i\jb} \bar{\chi}^\jb P_R \gamma^\mu \Gamma^i_{kT} \partial_\mu T \chi^k - \frac{1}{2} \delta_{i\jb} \bar{\chi}^i P_L \gamma^\mu \Gamma^{\jb}_{\bar{k}\bar{T}} \partial_\mu \bar{T} \chi^{\bar{k}}\,,
\eea
now with non vanishing contributions form the Christoffel symbols in the direction of $T$. In the limit of large volume we can find the corresponding Christoffel symbols from the Kähler potential \eqref{eq:barD3Kpot} approximately to be:
\be 
\Gamma^{i}_{jT} \approx \frac{\rmi}{2 \rmim(T)}\,.
\ee
Together with the derivatives acting on the Kähler potential all contributions combine correctly and match \eqref{eq:d3act}, again including coefficients.
\paragraph{In a similarly straightforward manner we can find the coupling of $\mathbf{\tau}$ to $\mathbf{\lambda}$} from the supergravity action. For the $D3$-brane the gauge kinetic function is $f(\tau) = -\rmi \tau$ and the kinetic term of the gaugino is normalized such that its prefactor is $\rmre(f) = \rmim(\tau) = \rme^{-\phi}$. Thus, we have to re-scale the fermion singlet like $\lambda = \rme^{-\phi/2} \lambda^\prime$, which does not lead to any new derivative terms since $\bar{\lambda} \gamma^\mu \lambda =0$ for $4d$ Majorana spinors. The supergravity action for the re-scaled singlet $\lambda^\prime$ is:
\bea 
\mathcal{L}^{\text{SUGRA}} \supset &-\frac{1}{2} \rmre(f) \bar{\lambda}^\prime \gamma^\mu \left( \partial_\mu + \frac{1}{4} [\partial_\mu \tau \partial_\tau K - \partial_\mu \bar{\tau} \partial_{\bar{\tau}} K ] \gamma_\star \right) \lambda^\prime\\
&+ \frac{\rmi}{4} \partial_\mu \rmim(f)\bar{\lambda}^\prime \gamma_\star \gamma^\mu \lambda^\prime\,,
\eea
which reproduces the coupling of $\lambda$ to $\partial_\mu \rmre(\tau)$ in the $D3$-brane action.
\paragraph{The terms coupling the gaugino to the modulus $\mathbf{T}$} do not involve any Christoffel symbols. Again, using $\lambda^\prime$ we can write the standard supergravity action and the terms, when compared to equation \eqref{eq:d3act}, match up to a different prefactor that is caused by the lack of the Christoffel symbols:
\be
\mathcal{L}^{\text{SUGRA}} \supset - \frac{1}{2} \rmre(f) \bar{\lambda}^\prime \gamma^\mu \left( \partial_\mu + \frac{1}{4} [\partial_\mu T \partial_T K - \partial_\mu \bar{T} \partial_{\bar{T}} K]\gamma_\star\right) \lambda^\prime\,.
\ee
\paragraph{The final terms to analyze contain the complex structure moduli $\mathbf{U^A}$.} The spin connection \eqref{eq:spincon} has two independent parts, that are proportional to $\partial_\mu U^A$ and $\partial_\mu \bar{U}^A$, individually. By an analogous reasoning as below equation \eqref{eq:spincon} we conclude that both terms are proportional to $\delta_{i\ib}$. Hence,  we can write:
\bea
\partial_{U^A} e^i_v 
&= - e^i_{w} (\partial_{U^A} e^{w\ib} ) e_{v\ib}\\
&= -\frac{1}{3} e^j_w (\partial_{U^A} e^{w\jb} ) \delta_{j\jb} \delta^{i\ib} e_{v\ib}\\
&= -\frac{1}{3} e^j_w (\partial_{U^A} e^{w\jb} ) \delta_{j\jb}  e^i_{v}\,,
\eea
where we used that $\partial_\mu e^i_v e^{v\ib} = 0$ and $v,w,\ldots = 1,2,3$ are curved and warped indices. We can related this to the $(3,0)$-form $\Omega$ of the Calabi-Yau manifold in the following way:
\bea
\partial_{U^A} \Omega &= \frac{1}{3! \cdot 3!} \partial_{U^A} \left( e^i_v e^j_w e^k_x \epsilon_{ijk} dz^v \wedge dz^w \wedge dz^x\right)\\
&= \frac{3}{3! \cdot 3!} \left(\partial_{U^A}  e^i_v\right) e^j_w e^k_x \epsilon_{ijk} dz^v \wedge dz^w \wedge dz^x\\
&= - \frac{1}{3! \cdot 3!} e^l_t \left(\partial_{U^A} e^{t\jb}\right) \delta_{l\jb} e^i_v e^j_w e^k_x \epsilon_{ijk} dz^v \wedge dz^w \wedge dz^x\\
&= -e^j_w \left(\partial_{U^A} e^{w\jb} \right) \delta_{j\jb} \Omega\,.
\label{eq:partomgexp}
\eea
In order to eventually match this with the component action of the $D3$-brane we will expand the 3-form $\Omega$ in a cohomology basis as:
\be 
\Omega = Z^K \alpha_K - F_K \beta^K\,,
\ee

with the 3-form basis given by $\alpha_K$ and $\beta^K$ that satisfy $\int \alpha_K \wedge \beta^L = \delta^L_K$. The $Z^K$ and $F_K$ are functions of the complex structure moduli $U^A$. The expansion of $\Omega$ together with \eqref{eq:partomgexp} tells us that:

\bea 
\partial_{U^A} Z^K &= - e^j_v \left(\partial_{U^A} e^{v\jb} \right) \delta_{j\jb} Z^K\,,\\
\partial_{U^A} F_K &= -e^j_v \left(\partial_{U^A} e^{v\jb}\right) \delta_{j\jb} F_K\,.
\eea

The Kähler potential for the complex structure moduli is given in terms of the 3-form $\Omega$ and can be written using the expansion above:

\bea 
K^{(U)} &= - \log \left[ -\rmi \int \Omega \wedge \bar{\Omega} \right] \\
&= - \log \left[ \rmi (Z^K \bar{F}_K - \bar{Z}^K F_K ) \right] \,.
\eea
Finally, we can use this to write the relevant part of the spin connection using the Kähler potential $K^{(U)}$ as:
\bea 
\omega_\mu^{i\ib} \delta_{i\ib} &= \delta_{i\ib} \left( e^{\bar{a}i} (\partial_{\bar{U}^A} e^{\ib}_{\bar{a}}) \partial_\mu \bar{U}^A - e^{a\ib} (\partial_{U^A} e^i_a )\partial_\mu U^A\right)\\
&= \partial_{U^A} K^{(U)} \partial_\mu U^A - \partial_{\bar{U}^A} K^{(U)} \partial_\mu \bar{U}^A\,.
\eea
Before we can use this result to match the component action \eqref{eq:d3act} with a standard supergravity expression we need to propose a fitting Kähler potential. 
\\\newpage One might naively propose:
\be
K = - \log\left[-\rmi (\tau-\bar{\tau})\right] - \log\left[-\rmi \int \Omega \wedge \bar{\Omega}\right] - 3 \log \left[-\rmi(T-\bar{T})  + k(H,\bar{H})\right]\,,
\ee
this, however, is incapable of reproducing the correct $D3$-brane action. In order to find the correct expression, we note that the coupling of the fermions $\lambda$ and $\chi^i$ to $\partial_\mu \rmre(U^A)$ has the same prefactor as the coupling to $\partial_\mu \rmre(T)$. Hence, the complex structure part has to couple in the same way to the chiral multiplets as the Kähler sector in the supergravity action. The Kähler potential:
\bea 
K =& - \log\left[-\rmi (\tau-\bar{\tau})\right] - 3 \log \left[-\rmi(T-\bar{T}) \left(-\rmi\int \Omega \wedge \bar{\Omega}\right)^{\frac{1}{3}} + k(H,\bar{H})\right]\\
=&  - \log\left[-\rmi (\tau-\bar{\tau})\right] - \log\left[-\rmi \int \Omega \wedge \bar{\Omega}\right] - 3 \log \left[-\rmi(T-\bar{T})  + \frac{k(H,\bar{H})}{\left(-\rmi\int \Omega \wedge \bar{\Omega}\right)^{\frac{1}{3}}}\right]\,,
\label{eq:deKpot}
\eea
does reproduce all the required couplings to $\partial_{\mu} \rmim(\tau)$, $\partial_\mu \rmim(T)$, $\chi^i$ and $\lambda^\prime$, via standard supergravity terms. Dropping, as above, the terms containing $k(H,\bar{H})$ the relevant supergravity pieces are:
\bea 
\mathcal{L}^{\text{SUGRA}} \supset &-\delta_{i\jb} \bar{\chi}^\jb P_R \gamma^\mu \left( \partial_\mu - \frac{1}{4} \left[\partial_\mu U^A \partial_{U^A} K - \partial_\mu \bar{U}^A \partial_{\bar{U}^A} K \right]\right)\chi^i\\
&- \frac{1}{2} \delta_{i\jb} \bar{\chi}^\jb P_R \gamma^\mu \Gamma^i_{kU^A} \partial_\mu U^A \chi^k-\frac{1}{2} \delta_{i\jb} \bar{\chi}^i P_L \gamma^\mu \Gamma^\jb_{\bar{k}\bar{U}^A} \partial_\mu \bar{U}^A \chi^{\bar{k}}\\
&-\frac{1}{2} \rmre(f)\bar{\lambda}^\prime \gamma^\mu \left( \partial_\mu + \frac{1}{4} \left[\partial_\mu U^A \partial_{U^A} K - \partial_\mu \bar{U}^A \partial_{\bar{U}^A} K\right]\gamma_\star\right)\lambda^\prime\,.
\eea
As a final note we remark that, in order to produce the correct kinetic terms involving the world-volume scalars $H^a$, we require:
\be
\partial_{H^a}\partial_{\bar{H}^{\bar{b}}} k(H,\bar{H}) \approx \frac{1}{6} \rme^{\frac{4}{3} (\mathcal{A} + u )} \left(-\rmi \int \Omega \wedge \bar{\Omega} \right)^{\frac{1}{3}} g_{a\bar{b}}\,.
\ee

\subsection[The Anti-$D3$-Brane Action in Supergravity]{The Anti-$\mathbf{D3}$-Brane Action in Supergravity}
\label{sec:sugraact}
We now move on to tackle the description of the supersymmetry breaking anti-$D3$-brane in the KKLT setup. This work is an extension of \cite{GarciadelMoral:2017vnz} to which we add the bosonic part of the action, the mixed terms of fermions and bosons as well as the $U(1)$ gauge sector with the vector multiplet. We will package all fields into the constrained multiplets of section \ref{sec:multiplets} which allows us to use the language of linear supergravity for a system with broken and hence non-linearly realized, supersymmetry. The fact that it is possible to do this, using constrained multiplets to write a system with broken supersymmetry in a seemingly linear way, aligns with the fact that supersymmetry is broken spontaneously.
\paragraph{We start by considering the Goldstino $\mathbf{\lambda}$ and the fermion triplet $\mathbf{\chi^i}$} which we will put into a nilpotent chiral field $X$ and into constrained chiral fields $Y^i$, respectively. The correct Kähler potential coupling the bulk moduli to the world-volume fermions is \cite{GarciadelMoral:2017vnz}:
\bea 
K = &- \log\left[\left(-\rmi(\tau-\bar{\tau})\right)\right] - 3 \log\left[\left(-\rmi(T-\bar{T})\right)f\left(U^A,\bar{U}^A\right)^{\frac{1}{3}}\right]\\
&- 3\log \Bigg[ 1- \frac{\rme^{-4\mathcal{A}} X\bar{X}}{3\left(-\rmi (\tau-\bar{\tau})\right)\left(-\rmi(T-\bar{T})\right)f\left(U^A,\bar{U}^A\right)}\\
&\qquad \qquad \;\,  - \frac{\rme^{-4\mathcal{A}}\delta_{i\jb}Y^i \bar{Y}^\jb}{3\left(-\rmi (\tau-\bar{\tau})\right)\left(-\rmi(T-\bar{T})\right)^2 f\left(U^A,\bar{U}^A\right)^{\frac{1}{3}}}\Bigg]\,,
\eea
where we used $f\left(U^A,\bar{U}^A\right) = -\rmi \int \Omega \wedge \Omega$. From here on out, we will only use $f\left(U^A,\bar{U}^A\right)$ as notation for the gauge kinetic function, in order to avoid confusion between the $(3,0)$ form $\Omega$ of the internal manifold and the fermions in the multiplets $X$ and $Y^i$. We now need to obtain the correct couplings of the multiplets $X$ and $Y^i$ to the bulk moduli:
\begin{itemize}
\item The coupling of $\tau$ is fixed by the modular invariance of the world-volume action. For the details on this see appendix \ref{app:modinv}.
\item For the Goldstino in $X$ we have to match the coupling using the scalar potential contribution in \eqref{eq:simpbraneact}. The kinetic term of $\lambda$ will not be matched, as the couplings of a Goldstino are not physical and hence can be set to zero. It is sufficient to write:
\be 
P_L \Omega = P_L \lambda + \ldots\,,
\ee
where the dots represent higher-order terms.
\item The couplings of the fermion triplet in $Y^i$ can be fixed by comparing with the kinetic terms of the massive spin-$1/2$ fields, as we did for the supersymmetric $D3$-brane in section \ref{sec:susyD3}. One finds that, suppressing higher-order terms, the $\Omega^i$ are related to the $\chi^i$ via the field redefinition:
\be 
P_L \Omega^i = 2 \rmi \rme^{4\mathcal{A}-\frac{\phi}{2}} f\left(U^A ,\bar{U}^A\right)^{\frac{1}{6}} P_L \chi^i + \ldots\,.
\ee
\end{itemize}
The superpotential of our model consists of the Gukov-Vafa-Witten flux-potential, non-perturbative contributions and an uplifting contribution that breaks supersymmetry:
\be 
W = W_{\text{GVW}} + W_{\text{np}} + M^2 X\,.
\label{eq:D3barsup}
\ee
$M$ is related to the scale at which SUSY breaks, the warped-down string scale. By re-scaling $X$ we set $M^2 = \sqrt{2}$ and the tension of the anti-$D3$-brane to $T_{\overline{D3}} = 2 \pi = M^4 \pi$. It is evident that the superpotential breaks SUSY via the inclusion of the nilpotent chiral multiplet $X$ which, in turn, supports our choice to identify the fermion in $X$ as the Goldstino.\\
As of now the fermions are massless in the supergravity description. A typical approach to give them masses would be to add a term $W_{\text{mass}} = h_{ij}Y^i Y^j$ to the superpotential \cite{Vercnocke:2016fbt,GarciadelMoral:2017vnz}. Unfortunately, with a dynamical axio-dilaton, this requires $h_{ij}\propto \bar{G}_3$ to be anti-holomorphic in $\tau$, which is incompatible with supersymmetry. Here, we instead choose to modify our Kähler potential in order to acquire the desired terms:
\bea 
K = &- \log\left[\left(-\rmi(\tau-\bar{\tau})\right)\right] - 3 \log\left[\left(-\rmi(T-\bar{T})\right)f\left(U^A,\bar{U}^A\right)^{\frac{1}{3}}\right]\\
&- 3\log \Bigg[ 1- \frac{\rme^{-4\mathcal{A}} X\bar{X}}{3\left(-\rmi (\tau-\bar{\tau})\right)\left(-\rmi(T-\bar{T})\right)f\left(U^A,\bar{U}^A\right)}\\
&\qquad \qquad \;\,  - \frac{\rme^{-4\mathcal{A}}\delta_{i\jb}Y^i \bar{Y}^\jb}{3\left(-\rmi (\tau-\bar{\tau})\right)\left(-\rmi(T-\bar{T})\right)^2 f\left(U^A,\bar{U}^A\right)^{\frac{1}{3}}}\\
&\qquad \qquad \;\,  +\frac{\rme^{-8\mathcal{A}}\left( m_{ij} \bar{X} Y^i Y^j + \bar{m}_{\ib\jb}X \bar{Y}^\ib\bar{Y}^\jb\right)}{6 M^2 \left(-\rmi (\tau-\bar{\tau})\right)^\frac{3}{2} \left(-\rmi(T-\bar{T})\right)^{\frac{3}{2}} f\left(U^A,\bar{U}^A\right)^{\frac{5}{6}}}    \Bigg]\,,
\label{eq:Kferm}
\eea
where $m_{ij}$ is the fermion mass matrix that we discussed above in equation \eqref{eq:fermmass}. This method of including the masses for the fermion triplet is consistent and compatible with modular invariance. Nevertheless, we will give an alternative description of these terms as part of the superpotential later on in this section. This relies on tools we will review when coupling the $U(1)$ sector, namely, the new $D$-term, described in section \ref{sec:newD}.

\paragraph{The world-volume scalars in the $\mathbf{H^a}$} give small fluctuations of the anti-brane around its rest position and come from the DBI part of the action. Hence, their kinetic terms are the same in the case of the anti-$D3$-brane as they are for the $D3$-brane that we discussed in section \ref{sec:susyD3}. As discussed there, the Kähler potential \eqref{eq:deKpot} accurately describes the $H^a$ for dynamical complex structure moduli and axio-dilaton. By putting the world-volume scalars into the constrained multiplets $H^a$ of section \ref{sec:constmult} we are able to modify the Kähler potential we have found thus far \eqref{eq:Kferm} in order to include the behavior of the scalars. For this we consider the warp factor $\mathcal{A}$ do depend on the $H^a$ and we shift the volume modulus $T$:
\be 
-\rmi (T-\bar{T}) \,\to \, -\rmi(T-\bar{{T}}) + \frac{k(H^a,\bar{H}^a)}{f(U^A,\bar{U}^A)^{\frac{1}{3}}}=\rme^{4u}\,.
\ee
Putting this into the \eqref{eq:Kferm} we find the Kähler potential that now also correctly reproduces the kinetic behavior of the world-volume scalars:
\bea
K = &-\left[\left(-\rmi(\tau-\bar{\tau})\right)\right]-3 \log \left[\left(-\rmi(T-\bar{T})\right)\,f\left(U^A,\bar{U}^A\right)^{\frac{1}{3}}+k\left(H^a,\bar{H}^a\right)\right]\\
&- 3 \log\Bigg[1-\frac{\rme^{-4\mathcal{A}(H^a,\bar{H}^a)-4u}}{3\left(-\rmi(\tau-\bar{t})\right)f\left(U^A,\bar{U}^A\right)}X\bar{X}\\
&\qquad\qquad\;\,-\frac{\rme^{-4\mathcal{A}(H^a,\bar{H}^a)-8u}}{3\left(-\rmi(\tau-\bar{t})\right)f\left(U^A,\bar{U}^A\right)^{\frac{1}{3}}}\delta_{i\jb}Y^i\bar{Y}^\jb\\
&\qquad\qquad\;\,+\frac{\rme^{-8\mathcal{A}(H^a,\bar{H}^a)-6u}\left(m_{ij}\bar{X}Y^iY^j+\bar{m}_{\ib\jb}X \bar{Y}^\ib\bar{Y}^\jb\right)}{6 M^2 \left(-\rmi(\tau-\bar{t})\right)^{\frac{3}{2}}f\left(U^A,\bar{U}^A\right)^{\frac{5}{6}}}\Bigg]\,.
\label{eq:KpotbarD3}
\eea
The shift in $-\rmi (T-\bar{T})$ also leads to a dependence of the non-perturbative superpotential, $W_{\text{np}}$ on the scalars $H^a$. However, these corrections are highly suppressed compared to our leading order contributions as the non-perturbative corrections originate from gaugino condensation on a stack of $D7$-branes in the bulk, while we are considering an anti-$D3$-brane at the bottom of a warped throat. Due to this fact we choose to neglect the contributions to the superpotential \eqref{eq:D3barsup}.

\paragraph{The vector field} in the $U(1)$ gauge sector of the anti-$D3$-brane action requires a kinetic term, coming from the DBI-part of the action, and contributions from the CS-part of the action. While the DBI part matches the considerations for the supersymmetric $D3$-brane exactly (see section \ref{sec:susyD3}) we have an issue with the CS action. There, a sign flip is present for the anti-$D3$-brane which is problematic since it would require an anti-holomorphic gauge kinetic function $f(\bar{\tau}) = \rmi \bar{\tau}$, which is forbidden, as discussed in section \ref{sec:D3barKKLT}. We will be able to correctly incorporate the CS action here because of the non-linear realization of supersymmetry and the utilization of constrained superfields and the new $D$-term of section \ref{sec:newD}.\\
For this we package the $U(1)$ gauge vector into a chiral field strength multiplet $P_L\Lambda$ with the constraint $X P_L\Lambda = 0$, which removes the gaugino (see section \ref{sec:constmult}). 
\\\newpage The standard supergravity action for a vector multiplet, up to fermionic terms, is \cite{Freedman:2012zz}:
\bea
-\frac{1}{4} \left[f(\tau)\bar{\Lambda}P_L\Lambda\right]_F = \int d^4x \sqrt{-g_4}\bigg[&-\frac{\rmre(f)}{4} F_{\mu\nu}F^{\mu\nu}\\
&+\frac{\rmim(f)}{8}\frac{\epsilon^{\mu\nu\rho\sigma}}{\sqrt{-g_4}}F_{\mu\nu}F_{\rho\sigma}+\frac{\rmre(f)}{2}D^2 + \dots \bigg]\,.
\eea
Here, $f(\tau) = -\rmi \tau$ and the term proportional to $\rmim(f)=-\rmre(\tau)$ has the opposite sign as compared to the component action of the anti-$D3$-brane in the KKLT setup \eqref{eq:simpbraneact}. Because of this we need to effectively flip the sign of this term by subtracting $2 \cdot \rmim(f) \epsilon^{\nu\nu\rho\sigma}F_{\mu\nu}F_{\rho\sigma}/(8 \sqrt{-g_4})$. In order to achieve this in a supersymmetric way we will utilize the new $D$-term as first introduced in \cite{Cribiori:2017laj} and discussed already in section \ref{sec:newD}. Here, we want to use it to include a typical $U(1)$ $\theta$-term:
\be 
S_{\theta} = -\frac{1}{4} \int d^4x \rmim(f)\epsilon^{\mu\nu\rho\sigma}F_{\mu\nu}F_{\rho\sigma} + \ldots \,,
\label{eq:Stheta}
\ee
into an existing supersymmetric action. For once, the $\ldots$ do not stand for fermionic contributions but generally for additional terms that are needed for supersymmetry. In fact, our goal is to find these terms. Usually, one would use the Noether method to determine them. For this, we would vary the action and add terms in order to cancel the total variation and then start over again until the procedure does not produce extra terms. The problem with this method is that it is not clear when it will stop. For non-linear supersymmetry the result can instead be obtained immediately by considering the multiplets $R_1$ and $R_2$ of section \ref{sec:newD}.\\
For our case here we take:
\bea 
R_1 &=\frac{\omega^2\bar{\omega}^2}{\Sigma(\bar{\omega}^2)\bar{\Sigma}(\omega^2)}\,,\\
R_2 &= \rmim(f) \epsilon^{\mu\nu\rho\sigma}F_{\mu\nu}F_{\rho\sigma}\,,
\eea
and view $R_2$ as a real multiplet with weights $(4,0)$. The gauge kinetic function $f(\tau)$ is then the lowest component of a chiral multiplet with weight $(0,0)$. Unfortunately, there is an issue that needs to be addressed. Supersymmetry in our setup is not broken by the auxiliary field of the vector multiplet but rather by the nilpotent chiral multiplet $X$. This means that the auxiliary field of the vector does not acquire a vacuum expectation value and the quantity $\bar{\Sigma}(\omega^2)$ vanishes. It can, however, be proven that we can formally replace $\omega^2$ via the identification \cite{Cribiori:2017laj}:
\be 
\frac{\omega^2 \bar{\omega}^2}{\Sigma(\bar{\omega}^2)\bar{\Sigma}(\omega^2)} = \frac{X^0 \bar{X}^0 X \bar{X}}{\Sigma (\bar{X}^0\bar{X})\bar{\Sigma}(X^0X)} = \frac{X^0 \bar{X}^0 \rme^{-\frac{K}{3}}X \bar{X}}{\Sigma (\bar{X}^0\rme^{-\frac{K}{6}}\bar{X})\bar{\Sigma}(X^0\rme^{-\frac{K}{6}}X)}\,,
\ee
which holds for as long the auxiliary field $F$ in the nilpotent chiral field $X$ is non-vanishing, which is always the case in our setup. Then, in the above expression, the right hand side never vanishes and we are able to write down the supersymmetric completion of the $\theta$-action \eqref{eq:Stheta} as:
\be
S_{\theta} = -\frac{1}{4} \left[\frac{X^0 \bar{X}^0 \rme^{-\frac{K}{3}}X \bar{X}}{\Sigma (\bar{X}^0\rme^{-\frac{K}{6}}\bar{X})\bar{\Sigma}(X^0\rme^{-\frac{K}{6}}X)}\rmim(f)\epsilon^{\mu\nu\rho\sigma}F_{\mu\nu}F_{\rho\sigma}\right]_D\,.
\ee
We go on to write $\epsilon^{\mu\nu\rho\sigma} F_{\mu\nu}F_{\rho\sigma}$ in terms of the field strength multiplet $P_L \Lambda$:
\be 
\frac{1}{4} X \bar{X} \frac{\epsilon^{\mu\nu\rho\sigma}}{\sqrt{-g_4}}F_{\mu\nu}F_{\rho\sigma} = X \bar{X} \left( X^0 \bar{X}^0 \rme^{-\frac{K}{2}}\right)^2 \frac{1}{2\rmi} \left[\bar{\Sigma}(\omega^2) - \Sigma(\bar{\omega}^2)\right]\,,
\ee
where we have multiplied both sides in order to remove fermionic contributions. We furthermore use that $ \left[\bar{\Sigma}(\omega^2) - \Sigma(\bar{\omega}^2)\right]/(2\rmi) = \rmim(\Sigma(\bar{\omega}^2))$ such that we can finally write the action for the $U(1)$ gauge sector as:
\bea 
S_{V} = &-\frac{1}{4} \left[ f(\tau) \bar{\Lambda} P_L \Lambda\right]_F\\
&+ \left[ \frac{X \bar{X} \left(X^0 \bar{X}^0 \rme^{-\frac{K}{3}}\right)^3}{\bar{\Sigma}\left(X^0 \rme^{\frac{K}{6}}X\right) \Sigma \left(\bar{X}^0 \rme^{-\frac{K}{6}} \bar{X}\right)} \rmim(f) \rmim(\Sigma(\bar{\omega}^2))\right]_D\,.
\label{eq:vecact}
\eea
Importantly, the expansion of the $D$-term in the above expression is exactly twice the contribution of the $\theta$-term that we wanted to obtain:
\be 
-\frac{1}{4} \int d^4x \, \rmim(f) \epsilon^{\mu\nu\rho\sigma} F_{\mu\nu}F_{\rho\sigma} + \ldots\,,
\ee
with the dots giving the fermionic terms required by supersymmetry. When going from the superconformal formalism to Poincar\'{e} supersymmetry by letting the compensator $X^0 = \kappa^{-1} \rme^{K/6}$ and setting $\kappa = 1$ we obtain the result for the bosonic sector:
\be 
S_{V,\text{ bos}} = \int d^4x \sqrt{-g_4} \left[-\frac{1}{4} \rmim(\tau) F^{\mu\nu}F_{\mu\nu} + \frac{1}{8} \rmre(\tau)\frac{\epsilon^{\mu\nu\rho\sigma}}{\sqrt{-g_4}}F_{\mu\nu}F_{\rho\sigma}\right]\,,
\ee
where we went on-shell by using the equations of motion for $D$ and the gauge kinetic function is $f(\tau) = -\rmi\tau$.
Expression \eqref{eq:vecact} contains higher-order fermionic terms that do not appear in the action of the anti-$D3$-brane as written in equation \eqref{eq:simpbraneact}. One might fear that this invalidates the match between the two expressions, however, due to the spontaneously broken supersymmetry, the fermions depend on the Goldstino and will vanish in unitary gauge. Since they can be set to zero via a gauge choice these couplings do not have a physical impact and hence the match holds. This is true for as long as the Goldstino is given only in terms of the fermion in our nilpotent chiral field $X$. If, for some reason, the Goldstino becomes a mixed expression of the fermion in $X$ and the world-volume fermions, this argument becomes invalid. \\
It is still possible to bring \eqref{eq:vecact} into a more familiar looking form. For this we need to make use of the constraints on the multiplets $X$ and $P_L \Lambda_\alpha$ in section \ref{sec:couplinggravity}. Furthermore, using $\Sigma(AB) = A \Sigma(B)$ \cite{Ferrara:2016een}, for a chiral field $A$ and a field $B$ with weights $(w,w-2)$, we observer that:
\bea 
\left[f(\tau) \bar{\Lambda}P_L\Lambda\right]_F &= \left[\frac{\Sigma\left(\bar{X}^0 \rme^{-\frac{K}{6}} \bar{X} f(\tau)\right)}{\Sigma\left(\bar{X}^0\rme^{-\frac{K}{6}}\bar{X}\right)}\bar{\Lambda}P_L\Lambda\right]_F\\
&= \left[\Sigma\left(\frac{\bar{X}^0 \rme^{-\frac{K}{6}} \bar{X} f(\tau)}{\Sigma \left(\bar{X}^0\rme^{-\frac{K}{6}}\bar{X}\right)}\right)\bar{\Lambda}P_L\Lambda\right]_F\,.
\eea
Using this, the fact that $P_L \Lambda_\alpha$ is constrained and $[C_D]=[\Sigma(C)]_F/2$ \cite{Ferrara:2016een} we can combine the two terms of the $U(1)$ gauge sector into \cite{Cribiori:2019hod}:
\bea 
S_V &= -\frac{1}{4} \left[\Sigma\left(\frac{\bar{X}^0 \rme^{-\frac{K}{6}} \bar{X} \bar{f}(\tau)}{\Sigma \left(\bar{X}^0\rme^{-\frac{K}{6}}\bar{X}\right)}\right)\bar{\Lambda}P_L\Lambda\right]_F\\
:&= -\frac{1}{4} \left[\hat{f}_{\overline{D3}}(\bar{\tau},\bar{X})\bar{\Lambda}P_L\Lambda\right]_F\,,
\eea 
where we defined the generalized gauge kinetic function for the anti-$D3$-brane:
\bea 
\hat{f}_{\overline{D3}}(\bar{\tau},\bar{X}) := \Sigma\left(\frac{\bar{X}^0 \rme^{-\frac{K}{6}} \bar{X} \bar{f}(\tau)}{\Sigma \left(\bar{X}^0\rme^{-\frac{K}{6}}\bar{X}\right)}\right)\,,
\label{eq:gengaugekin}
\eea
with $\bar{f} = \rmre(f)-\rmi \rmim(f)$. In this case the gauge kinetic function is a chiral multiplet in $\tau$ and contains Goldstino interactions that are required for the non-linear realization of supersymmetry. Importantly, the lowest component of $\hat{f}_{\overline{D3}}$ is just $\bar{f}(\bar{\tau})$ with higher-order fermionic terms that vanish in unitary gauge. It is possible to put this into the language of constrained multiplets where $\hat{f}_{\overline{D3}}$ is a chiral multiplet with the constraint:
\be 
X\bar{X}\hat{f}_{\overline{D3}} = X \bar{X}\bar{f}(\bar{\tau})\,.
\ee
\paragraph{A superpotential description of the fermion masses} requires a way to consistently put $\bar{\tau}$ into the holomorphic superpotential $W$. In order to achieve this we follow the logic we used when constructing the generalized gauge kinetic function $\hat{f}_{\overline{D3}}$ \eqref{eq:gengaugekin}. We define the chiral multiplet:
\be 
\hat{M}_{ij} = \Sigma \left(\frac{\left(\bar{X}^0 \rme^{-\frac{K}{6}}\bar{X}\right)\rme^{-4\mathcal{A}-2u}\,m_{ij}}{\Sigma\left(\bar{X}^0 \rme^{-\frac{K}{6}}\bar{X}\right)\left(-\rmi (\tau-\bar{\tau})\right)^{\frac{1}{2}}f\left(U^A,\bar{U}^A\right)^{-\frac{1}{6}}}\right)\,,
\ee
which one could call the chiral mass multiplet. Here, $m_{ij}$ are the fermion masses given at the beginning of this section in equation \eqref{eq:fermmass}. Again, the component expansion reads:
\be 
\hat{M}_{ij} = \frac{\rme^{-4\mathcal{A}-2u}}{\left(-\rmi (\tau-\bar{\tau})\right)^{\frac{1}{2}}f\left(U^A,\bar{U}^A\right)^{-\frac{1}{6}}} \; m_{ij} + \text{fermions}\,,
\ee
where the fermionic terms depend on the Goldstino and vanish in unitary gauge. Using a constraint to write this we find:
\be 
X\bar{X} \hat{M}_{ij} = X \bar{X} \left( \frac{\rme^{-4\mathcal{A}-2u}}{\left(-\rmi (\tau-\bar{\tau})\right)^{\frac{1}{2}}f\left(U^A,\bar{U}^A\right)^{-\frac{1}{6}}}\; m_{ij} \right)\,.
\ee 
The superpotential contribution that we would add to \eqref{eq:D3barsup} then is:
\be 
W_m(\hat{M},Y) = \frac{1}{2} \hat{M}_{ij}Y^iY^j\,.
\ee
This gives the correct mass terms to the fermions contained in the multiplets $Y^i$. Once again, this crucially depends on spontaneously broken supersymmetry, described by the nilpotent chiral multiplet $X$ that has a non-vanishing auxiliary field $F$.

\paragraph{The scalar potential $\mathbf{V}$ of our setup} can now be found using the typical supergravity formula and is of the form:
\be 
V = V_{KKLT} + V_{\overline{D3}}\,.
\ee
Here $V_{KKLT}$ contains the bulk fields of supergravity and for the anti-$D3$-brane contribution one needs to make sure to only use one description at a time and to set the constrained multiplets to zero at the end of the calculation as to eliminate fermionic contributions. We have chosen the Kähler potential \eqref{eq:KpotbarD3} and superpotential \eqref{eq:D3barsup} which gives the uplifting contribution to the scalar potential as:
\bea
V_{\overline{D3}} &= \frac{M^4 \rme^{4\mathcal{A}(H^a,\bar{H}^a)}}{\left[\left(-\rmi(T-\bar{T})\right) + k(H^a,\bar{H}^a)f\left(U^A,\bar{U}^A\right)^{-\frac{1}{3}}\right]^2}\\
\Rightarrow V_{\overline{D3}} &= 2 \rme^{4\mathcal{A}(H^a,\bar{H}^a)-8u}\,.
\eea
This matches correctly the expression \eqref{eq:barD3scalpotorig} of section \ref{sec:D3barKKLT} but comes from a Kähler and superpotential, as we desired.
\subsection[Anti-$D3$-brane Action - Formulas]{Anti-$\mathbf{D3}$-brane Action - Formulas}
For practical purposes we will condense the results of the previous subsection here in order to collect all relevant formulas in one place. The motivation and explanation for these expressions are found in the prior subsections. The goal of this section was to find a description of the anti-$D3$-brane in the KKLT setup in terms of a Kähler and superpotential. For this we utilized the fact that the anti-brane breaks supersymmetry spontaneously and we used constrained multiplets in order to obtain the desired description. We stress once more that constrained superfields are a convenient description we used here but other ways to write these formulas exist. Furthermore, the specific choice of constrained multiplets is not unique.\\
Up to terms quadratic in fermions we can write the anti-$D3$-brane action \eqref{eq:simpbraneact} in $4d$, $\mathcal{N}=1$ supergravity as:
\be 
S_{\overline{D3}}=\left[\hat{f}_{\overline{D3}}(\bar{\tau},\bar{X}) \bar{\Lambda}P_L\Lambda\right]_f + \left[-3X^0 \bar{X}^0\rme^{-\frac{K}{3}}\right]_D + \left[\left(X^0\right)^3 W\right]_F\,,
\label{eq:D3barfinalact}
\ee
where we remind the reader that we are using the conventions of \cite{Freedman:2012zz}. The generalized gauge kinetic function for the anti-$D3$-brane is defined as:
\be 
\hat{f}_{\overline{D3}} = \Sigma\left( \frac{\bar{X}^0 \rme^{-\frac{K}{6}}\bar{X}\bar{f}(\bar{\tau})}{\Sigma \left( \bar{X}^0 \rme^{-\frac{K}{6}}\bar{X}\right)}\right)\,,
\ee
with $\bar{f}(\bar{\tau}) = \rmi \bar{\tau}$. 
\\\newpage The multiplets that contain the world-volume fields are constrained in the following way:
\be
\begin{pmatrix}
X & \quad & X^2=0 & \quad & \text{Goldstino}\\
Y^i & & X Y^i = 0& & \text{massive fermion triplet}\\
P_L \Lambda_\alpha & & X P_L \Lambda_\alpha = 0 & & U(1) \text{ gauge vector}\\
H^a & & X\bar{H}^a = \text{chiral} & & \text{3 complex scalars}
\end{pmatrix}\,.
\label{eq:D3barmultfinal}
\ee

\paragraph{When generating the masses for the fermions from the Kähler potential} we write:
\bea 
K_{K} = & -\log\left(-\rmi(\tau-\bar{\tau})\right) - 3 \log \left[\left(-\rmi (T-\bar{T})\right) f\left(U^A,\bar{U}^A \right)^{\frac{1}{3}} + k\left(H^a,\bar{H}^a\right) \right]\\
& - 3 \log \left[ 1 - a X\bar{X} - b \delta_{ij} Y^i \bar{Y}^\jb + c\left(m_{ij} \bar{X} Y^i Y^j + \bar{m}_{\ib\jb} X \bar{Y}^\ib \bar{Y}^\jb \right)\right]\,,
\label{eq:D3barKKfinal}
\eea
with the gauge kinetic function given as:
\be 
f\left(U^A ,\bar{U}^A\right) = -\rmi \int \Omega \wedge \bar{\Omega}\,,
\label{eq:D3barKinFuncfinal}
\ee
the $a$, $b$, and $c$ are shorthand for:
\bea 
a &= \frac{\rme^{-4\mathcal{A}(H^a,\bar{H}^a)}}{3\left(-\rmi (\tau-\bar{\tau})\right) \left[\left(-\rmi(T-\bar{T})\right) + k\left(H^a,\bar{H}^a\right) f\left(U^A,\bar{U}^A\right)^{-\frac{1}{3}}\right] f\left(U^A,\bar{U}^A \right)}\,,\\
b &= \frac{\rme^{-4\mathcal{A}(H^a,\bar{H}^a)}}{3\left(-\rmi (\tau-\bar{\tau})\right) \left[\left(-\rmi(T-\bar{T})\right) + k\left(H^a,\bar{H}^a\right) f\left(U^A,\bar{U}^A\right)^{-\frac{1}{3}}\right]^2 f\left(U^A,\bar{U}^A \right)^{\frac{1}{3}}}\,,\\
c &= \frac{\rme^{-8\mathcal{A}(H^a,\bar{H}^a)}}{6 M^2 \left(-\rmi (\tau-\bar{\tau})\right)^{\frac{3}{2}} \left[\left(-\rmi(T-\bar{T})\right) + k\left(H^a,\bar{H}^a\right) f\left(U^A,\bar{U}^A\right)^{-\frac{1}{3}}\right]^{\frac{3}{2}} f\left(U^A,\bar{U}^A \right)^{\frac{5}{6}}}\,,
\label{eq:D3barparafinal}
\eea
and finally the fermion mass matrix is (see also \eqref{eq:fermmass} and the discussion there):
\be
m_{ij} = \rmi \,\frac{\sqrt{2}}{8} \rme^{2\mathcal{A}-4 u}\, \rme^{\frac{\phi}{2}} \left(e_i^c e_j^d + e_j^c e_i^d \right) \Omega_{abc}g^{a\bar{a}}g^{b\bar{b}} \bar{G}_{d\bar{a}\bar{b}}\,.
\label{eq:D3barfermmassfinal}
\ee
The correct superpotential for this description is:
\be 
W_K = W_{\text{GVW}} + W_{\text{np}} + M^2 X\,,
\label{eq:D3barWKfinal}
\ee
with $M^2 = \sqrt{2}$ here and also below. Note once more that the fermion mass terms do not appear in this superpotential.
\paragraph{Writing the fermion masses as a superpotential  contribution} we have a Kähler potential:
\bea 
K_M = & - \log \left(-\rmi (\tau-\bar{\tau})\right) - 3 \log \left[ \left(-\rmi (T-\bar{T})\right) f\left(U^A,\bar{U}^A \right)^{\frac{1}{3}}+k\left(H^a,\bar{H}^a\right)\right]\\
 & - 3 \log \left[1-a X \bar{X} - b \delta_{i\jb}Y^i \bar{Y}^\jb \right]\,,
 \label{eq:D3barKMfinal}
 \eea
 with the same $a$ and $b$ as above. The fermion masses are now included in the superpotential:
 \be
 W_M = W_{\text{GVW}} + W_{\text{np}} + M^2 X + \frac{1}{2} \hat{M}_{ij} Y^i Y^j\,,
 \label{eq:D3barWMfinal}
 \ee
with the chiral mass multiplet:
\be 
\hat{M}_{ij} = \Sigma \left(\frac{\left(\bar{X}^0 \rme^{-\frac{K}{6}}\bar{X}\right)\rme^{-4\mathcal{A}-2u}\,m_{ij}}{\Sigma\left(\bar{X}^0 \rme^{-\frac{K}{6}}\bar{X}\right)\left(-\rmi (\tau-\bar{\tau})\right)^{\frac{1}{2}}f\left(U^A,\bar{U}^A\right)^{-\frac{1}{6}}}\right)\,.
\label{eq:D3barmassmultfinal}
\ee

\subsection[The anti-$D3$-brane in KKLT - Interim Summary]{The anti-$\mathbf{D3}$-brane in KKLT - Interim Summary}
In this section we studied the anti-$D3$-brane at the bottom of a warped throat in the KKLT scenario. We found a description that includes all world-volume fields and couplings using the language of non-linear supersymmetry. The action is supersymmetric and the symmetry gets broken spontaneously by the anti-brane. With the results presented here and originally in \cite{Cribiori:2019hod}, we have found a complete description that should find ample use in further studies of KKLT-like models and string phenomenology in general. In particular, in \cite{Parameswaran:2020ukp} an anti-$D3$-/$D7$ setup was discussed using similar methods as described here. Furthermore, it should be possible to extend the description here to stacks of anti-$D3$-branes by promoting the world-volume fields to have $SU(N_{\overline{D3}})$-symmetry \cite{McGuirk:2012sb}. Such a setup in the KKLT scenario leads, however, to the polarization of the stack of anti-$D3$-branes into an $NS5$-brane \cite{Kachru:2002gs}. The resulting theory \cite{Aalsma:2017ulu,Aalsma:2018pll} has been studied but a description in terms of constrained superfields has not yet been found. Another interesting extension of our work here regards the large volume scenario (LVS) \cite{Balasubramanian:2005zx,Conlon:2005ki}. There, SUSY is broken already in anti-de Sitter space before the uplifting brane is introduced. This leads to a Goldstino that is composed of the world-volume fermions and the closed string fermions. It should be possible to re-write the action of this situation in a similar manner as we did here for KKLT. This would be interesting as LVS is one of the main competitors to the KKLT setup.

\chapter{KKLT-like Constructions} 
\label{sec:kkltconstr}
In this chapter we review extensions of the class of de Sitter constructions based on the KKL(MM)T model \cite{Kachru:2003aw,Kachru:2003sx} that we published in \cite{Cribiori:2019bfx,Cribiori:2019drf,Cribiori:2019hrb}. After briefly reviewing the original setup, we first show a way to translate the setup to type IIA models \cite{Cribiori:2019bfx} where one uses and anti-$D6$-brane instead of the anti-$D3$-brane of the original models that are based in type IIB theory. Importantly, we rely on non-perturbative corrections in order to achieve the uplift. We go on by introducing the ''mass production method" \cite{Cribiori:2019drf} in section \ref{sec:massprod} that allows for the easy construction de Sitter vacua without the fear of tachyonic directions. Finally, based on \cite{Cribiori:2019hrb}, we consider more general, M-theory motivated, examples that allow for different kinds of non-perturbative corrections to be absent.
\section{KKL(MM)T}
\label{sec:kklt}
\begin{figure}[H]
     \centering
     \includegraphics[trim=0 0 100 20,clip,width=0.8\textwidth]{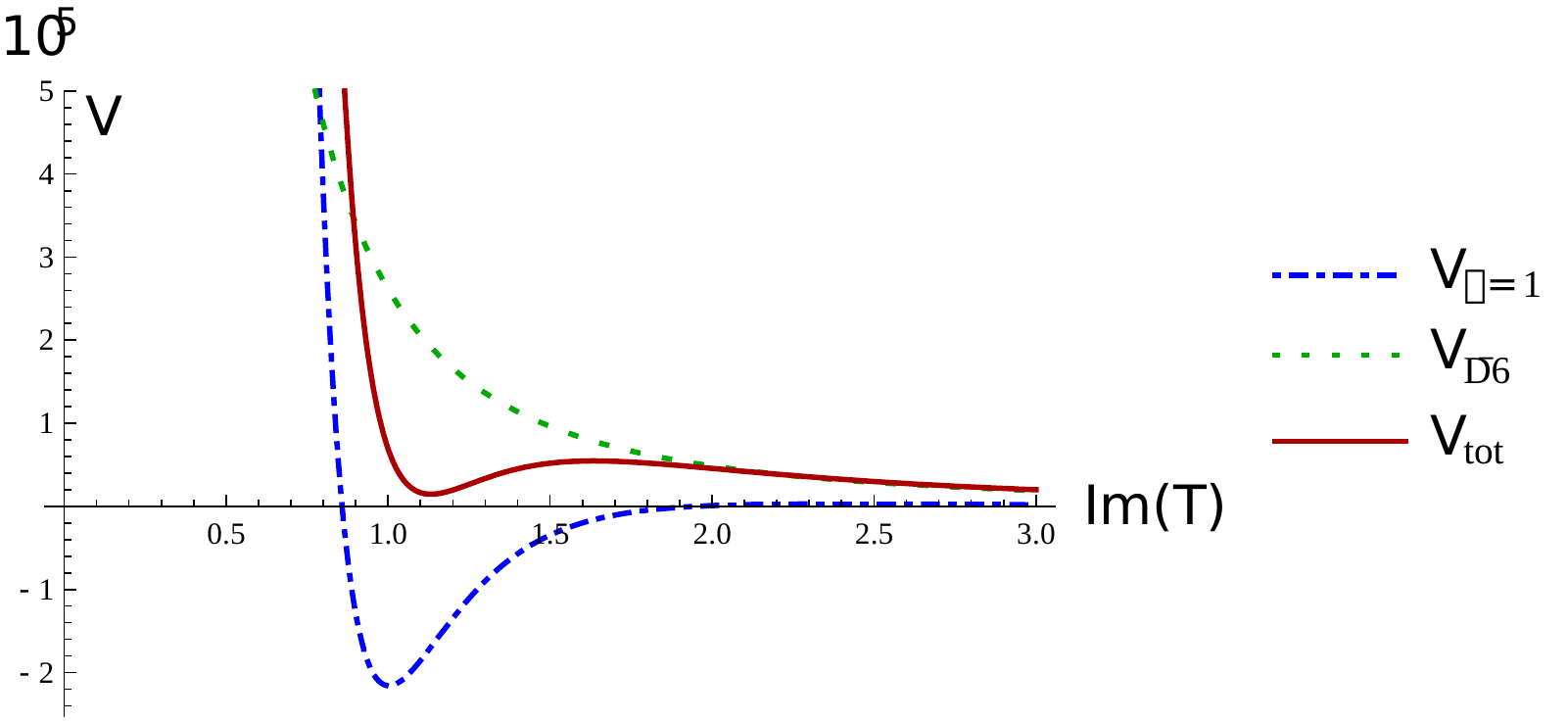}
     \caption{Example depiction of the three scalar potentials in KKLT. The dash-dotted (blue) line is the stable AdS progenitor that gets uplifted by the dashed (green) positive energy contribution from the anti-$D3$-brane, in this example placed at the bottom of a warped throat. The result is a (meta-) stable de Sitter vacuum in the solid (red) line.}
     \label{fig:KKLTscalpot}
\end{figure}

The KKLT-scenario \cite{Kachru:2003aw,Kachru:2003sx} is a potentially possible way to obtain meta-stable de Sitter spaces from string theory. The initial setup is based on a simple type IIB model with 3 moduli. Here, $S$ is the axio-dilaton, $T$ the Kähler-modulus and $U$ the complex structure modulus. Note that in more general setups the moduli $T$ and $U$ can split into many different, independent scalars each, depending on the Hodge-numbers $h^ {1,1}$ and $h^ {2,1}$ of the internal manifold. In the simplest model these are taken to be the same and the model simplifies. Such an $STU$-model has the following Kähler- and Superpotential in IIB supergravity:
\bea
K &= - \log \left( -\rmi (S- \bar{S}\right) - 3 \log\left(-\rmi (T- \bar{T}) \right) - 3 \log\left(-\rmi (U- \bar{U}) \right)\,, \\
W &= W_{\text{flux}} + W_{\text{np}}\,.
\label{eq:KWkklt}
\eea
In the superpotential the first term comes from introducing $p$-form fluxes on the internal manifold and the second term from non-perturbative contributions. In the KKLT-scenario one follows a three-step process:
\begin{enumerate}
\item Use the flux superpotential in order to stabilize the axio-dilaton $S$ and the complex structure moduli $U$.
\item Introduce a non-perturbative contribution of the form:
\begin{equation}
W_{\text{np}} = A \rme^{\rmi a T}\,,
\end{equation}
in order to stabilize the Kähler moduli and obtain a stable AdS space.
\item Add an anti-$D3$-brane at the bottom of a warped throat, effectively giving a positive energy contribution to the scalar potential, in order to lift the minimum of the scalar potential to positive energy.
\end{enumerate}
In figure \ref{fig:KKLTscalpot} a rough example is given of how the stable AdS space from step 2, the contribution of the anti-$D3$-brane and the resulting meta-stable dS space are plotted.
The cosmological constant is given by the value of the scalar potential:
\bea
V\; &= V_{0} + V_{\overline{D3}} \qquad \text{where:}
\eea
\bea
V_{0}\, &= \rme^{-K} \left( K^{I\bar{J}} D_I W \overline{D_J W} - 3 |W|^2\right)\,,\\
V_{\overline{D3}} &= \frac{\mu}{(T + \bar{T})^p} \quad \text{with} \quad p=2,3\,,
\eea
at the minimum. The value of $p$ depends on whether the brane is located in te bulk ($p=2$) or at the bottom of a warped throat $p=3$. The idea is that one can tune the amount of uplift in order to, in principle, match the cosmological constant. The description of the anti-$D3$-brane in terms of Kähler- and superpotential has been discussed in section \ref{sec:D3barKKLTmain} and relies on constrained superfields.\\
This model has been investigated thoroughly since its inception and many questions have been raised, for an overview see \cite{Andriot:2019wrs}. A main concern regards the nature of the non-perturbative corrections. These can arise in 4d supergravity from gaugino condensation on $D7$-branes. However, this is a 4d effect that lacks an obvious physical counterpart in $10d$ and the description is likewise not immediately clear. Recently it has been shown that one can indeed match these effects to fermionic coupling terms in $10d$ \cite{Kachru:2019dvo}, see \cite{Moritz:2017xto,Hamada:2018qef,Gautason:2018gln,Carta:2019rhx,Gautason:2019jwq,Hamada:2019ack,Bena:2019mte,Koerber:2007xk,Baumann:2010sx,Dymarsky:2010mf} for earlier works on this and related issues. Another topic under investigation is the stability and possible singularities of the anti-brane setup at the bottom of a warped throat \cite{Armas:2018rsy,Blaback:2012nf,Bena:2009xk,Bena:2012tx,McGuirk:2009xx,Dymarsky:2011pm,Polchinski:2000uf,Blaback:2019ucp,Danielsson:2016cit,Bena:2013hr,Blaback:2014tfa,Polchinski:2015bea,Cohen-Maldonado:2015ssa,Gautason:2015tla,Kuperstein:2014zda,Danielsson:2014yga}. Currently, the consensus is that such setups can be meta-stable. A more recent, issue regards the sizes of the bulk of the internal manifold and the warped throat. It has been claimed \cite{Gao:2020xqh} that the warped throat needs to be bigger than the rest of the bulk in order for the model to work, which goes against common intuition about the internal manifold in typical setups.\\
Nevertheless, the KKLT scenario remains one of the more promising candidates for the construction of dS space from string theory.

\section{Uplifting in Type  IIA}
\label{sec:IIAuplift}
The KKLT-scenario \cite{Kachru:2003aw,Kachru:2003sx}, as described in section \ref{sec:kklt}, gives a possible way of constructing dS vacua from type IIB string theory using an uplifting anti-$D3$-brane. One may wonder if something similar is possible in type IIA as well. First off, since we are going to consider type IIA theory on $SU(3)$ structure manifolds, only anti-$D6$-branes can appear because $Dp$-branes wrap $p-3$ cycles in the internal manifold and no non-trivial 1- or 5-cycles exist for our choice. This leaves us with an unique choice for the uplifting brane. An anti-$D6$-brane in a similar setup was used first in order to stabilize an existing but unstable vacuum in \cite{Kallosh:2018nrk}. The present section is based on \cite{Cribiori:2019bfx} where we instead were interested in uplifting a stable AdS point with the contribution from the anti-$D6$-brane, just like in the KKLT model.

\subsection{The Type IIA STU-Model}
\label{sec:antibraneupSTU}
For an explicit construction we will employ a simple 3 moduli setup with $S$ being the axio-dilaton, $T$ the complex structure modulus and $U$ the Kähler modulus \cite{Dibitetto:2011gm,Danielsson:2013rza}. Note that $T$ and $U$ exchange their physical meaning when going from type IIB to IIA, due to T-duality. 
The model is a compactification to 4 dimensions on $T^6/(\mathbb{Z}_2 \times \mathbb{Z}_2)$ with added anti-$D6$-branes wrapping 3-cycles. These branes can either wrap only on one such 3-cycle ($N_{\overline{D6}}^\parallel$) or wrap directions along both cycles of the geometry ($N_{\overline{D6}}^\perp$). The Kähler- and Superpotential we will consider are:
\bea
K &= -\log{(-\rmi (S- \bar{S}))} - 3\log{(-\rmi (T- \bar{T}))} - 3\log(-\rmi (U- \bar{U}))\,,\\
W &= f_6 + W_{np}\,.
\eea
$f_6$ corresponds to a 6-flux and we assume that all other $p$-form fluxes are turned off. This is not only due to simplicity but it turns out that including those fluxes leads to tachyons in the initial AdS space. The non-perturbative part $W_{np}$ is given as:
\be 
W_{\text{np}} = A_S \rme^{\rmi a_S S} + A_T \rme^{\rmi a_T T} + A_U \rme^{\rmi a_U U}\,,
\label{eq:Wnonpert}
\ee
where we take all parameters appearing to be real and constant. Note that, in principle, the parameters $A_S$, $A_T$ and $A_U$ can be moduli dependent. For example of one field $\Phi$ and since we are interested in $\rmim(\Phi)$ as our modulus, we have: $A(\rme^\Phi) \cong A(0) + A^\prime (0) \rme^{-\rmim (\Phi)} + \ldots$ which we will approximate by a constant expression for $\rme^{-\rmim ( \Phi)}\ll 1$. For this reason we will require the last condition throughout our analysis. This corresponds to suppressed higher-order perturbative terms and $\alpha^\prime$ corrections, which is a typical requirement for most string constructions. \\
Further stringy requirements are flux quantization and the non-trivial Bianchi identity including the anti-$D6$-brane charges. The first condition is easily satisfied in our model as we can set $f_6$ to any value we want. In fact, due to a scaling symmetry of the model, this can even be achieved after finding a model with non-integer flux by shifting the position in moduli space and  changing the parameters in the superpotential in the following way:
\bea 
S &\to \lambda_S S\,, \qquad T \to \lambda_T T\,, \qquad U\to \lambda_U U\,,\\
a_S &\to \frac{a_S }{\lambda_S}\,,\qquad\, a_T \to \frac{a_T }{\lambda_T}\,,\qquad\, a_U \to \frac{a_U }{\lambda_U}\,,\\
A_i &\to \sqrt{\lambda_S \lambda_T^3 \lambda_U^3}\, A_i\,,\qquad f_6 \to \sqrt{\lambda_S \lambda_T^3 \lambda_U^3}\, f_6\,.
\eea
The Kähler potential shifts by a constant $-\log (\lambda_S \lambda_T^3 \lambda_U^3)$, which can be compensated via a Kähler transformation that does not change the physics.\\
The Bianchi identity:
\be 
\int dF_2 - F_0 H = -2 N_{O6} + N_{D6} - N_{\overline{D6}}\,,
\ee
on the other hand, needs to be satisfied for each 3-cycle independently. Since the only non-zero flux we have is $F_6$, we can satisfy this condition only by adding $D6$-branes in order to cancel the charges of our uplifting anti-$D6$-branes and any potential $O6$-planes. In principle having $D6$- and anti-$D6$-branes on the same cycle could prove problematic, since they can annihilate, but if one chooses the correct geometry this can work out \cite{Retolaza:2015sta}. Other possibilities to satisfy the Bianchi identity would require exotic ingredients, like anti-$O6$-planes, the study of which is not the subject of this work.

\subsection{Origin of the non-perturbative Corrections}
The non-perturbative corrections in the superpotential \ref{eq:Wnonpert} can arise due to different effects \cite{Palti:2008mg}. For $S$ and $T$ one possibility that is often considered is gaugino condensation. The terms then arise from a Yang-Mills (YM) theory on the D6-branes. Depending on the brane orientation the coupling constants of the YM-theory are related to the moduli via \cite{Danielsson:2013rza}:
\be
\left(g_{YM}^\parallel \right)^{-2} \sim \rmim (S)\,, \qquad \left(g_{YM}^\perp \right)^{-2} \sim \rmim (T)\,.
\ee
An alternative explanation for the appearance of non-perturbative contributions in the superpotential would be Euclidean D2-branes wrapping 3-cycles of the internal manifold. Such branes would give rise to the wanted terms with $a_S =a_T= 2\pi$.\\
Thus far the exponent for the volume modulus $U$ has not been motivated. Generally, such a non-perturbative correction is less established than the other two. Nevertheless, there are possibilities for this contribution to arise.\\
Here we argue that string theory U-duality should exist. String theories exhibit $S$- and $T$-duality. In M-theory it is expected that the two combine into U-duality \cite{Hull:1994ys,Schwarz:1996bh}. The discrete U-duality  contains S- and T-duality as subgroups: $E_7(\mathbb{Z}) \supset SL(2,\mathbb{Z}) \times O(6,6;\mathbb{Z})$. This suggests that non-perturbative corrections in the $U$-direction can arise since they appear on equal footing in M-theory \cite{Acharya:2007rc}. One physical motivation for this claim comes from STU black holes, which exhibit a feature called string triality, see \cite{Behrndt:1996hu}. Still, the question remains which phenomenon in type IIA string theory can produce the required terms in the superpotential. One possible explanation would be open  string worldsheet instantons in $\mathcal{N} = 1$ orientifold compactifications \cite{Kachru:2000ih,Blumenhagen:2009qh}.

\subsection[The uplifting Anti-$D6$-Brane]{The uplifting Anti-$\mathbf{D6}$-Brane}
The final step in the procedure is the lift from anti-de Sitter space to de Sitter via the inclusion of anti-$D6$-branes. This corresponds to an additional term in the scalar potential of the following form:
\be 
V_{\overline{D6}} = \frac{\mu_1^4}{\rmim (T)^3} + \frac{\mu_2^4}{\rmim(T)^2\rmim(S)}\,.
\ee
Here, $\mu_1^4 = 2 \rme^{\mathcal{A}_1} N_{\overline{D6}}^\parallel$ and $\mu_2^4 = 2 \rme^{\mathcal{A}_2} N_{\overline{D6}}^\perp$ describe $N_{\overline{D6}} = N_{\overline{D6}}^\parallel + N_{\overline{D6}}^\perp$ number of anti-$D6$-branes wrapping, potentially warped, 3-cycles. The warp factors are given by $\rme^{\mathcal{A}_1}$ and $\rme^{\mathcal{A}_2}$, respectively. The total scalar potential then is:
\be 
V_{\text{tot}} = \rme^K \left( K^{I\bar{J}} D_{I}W \overline{D_{J} W} - 3 W \overline{W} \right) + \frac{\mu_1^4}{\rmim (T)^3} + \frac{\mu_2^4}{\rmim(T)^2\rmim(S)}\,.
\label{eq:IIAupliftVtot}
\ee
Adding the contribution of the anti-branes directly to the scalar potential might seem strange and ill-motivated. However, it is possible to include them in the 4 dimensional supergravity directly by including contributions in the Kähler- and superpotential. This can be achieved by using constrained superfields. In particular, for this case, one can use a nilpotent chiral field $X = \phi + \sqrt{2} \theta \chi + F \theta^2$ with $X^2=0$. Here, $\phi$ is a scalar, $\chi$ a fermion, $F$ and auxiliary field and $\theta$ the superspace coordinates. Importantly, after enforcing the nilpotency condition, the only remaining degree of freedom will be $\chi$. Using this field, we can include the contribution of the anti-$D6$-brane into our supergravity potentials in the following way \cite{Kallosh:2018nrk}:
\bea 
K = &- \log \left( -\rmi(S-\bar{S}\right) -  \log \left( [-\rmi (T-\bar{T})]^3\right) \\&- \log \left( [-\rmi(U-\bar{U})]^3 - \frac{X \bar{X}}{\rme^{\mathcal{A}_1} N_{\overline{D6}_1} (-\rmi (S-\bar{S})) +\rme^{\mathcal{A}_2} N_{\overline{D6}_2} (-\rmi (T-\bar{T}))}\right)\,,\\
W = &f_6 + A_S \rme^{\rmi a_S S}+ A_T \rme^{\rmi a_T T}+ A_U \rme^{\rmi a_U U} + \mu^2 X\,,
\eea 
where we identify $\mu^4_{1/2} = 1/8\,\mu^4 \rme^{\mathcal{A}_{1/2}} N_{\overline{D6}_{1/2}}$.
Using these two potentials and the formula for the scalar potential in supergravity: $V = \rme^K \left( K^{I\bar{J}} D_{I}W \overline{D_{J} W} - 3 W \overline{W} \right)|_{X=0}$, returns the complete $V_{tot}$ given in \eqref{eq:IIAupliftVtot}. \\
Note that under the scaling symmetry that was discussed above, we need to let $ \mu_1^4 \to \lambda_T^3 \mu_1^4 $ and $\mu_2^4 \to \lambda_T^2 \lambda_S\mu_2^4 $ in order to leave the model invariant when performing a re-scaling.

\subsection{Explicit Models}
\label{sec:uplfitbraneexamples}
\begin{figure}[H]
\includegraphics[scale=0.5]{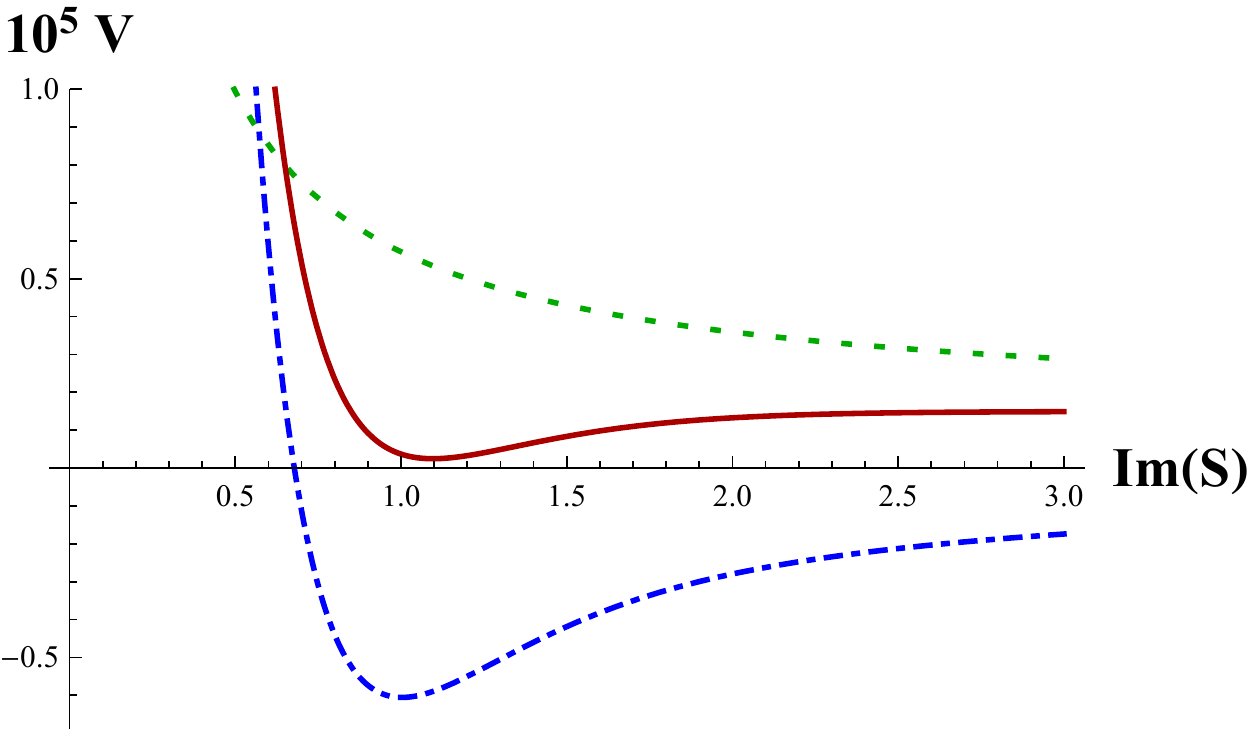}\hspace{10pt} \includegraphics[scale=0.5]{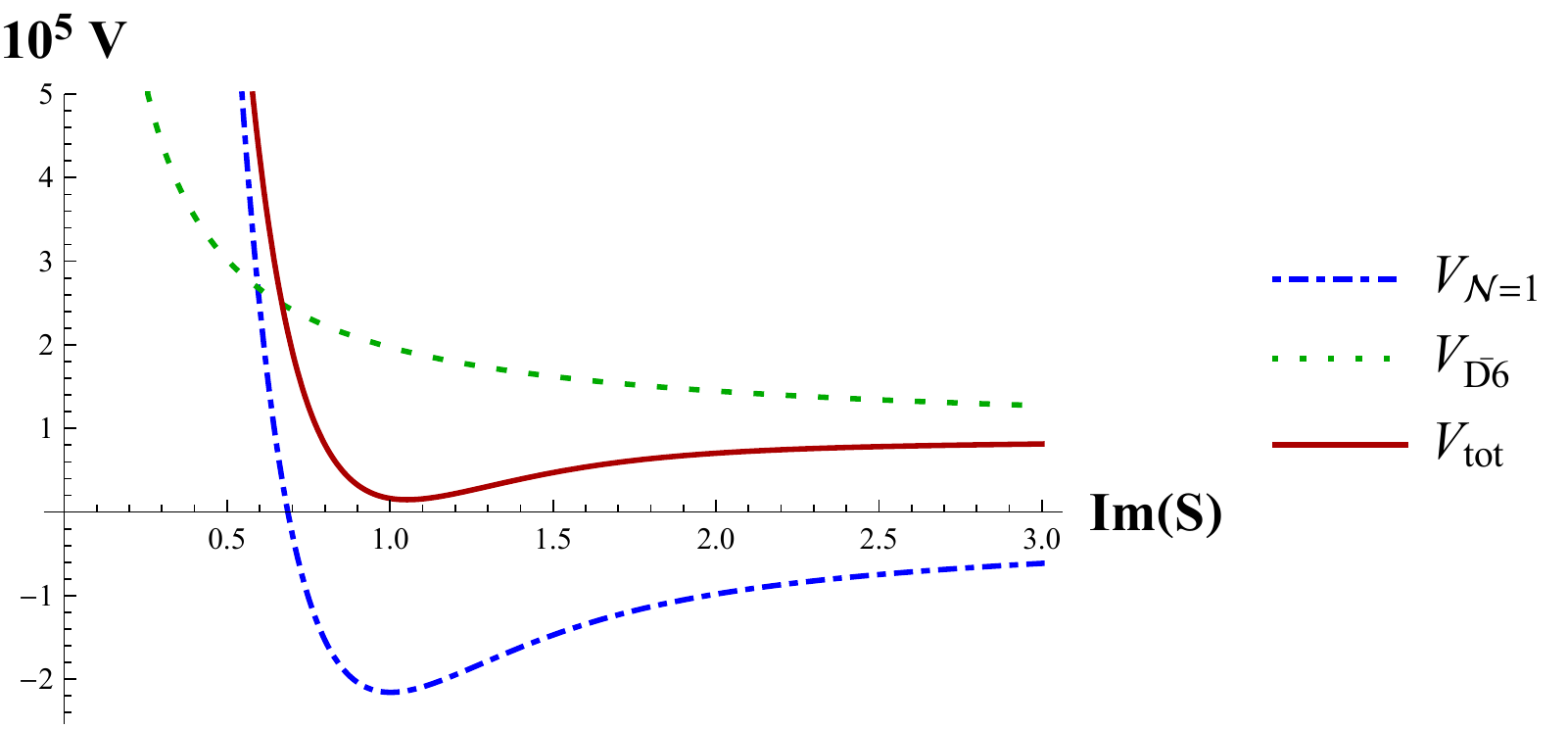}\vspace{15pt}\\
\includegraphics[scale=0.5]{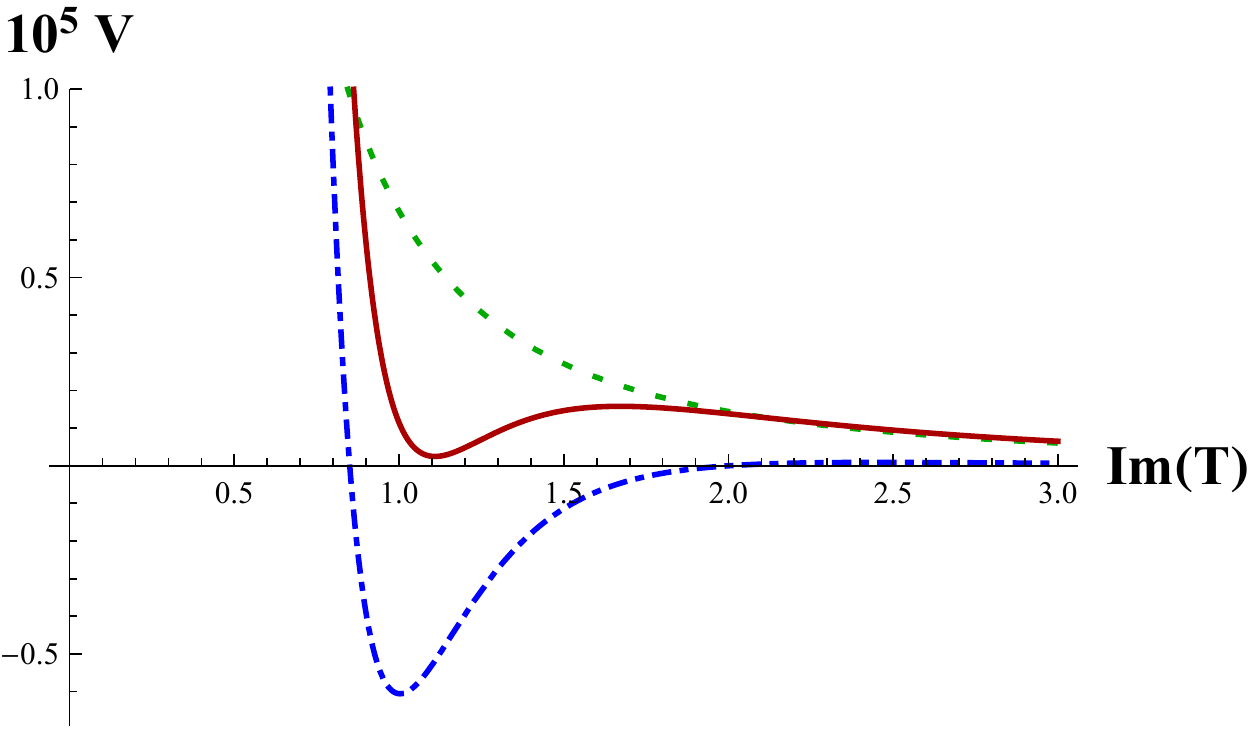}\hspace{10pt} \includegraphics[scale=0.5]{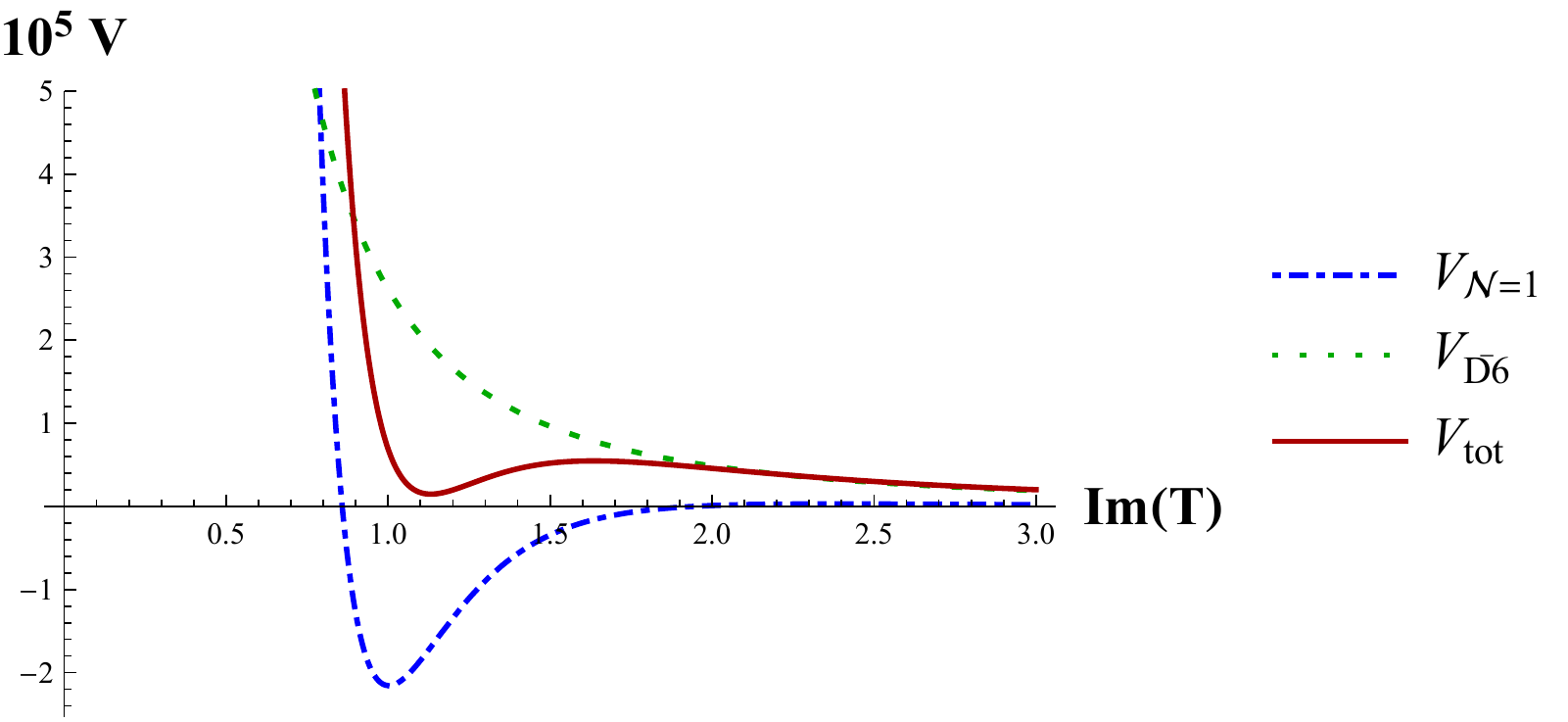}\vspace{15pt}\\
\includegraphics[scale=0.5]{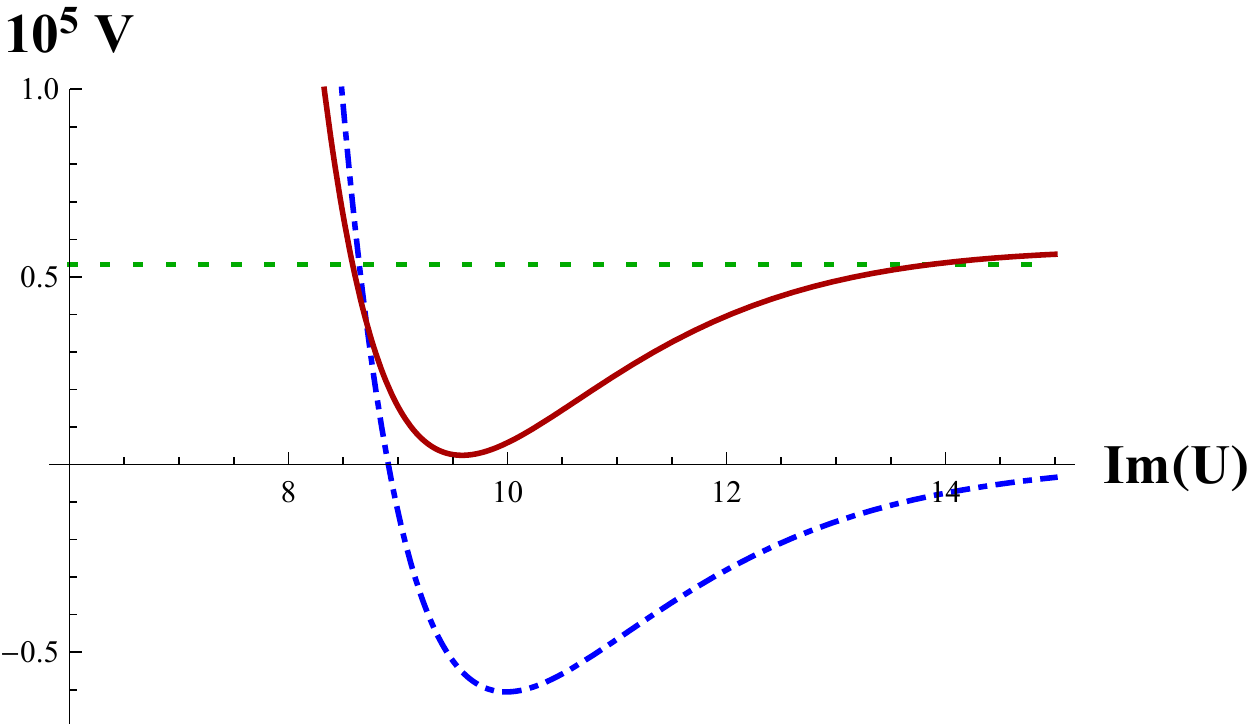}\hspace{10pt} \includegraphics[scale=0.5]{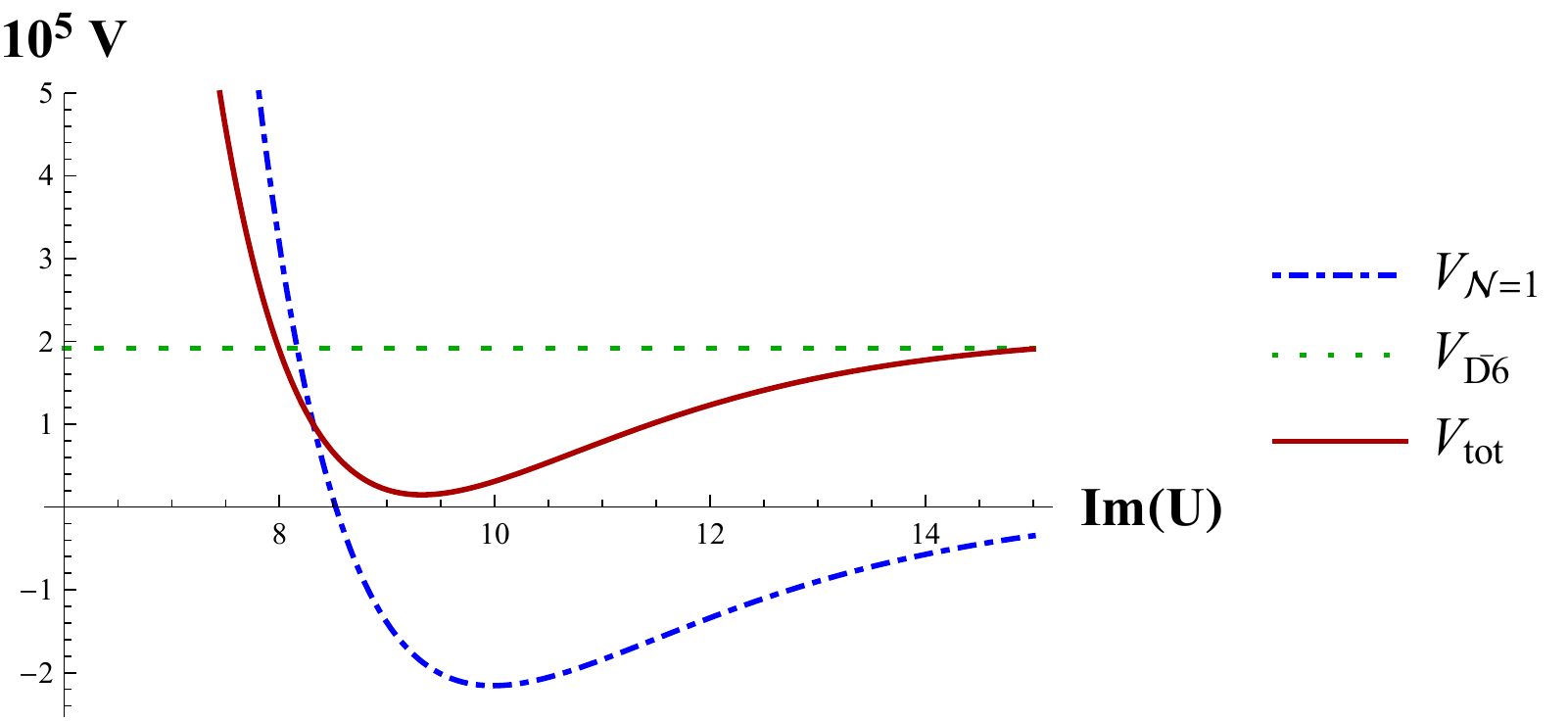}
\caption{Two dimensional plots showing the scalar potential in the AdS minimum (blue, dash-dotted), the contribution from the anti-$D6$-brane (green, dotted) and the resulting, total dS scalar potential (red, solid) for each direction in moduli space. The left side is for Set 1 and the right for Set 2.}
\label{fig:IIAupliftplot}
\end{figure}

\begin{figure}[H]
\hspace{100pt}
\includegraphics[scale=0.53]{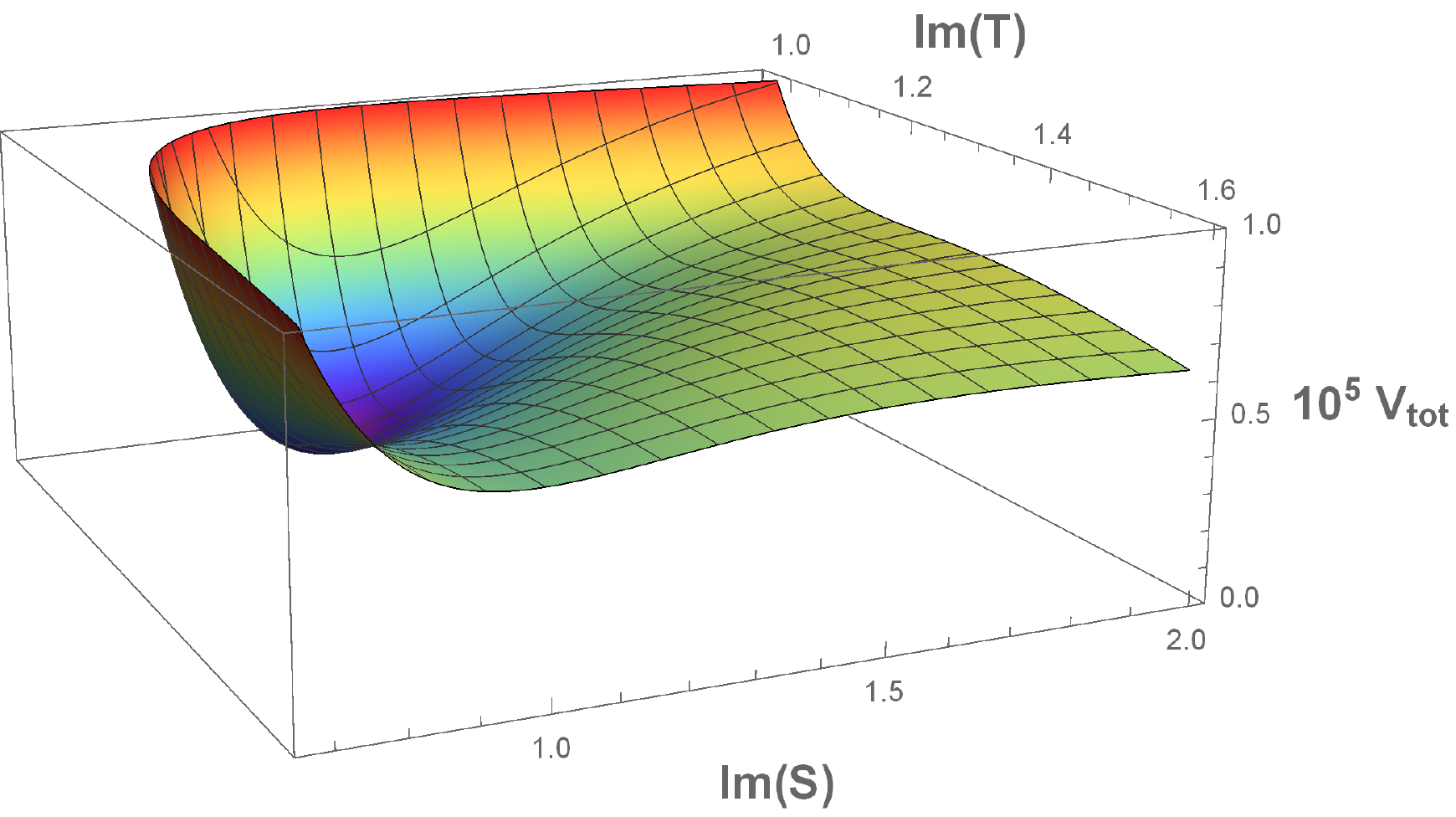}\vspace{15pt}\\
\includegraphics[scale=0.53]{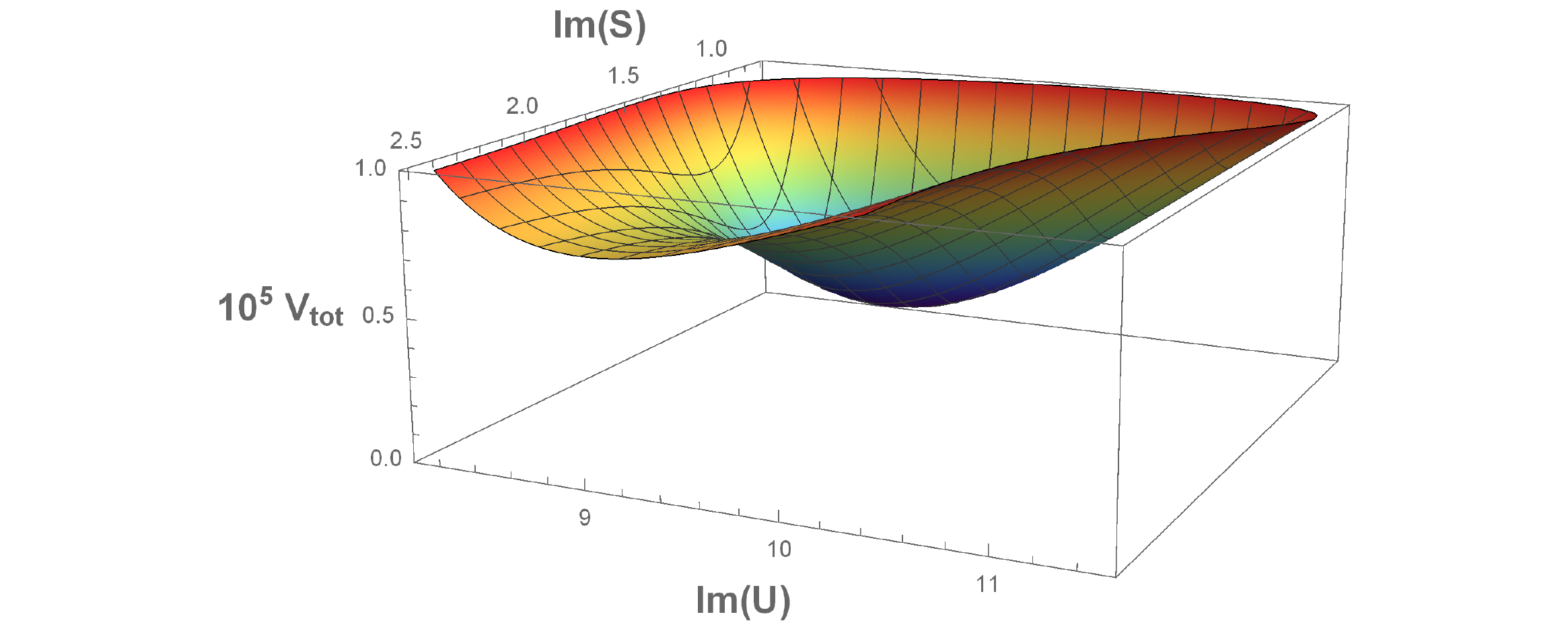}\vspace{15pt}\\
\hbox{\hspace{50pt}\includegraphics[scale=0.53]{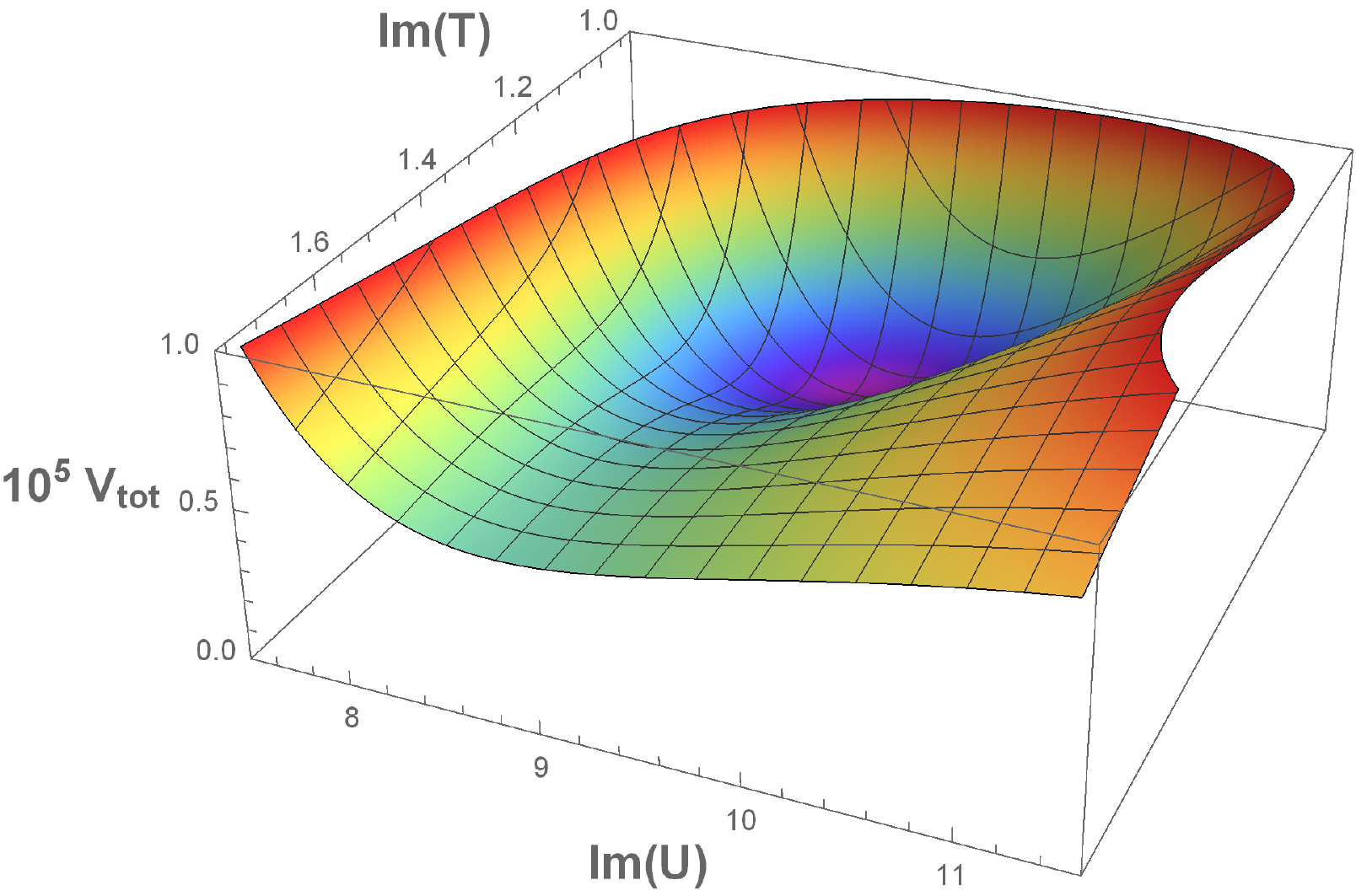}}
\caption{Three dimensional plots for Set 2, showing the (meta-) stability of the scalar potential. Top: Im$(S)$ and Im$(T)$, Middle: Im$(S)$ and Im$(U)$, Bottom: Im$(U)$ and Im$(T)$. }
\label{fig:IIAupliftplot3d}
\end{figure}

The first step in the procedure to find a de Sitter vacuum via an anti-$D6$-brane uplift in type IIA is to generate a stable anti-de Sitter minimum. For this we solve the equations:
\be 
D_I W = 0 \quad \text{where} \quad I = \{S,T,U\}\,,
\ee
and tune the free parameters such that all masses are positive. In principle, in anti-de Sitter, it would suffice to find masses above the Breitenlohner-Freedman bound, however, we choose to focus on strictly positive masses. Solving $D_I W = 0$ implies $\partial_I V=0$\footnote{Note that $\partial_I V=0$ does not imply $D_I W = 0$.}. One simplification we can make is to set the axions $\rmre(S)$, $\rmre(T)$ and $\rmre(U)$ to zero. This can consistently be done as long as all moduli masses are positive.\\
We furthermore choose to fix the position of the minimum at $\rmim (S) = S_0$, $\rmim (T) = T_0$ and $\rmim (U) = U_0$. Then, we have to solve the equations $D_IW = 0$ with $W$, given in equation \eqref{eq:Wnonpert}, in terms of the parameters $A_I$ with $I = \{S,T,U\}$. Importantly, $f_6$ remains a free parameter, making flux quantization trivial. After a solution is found one needs to check the mass matrix:
\be 
m_J^I = \frac{1}{2} K^{J\bar{K}} \nabla_{\bar{K}} \partial_I V\,,
\ee 
to ensure that all eigenvalues are positive at the minimum\footnote{This expression has to be considered with the Kähler metric in real coordinates.}. The mass matrix now depends on several free parameters:
\be 
m_{IJ} = m_{IJ}(f_6,a_S,a_T,a_U,S_0,T_0,U_0)\,,
\ee
that can be used to tune the values of the masses. One important restriction concerns the values of the $a_I$. Those have to be chosen in order to guarantee that $\rme^{-a_I \rmim (I)}$ is small, such that higher-order corrections can safely be neglected. In particular, $\rme^{-a_I \rmim (I)} < \mathcal{O}(10^{-1})$ was chosen as a numerical bound. Another condition that has to be met is that, in order to trust the supergravity approximation, we need to have a large internal volume. For this, we choose the minimum of the volume modulus to be $U_0 \sim \mathcal{O}(10)$. Similarly, we require for the axio-dilaton $S_0 \sim \mathcal{O}(1)$, such that we can neglect string loop corrections.\\
It is relatively easy to find solutions to $D_IW=0$, even with all conditions mentioned above. Here, two such solutions are presented. For both of them our choice for the minimum in moduli space is:
\be 
S_0 = T_0=1 \qquad \text{and}\qquad U_0=10\,.
\ee
The parameters $A_I$ are set by the solution of the equations and for the remaining ones two choices can be seen in table \ref{tab:IIAupliftpar}. The parameter $a_U$ is a magnitude smaller than the other ones which is due to the fact that we have chosen $U_0=10$ in order to obtain a large volume.
\begin{table}[H]
\center
\begin{tabular}{|c|c|c|c|c|}\hline
 & $f_6$ & $a_S$ & $a_T$ & $a_U$ \\\hline
Set 1 & $1$ & $3$ & $3$ & $0.5$ \\\hline
Set 2 &$2$ & $3.1$ & $3.3$ & $0.32$ \\\hline
\end{tabular}
\caption{Two possible choices of parameters for an anti-$D6$-uplift in type IIA.}
\label{tab:IIAupliftpar}
\end{table}
These choices of parameters lead to the masses given in table \ref{tab:IIAupliftmass} that are all positive. Furthermore, as can be seen from table \ref{tab:IIAupliftreq}, we satisfy the condition required in order to neglect the moduli dependence of the parameters $A_I$.
\begin{table}[H]
\center
\begin{tabular}{|c|c|c|c||c|c|c|}\hline
 & $A_S$ & $A_T$ & $A_U$  & $e^{-a_S {\rm Im} S}$ & $e^{-a_T {\rm Im} T}$ & $e^{-a_U {\rm Im} U}$  \\\hline
Set 1 & $-1.70$ & $-5.11$ & $-22.6$ & $0.0498$ & $0.0498$ & $0.00674$  \\\hline
Set 2 & $-3.43$ & $-11.8$ & $-11.0$ & $0.0450$ & $0.0369$ & $0.0408$  \\\hline
\end{tabular}
\caption{The parameters $A_I$ and the conditions for them to be moduli-independent.}
\label{tab:IIAupliftreq}
\end{table}
With stable anti-de Sitter minima found, we can now attempt to lift them to de Sitter by introducing anti-$D6$-branes according to equation \eqref{eq:IIAupliftVtot}. For this we choose the following values for $\mu_1$ and $\mu_2$:
\bea 
\text{Set 1} \qquad \mu_1^4 &= 2.01 \cdot 10^{-6}, \qquad \mu_2 = 5.21 \cdot 10^{-6},\\
\text{Set 2} \qquad \mu_2^4 &= \mu_2^4 = 1.34 \cdot 10^{-5}\,.
\eea
These values were chosen for visibility of the behavior in the plots \ref{fig:IIAupliftplot}. In principle they can be tuned to achieve any desired value. A physical motivation would be to match the cosmological constant.\\
With these uplift values one obtains new masses after the uplift, given in table \ref{tab:IIAupliftmass}. Importantly, the masses stay positive, which means that the resulting de Sitter space is meta-stable. Two dimensional plots of all directions are shown in figure \ref{fig:IIAupliftplot}, where it can be seen how we lift the potential from AdS to dS using the anti-$D6$-brane. The scalar potential of the brane does not depend on the $U$-direction and hence the contribution is constant for the $U$-slice. In figure \ref{fig:IIAupliftplot3d} three dimensional plots for all slices of two moduli are presented, visualizing the (meta-)stability of the model. 

\begin{table}[H]
\center
\begin{tabular}{|c|c|c|c|c|c|c|}\hline
Set 1 & $m_{1}^{\,2}$ & $m_{2}^{\,2}$  & $m_{3}^{\,2}$ & $m_{4}^{\,2}$ & $m_{5}^{\,2}$ & $m_{6}^{\,2}$ \\\hline
AdS & $4.36\cdot10^{-4}$& $3.79\cdot 10^{-4}$& $1.01\cdot 10^{-4}$ & $7.37\cdot 10^{-5}$& $5.66 \cdot 10^{-5}$ & $3.64 \cdot 10^{-5}$ \\\hline
dS & $3.43\cdot10^{-4}$& $3.38\cdot 10^{-4}$& $6.46\cdot 10^{-5}$ & $5.40\cdot 10^{-5}$& $4.15 \cdot 10^{-5}$ & $3.47 \cdot 10^{-5}$ \\\hline\hline
Set 2 & $m_{1}^{\,2}$ & $m_{2}^{\,2}$  & $m_{3}^{\,2}$ & $m_{4}^{\,2}$ & $m_{5}^{\,2}$ & $m_{6}^{\,2}$ \\\hline
AdS & $1.19 \cdot 10^{-3}$ & $1.01\cdot 10^{-3}$& $2.43 \cdot 10^{-4}$ & $2.20 \cdot 10^{-4}$& $1.64 \cdot 10^{-4}$ & $1.45\cdot 10^{-4}$  \\\hline
dS & $8.00 \cdot 10^{-4}$ & $7.40 \cdot 10^{-4}$ & $1.76\cdot 10^{-4}$ & $1.63 \cdot 10^{-4}$ & $1.61 \cdot 10^{-4}$& $1.50 \cdot 10^{-4}$  \\\hline
\end{tabular}
\caption{All moduli masses before and after the uplift.} 
\label{tab:IIAupliftmass}
\end{table}

Two final remarks are in order. First, the parameters are not particularly fine-tuned, meaning that it was easy to find the two sets presented here. In fact, it is not hard to find even more examples. The only requirement is to keep the consistency conditions in mind. Second, this can be extended to a more general 7-moduli model. For this the moduli $T$ and $U$ split into three independent fields each. When having 7 independent directions the Kähler- and superpotential change to represent that fact in the following way:
\bea 
K&=  - \log\left(-\rmi (S-\bar{S})\right) - \sum_{i=1}^3  \log\left(-\rmi (T_i-\bar{T}_i)\right) - \sum_{i=1}^3 \log\left(-\rmi (U_i-\bar{U}_i)\right)\,,\\
W& = f_6 + A_{S}\, e^{\rmi a_S S} +  \sum_{i=1}^3 A_{T,i}\, e^{\rmi a_{T,i} T_i}+ \sum_{i=1}^3 A_{U,i}\, e^{\rmi a_{U,i} U_i}\,.
\eea
Furthermore, the contribution of the uplifting anti-$D6$-branes changes to represent the fact that there are now more cycles that can be wrapped:
\bea
V_{\overline{D6}} =\; &\frac{\mu_1^{\,4}}{Im(T_1)  Im(T_2)  Im(T_3)} +\frac{\mu_2^{\,4}}{Im(S)  Im(T_2)  Im(T_3)} \\+&\frac{\mu_3^{\,4}}{Im(S)  Im(T_1)  Im(T_3)} +\frac{\mu_4^{\,4}}{Im(S)  Im(T_1)  Im(T_2)}\,.
\eea
The procedure is otherwise analogous to what was done before. It is necessary to solve the 7 equations $D_IW=0$ in terms of the parameters $A_S$, $A_{T,i}$ and $A_{U,i}$. Then, using the remaining free coefficients, $S_0$, $T_{i,0}$, $U_{i,0}$, $f_6$, $a_S$, $a_{T,i}$ and $a_{U,i}$, one builds a stable anti-de Sitter minimum which gets lifted by the choice of the parameters $\mu_i$ ($i=1,2,3,4$) that correspond to the anti-$D6$-branes. Explicit examples are given in \cite{Cribiori:2019bfx} but are left out here since they do not provide any new insights.

\section{Mass Production of de Sitter Vacua in Type IIA and IIB}
\label{sec:massprod}
The method of the previous section is an easy way to obtain de Sitter vacua in type IIA supergravity models. However, there is no guarantee that one will arrive at a stable model following this procedure. In \cite{Kallosh:2019zgd} it was noted that it is possible to predict that the masses will stay positive during the procedure if one performs a detour by first going to Minkowski space. This allows to generate many meta-stable de Sitter vacua via an easy process. In this section we review the work done in \cite{Cribiori:2019drf}, where this procedure was supplied with explicit examples and generalized for use not only in IIA but also IIB settings.

\subsection{The mass Production Procedure}
The mass production procedure of meta-stable de Sitter vacua follows an easy 3-step process in $4d$, $\mathcal{N}=1$ supergravity:
\begin{enumerate}
\item Find a Minkowski progenitor model.
\item Downshift the model to anti-de Sitter via a perturbation to the superpotential.
\item Introduce anti-branes in order to perform an uplift to de Sitter.
\end{enumerate}
Following this path guarantees that the resulting de Sitter point is meta-stable under certain conditions.\\
Before going into the details of the construction some remarks are in order. Finding a Minkowski minimum requires us to solve not only the equations $D_I W =0$, as we did before, but also $W=0$. This then leads to $\partial_I V=0$ and $V=0$ at the minimum. Not every possible superpotential allows for such solutions. Importantly, the model we investigated in the previous section, with $W = f_6 + \sum_I A_I \rme^{\rmi a_I \Phi_I}$ ($I = \{S,T,U\}$), cannot be  solved for Minkowski space. The models of \cite{Kallosh:2004yh,BlancoPillado:2005fn,Kallosh:2011qk}, however, are a generalization of our previous superpotential that do allow for Minkowski vacua. These are called racetrack models and are schematically of the following form:
\be 
W = W_0 + \sum_I \left( A_I \rme^{\rmi a_I  \Phi^I} - B_I \rme^{\rmi b_I  \Phi^I}\right)\,.
\label{eq:racetrack}
\ee
Here we generalized the $f_6$ term of our type IIA models for use in general setups and the $\Phi^I$ represent all appearing moduli. It has been shown that models with such a superpotential can in fact produce Minkowski minima, see for example \cite{Kallosh:2019zgd,Kallosh:2004yh}. Thus, the superpotential \eqref{eq:racetrack} will be the basis of our investigations in this section. We will go into detail about the Kallosh-Linde racetrack in section \ref{sec:KLmodel}.\\
The second step in our process, the AdS downshift, can be achieved via adding a small perturbation to the superpotential of the form $W_0 \to W_0 + \Delta W$. This will result in pushing a Minkowski vacuum with vanishing scalar potential at the minimum to an AdS vacuum with negative vacuum energy. For small downshifts this is expected to not change the position of the minimum too much and will not make the vacuum unstable.\\
Finally, the uplift follows as before and will lift the vacuum to de Sitter with positive vacuum energy. Note that we could have lifted starting from Minkowski space, however, this comes with an issue. Usually, one wants to match the cosmological constant in semi-realistic constructions. Since the cosmological constant is of order $10^{-120}$ an uplift from Minkowski to de Sitter would require an extreme amount of fine tuning but also leads to a state that is very close to a supersymmetric one. In fact, the uplift scale and SUSY breaking scale are related: $E_{\text{SUSY breaking}} \sim \Lambda^ {1/4} \sim 10^ {-30} M_P \sim 10^ {-3} eV$. This problem can be alleviated by first going to anti-de Sitter and thus disentangling the uplift scale from the SUSY breaking scale. 

\subsection{The Kallosh-Linde Racetrack Potential}
\label{sec:KLmodel}
The simplest Kähler- and superpotential we can write down in order to explain the procedure can be written as follows:
\bea 
K &=-3\log \left(-\rmi (\Phi - \bar{\Phi})\right) + X \bar{X}\,,\\
W &=W_0 + \Delta W +  A \rme^{\rmi a \Phi} - B \rme^{\rmi b \Phi} + \mu^2 X\,,
\label{eq:KLWandK}
\eea 
where $X$ is again a nilpotent chiral field that is used to include the uplifting contributions of anti-$Dp$-branes. The Minkowski solution is found by only considering $W_{KL} = W_0 +  A \rme^{\rmi a \Phi} - B \rme^{\rmi b \Phi}$.\\
To begin with, let us start by considering the Kallosh-Linde racetrack potential where we write the single field as $\Phi = \phi + \rmi \theta$. The solution to $W_{KL} = 0$ and $DW=0$ is given by:
\bea
\Phi &= \phi = \frac{\rmi}{a-b} \log \frac{a\, A}{b\, B} \qquad \text{with}\\
W_0 &= -A\left(\frac{a\, A}{b\, B}\right)^{\frac{a}{b-a}} + B\left(\frac{a\, A}{b\, B}\right)^{\frac{b}{b-a}}\,,
\eea
where the axions have once again been set to zero. A solution to these equations exists for as long as $a>b$ and $a\cdot A > b\cdot B$. The mass of the field at the minimum evaluates to:
\be 
m_\Phi^2 = \frac{2\phi_0}{9} \left( W'' \right)^2 = \frac{2}{9}\, a\,A\,b\,B\,(a-b) \left( \frac{a\,A}{b\,B}\right)^{-\frac{a+b}{a-b}} \log \frac{a\, A}{b\,B}\,.
\ee
In order to go from Minkowski to AdS we now introduce a shift via $W_0 \to W_0 + \Delta W$. This changes the value of the scalar potential $V$ to:
\be 
V_{AdS} = - \frac{3(\Delta W)^2}{8 \phi_0^3} = -\frac{3}{8} \left(\frac{a-b}{\log \frac{a\, A}{b\,B}}\right)^3 (\Delta W)^2\,.
\ee
At leading order, this modification shifts the position of the vacuum in moduli space only by:
\be 
\Delta \Phi = - \frac{K_\Phi \Delta W}{\partial_\Phi^2 W}\,.
\ee
Importantly, the changes, both in mass and position, depend on the tunable and small parameter $\Delta W$.\\
Now we can attempt to lift the minimum, with a negative value for the cosmological constant, to a de Sitter point with nearly vanishing cosmological constant. This requires an uplift parameter of approximately:
\be 
\mu^4 \approx 3 (\Delta W)^2\,.
\ee
Then, the gravitino mass in dS is given as:
\be 
m_{3/2}^2 = \rme^K (\Delta W)^2 = \frac{(\Delta W)^2}{8 \phi_0^3} =  \frac{\mu^4}{24\phi_0^3}\,.
\ee
This model readily generalizes to more moduli and more exponents for each modulus. In the following we will use the model as given here but for $2$, $3$ or $7$ moduli.

\subsection{The Mass Matrix during the mass Production Process}
\label{sec:massmass}
The most important claim of this section is that, given a Minkowski progenitor, we have a deterministic way to arrive at a meta-stable de Sitter vacuum. An important part of that claim is that masses in a supersymmetric Minkowski minimum are non-negative. This is discussed in detail in \cite{Kallosh:2019zgd}, based on earlier work from \cite{BlancoPillado:2005fn}. Here, the arguments of this claim will briefly be summarized.\\
For this discussion it is useful to define a covariant, holomorphic superpotential, also called the complex gravitino mass as:
\be 
\rme^{K/2} W =: m(z^a,\bar{z}^{\bar{a}})\,,
\ee
which depends on an arbitrary number of chiral matter superfields $z^a$. In terms of this quantity the real gravitino mass is $M_{3/2} = \sqrt{|m\bar{m}|}$. The complex masses of the chiral fermions are then given to be:
\be 
D_a D_b m =: m_{ab}\,,\qquad \bar{D}_{\bar{a}}\bar{D}_{\bar{b}}\bar{m} =: \bar{m}_{\bar{a}\bar{b}}\,.
\ee
Alternatively, this can be written as \cite{Freedman:2012zz}:
\be 
m_{ab} = \rme^{K/2} \left(\partial_a + K_a\right) D_b W - \rme^{K/2} \Gamma^{c}\,_{ab} D_c W\,.
\ee
\paragraph{For a supersymmetric Minkowski minimum,} $W = 0$ and $D_aW=0$, the expression for the fermion masses simplifies to:
\be 
m_{ab}^{\text{Mink}} = \rme^{K/2} \partial_a \partial_b W\,.
\ee
Rewriting the scalar potential $V=\rme^K \left(|DW|^2 - 3 |W|\right)$ with this notation we find:
\be 
V = m_a K^{a\bar{b}}\bar{m}_{\bar{b}} - 3 |m|^2 =: |m_a|^2 - 3 |m|^2\,,
\ee
where the Kähler metric is defined as $K_{a\bar{b}} = \partial_a \partial_{\bar{b}}K$. This scalar potential has an extremum for:
\be 
\partial_a V = -2 m_a \bar{m} + m_{ab} K^{b\bar{b}} \bar{m}_{\bar{b}}=0\,,
\ee
and it is supersymmetric if $m_a = \bar{m}_{\bar{b}}=0$. Note that non-supersymmetric extrema are possible as well for $2 m_a \bar{m} = m_{ab} K^{b\bar{b}} \bar{m}_{\bar{b}}$ with $m_a \neq 0$ and $\bar{m}_{\bar{b}} \neq 0$.\\
Writing the scalar mass-matrix as:
\be 
\mathcal{M}^2 = \begin{pmatrix} V_{a\bar{b}} & V_{ab}\\V_{\bar{a}\bar{b}} & V_{\bar{a}b}\end{pmatrix}\,,
\ee
we can go on about our study of the behavior of the masses under our mass production procedure. In a supersymmetric Minkowski vacuum the mass matrix becomes block diagonal:
\be 
\left(\mathcal{M}^2\right)^{\text{Mink}} = \begin{pmatrix} V_{a\bar{b}}^{\text{Mink}} &0\\0 & V_{\bar{a}b}^{\text{Mink}}\end{pmatrix}\,,
\ee
with $V_{a\bar{b}}^{\text{Mink}} = m_{ac} K^{c\bar{c}}\bar{m}_{\bar{c}\bar{b}}$. Since the diagonal blocks are positive definite we conclude that all eigenvalues, corresponding to the masses of the scalars in our model, are non-negative. This can be further underlined by considering an arbitrary vector $\phi$:
\be 
\phi^a V_{a\bar{b}}^{\text{Mink}} \bar{\phi}^{\bar{b}} = \phi^a m_{ac} K^{c\bar{c}}\bar{m}_{\bar{c}\bar{b}} \bar{\phi}^{\bar{b}} = \Phi_c K^{c\bar{c}} \bar{\Phi}_{\bar{c}}\,,
\label{eq:MinkPosDef}
\ee
with $\Phi_c = \phi^a m_{ac}$. Since the Kähler metric $K^{c\bar{c}}$ is positive definite the same holds for the above combination. For the special choice of a racetrack superpotential \eqref{eq:KLWandK} the masses will in fact be strictly larger than zero and this property, as per the following discussion, will transfer to the resulting de Sitter minimum. 
\paragraph{Performing the downshift to anti-de Sitter space} is done via the inclusion of a small shift in $W_0$: $W_0 \to W_0 +\Delta W$. Including such a term in the superpotential will yield an AdS minimum ($V<0$). This means that $D_aW=0$ will still be satisfied but $W=0$ will no longer be the case. Likewise, the complex gravitino mass will no longer be zero:
\be 
m = \rme^K W \neq 0\,,
\ee 
while we still have 
\be 
m_a = 0\,.
\ee
The fermion mass matrix loses its block diagonality but due to the setup we can still quantify the change for a small shift $\Delta W$. Under this change the Kähler potential will be unaffected. Going forward we will distinguish quantities by superscripts Mink and AdS when necessary. To investigate the AdS masses we consider the change of the position of the minimum: $z^{a,\,\text{Mink}} \to z^{a,\,\text{AdS}}$. Since we still have unbroken supersymmetry which leads, at leading order, to:
\be 
\delta z^a = - (m_{ab})^{-1} K_b \Delta m + \ldots\,.
\ee
The change in the complex gravitino mass is $\Delta m = \rme^{K/2} \Delta W$. For the position of the minimum the change is small if $\Delta m$ is smaller than the smallest eigenvalue $m_\chi$ of the fermion mass matrix. The fermion mass matrix in AdS at the point $z + \delta z$ can be written as:
\be 
V_{a\bar{b}}^{\text{AdS}}=m_{ac}K^{c\bar{c}}-2K_{a\bar{b}} m \bar{m}\,.
\ee
The first part is still positive definite, as in the Minkowski case. We immediately see that the AdS minimum will be (meta-)stable if the gravitino mass is smaller than the lightest eigenvalue of the Minkowski fermion mass matrix:
\be 
m_{\chi} \gg m_{3/2}\,.
\ee
If this condition is satisfied we are certain to find a stable, supersymmetric anti-de Sitter vacuum by performing a shift in the superpotential from a previously known Minkowski progenitor.
\paragraph{Under the uplift to de Sitter} we likewise need to check what happens to the position of the minimum and the masses of the model. The lift to dS is performed by including a nilpotent, chiral field $X$ into the Kähler- and superpotential as in equations \eqref{eq:KLWandK}. Because of $X$ we now have new contributions to the scalar potential and mass matrix. We will include them by using the index $I = \{a,X\}$. The de Sitter scalar potential is
\be 
V^{\text{dS}} = \rme^K \left(|D_IW|^2-3|W|^2\right) = |m_I|^2 - 3 |m|^2 >0\,.
\ee
All new contributions have been included by our change of the index. For a successful uplift the supersymmetry breaking terms in $|m_I|^2$ need to be larger than the gravitino mass. 
\\ The holomorphic/anti-holomorphic part of the mass matrix now evaluates to \cite{Denef:2004ze,Kallosh:2014oja}:
\be 
V_{a\bar{b}}^{\text{dS}}=m_{aI} K^{I\bar{J}} \bar{m}_{\bar{J}\bar{b}} - 2 K_{a\bar{b}}m_I \bar{m}^I - R_{a\bar{b}I\bar{I}} \bar{m}^I m^{\bar{I}} - m_a \bar{m}_{\bar{b}}\,.
\label{eq:dSmassmatrix}
\ee
Here, we have not considered the $X$ direction in moduli space since they are fundamentally fermionic in nature. Furthermore, $R_{a\bar{b}I\bar{I}}$ is the moduli space curvature tensor. The holomorphic/holomorphic part of the mass matrix in de Sitter reads:
\be 
V_{ab}^{\text{dS}} = -m_{ab}\bar{m} + m_{abI} \bar{m}^I\,.
\ee
Once again, we will consider the mass change due to the introduction of the new contribution in terms of the shift of the minimum. For this we consider the AdS minimum at $z^{a,\,\text{AdS}} = z^{a,\,\text{AdS}}_{r} + \rmi z^{a,\,\text{AdS}}_{i}$ and then the change will be due to the uplifting contribution in the scalar potential:
\be 
V^{\text{up}} = \mu^4 \rme^K K_{X\bar{X}}\,.
\ee
The de Sitter minimum is located at:
\be 
\partial_{z^ a_{\alpha}} \left[V^{\text{AdS}} + V^{\text{up}}\right] =0\,,
\ee
where $\alpha = \{i,\,r\}$. Using our knowledge about the AdS minimum and the uplift potential we find for the two terms individually:

\be 
\partial_{z^a_{\alpha}} V^{\text{AdS}}=\left(\partial_{z^a_\alpha } \partial_{z^b_\beta} V^{\text{AdS}} \right) \delta z^b_\beta\,, \qquad \partial_{z^a_{\alpha}} V^{\text{up}} = \mu^4 \partial_{z^a_\alpha} \left(\rme^K K_{X\bar{X}}\right)\,.
\ee
With this we can express the shift of the minimum, when going from AdS to dS as:
\be 
\delta z^b_\beta = -\left( \partial_{z_\alpha^a}\partial_{z^b_\beta} V^{\text{AdS}}\right)^{-1} \mu^4 \partial_{z^a_\alpha} \left(\rme^K K_{X\bar{X}}\right)\,.
\ee
Since the uplift is of the same order as the downshift we know that the anti-de Sitter masses $\partial_{z_\alpha^a}\partial_{z^b_\beta} V^{\text{AdS}}$ are larger than the scale of the uplift and thus the shift of the position of the minimum will likewise be small. Using this result we furthermore find that the amount of supersymmetry breaking in the direction of the unconstrained moduli is small compared to the nilpotent one \cite{Linde:2011ja}:
\be 
|m_a|^2 \ll |m_X|^2\,.
\ee
For stability of the de Sitter minimum we require the amount of supersymmetry breaking to be small compared to the chiral masses. We thus have, together with our earlier requirement that the gravitino mass is parametrically small, for a positive value of the potential at the minimum the following conditions:
\be 
m_\chi \gg |m_I|^2\qquad m_\chi^2 \gg m^2_{3/2} \qquad |m_I|^2 > 3 m_{3/2}^2\,,
\label{eq:dSstabcond}
\ee
where $m_\chi$ is once again the smallest eigenvalue of the mass matrix. As observed before, supersymmetry breaking in the chiral directions is very small and thus SUSY breaking in the nilpotent direction is of the order of the gravitino mass for an almost vanishing cosmological constant.\\
Finally, we have to consider the mass matrix \eqref{eq:dSmassmatrix}. The first term is positive definite and will be strictly positive if the progenitor Minkowski space has no flat directions. Because of the conditions in \eqref{eq:dSstabcond}, all other terms in \eqref{eq:dSmassmatrix} are parametrically small and thus we conclude that the de Sitter mass matrix will be positive definite, or in the case of a Minkowski progenitor without flat directions, it will have strictly  positive eigenvalues. With this we have shown how to obtain a (meta-) stable dS vacuum from our mass production procedure.

\subsection{Model Choices in Type IIA and IIB}
The discussion up until this point has been about fairly general Kähler- and superpotentials. By considering specific choices we can find stronger bounds on the masses in our de Sitter model. In the following sections we will show explicit examples in detail that are based on the choices of $K$ and $W$ we discuss in this section. 
\paragraph{For type IIA} we consider models based on the superpotentials that were already studied in \cite{Kallosh:2019zgd}:
\bea 
K &= - \sum_{I=1}^n N_I \log \left( -\rmi (\Phi^I - \bar{\Phi}^I )\right)\,,\\
W &= W_0 + \sum_{I=1}^n \left( A_I \rme^{\rmi a_I \Phi^I} - B_I \rme^{\rmi b_I \Phi^I} \right)\,.
\label{eq:IIAKW}
\eea 
Here, $\Phi^i$ runs over all the moduli we are considering and $W_0$, $a_I$, $b_I$, $A_I$, $B_I$ and $N_I$ are real parameters. For this class of models we find that the mass matrix is diagonal at the Minkowski minimum when splitting the moduli as $\Phi^I = \phi^I + \rmi \theta^I$ with the $\phi^I$ and $\theta^I$ real:
\be 
\begin{pmatrix}
V_{\phi^I\phi^J} & V_{\phi^I \theta^J}\\
V_{\theta^I\phi^J} & V_{\theta^I \theta^J}
\end{pmatrix}\Bigg|_{\text{Mink}}
=
\begin{pmatrix}
m^2_{\phi^I\phi^I} & 0\\
0 & m^2_{\theta^I \theta^I}
\end{pmatrix}\,.
\ee
Furthermore, there is a mass degeneracy between the scalars and pseudo-scalars:
\be 
m^2_{\phi^I\phi^I} = m^2_{\theta^I \theta^I}\,.
\ee
When perturbing the model in order to go to an anti-de Sitter minimum by a small $\Delta W$, the mass matrix will remain block-diagonal:
\be 
\begin{pmatrix}
V_{\phi^I\phi^J} & V_{\phi^I \theta^J}\\
V_{\theta^I\phi^J} & V_{\theta^I \theta^J}
\end{pmatrix}\Bigg|_{\text{AdS}}
=
\begin{pmatrix}
V^{\text{Mink}}_{\phi^I\phi^J} + \Delta_{\phi^I \phi^J}& 0\\
0 & V^{\text{Mink}}_{\theta^I \theta^J}+\Delta_{\theta^I \theta^J}
\end{pmatrix}\,.
\ee
Note that the terms $\Delta_{\phi^I \phi^J}$ and $\Delta_{\theta^I \theta^J}$ can appear in the off-diagonals of the blocks but, as per our previous analysis, they are parametrically small compared to the diagonal terms. Now, if we introduce the uplift, as discussed above, we will change the mass matrix one more time, however, by another parametrically small amount:

\be 
\begin{pmatrix}
V_{\phi^I\phi^J} & V_{\phi^I \theta^J}\\
V_{\theta^I\phi^J} & V_{\theta^I \theta^J}
\end{pmatrix}\Bigg|_{\text{dS}}
=
\begin{pmatrix}
V^{\text{Mink}}_{\phi^I\phi^J} + \Delta_{\phi^I \phi^J}+ \tilde{\Delta}_{\phi^I \phi^J}& 0\\
0 & V^{\text{Mink}}_{\theta^I \theta^J}+\Delta_{\theta^I \theta^J}+\tilde{\Delta}_{\theta^I \theta^J}
\end{pmatrix}\,.
\ee
Since all the contributions to the mass matrix during the mass production procedure are small we conclude that we should find $V^{\text{dS}}_{IJ} \approx V^{\text{Mink}}_{IJ}$, where $I$ and $J$ symbolically stand for the fields $\phi^I$ and $\theta^I$. Furthermore, due to the the mass degeneracy in Minkowski space we also expect that in AdS and dS at least an approximate mass degeneracy is present.
\paragraph{In type IIB} the superpotential will be the same as in the type IIA case \eqref{eq:IIAKW}. The Kähler potential, on the other hand, will differ and in general be of the form:
\be 
K = K\left(-\rmi ( \Phi^I - \bar{\Phi}^I )\right)\,.
\ee 
One immediate consequence of this is that already the Minkowski mass matrix is now only block-diagonal, instead of diagonal:
\be 
V_{IJ}|_{\text{Mink}}=
\begin{pmatrix}
V_{\phi^I\phi^J} & 0\\
0 & V_{\theta^I \theta^J}
\end{pmatrix}\,.
\ee
Still, due to \eqref{eq:MinkPosDef}, this matrix is positive definite and there will be still a mass degeneracy between the scalars and pseudo-scalars. Nevertheless, the same arguments as in type IIA hold and the changes in the masses in the process of going from Minkowski space to de Sitter will be small.

\subsection{Explicit Examples in Type IIA}
\label{sec:massIIAexampel}
In this section two explicit examples in type IIA will be presented. We will discuss the seven moduli example and its simplification, the STU model. More details can be found in \cite{Cribiori:2019drf}, from which the numerical data and plots were taken as well.
\paragraph{The seven-moduli model in IIA} exhibits an $SL[(2,\mathbb{R})]^7$ symmetry and is closely related to M-theory models and $d=4$, $\mathcal{N}=8$ supergravity. These models were constructed by compactifying from $10$ to $4$ on $T^6/(\mathbb{Z}_2 \times \mathbb{Z}_2)$ in \cite{Derendinger:2004jn,Villadoro:2005cu} and are in particular interesting for future B-mode experiments as discussed in \cite{Ferrara:2016fwe,Kallosh:2017ced}. The model includes the seven moduli:
\be 
\Phi^I = \{S,T_1,T_2,T_3,U_1,U_2,U_3\}\,,
\ee
where $S$ is the axio-dilaton, the $T_i$ ($i=1,2,3$) are the complex structure moduli and the $U_i$ are the Kähler moduli. The $\Phi^I$, in a sense, can be viewed as the coordinates of the $[SL(2,\mathbb{R})/U(1)]^7$ coset. The Kähler- and superpotential for our example are:
\bea 
\label{eq:massIIA7modpot}
K &= - \sum_I \log \left(-\rmi (\Phi^I -\bar{\Phi}^I)\right)\,,\\
W &= f_6 + \sum_I \left( A_I \rme^{\rmi a_I \Phi^I} - B_I \rme^{\rmi b_I \Phi^I} \right)\,,
\eea 
where we will utilize the Kallosh-Linde racetrack superpotential. Once again, we do not consider $p$-fluxes other than $F_6$. The shift to anti-de Sitter is performed by letting $f_6 \to f_6 + \Delta f_6$. The contribution of the uplift to the scalar potential due to the anti-$D6$-brane uplift is:
\bea
V_{\overline{D6}}^{\text{up}} =\; & \frac{\mu_0^4}{\rmim(T_1)\rmim(T_2)\rmim(T_3)} + \frac{\mu_1^4}{\rmim(S)\rmim(T_2)\rmim(T_3)} \\+ & \frac{\mu_2^4}{\rmim(S)\rmim(T_1)\rmim(T_3)} + \frac{\mu_3^4}{\rmim(S)\rmim(T_1)\rmim(T_2)}\,,
\eea
which can, again, can be obtained using a nilpotent field $X$ \cite{Kallosh:2018nrk,Cribiori:2019bfx}. As is evident, we have now several cycles around which the anti-branes can be wrapped and the parameters $\mu_i^4$ can, but do not necessarily need to be, tuned individually.\\
In order to find solutions to the equations $D_I W =0$ and $W=0$, which will give a Minkowski minimum, we use the parameters $B_I$ and $f_6$. Thus, we can freely choose the point of the extremum in moduli space and the parameters $A_I$, $a_I$ and $b_I$. The choices are given in table \ref{tab:7modchoice}. Once again, we set the axions to zero, which will not produce any problems as discussed in section \ref{sec:uplfitbraneexamples}. The position of the Minkowski minimum will thus be at $S=\rmi \,S_0$, $T_i = \rmi \,T_{i,0}$ and $U_i = \rmi \,U_{i,0}$ where $i=1,2,3$.\\
In choosing the parameters we had to keep the same requirements in mind as in section \ref{sec:antibraneupSTU}.  Namely, $a_I \Phi^I$ needs to be small such that higher-order corrections can safely be neglected and we choose the $U_{i,0}$ large in order to obtain a large internal volume. Other than that, no particular care is required in choosing the values found in \ref{tab:7modchoice}. Even with these conditions it is relatively easy to find positive masses, meaning that the parameters are not particularly fine-tuned. For the downshift parameter we choose:
\be 
\Delta f_6 = -10^{-5}\,.
\ee 
The sign of this parameter does matter and can change the behavior but, for small downshifts, the qualitative effect will be independent of the sign. For the uplift we set all the parameters to be equal:
\be 
\mu_0 ^4 = \mu_1 ^4 = \mu_2 ^4 = \mu_3 ^4 = 5.49028 \cdot 10^{-15}\,.
\ee
\begin{table}[H]
\centering
\begin{tabular}{|c|c|c|c|c|c|c|}\hline
$A_S = 1$ & $A_{T_1} = 3.1$& $A_{T_2} = 3.2$& $A_{T_3} = 3.3$ & $A_{U_1} =11$& $A_{U_2} =12$& $A_{U_3} =13$\\\hline
$a_S = 2$ & $a_{T_1} = 2.1$& $a_{T_2} = 2.2$& $a_{T_3} = 2.3$ & $a_{U_1} =0.41$& $a_{U_2} =0.42$& $a_{U_3} =0.43$\\\hline
$b_S = 3$ & $b_{T_1} = 3.1$& $b_{T_2} = 3.2$& $b_{T_3} = 3.3$ & $b_{U_1} =1.1$& $b_{U_2} =1.2$& $b_{U_3} =1.3$\\\hline
$S_0 = 1$ & $T_{1,\,0} = 1.1$  & $T_{2,\,0} = 1.2$  & $T_{3,\,0} = 1.3$ & $U_{1,\,0} = 5.1$& $U_{2,\,0} = 5.2$& $U_{3,\,0} = 5.3$\\\hline
\end{tabular}
\caption{Choices for the position in moduli space and parameters in the 7-moduli example in type IIA.}
\label{tab:7modchoice}
\end{table}
The value chosen here, and in the following, have not been selected to match the observed cosmological constant but rather for illustrative purposes. With the choices in table \ref{tab:7modchoice} one obtains a Minkowski minimum with strictly positive values for the masses and thus, as by our prior discussion, the same should hold for the de Sitter masses. We give both, the Minkowski and de Sitter, masses in table \ref{tab:7modmass}. The change in masses is noticeable but small, as predicted. Furthermore, the mass degeneracy is  broken in de Sitter, to a minute degree. In addition to its eigenvalues, below the upper right corner block of the dS mass matrix is shown. It is evident that the off-diagonal terms are much smaller than the diagonal entries, visualizing our discussion from section \ref{sec:massmass}.\\
Due to the sheer amount of plots necessary do properly represent the 7-moduli example we instead will use the related STU-model in order  to visualize the behavior.

{\tiny
\begin{equation}
 \left(
\begin{array}{ccccccc}
1.89809 \cdot 10^{-5} & -6.19837 \cdot 10^{-10} &- 5.36624 \cdot 10^{-10}& -4.56280 \cdot 10^{-10} & -1.30375 \cdot 10^{-9}& -1.52470 \cdot 10^{-9} & -1.74268 \cdot 10^{-9}\\
\\
-6.19837 \cdot 10^{-10} & 1.30911 \cdot 10^{-4} & -7.38721 \cdot 10^{-10} & -6.46383 \cdot 10^{-10} & -1.24516 \cdot 10^{-9} & -1.44398 \cdot 10^{-9} & -1.64104 \cdot 10^{-9}\\
\\
-5.36624 \cdot 10^{-10} & -7.38721 \cdot 10^{-10} & 9.41667 \cdot 10^{-5} & -5.68241 \cdot 10^{-10} & -1.13520 \cdot 10^{-9} &-1.31758 \cdot 10^{-9} &-1.49834 \cdot 10^{-9}\\
\\
-4.56280 \cdot 10^{-10} & -6.46383 \cdot 10^{-10} & -5.68241 \cdot 10^{-10} &6.37888 \cdot 10^{-5}&-1.04022 \cdot 10^{-9}& -1.20871 \cdot 10^{-9}&-1.37571 \cdot 10^{-9}\\
\\
-1.30475 \cdot 10^{-9} & -1.24516 \cdot 10^{-9} & -1.13520 \cdot 10^{-9} & -1.04022 \cdot 10^{-9} & 9.96472 \cdot 10^{-4} &-5.36645 \cdot 10^{-10} & -5.74900 \cdot 10^{-10}\\
\\
-1.52470 \cdot 10^{-9}&-1.44398 \cdot 10^{-9}& -1.31758 \cdot 10^{-9}& -1.20871 \cdot 10^{-9 }&-5.36646 \cdot 10^{-10} & 1.37262 \cdot 10^{-3}& -6.10079 \cdot 10^{-10}\\
\\
-1.74268 \cdot 10^{-9}& -1.64104 \cdot 10^{-9}& -1.49834 \cdot 10^{-9} & -1.37571 \cdot 10^{-9}& -5.74900 \cdot 10^{-10} &-6.10079 \cdot 10^{-10}&1.80465 \cdot 10^{-3}
\end{array}
\right)\nonumber
\end{equation}
}

\begin{table}[htb]
\centering
\begin{tabular}{|c|c|c|c|}\hline
& Minkowski  & de Sitter \\\hline
$m_1^{\,2}$   & $\;1.80473 \cdot 10^{-3}\;$  & $\;1.80465 \cdot 10^{-3}\;$\\\hline
$m_2^{\,2}$   & $1.80473 \cdot 10^{-3}$  & $1.80465 \cdot 10^{-3}$\\\hline
$m_3^{\,2}$   & $1.37269 \cdot 10^{-3}$  & $1.37262 \cdot 10^{-3}$\\\hline
$m_4^{\,2}$   & $1.37269 \cdot 10^{-3}$  & $1.37262 \cdot 10^{-3}$\\\hline
$m_5^{\,2}$   & $9.96519 \cdot 10^{-4}$  & $9.96472 \cdot 10^{-4}$\\\hline
$m_6^{\,2}$   & $9.96519 \cdot 10^{-4}$  & $9.96471 \cdot 10^{-4}$\\\hline
$m_7^{\,2}$   & $1.30924 \cdot 10^{-4}$  & $1.30911 \cdot 10^{-4}$\\\hline
$m_8^{\,2}$   & $1.30924 \cdot 10^{-4}$  & $1.30911 \cdot 10^{-4}$\\\hline
$m_9^{\,2}$   & $9.41773 \cdot 10^{-5}$  & $9.41667 \cdot 10^{-5}$\\\hline
$m_{10}^{\,2}$& $9.41773 \cdot 10^{-5}$  & $9.41660 \cdot 10^{-5}$\\\hline
$m_{11}^{\,2}$& $6.37973 \cdot 10^{-5}$  & $6.37888 \cdot 10^{-5}$\\\hline
$m_{12}^{\,2}$& $6.37973 \cdot 10^{-5}$  & $6.37883 \cdot 10^{-5}$\\\hline
$m_{13}^{\,2}$& $1.89843 \cdot 10^{-5}$  & $1.89809 \cdot 10^{-5}$\\\hline
$m_{14}^{\,2}$& $1.89843 \cdot 10^{-5}$  & $1.89806 \cdot 10^{-5}$\\\hline
\end{tabular}
\caption{  The masses in Minkowski and de Sitter for the 7 moduli IIA example. The changes during the procedure are small but noticeable. A very small deviation from the mass degeneracy, originally present in Minkowski space, is visible. }
\label{tab:7modmass}
\end{table}

\begin{figure}[htb]
\center
\includegraphics[scale=0.52]{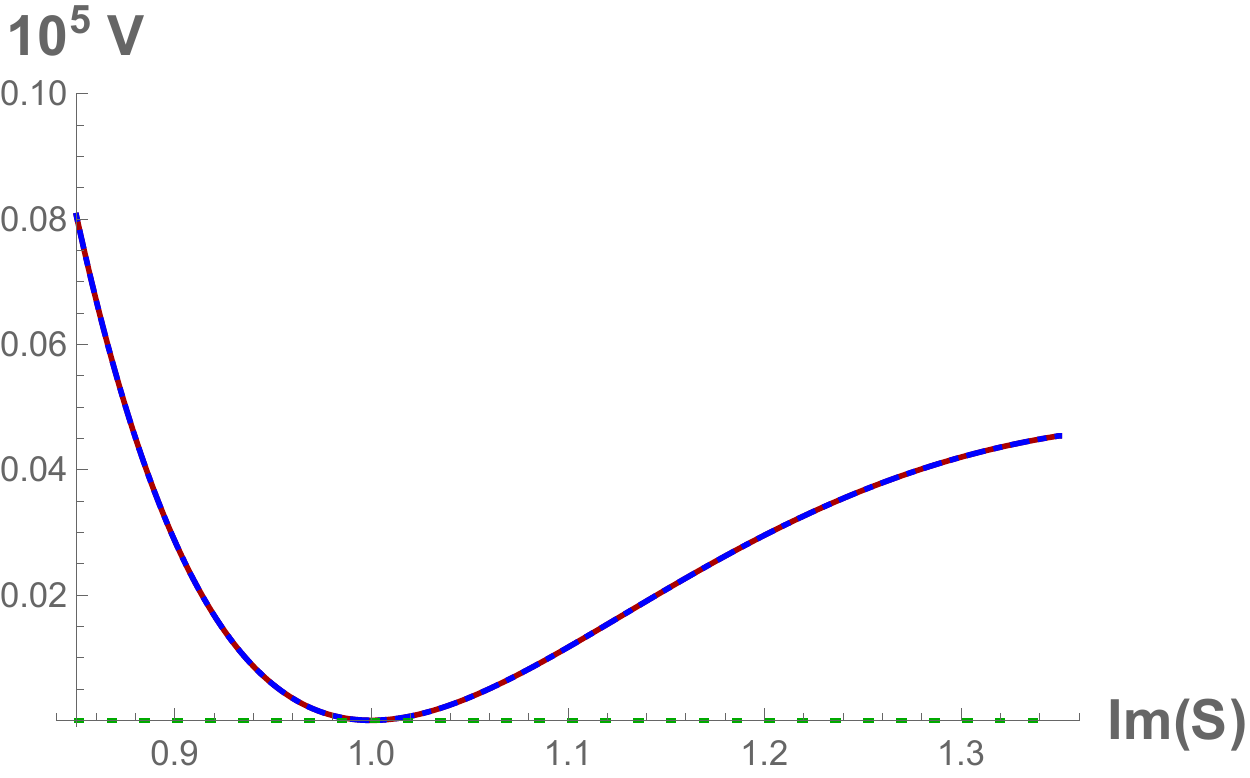}\qquad
\includegraphics[scale=0.52]{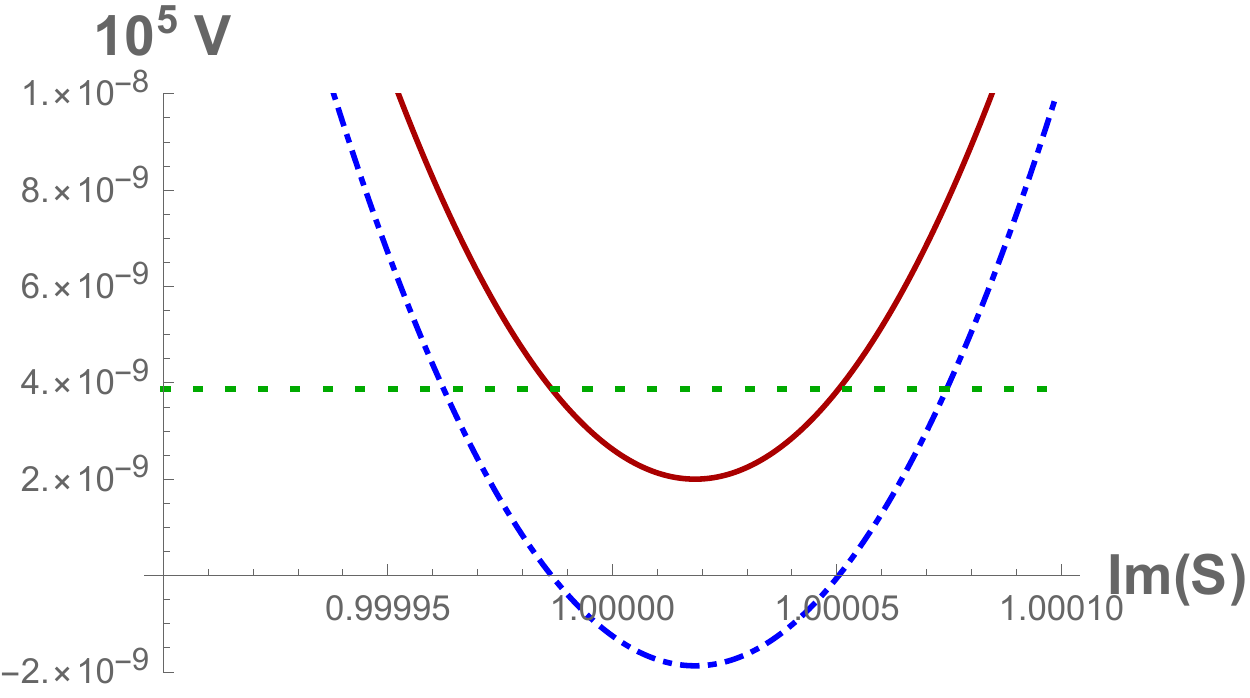}
\caption{ The shift of the position of the minimum in the IIA STU-model when going from AdS to dS for the $\rmim(S)$ direction. On the left the AdS and dS scalar potential visually overlap while the anti-brane contribution seems to be at zero. In the close-up on the right the differences become visible. The anti-$D6$-brane gives a flat contribution in this direction (dotted, green) and it is evident that the position of the minimum moves slightly when going from AdS (dash-dotted, blue) to dS (solid, red).}
\label{fig:stushift}
\end{figure}

\begin{figure}[H]\hskip 2cm
\includegraphics[scale=0.6]{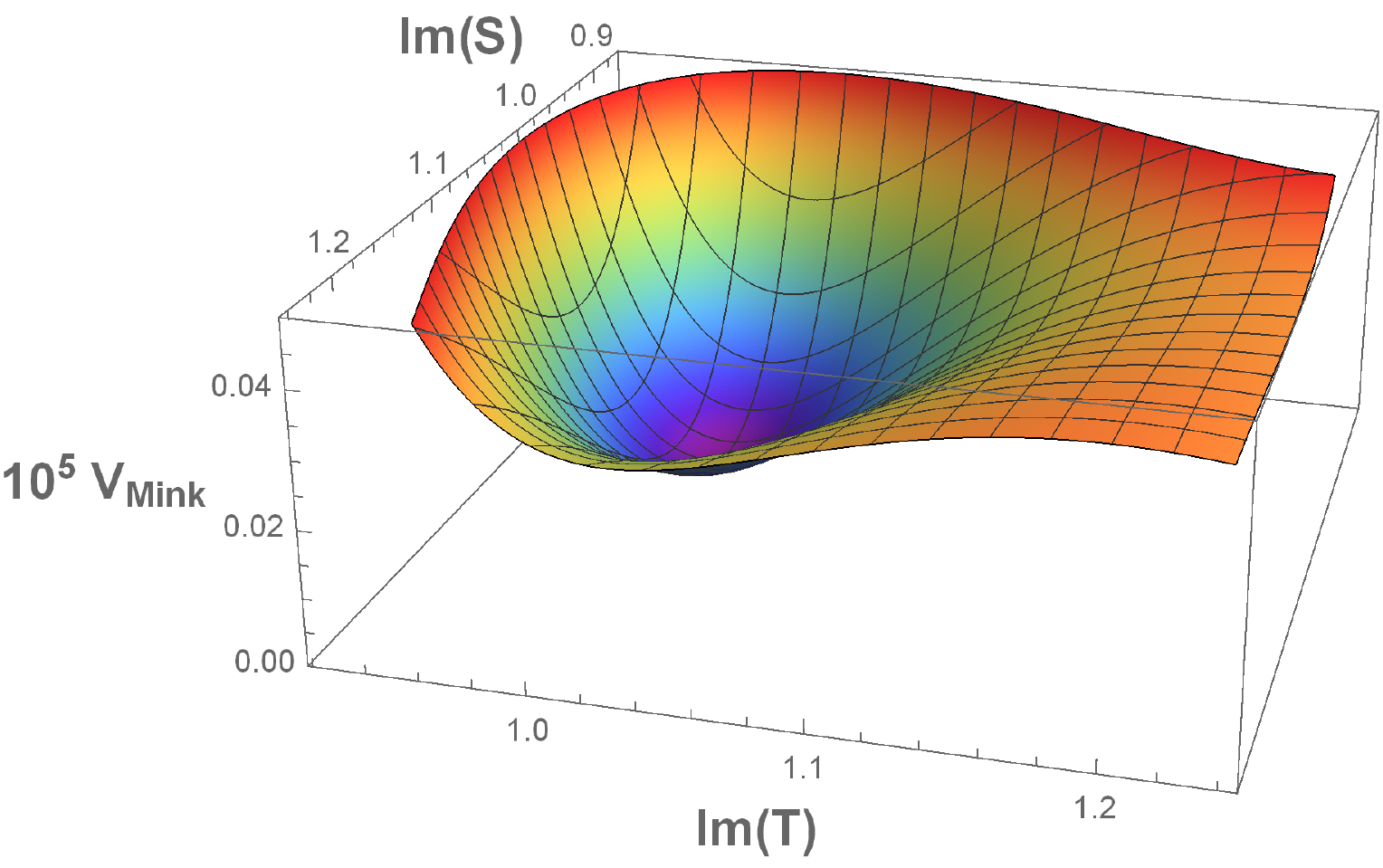}
\caption{ The $\rmim(S)$ and $\rmim(T)$ 3D slice of the scalar potential in the IIA STU-model. The potential shows a meta-stable behavior around the minimum. The form of this  potential and the other possible slices do not change significantly during the 3 steps of the procedure.}
\label{fig:stu3Dlarge}
\end{figure}

\begin{figure}[H]
\includegraphics[scale=0.457]{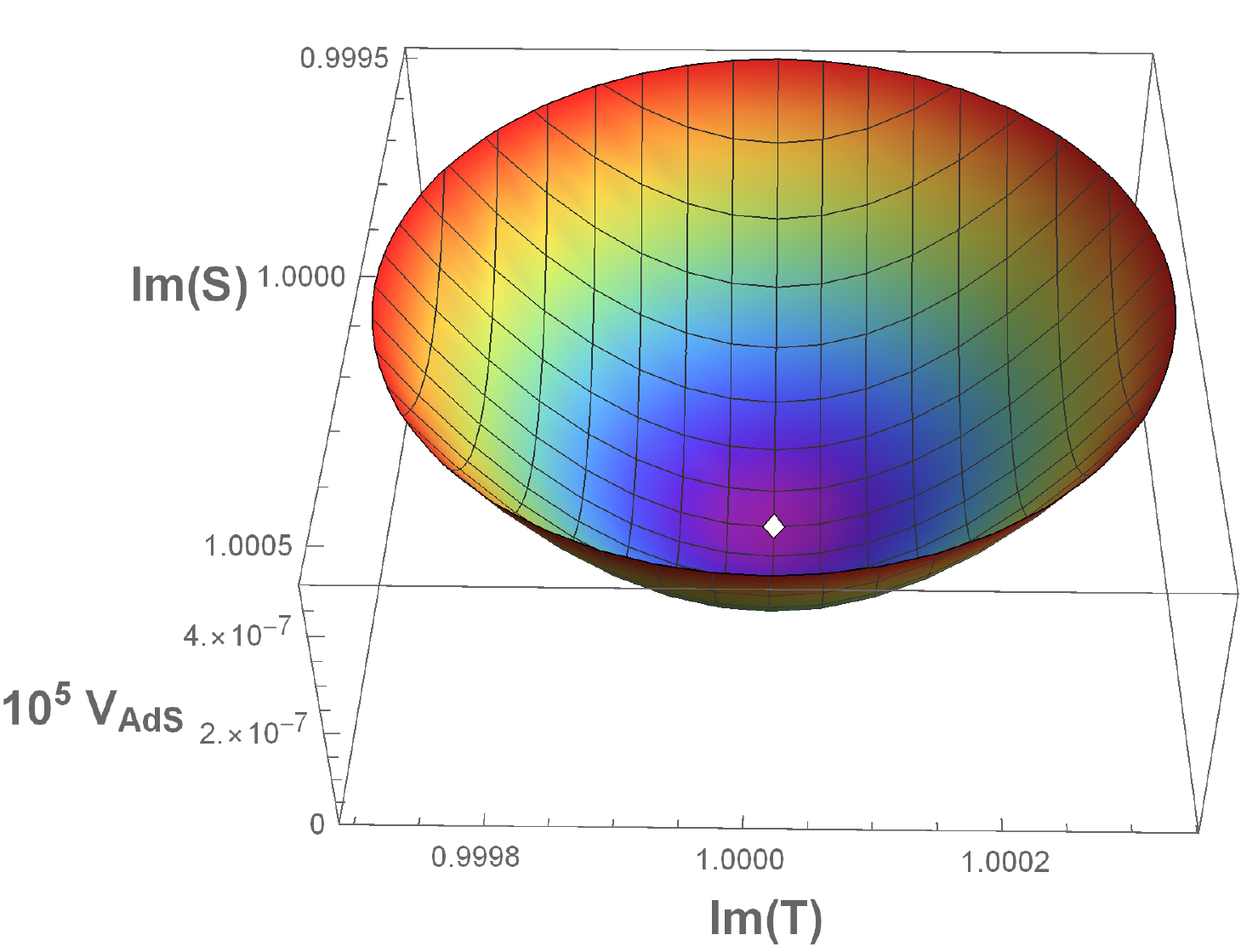} 
\includegraphics[scale=0.46]{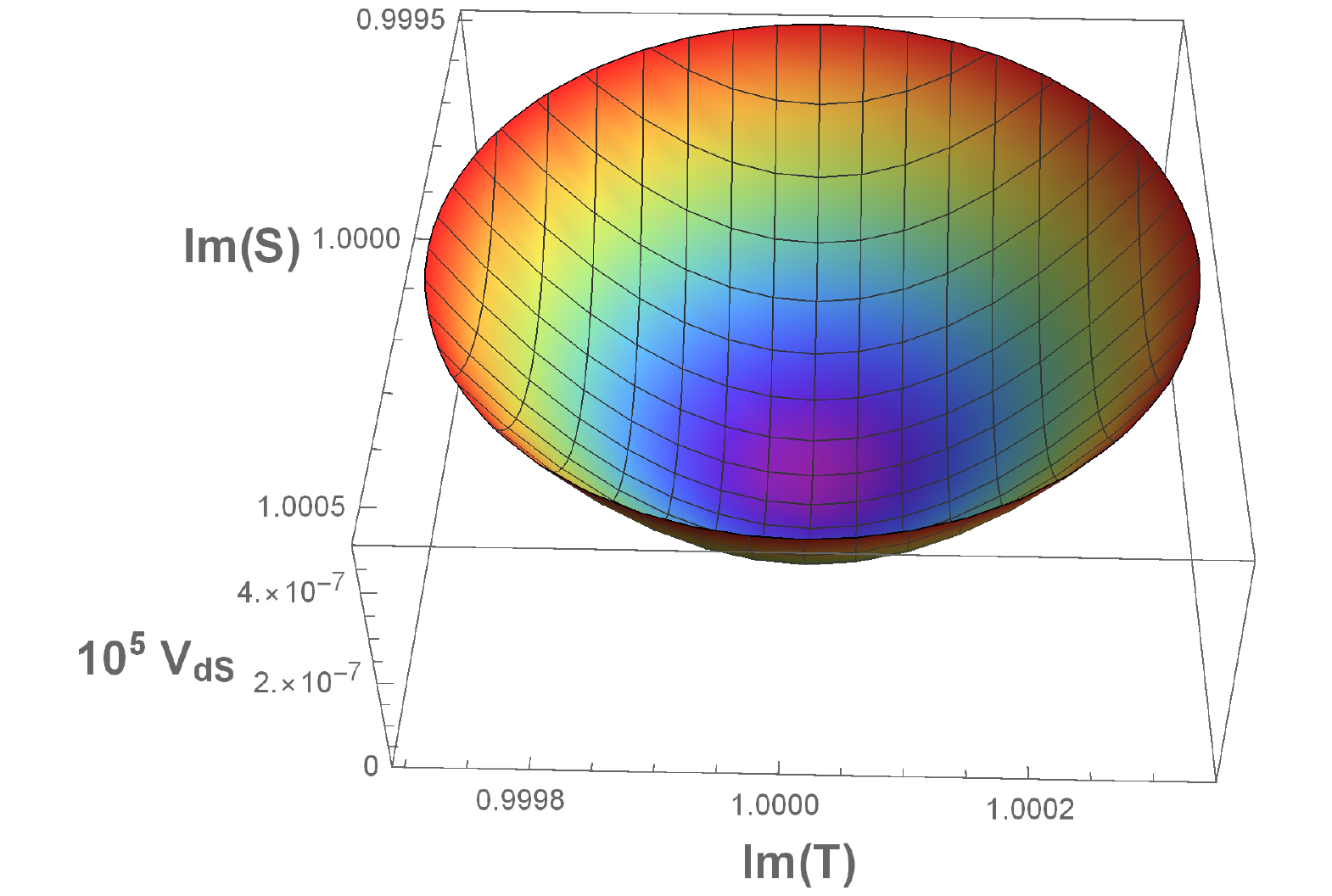}
\caption{  The minimum of the scalar potential in the $\rmim(S)$ and $\rmim(T)$ slice of the IIA STU-model. On the left the parts of the potential with negative values is visible as the hole. On the right the whole of the potential is now positive due to the uplift term. }
\label{fig:stu3Dclose}
\end{figure}

\FloatBarrier
\paragraph{The IIA STU model} is a simplification of the above case where one identifies the different complex structure and Kähler moduli:
\be 
T_i \to T\,, \qquad U_i \to U\,, \qquad \text{for }\,i=1,2,3\,.
\ee
The model is then very similar to the model of section \ref{sec:antibraneupSTU} but with the racetrack superpotential. The Kähler- and superpotential are:
\bea 
K &= -\log \left( -\rmi (S-\bar{S})\right) - 3 \log \left( -\rmi (T-\bar{T})\right)- 3 \log \left( -\rmi (U-\bar{U})\right)\,,\\
W &= f_6 \,+ \sum_{\Phi = S,T,U} \left( A_\Phi \rme^{\rmi a_\Phi \Phi} - B_\Phi \rme^{\rmi b_\Phi \Phi}\right)\,.
\eea
The uplift contribution to the scalar potential in this model likewise simplifies to
\be 
V_{\overline{D6}}^{\text{up}} =  \frac{\mu_0^4}{\rmim(T)^3} + \frac{\mu_1^4}{\rmim(S)\rmim(T)^2} \,.
\ee
Finding solutions to this proceeds as before by first setting the axions to zero and then solving in terms of $f_6$ and the $B_I$, while the other parameters remain free and are listed in table \ref{tab:stupara}.
\begin{table}[H]
\centering
\begin{tabular}{|c|c|c|}\hline
$A_S = 1$ & $A_T = 3$ & $A_U =11$\\\hline
$a_S = 2$ & $a_T = 2.1$ & $a_U = 1$\\\hline
$b_S = 3$ & $b_T = 3.1$ & $b_U = 1.2$\\\hline
$S_0 = 1$ & $T_0 = 1$ & $U_0 = 5$\\\hline
\end{tabular}
\caption{   Our choice of parameters for the 3 moduli IIA example.}
\label{tab:stupara}
\end{table}
The downshift, once again, is:
\be 
\Delta f_6 = -10^{-5}\,,
\ee 
while the uplift is:
\be 
\mu_0^4 = \mu_1^4 = 1.93753 \cdot 10^{-14}\,.
\ee
The resulting masses are given in table \ref{tab:stumass}, where we see once more that the masses change only slightly and the mass degeneracy breaks in de Sitter space.
\begin{table}[htb]
\centering
\begin{tabular}{|c|c|c|c|}\hline
&  Minkowski  & de Sitter \\\hline
$m_1^{\,2}$ & $\; 9.91957 \cdot 10^{-5} \;$ & $\; 9.91570 \cdot 10^{-5} \;$ \\\hline
$m_2^{\,2}$ & $\; 9.91957 \cdot 10^{-5} \;$ & $\; 9.91551 \cdot 10^{-5} \;$ \\\hline
$m_3^{\,2}$ & $\; 3.66313 \cdot 10^{-5} \;$ & $\; 3.66189 \cdot 10^{-5} \;$ \\\hline
$m_4^{\,2}$ & $\; 3.66313 \cdot 10^{-5} \;$ & $\; 3.66183 \cdot 10^{-5} \;$ \\\hline
$m_5^{\,2}$ & $\; 9.15565 \cdot 10^{-7} \;$ & $\; 9.14587 \cdot 10^{-7} \;$ \\\hline
$m_6^{\,2}$ & $\; 9.15565 \cdot 10^{-7} \;$ & $\; 9.14550 \cdot 10^{-7} \;$ \\\hline
\end{tabular}
\caption{  Values for the masses in Minkowski and de Sitter for the STU-model. As in the 7-moduli case the change in  going from Minkowski to de Sitter is evident as well as the breaking of the mass degeneracy.}
\label{tab:stumass}
\end{table} For this model we illustrate the behavior via some exemplary plots. In figure \ref{fig:stushift} one can see that the position of the minimum shifts slightly under the procedure. In figure \ref{fig:stu3Dlarge} the form of the potential for the $\rmim(S)$ and $\rmim(T)$ directions is shown. The meta-stability of the minimum in these directions is evident. Similar plots can be obtained for the other directions. Finally, in figure \ref{fig:stu3Dclose} a close-up of the minimum, both in AdS and dS, for the same directions is depicted. The hole seen there is the part in AdS with negative values for the potential.

\subsection{Explicit Examples in Type IIB}
In type IIB we focus on models already investigated in \cite{Cicoli:2008va,Cicoli:2008gp,Burgess:2016owb,Kallosh:2017wku,Bobkov:2010rf}\footnote{Note that there is difference in conventions between \cite{Cicoli:2008va,Cicoli:2008gp,Burgess:2016owb,Kallosh:2017wku,Bobkov:2010rf} and \cite{Cribiori:2019drf}. As is discussed in \cite{Cribiori:2019drf} this does not have any physical effects.}  for their applications in cosmology. There, they were studied using either LVS or KKLT type uplifts. The difference for us here is that we use our mass production method with the racetrack superpotential \eqref{eq:racetrack}.\\
Flux compactifications in IIB are slightly different than in IIA. The 4 dimensional $\mathcal{N}=1$ supergravity theory in IIB is given by the Kähler- and superpotential \cite{Grimm:2004uq}:
\bea 
K &= -\log\left(\rmi \int \Omega \wedge \bar{\Omega} \right) - \log\left(-\rmi (\tau -\bar{\tau})\right) - 2 \log \left(\mathcal{V}_6\right)\,,\\
W &= \int G_3 \wedge \Omega\,.
\eea
Here, $\Omega$ is a function of the complex structure moduli, $\tau$ is the axio-dilaton and $G_3$ is a complex 3-form flux. For our purpose in this part we will only focus on the Kähler moduli, meaning that the complex structure moduli and the axio-dilaton are stabilized at an earlier stage which is typical for type IIB compactifications. $\mathcal{V}_6$ is the six dimensional, internal volume given by:
\be 
\mathcal{V}_6 = \frac{1}{3!} \int J \wedge J \wedge J=\frac{1}{3!} r_{ijk}t_i t_j t_k\,.
\ee
The $r_{ijk}$ are the Calabi-Yau intersection numbers and the $t_i$ are volumes of 2-cycles. We eventually have to make the connection to the notation we used up until now. For this we use complexified Kähler moduli:
\be 
T_i = \chi_i + \rmi \tau_i \qquad \text{where} \qquad \tau_i = \frac{\partial \mathcal{V}}{\partial t_i}\,.
\ee
The $\tau_i$ are thus volumes of 4-cycles. This allows us to write $\mathcal{V}_6$ in terms of the 4-cycles.\\
The uplift in type IIB is facilitated via anti-$D3$-branes, like in the KKLT scenario. The brane can be placed either in the bulk of the internal space or at the bottom of a warped throat \cite{Kallosh:2017wku}. Depending on the placement of the brane both the description in terms of the 4 dimensional Kähler potential as well as the effective contribution to the scalar potential change. The Kähler potentials are:
\bea
K_{\text{bulk}} &= -2 \log \left( \mathcal{V}_6 (\tau_i)\right) + X\bar{X}\,,\\
K_{\text{throat}} &= -3 \log \left( \mathcal{V}^ {\frac{2}{3}}_6 (\tau_i) - \frac{1}{3} X\bar{X}\right)\,,
\eea
where $X$ is our familiar nilpotent chiral multiplet. Note that from here on we will use the letters $S$, $T$ and $U$ to label different Kähler moduli, as opposed to type IIA, where they labeled different types of moduli. With this changed convention we can use the racetrack superpotential \eqref{eq:racetrack} to compute the scalar potential and find:
\bea 
V_{\overline{D3},\,\text{bulk}} &= \rme^{K_{\text{bulk}}} D_X W K^{X\bar{X}} \overline{D_X W}\big|_{X=0} = \frac{\mu^4}{\mathcal{V}_6^{\,2}}\,,\\
V_{\overline{D3},\,\text{throat}} &=  \rme^{K_{\text{throat}}} D_X W K^{X\bar{X}} \overline{D_X W}\big|_{X=0} = \frac{\mu^4}{\mathcal{V}_6^{\,4/3}}\,.
\eea
Armed with these formulas we are ready to tackle explicit examples.

\paragraph{The 2 parameter K3-fibration in type IIB} will be our first model that is not set in type IIA. This model depends on two moduli only and the internal volume is given as \cite{Cicoli:2008va}:
\be 
\mathcal{V}_6 (\tau_i) = \frac{1}{2} \sqrt{\tau_1} \left[\tau_2 - \frac{2}{3} \tau_1 \right]\,.
\ee
Importantly, this model does not allow for an LVS-type stabilization \cite{Balasubramanian:2005zx}. In our conventions the above expression translates to:
\be 
\mathcal{V}_6(S,T) = \frac{1}{2} \sqrt{-\rmi (S-\bar{S})} \left[ \left( -\rmi (T-\bar{T}) \right) - \frac{2}{3} \left( -\rmi (S-\bar{S})\right) \right]\,.
\ee
Again, we will use the racetrack type potential \eqref{eq:racetrack}, set the axions to zero and use $S=\rmi S_0$, $T=\rmi T_0$ (and in the following $U=\rmi U_0$) as in the type IIA case. The parameters we pug in are given in table \ref{tab:2modK3para} and for the downshift and uplift parameters we plug in:
\bea 
\Delta W_0 &= -10^{-5}\,,\\
\mu_{\text{bulk}}^4 &= 3.61516 \cdot 10^{-10}\,.
\eea
\begin{table}[htb]
\centering
\begin{tabular}{|c|c|}\hline
$A_S = 1.1$ & $A_T = 1.2$ \\\hline
$a_S = 2.1$ & $a_T = 2.2$ \\\hline
$b_S = 3.1$ & $b_T = 3.2$ \\\hline
$S_0 = 1$ & $T_0 = \pi$ \\\hline
\end{tabular}
\caption{ The chosen parameters for the two-moduli K3 fibration.}
\label{tab:2modK3para}
\end{table}

\begin{figure}[H]
\centering
\includegraphics[scale=0.72]{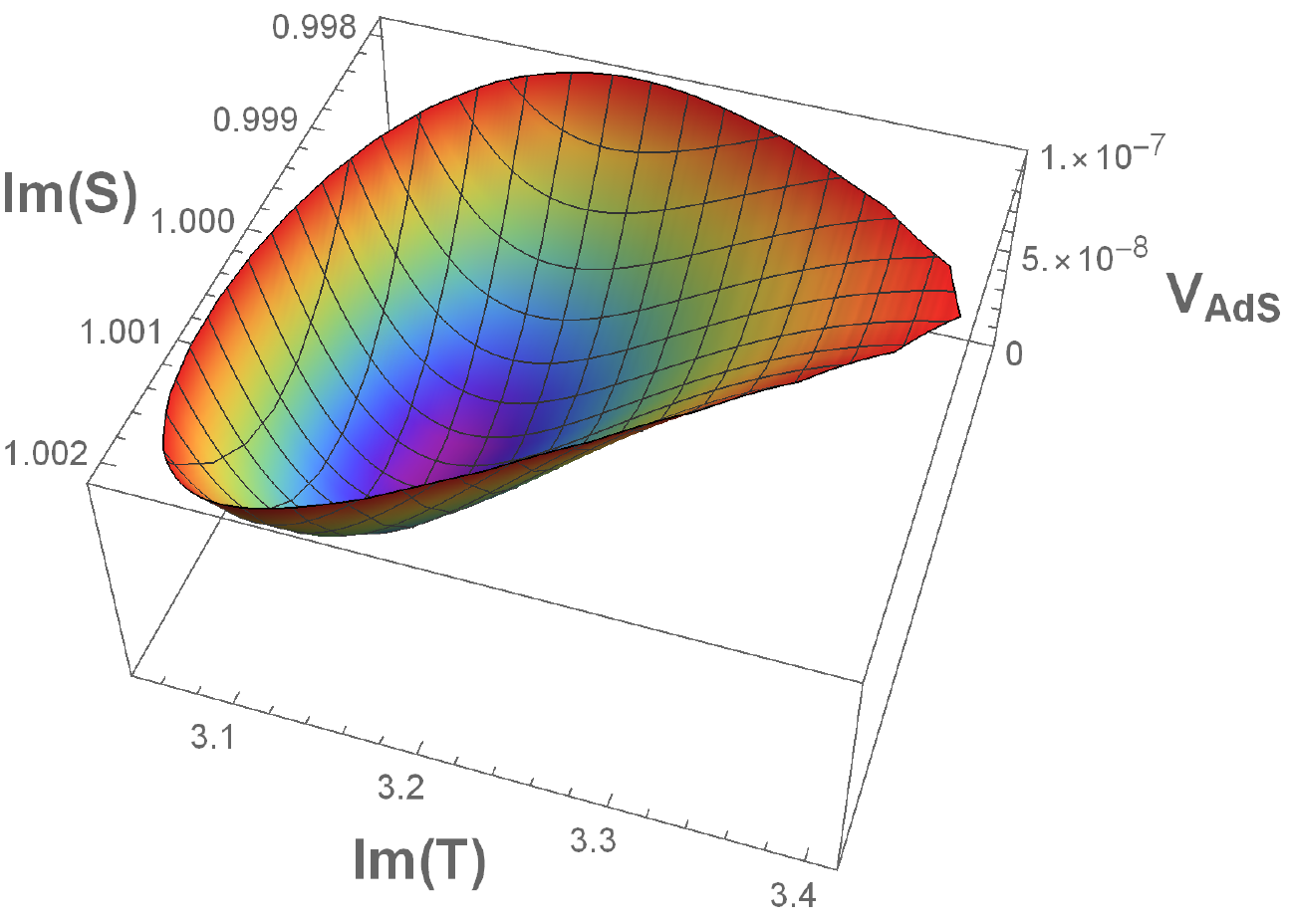}
\caption{The overall shape of the scalar potential around the minimum of the 2-moduli K3 fibration. While stable locally, globally the potential is only meta-stable. At this scale the difference between dS and AdS would not be visible.}
\label{fig:K3large}
\end{figure}

\begin{figure}[H]
\centering
\includegraphics[scale=0.54]{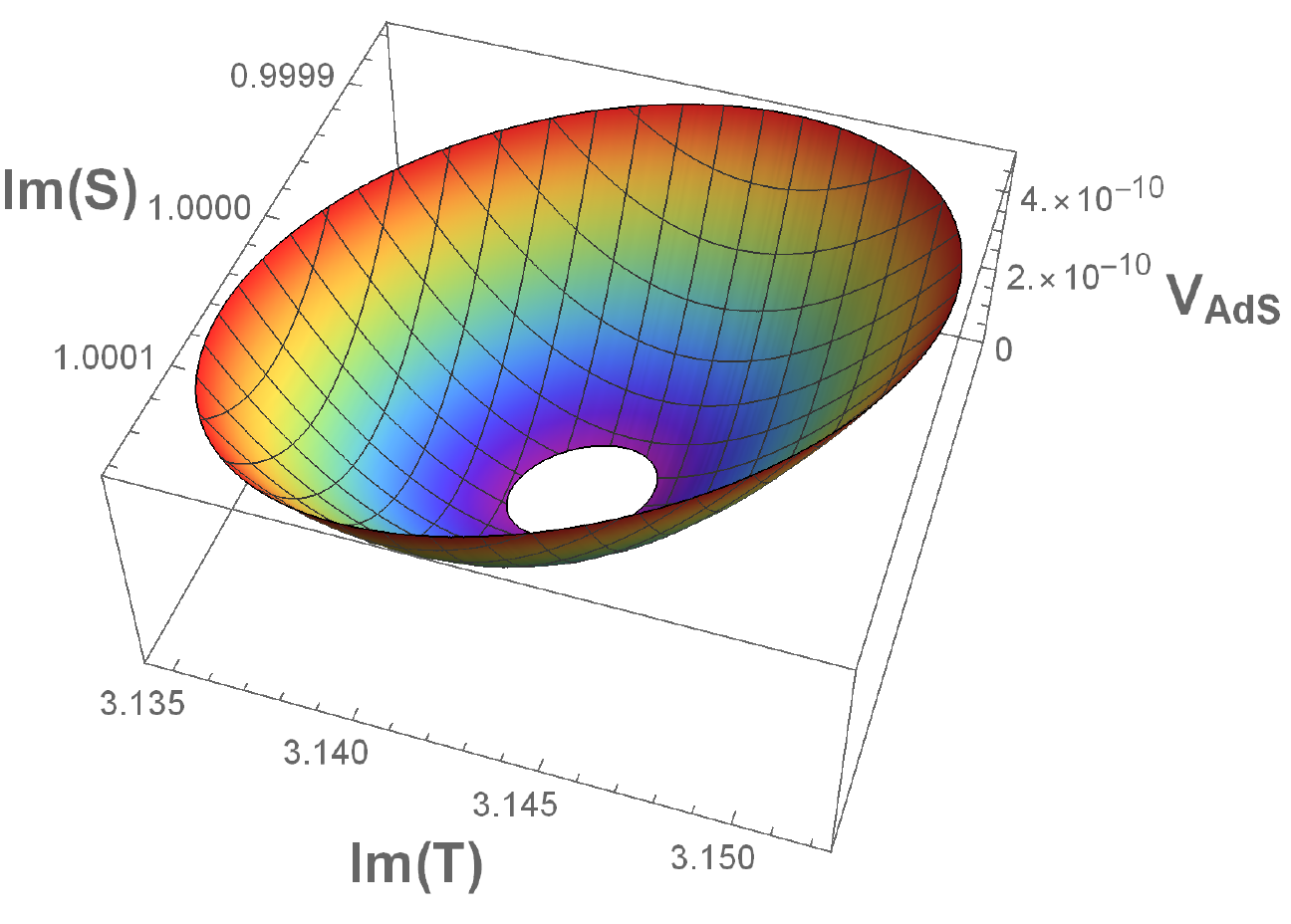} \qquad
\includegraphics[scale=0.52]{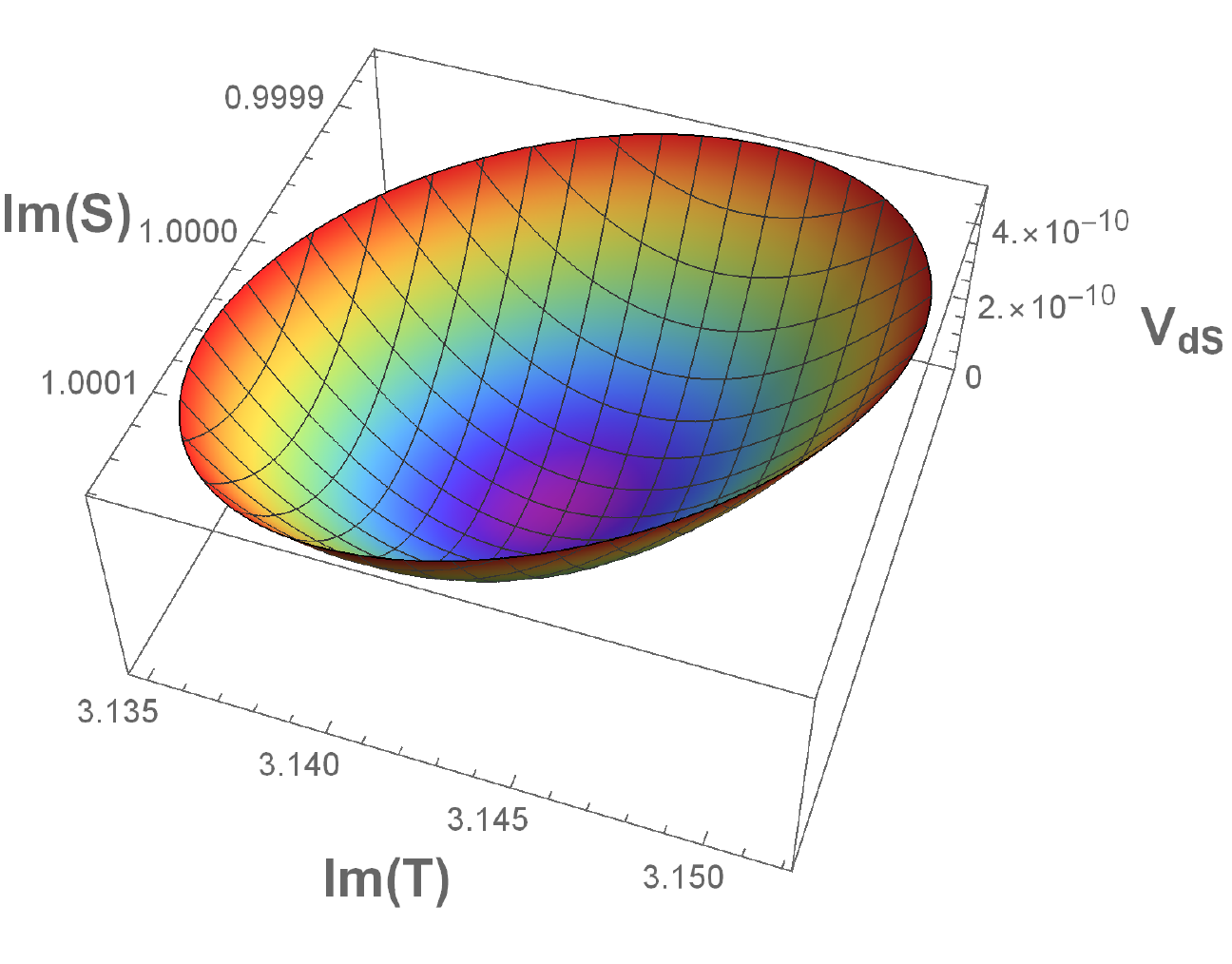}
\caption{Close up of the scalar potential of the K3 fibration around the minimum for AdS (left) and dS (right).}
\label{fig:2modK33d}
\end{figure}

\begin{figure}[H]
     \includegraphics[scale=0.52]{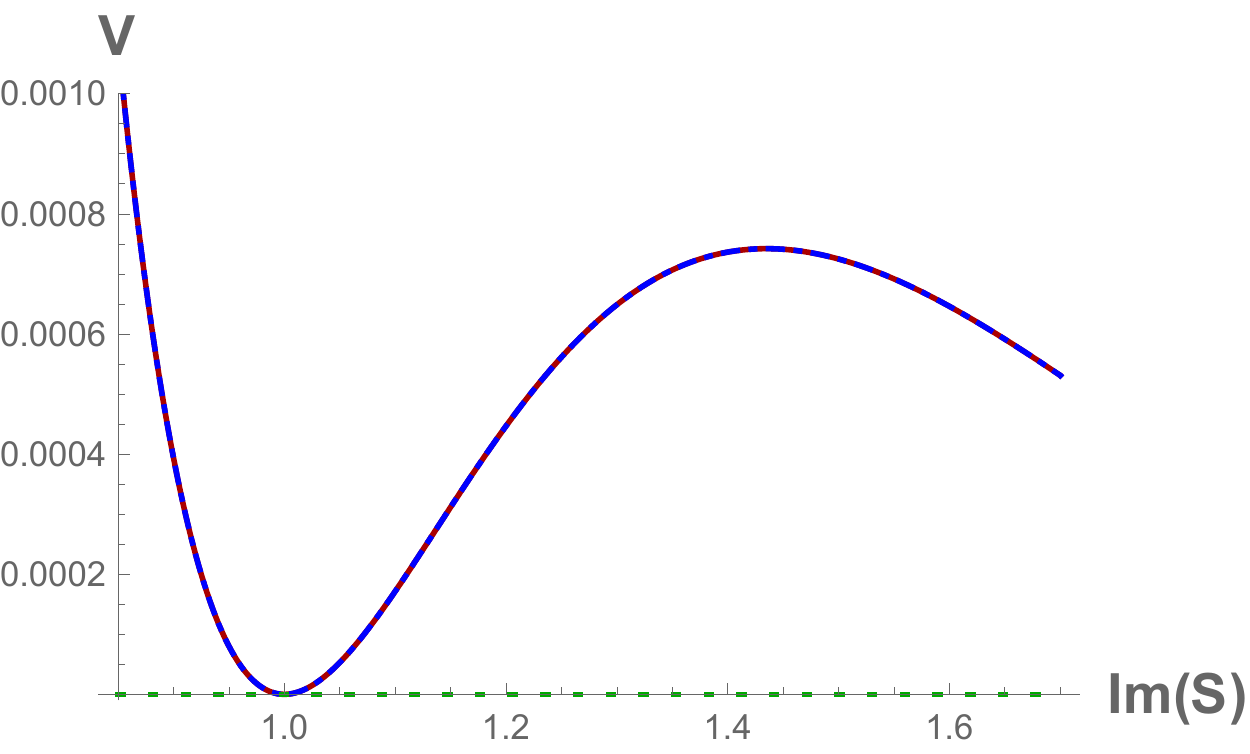}\qquad\includegraphics[scale=0.58]{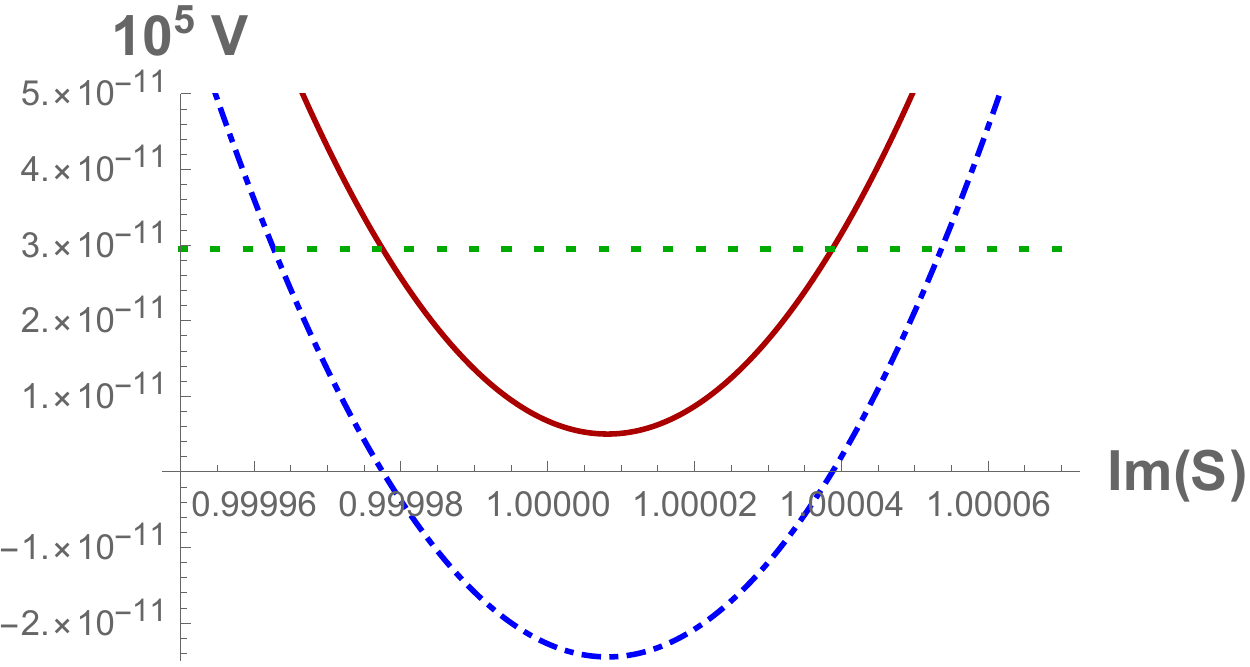}
     \includegraphics[scale=0.52]{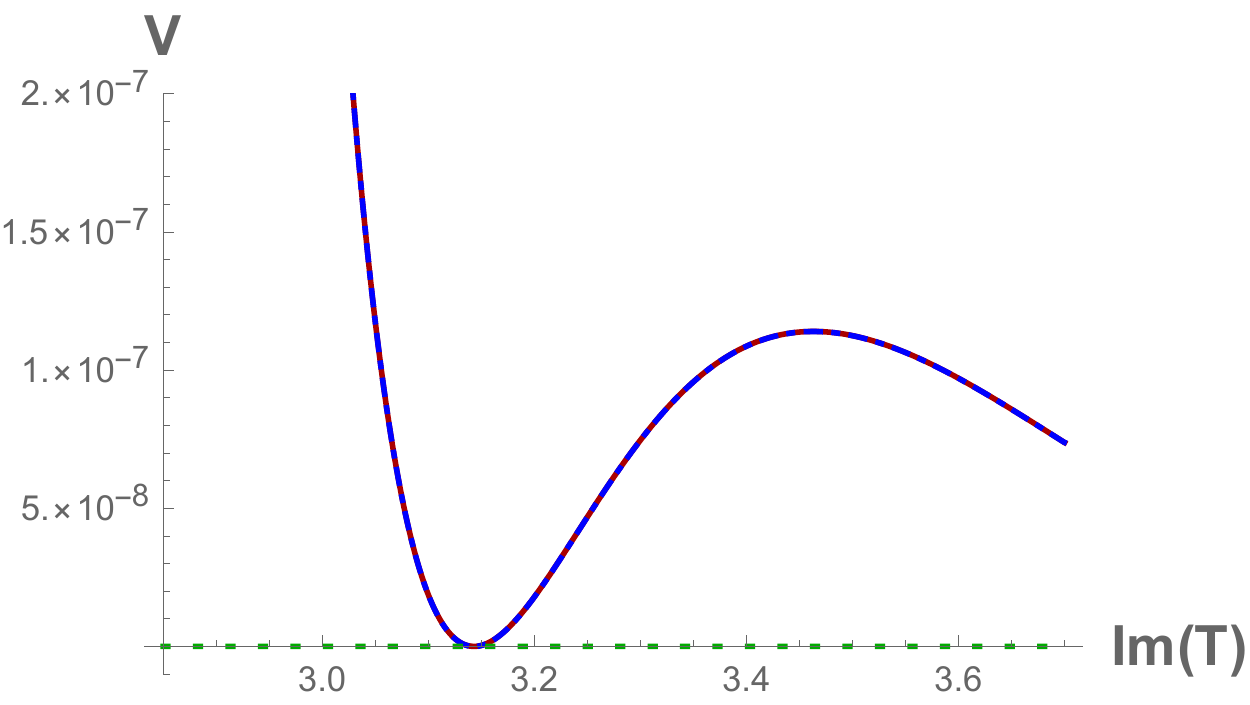}\qquad\includegraphics[scale=0.58]{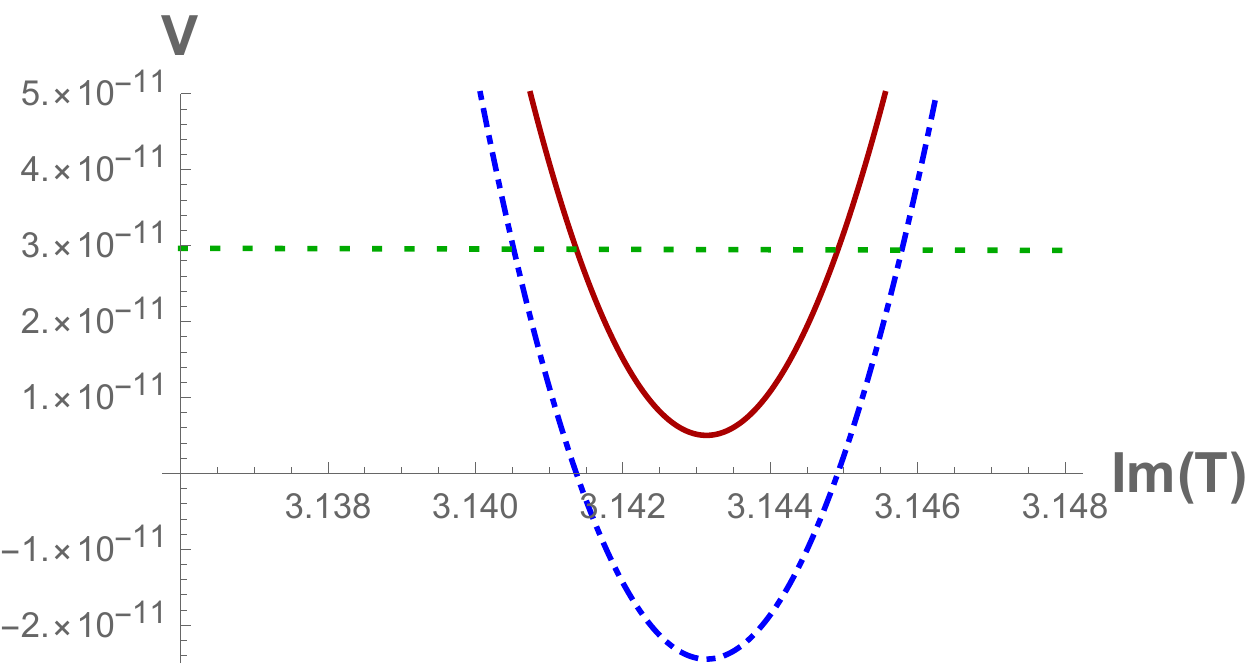}
     \caption{2d plots of the scalar potential for both moduli directions in the 2 moduli K3-fibration model. In the large-scale plot on the left, dS and AdS are not distinguishable. The close-ups, on the right hand side, show the AdS potential (blue, dash-dotted), the de Sitter potential (red, solid) and the uplifting contribution from the anti-$D3$-brane (green, dotted). The shift in the position is barely noticeable with our choice of parameters.}
     \label{fig:2modK32d}
     \end{figure}

For brevity and because the values of the masses barely change at all, we will only focus on the case where the brane is in the bulk of the internal space. As the model only has two moduli we are able to give all relevant plots. In the plot \ref{fig:K3large} the overall shape of the scalar potential is shown while in figure \ref{fig:2modK33d} a closee up of the potential around the minimum is depicted. The overall shape is nearly identical for AdS and dS, while in the close up the hole, where the values of the potential are below zero, is clearly visible. At the same point in de Sitter the potential is above zero. In figure \ref{fig:2modK32d} the scalar potential in AdS and dS can be seen for both moduli. As in the type IIA examples, the position of the minimum shifts only slightly.

Finally, the same basic picture for the masses unfolds as in type IIA, evident from table \ref{tab:2modK3mass}. There we give the masses in Minkowski space and de Sitter. The masses change slightly during the process and in de Sitter the loss of the degeneracy between scalars and pseudo-scalars can be seen.
\begin{table}[H]
\centering
\begin{tabular}{|c|c|c|}\hline
& Minkowski  & de Sitter   \\\hline
$m_1^{\,2}$ & $\; 5.22564 \cdot 10^{-2} \;$ & $\; 5.21884 \cdot 10^{-2} \;$\\\hline
$m_2^{\,2}$ & $\; 5.22564 \cdot 10^{-2} \;$ & $\; 5.21875 \cdot 10^{-2} \;$\\\hline
$m_3^{\,2}$ & $\; 1.38346 \cdot 10^{-5} \;$ & $\; 1.36014 \cdot 10^{-5} \;$\\\hline
$m_4^{\,2}$ & $\; 1.38346 \cdot 10^{-5} \;$ & $\; 1.35926 \cdot 10^{-5} \;$\\\hline
\end{tabular}
\caption{ The squared masses for the Minkowski and de Sitter case. Both the changes when going from Minkowski to de Sitter and the splitting of the degeneracy of scalars and axionic partners can be seen.}
\label{tab:2modK3mass}
\end{table}
\paragraph{A K3 fibered Calabi-Yau model} can be used for so-called fibre inflation \cite{Cicoli:2008gp,Burgess:2016owb,Kallosh:2017wku}. For this example, we first consider a generalization of the 2 moduli K3 fibration we just discussed by including a blow-up mode $\tau_3$. This leads to an internal volume \cite{Cicoli:2008gp}:
\be 
\mathcal{V}_6 (\tau_i) = \alpha \left( \sqrt{\tau_1} (\tau_2 - \beta \tau_1 ) - \gamma \tau_3^{3/2} \right)\,.
\ee
The parameters $\alpha$, $\beta$ and $\gamma$ are positive and model dependent. An example for this general case is presented in detail in \cite{Cribiori:2019drf}. Here, instead, we immediately focus on the special case where $\beta = 0$, which gives:
\be 
\mathcal{V}_6 (\tau_i) = \alpha \left( \sqrt{\tau_1} \tau_2  - \gamma \tau_3^{3/2} \right)\,,
\ee
or in our conventions, using $S$, $T$ and $U$ for the three complex Kähler moduli:
\be 
\mathcal{V}_6 (S,T,U) = \alpha \left( \sqrt{\left(-\rmi ( S- \bar{S})\right)} \left(-\rmi (T-\bar{T})\right) - \gamma \left(-\rmi (U-\bar{U})\right)^{3/2} \right)\,.
\ee
The physical meaning of the moduli here is as follows:
\begin{itemize}
\item $\rmim(S)$ controls the volume of the K3 fiber.
\item $\rmim(T)$ is proportional to the overall volume of the compactification manifold.
\item $\rmim(U)$ corresponds to the blow-up volume.
\end{itemize}
The special case with $\beta=0$ can be used for fibre inflation that usually relies on an LVS type uplift \cite{Balasubramanian:2005zx}. It is not our intent to construct an inflation model here but we are still inclined to investigate models that have potential for applications. Instead of the usual LVS uplift we will employ our mass production procedure with the racetrack superpotential \eqref{eq:racetrack} and the parameters given in table \ref{tab:fibrepara}. Note that we also have to set the model specific parameters $\alpha$, $\beta$ and $\gamma$.
\begin{table}[htb]
\centering
\begin{tabular}{|c|c|c|}\hline
$A_S = 1.1$ & $A_T = 1.2$ & $A_U =1.3$\\\hline
$a_S = 2.1$ & $a_T = 2.2$ & $a_U = 2.3$\\\hline
$b_S = 3.1$ & $b_T = 3.2$ & $b_U = 3.3$\\\hline
$S_0 = 1$ & $T_0 = 1$ & $U_0 = 1$\\\hline
$\alpha = 1$ & $\beta=0$ & $\gamma = \frac{1}{2} $ \\\hline
\end{tabular}
\caption{  Chosen parameters for the 3-moduli K3 fibration with $\beta = 0$.}
\label{tab:fibrepara}
\end{table}

\begin{figure}[H]
          \centering
     \includegraphics[width=0.4\textwidth]{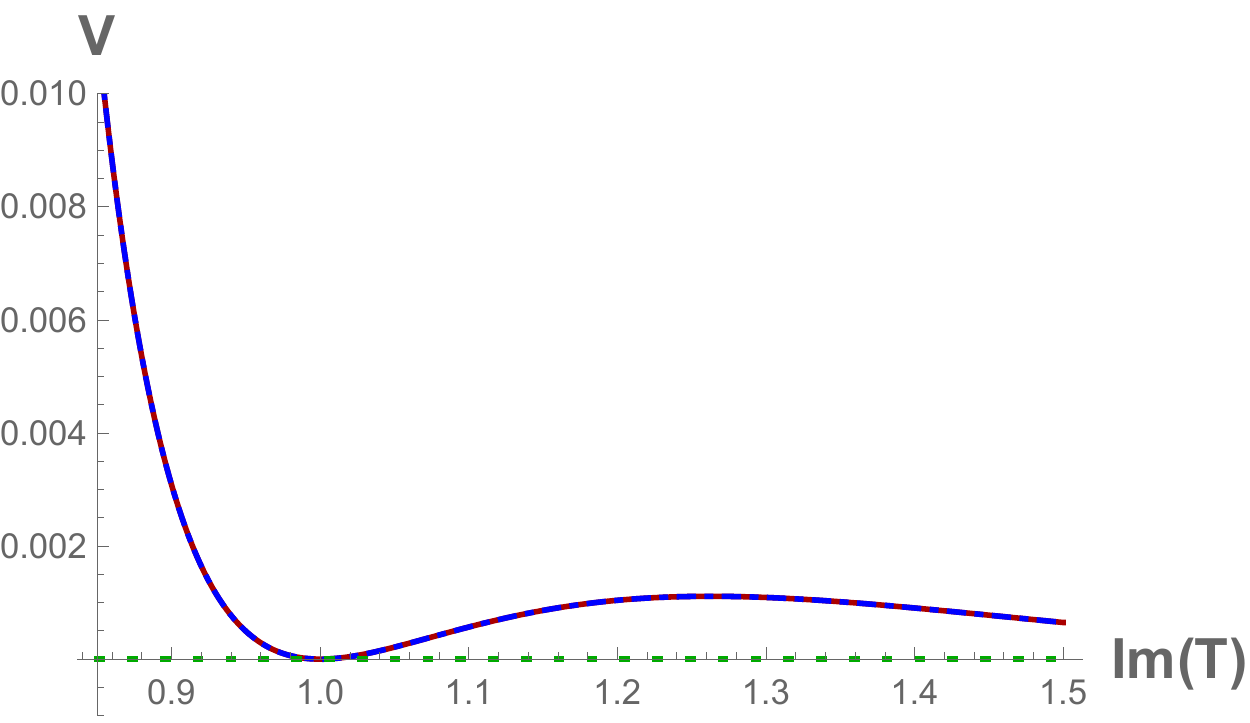} \hspace{20pt} \includegraphics[width=0.45\textwidth]{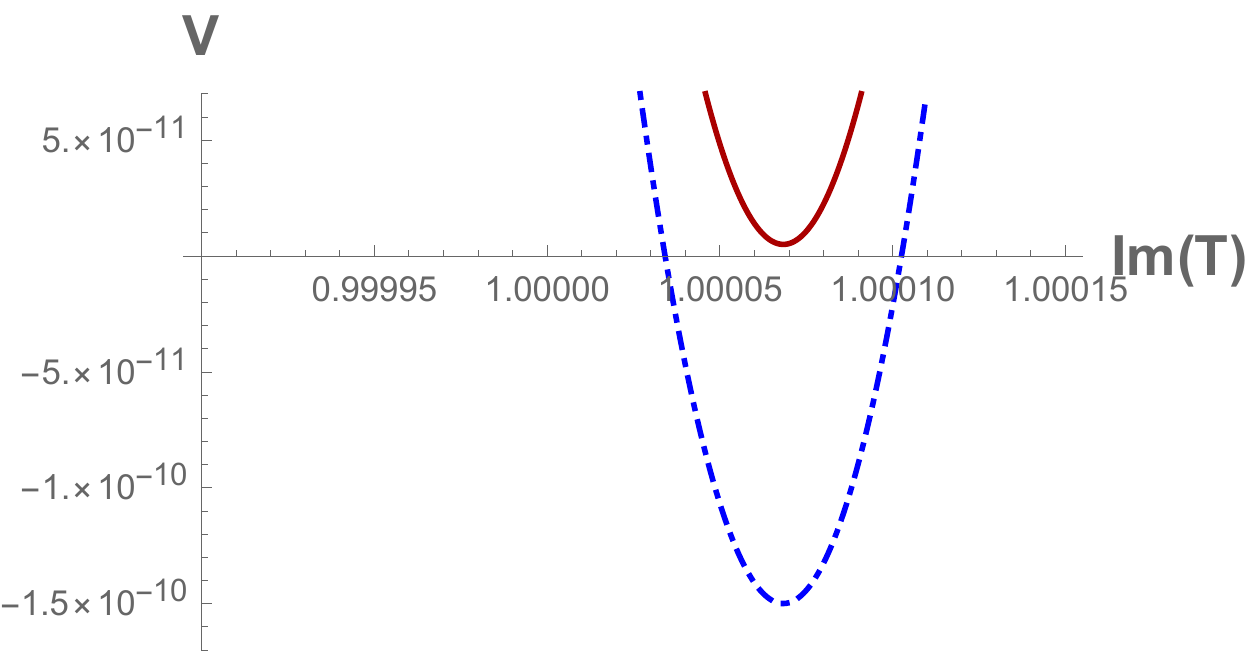}\\
     \includegraphics[width=0.4\textwidth]{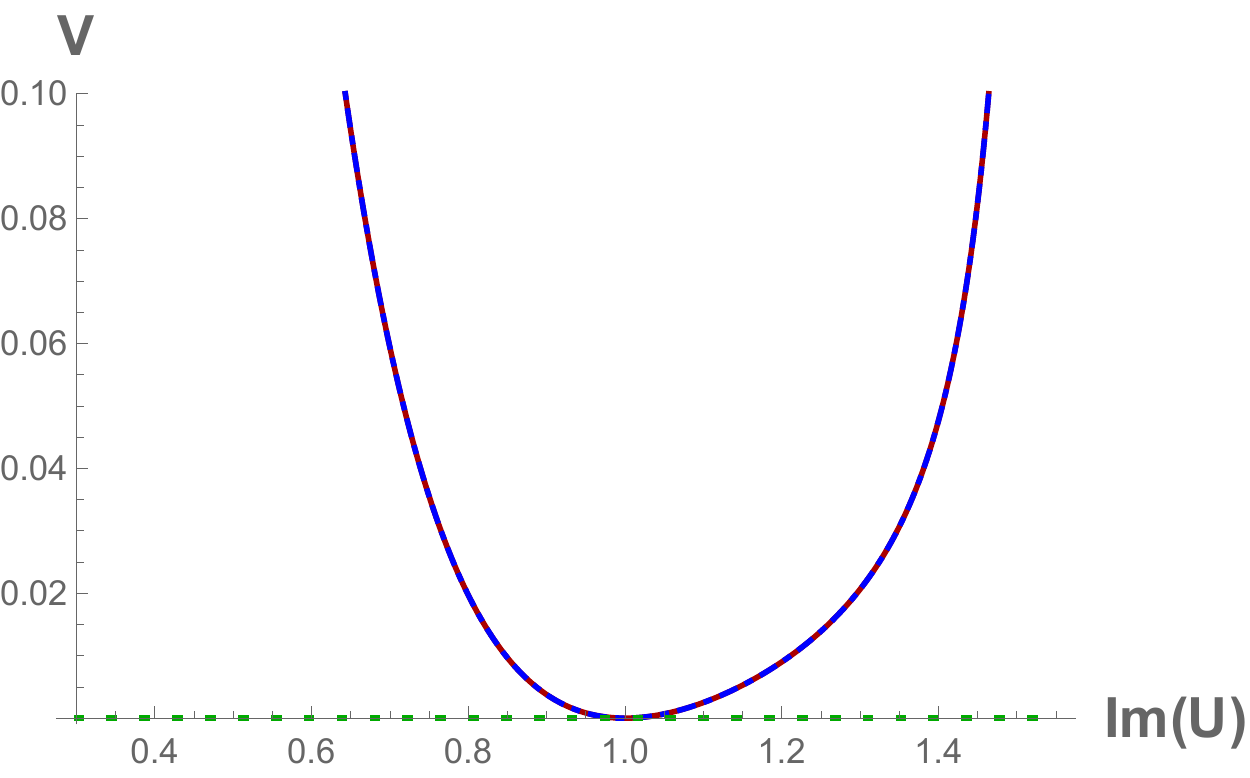} \hspace{20pt} \includegraphics[width=0.45\textwidth]{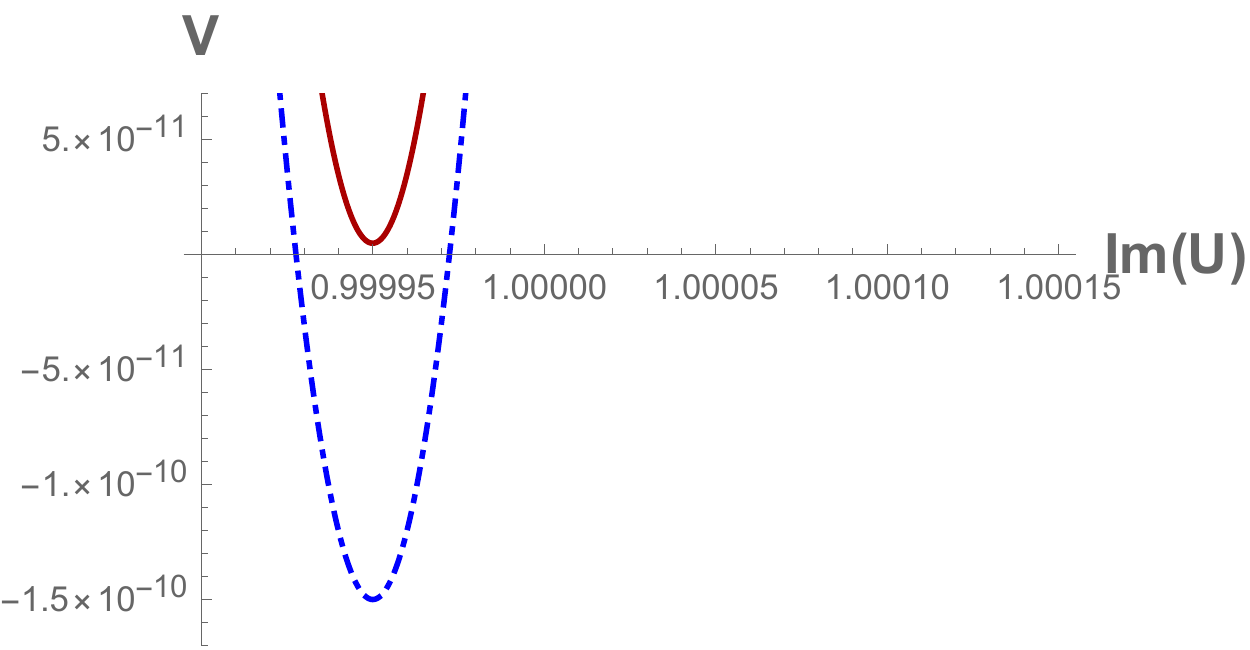}
\caption{2-dimensional plots of the minimum of the K3-fibration model used for fibre inflation. The large scale pictures on the left show the AdS and dS scalar potential overlapping. On the right the close up of the minimum depicts the AdS minimum (blue, dash-dotted) being lifted to a dS minimum (red, solid). The anti-$D3$-brane contribution (green, dotted) is above the close up of the minimum in this case. }
\label{fig:fibre2d}
\end{figure}

\begin{figure}[H]
     \centering
     \includegraphics[width=0.6\textwidth]{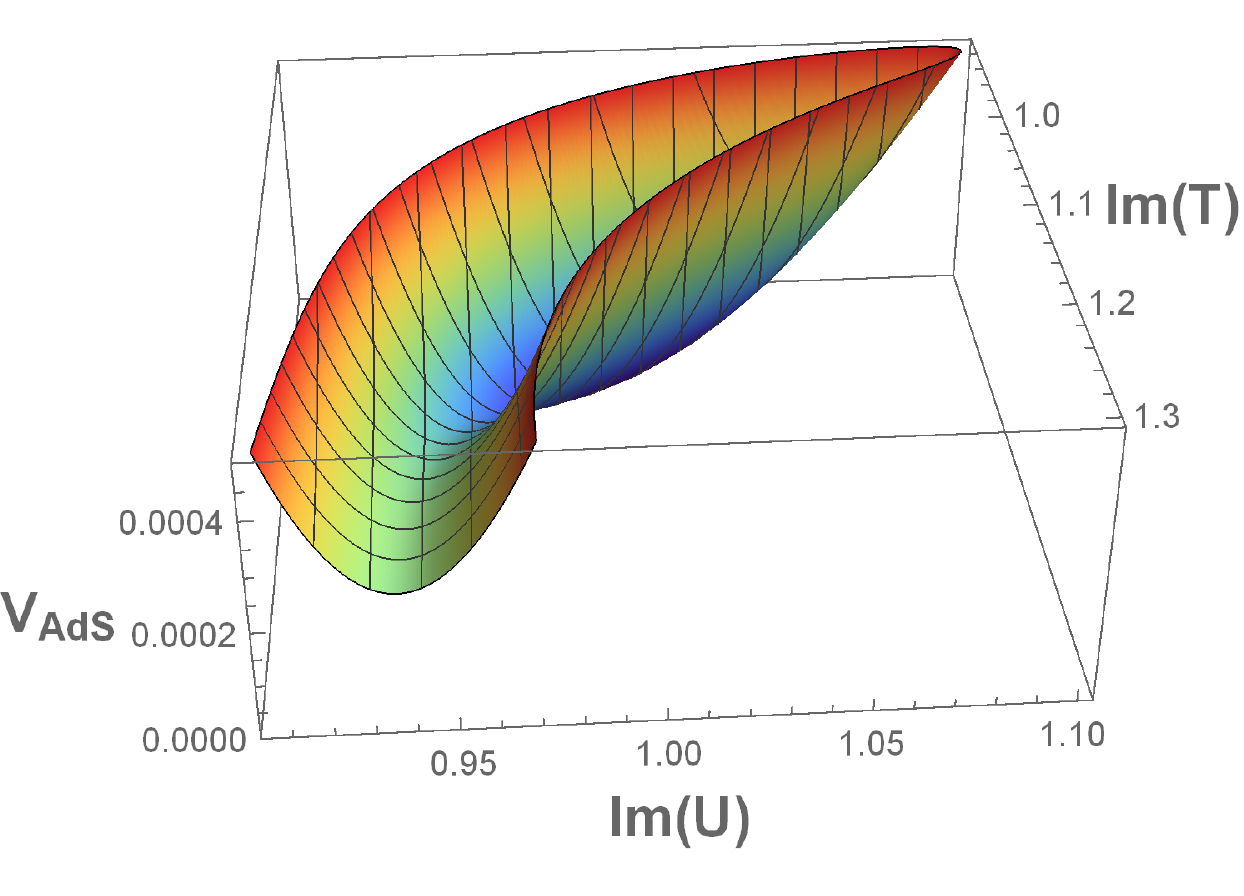}
     \caption{Form of the scalar potential around the minimum of the fibre inflation model in the $\rmim(T)$ and $\rmim(U)$ slice in 3d.}
     \label{fig:fibre3d}
\end{figure}

For the downshift and uplift parameters we choose values:
\bea
\Delta W_0 &= - 10^{-5}\,,\\
\mu^4_{\text{bulk}} &= 3.10079 \cdot 10^{-10}\,,
\eea
where we once again will focus on the placement of the anti-$D3$-branes in the bulk. Placing the branes at the bottom of a warped throat and using $\mu^4_{\text{throat}} = 2.46069 \cdot 10^{-10}$ will produce similar results. The resulting potential is visualized in figures \ref{fig:fibre2d} and \ref{fig:fibre3d}.\\
The values of the eigenvalues of the mass matrix are given in table \ref{tab:fibremass}, for Minkowski and de Sitter. Once again, we find the usual picture that they change slightly and that the degeneracy of scalars and pseudo-scalars is broken in dS.
\begin{table}[H]
\centering
\begin{tabular}{|c|c|c|}\hline
&  Minkowski  & de Sitter \\\hline
$m_1^{\,2}$ & $\; 1.01997 \cdot 10^{\;0} \,\;$ & $\; 1.01957 \cdot 10^{\;0} \,\;$\\\hline
$m_2^{\,2}$ & $\; 1.01997 \cdot 10^{\;0} \,\;$ & $\; 1.01957\cdot 10^{\;0}  \,\;$\\\hline
$m_3^{\,2}$ & $\; 1.31424 \cdot 10^{-1} \;$ & $\; 1.31344 \cdot 10^{-1} \;$\\\hline
$m_4^{\,2}$ & $\; 1.31424 \cdot 10^{-1} \;$ & $\; 1.31338 \cdot 10^{-1} \;$\\\hline
$m_5^{\,2}$ & $\; 2.44807 \cdot 10^{-2} \;$ & $\; 2.44724 \cdot 10^{-2} \;$\\\hline
$m_6^{\,2}$ & $\; 2.44807 \cdot 10^{-2} \;$ & $\; 2.44665 \cdot 10^{-2} \;$\\\hline
\end{tabular}
\caption{The squared masses for the fibre inflation model. The masses change minutely when going from Minkowski to de Sitter and the degeneracy is broken in de Sitter.}
\label{tab:fibremass}
\end{table}

\begin{figure}[H]
     \centering
     \includegraphics[width=0.4\textwidth]{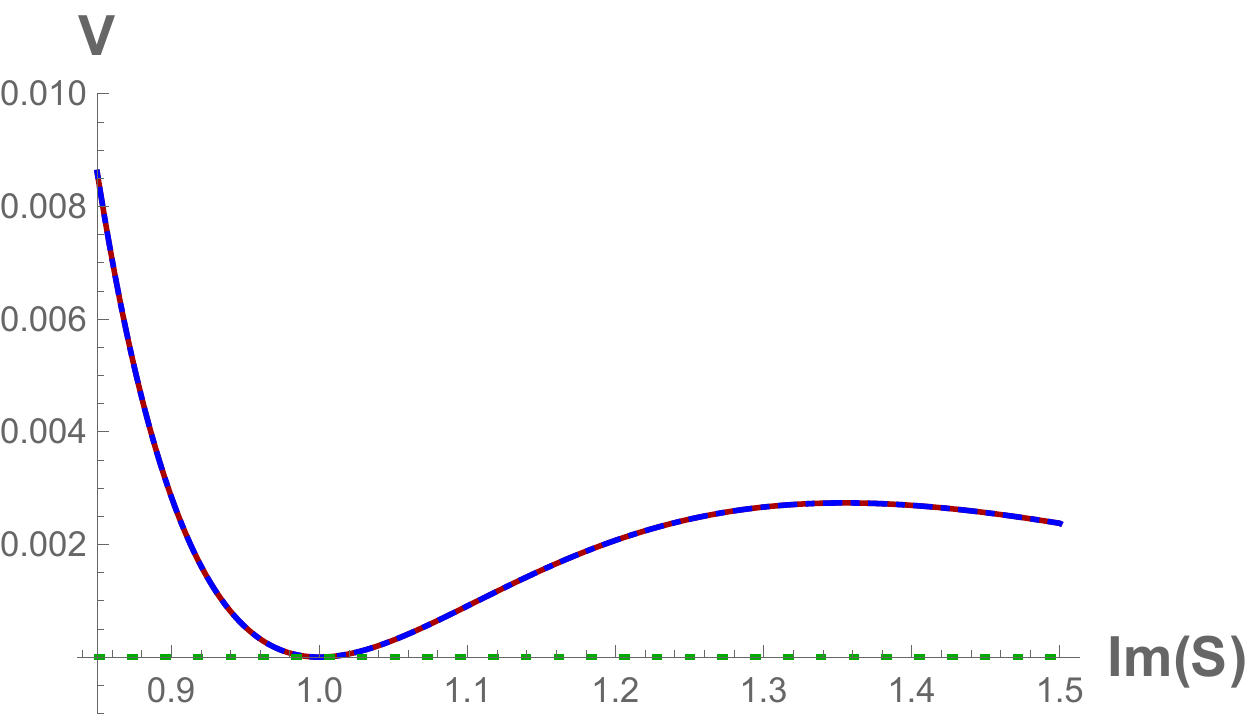} \hspace{20pt} \includegraphics[width=0.45\textwidth]{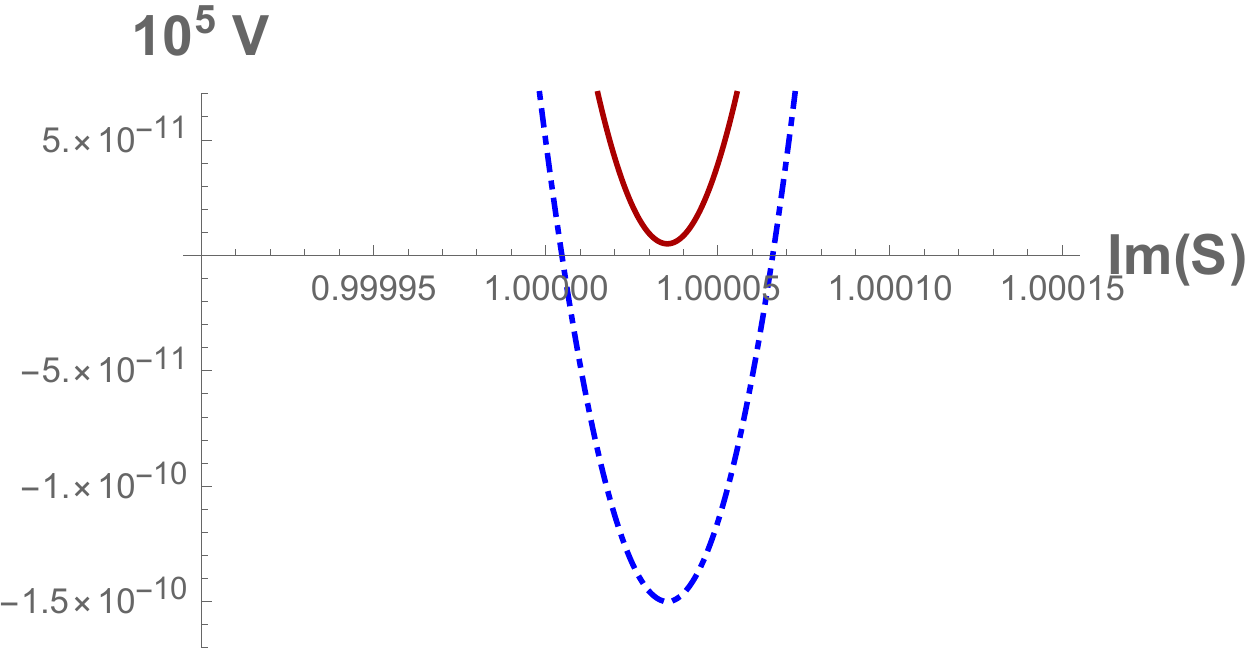}
\end{figure}

\begin{figure}[H]
\centering
\includegraphics[scale=0.55]{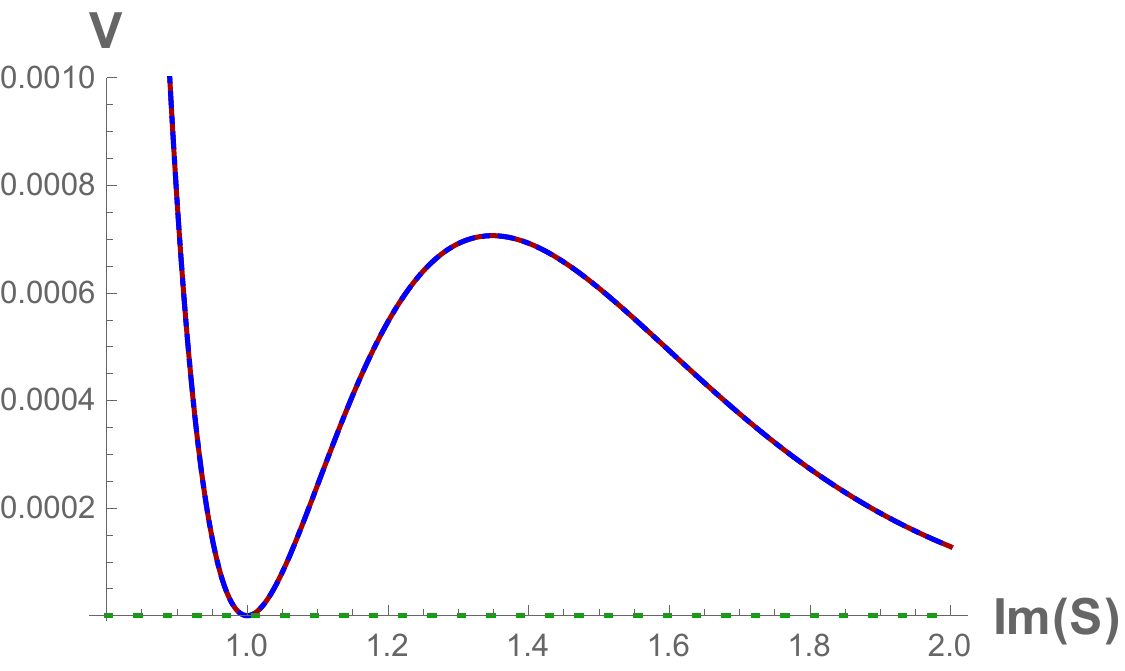}\qquad\includegraphics[scale=0.59]{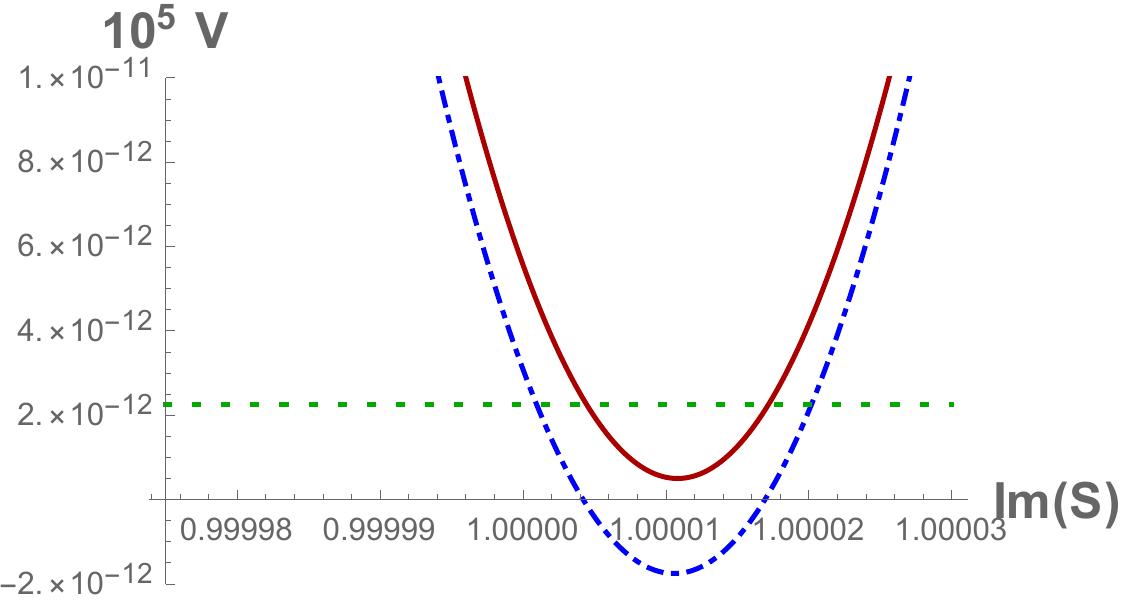}\\
\includegraphics[scale=0.55]{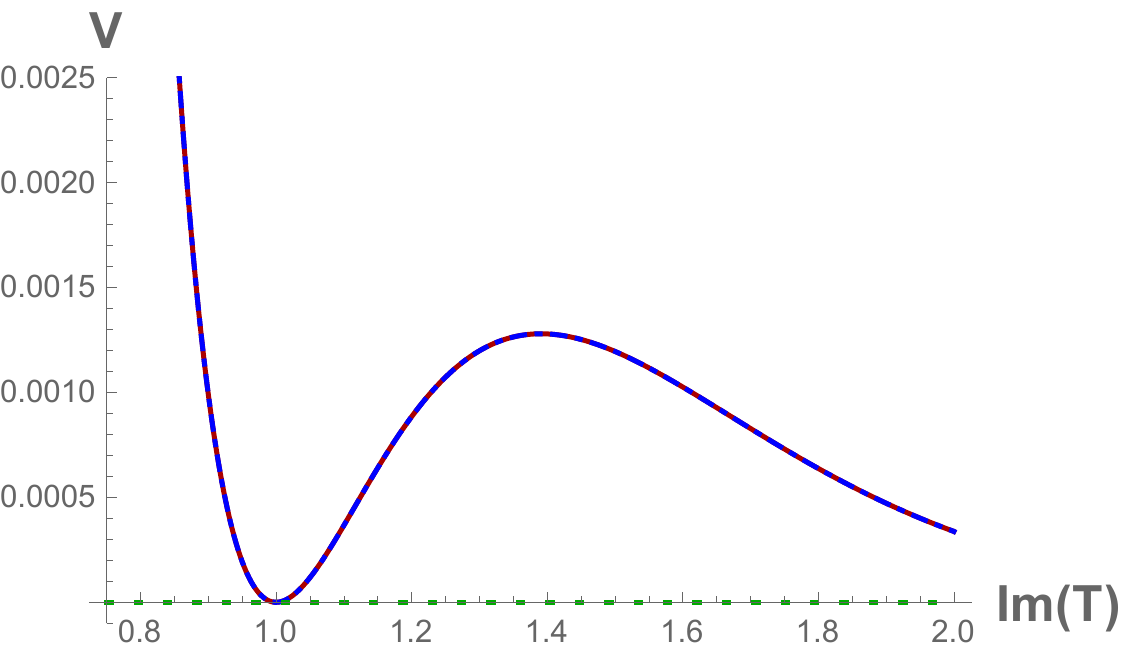}\qquad\includegraphics[scale=0.59]{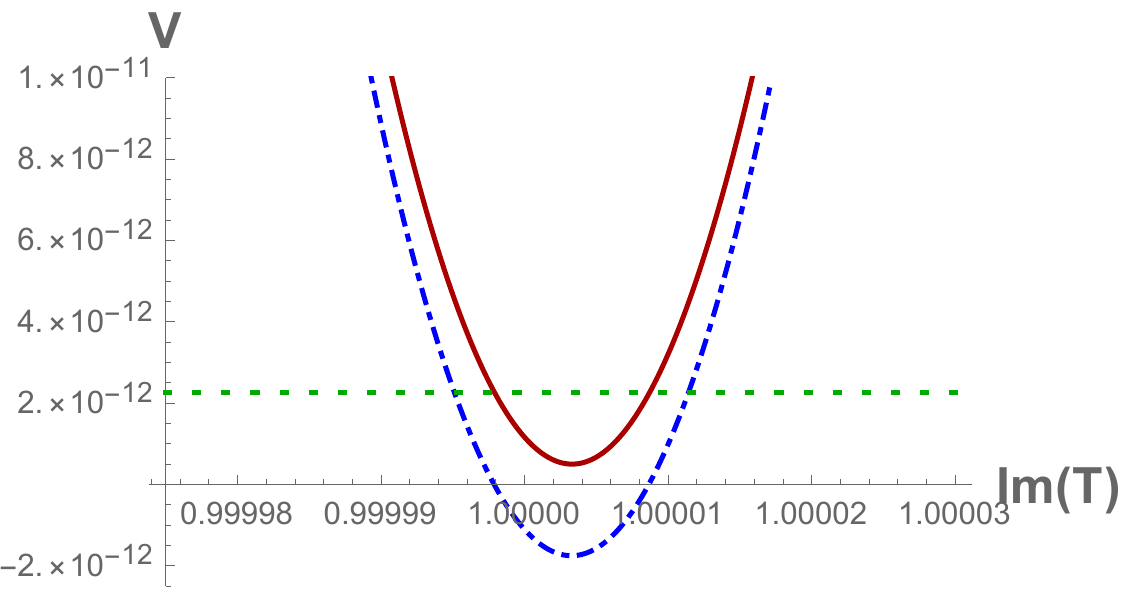}\\
\includegraphics[scale=0.55]{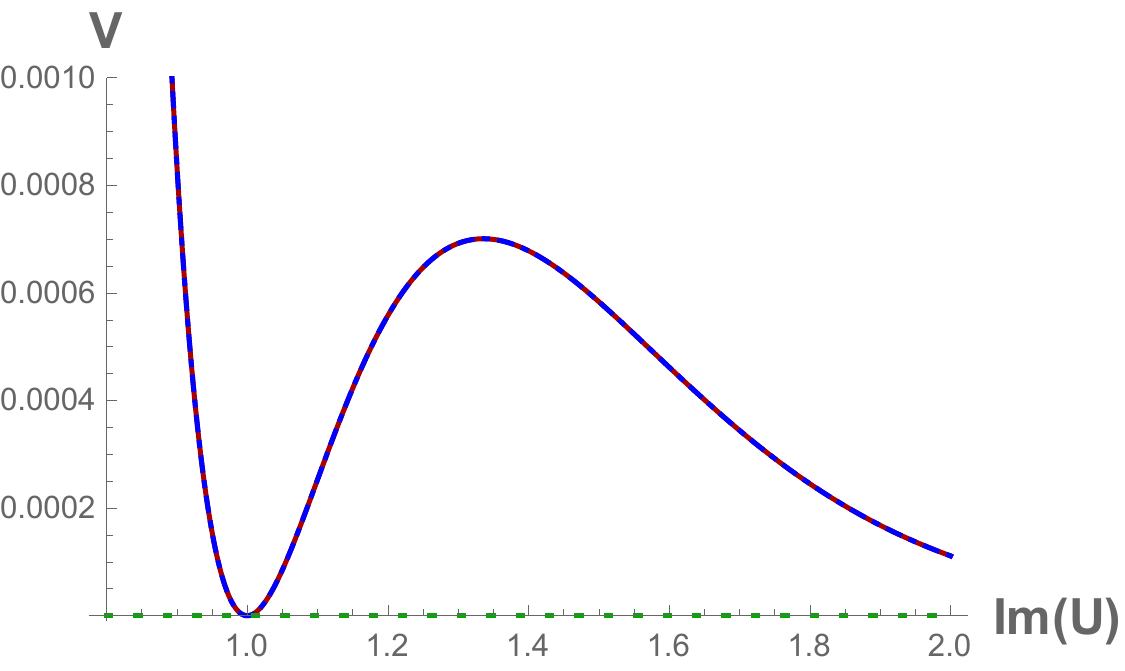}\qquad\includegraphics[scale=0.59]{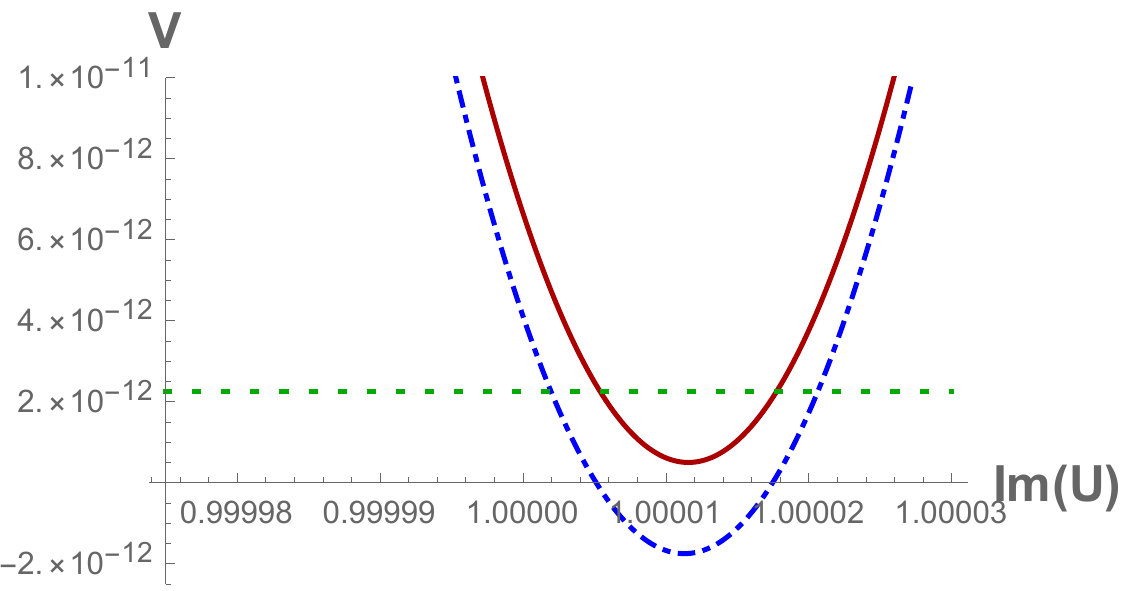}
\caption{2d plots of the scalar potential in the Fano three-fold model. All 3 directions show a similar meta-stable behavior. The AdS potential (blue, dash-dotted) gets lifted to dS (red, solid) via the contribution of the ant-D3-brane (green, dotted).}
\label{fig:swiss2d}
\end{figure}

\begin{figure}[H]
     \centering
     \includegraphics[scale=0.78]{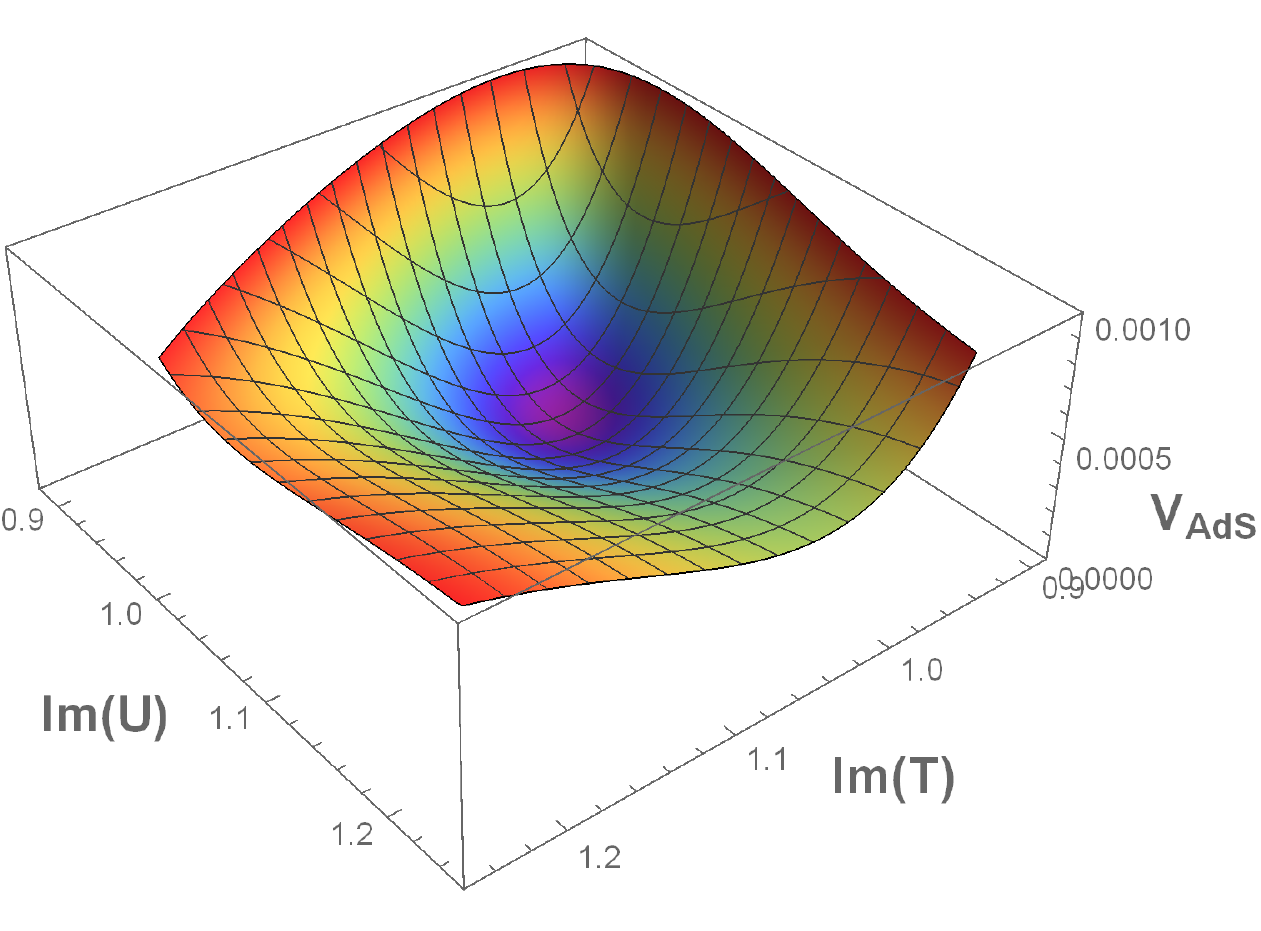}
     \caption{3d plot of the $\rmim(T)$ and $\rmim(U)$ slice in the Fano three-fold model with the meta-stable minimum clearly visible.}
     \label{fig:swiss3d}
\end{figure}

\paragraph{Our final example is based on the Fano three-fold $\bm{\mathcal{F}_{11}}$} and is sometimes called a multi-hole Swiss-cheese model. This manifold is topologically equivalent to a Calabi-Yau three-fold with Hodge numbers $h^{1,1} = 3$ and $h^{2,1} =111$. It has been studied in detail in \cite{Denef:2004dm}. The internal volume can be described by \cite{Cicoli:2008gp}:
\be 
\mathcal{V}_6 (\tau_i) = \frac{1}{3 \sqrt{2}} \bigg(2[\tau_1  + \tau_2 + 2 \tau_3]^{3/2} - [\tau_2 + 2 \tau_3]^{3/2} - [\tau_2]^{3/2}\bigg)\,.
\ee
Translating this to our complex Kähler moduli we obtain:
\bea
\mathcal{V}_6(S,T,U) = &\frac{1}{3 \sqrt{2}} \bigg( 2 \left[ \left(-\rmi (S-\bar{S})\right) +\left(-\rmi (T-\bar{T})\right) +2 \left(-\rmi (U-\bar{U})\right)\right]^{3/2}\\
&- \left[\left(-\rmi (T-\bar{T})\right) + 2\left(-\rmi (U-\bar{U})\right) \right]^{3/2} - \left[-\rmi (T-\bar{T})\right]^{3/2}\bigg)\,.
\eea
The mass production procedure continues as usual with:
\bea
\Delta W &= - 5 \cdot 10^{-6}\,,\\
\mu^4_{\text{bulk}} &= 9.62862 \cdot 10^{-11}\,,
\eea
and the other parameters given in table \ref{tab:swisspara}. The masses obtained with these parameters are given in table \ref{tab:swissmass}. As before, we show 2d and 3d plot of the minimum in figures \ref{fig:swiss2d} and \ref{fig:swiss3d}.
\begin{table}[H]
          \centering
          \begin{tabular}{|c|c|c|}\hline
          $A_S = 1.1$ & $A_T = 1.2$ & $A_U =1.3$\\\hline
          $a_S = 2.1$ & $a_T = 2.2$ & $a_U = 2.3$\\\hline
          $b_S = 3.1$ & $b_T = 3.2$ & $b_U = 3.3$\\\hline
          $S_0 = 1$ & $T_0 = 1$ & $U_0 = 1$\\\hline
          \end{tabular}
          \caption{ Parameter choices for the model based on the Fano 3-fold $\mathcal{F}_{11}$.}
          \label{tab:swisspara}
\end{table}
\begin{table}[H]
     \centering
     \begin{tabular}{|c|c|c|}\hline
     &  Minkowski  & de Sitter \\\hline
     $m_1^{\,2}$ & $\; 1.85578 \cdot 10^{-1} \;$ & $\; 1.85557 \cdot 10^{-1} \;$\\\hline
     $m_2^{\,2}$ & $\; 1.85578 \cdot 10^{-1} \;$ & $\; 1.85557 \cdot 10^{-1} \;$\\\hline
     $m_3^{\,2}$ & $\; 1.00760 \cdot 10^{-1} \;$ & $\; 1.00753 \cdot 10^{-1} \;$\\\hline
     $m_4^{\,2}$ & $\; 1.00760 \cdot 10^{-1} \;$ & $\; 1.00753 \cdot 10^{-1} \;$\\\hline
     $m_5^{\,2}$ & $\; 1.30646 \cdot 10^{-2} \;$ & $\; 1.30627 \cdot 10^{-2} \;$\\\hline
     $m_6^{\,2}$ & $\; 1.30646 \cdot 10^{-2} \;$ & $\; 1.30626 \cdot 10^{-2} \;$\\\hline
     \end{tabular}
     \caption{Squared masses of the Fano three-fold model in Minkowski space and de Sitter.}
     \label{tab:swissmass}
\end{table}

\FloatBarrier
\subsection{Models with large Downshifts and Uplifts}
\begin{figure}[htb] 
\includegraphics[width=0.9 \textwidth]{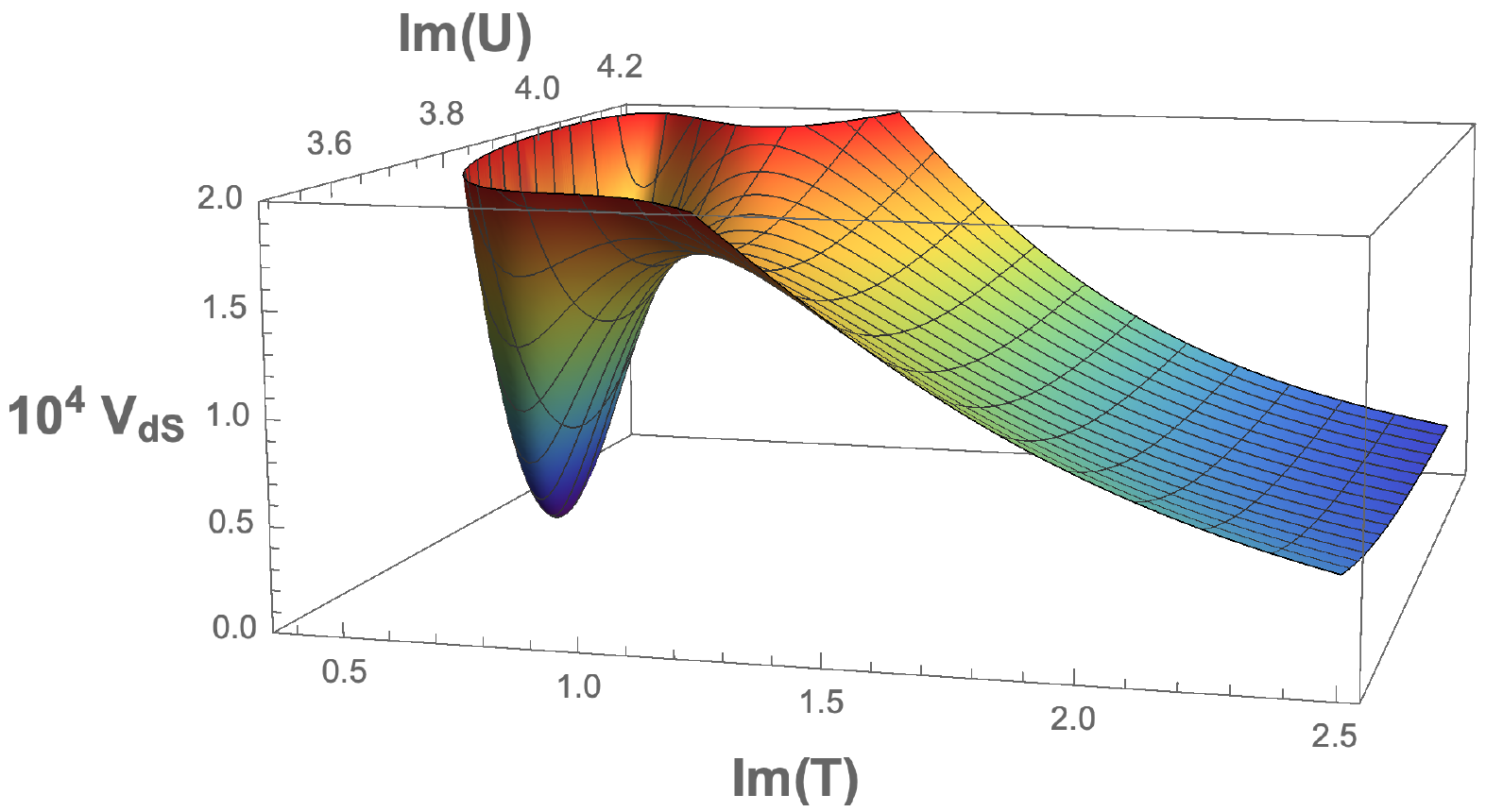}
\caption{3d plot of the STU model with large downshift and uplift in to dS. It is evident that a de Sitter vacuum with small cosmological constant can be obtained after a large downshift and subsequent uplift. The other possible slices show similar behavior. }
\label{fig:3dlargeshift}
\end{figure}

Up until now we looked at de Sitter models that were obtained by performing small deviations from Minkowski space. This is an intuitive safe bet if one hopes to be successful. The way to quantify what is considered small is by requiring that the gravitino mass is small compared to the lightest field in the mass matrix. During the investigation of all models presented in \cite{Cribiori:2019drf} it was found that this condition might not necessarily be required. \\
First, let us mention that for small $\Delta W$ the sign did not matter too much as the downshift is proportional to its square. For large downshift the linear term becomes relevant and it is important to use $\Delta W >0$. Then it is possible to use a large downshift and a large uplift in order to obtain a stable de Sitter minimum. During this procedure the masses will now change significantly but they will stay positive, as will be shown in the example in the following. Another interesting feature ist that large uplifts allow to control the degree of supersymmetry breaking, which might be useful for certain applications.\\
As an example for a model with large shifts we return to our familiar type IIA STU model\footnote{Here we use our type IIA notation with $S$ being the axio-dilaton, $T$ the complex structure modulus and $U$ the Kähler modulus.} of section \ref{sec:massIIAexampel}. The relevant potentials \eqref{eq:massIIA7modpot} are utilized in the exact same way as before with the parameters in table \ref{tab:largpara}. Here we solved for $A_S$, $B_S$, $B_T$ and $B_U$. The downshift and subsequent uplift are given by the parameters:
\be 
\Delta f_6 = 1 \qquad \text{and} \qquad \mu_1^4 = \mu_2^4 = 8.1479 \cdot 10 ^{-5}\,.
\ee
These values are several orders of magnitude larger than what we have used thus far. Nevertheless, as can be seen from figure \ref{fig:3dlargeshift} we obtain (meta-) stable minima in AdS and dS. The position of the minimum does shift significantly when going from Minkowski do anti-de Sitter and then de Sitter:
\bea 
\rmim(S):& \qquad 1 \; \to \; 1.03367 \; \to \; 1.11369 \,,\\
\rmim(T):& \qquad 1 \; \to \; 0.13319 \; \to \; 0.51958 \,,\\
\rmim(U):& \qquad 5 \; \to \; 4.58783 \; \to \; 3.88815 \,.
\eea
While  one does need to be careful to follow this minimum properly during the procedure there are no obstacles to do so and the procedure works out properly.
\begin{table}[htb]
\centering
\begin{tabular}{|c|c|c|}\hline
$a_S = 1$ & $a_T = 3/4$ & $a_U = 1/2$\\\hline
$b_S = 3/2$ & $b_T =5/3$ & $b_U = 2/3$\\\hline
$f_6 = 1 $ & $A_T = 3$ & $A_U =20$\\\hline
$S_0 = 1$ & $T_0 = 1$ & $U_0 = 5$\\\hline
\end{tabular}
\caption{ The parameters for the IIA STU model with large downshift and uplift.}
\label{tab:largpara}
\end{table}
The eigenvalues of the mass matrix do change significantly during the procedure but the minimum does not become unstable. Indeed, as can be seen from table \ref{tab:largemass} the fields become heavier. The deviation from the mass degeneracy also becomes more distinct. Still, there are no significant problems in the procedure and we obtain the desired result.
\begin{table}[htb]
\centering
\begin{tabular}{|c|c|c|}\hline
&  Minkowski  & de Sitter \\\hline
$m_1^{\,2} $ & $\; 1.07895 \cdot 10^{-2} \;$ & $\; 1.32013 \cdot 10^{-1} \;$  \\\hline
$m_2^{\,2} $ & $\; 1.07895 \cdot 10^{-2} \;$ & $\; 1.22573 \cdot 10^{-1} \;$ \\\hline
$m_3^{\,2} $ & $\; 1.29976 \cdot 10^{-3} \;$ & $\; 9.61218 \cdot 10^{-2} \;$ \\\hline
$m_4^{\,2} $ & $\; 1.29976 \cdot 10^{-3} \;$ & $\; 9.02659 \cdot 10^{-2} \;$ \\\hline
$m_5^{\,2} $ & $\; 1.05464 \cdot 10^{-4} \;$ & $\; 2.14932 \cdot 10^{-3} \;$ \\\hline
$m_6^{\,2} $ & $\; 1.05464 \cdot 10^{-4} \;$ & $\; 1.87824 \cdot 10^{-3} \;$ \\\hline
\end{tabular}
\caption{  Squared masses in the type IIA STU model with large downshift and uplift. }
\label{tab:largemass}
\end{table}

The idea of large shifts was further investigated in \cite{Linde:2020mdk} where it was shown that it is potentially possible to uplift to de Sitter even from models that do not have an AdS minimum but rather a runaway potential. 

\section{de Sitter Minima from M-theory and String Theory}
\label{sec:mtheory}
This section is based on \cite{Cribiori:2019hrb} where we considered moduli stabilization on a twisted seven torus in M-theory. Generally, this is a special case of a manifold with $G_2$-structure as $\mathbb{Z}_2 \times \mathbb{Z}_2 \times \mathbb{Z}_2 \subset G_2$. The twisted seven torus is obtained as a toroidal orbifold $X_7 = \mathbb{T}^7/(\mathbb{Z}_2 \times \mathbb{Z}_2 \times \mathbb{Z}_2)$ \cite{DallAgata:2005zlf,Duff:2010vy,Derendinger:2014wwa,Ferrara:2016fwe} with Betti numbers $(b_0,b_1,b_2,b_3) = (1,0,0,7)$. The quotient can be made non-singular by a choice of a free orbifold action. The resulting theory corresponds to a maximal rank reduction on the seven torus and gives $4d, \; \mathcal{N}=1$ supergravity with seven moduli. Finally, we introduce a twist which can be viewed as Scherk-Schwarz reduction \cite{Scherk:1978ta,Scherk:1979zr,Mueller-Hoissen:1987cwl,DallAgata:2005zlf,Grana:2013ila} of the initial torus.\\
In 11d one can extend the action of 11d supergravity to a form where the potentials and dual curvatures appear at the same time. This democratic form results in a pseudo-action and was studied in \cite{DallAgata:2005zlf}. In 10d the corresponding pseudo-action was proposed in \cite{Bergshoeff:2001pv}. Here, we follow \cite{DallAgata:2005zlf,Derendinger:2014wwa} in order to use these pseudo actions for moduli stabilization in 4d.\\
Going one step further, one can consider the generalized twisted seven torus \cite{Derendinger:2014wwa,Blaback:2018hdo,Villadoro:2007yq}. In this model the inclusion of KK5 and KKO5-planes in 10d or KK6 and KKO6-planes in 11d allows to lift some restrictions, like the tadpole conditions.\\
Our goal here is to use models from the generalized twisted seven torus in order to find Minkowski vacua that then can be used for the mass production procedure we already discussed \cite{Kallosh:2019zgd,Cribiori:2019drf}. In this section we will always have 7 complex moduli, corresponding to 14 real fields. Previously, we used a superpotential that prominently featured non-perturbative corrections in all directions and lacked tree-level flux terms. Moreover, in the mass production procedure a Kallosh-Linde type double exponent was used. Here, we will construct models that include quadratic flux terms in the superpotential and we will find not only that it is possible to have single exponential terms in the superpotential, but also models where it is not necessary to have exponents in all directions. The connection to type IIA on a six torus can be made for some of these models. Finally, by including conjectured non-geometric fluxes \cite{Aldazabal:2006up}, we also present a model that is related to type IIB that does not rely on any non-perturbative corrections. The origin of the required terms is not clear but can be motivated by S-duality.

\subsection{The Generalized twisted Seven Torus}
The Kähler- and superpotential that we will use for the generalized twisted seven torus model are \cite{DallAgata:2005zlf,Derendinger:2014wwa}:
\bea 
\label{eq:mtheorypots}
K &= - \sum_{I=1}^7 \log \left( -\rmi (\Phi^I - \bar{\Phi}^I)\right)\,,\\
W &=  g_7 + \frac{1}{2} M_{IJ} \Phi^I \Phi^J\, + \sum_{i=1}^7 A_I \rme ^{\rmi a_I \Phi^I},
\eea
where we already made some modifications to the superpotential. The complete perturbative part of $W$ is:
\be
W_{pert} =  g_7 + G_I \Phi^I + \frac{1}{2} M_{IJ} \Phi^I \Phi^J\,,
\ee
where all parameters are real. We have set the linear terms to zero via $G_I =0$ and introduced a single non-perturbative contribution for each field. Later on we will set some of those to zero as well, by enforcing the corresponding $A_I$ to be zero. For us, the relevant contributions are a seven flux $g_7$ and the terms quadratic in the moduli that come from geometric fluxes. All parameters are real and the matrix $M_{IJ}$ has a vanishing entries on the diagonal and is symmetric. Thus, it gives 21 parameters in total. In this model, the non-perturbative contributions may arise from wrapping M2-branes around 3-cycles \cite{Harvey:1999as}. For $\mathbb{T} ^7/(\mathbb{Z}_2 \times \mathbb{Z}_2 \times \mathbb{Z}_2)$ seven 3 cycles are available, allowing for non-perturbative corrections in all directions.\\
The first step in the mass production procedure is to find a stable, supersymmetric vacuum state with vanishing scalar potential. This is done by solving $W = 0$ and $\partial_I W = 0$. Once, again we split the complex moduli $\Phi^I = \theta^I + \rmi \phi^I$ and set $\theta^I=0$, which is consistent for as long as the masses remain positive. The seven equations $\partial_I W =0$ can be solved in terms of the coefficients $A_I$ by setting:
\be 
A_I = \rmi a^{-1}_I \rme^{-\rmi a_I \Phi^I} M_{IJ} \Phi^J\,.
\ee
Then, we use $g_7$ to solve the remaining condition $W=0$, giving the desired Minkowski vacuum state. In our examples, we will also consider models where no exponent appears in one or more directions. Then, the solutions to the supersymmetry equations are obtained in terms of flux parameters. By inclusion of some conjectured S-dual fluxes it is even possible to find a model that does not rely on any non-perturbative corrections. It is highly non-trivial and unexpected that such a solution can been found. In order to employ the mass production procedure we have to keep in mind that we require there to be no flat directions. We thus have to make sure that mass matrix in Minkowski space, given to be:
\be 
V_{I\bar{J}}^{\text{Mink}} = m_{IL} K^{L\bar{L}} m_{\bar{L}\bar{J}} = \rme^K W_{IL} K^{L\bar{L}} \overline{W}_{\bar{L}\bar{J}}\,,
\ee
has no zero eigenvalues. This is different from the models considered in section \ref{sec:massprod}, since there the double exponent superpotential guarantees that we will be able to find solutions with no flat directions. Still, in constructing the explicit examples in \cite{Cribiori:2019hrb} and in the following, no particular tedious fine-tuning of the free parameters was necessary in order to find Minkowski solutions without flat directions.\\
For convenience, we will use our type IIA notation where:
\be 
\Phi^I = \{S,T_i,U_i\}\,,\qquad i=1,2,3\,,
\ee
but note that the physical interpretation of these fields is not necessarily the same as before. We also decompose the quadratic terms in the superpotential as \cite{Derendinger:2014wwa}:
\be 
\label{eq:mtheoryW}
\frac{1}{2} M_{IJ} \Phi^I \Phi^J = S b^k U_k + U_i C^{ij} T_j + a^i \frac{U_1 U_2 U_3}{U_i} + c^i \frac{T_1 T_2 T_3}{T_i} + S d^k T_k\,.
\ee
The 21 independent components of $M_{IJ}$ are translated to $a^i$, $b^k$, $c^i$, $d^k$ and the $C^{ij}$. It is not necessary to always use all of these terms in order to obtain a Minkowski minimum and several cases will be explored in section \ref{sec:mtheoryexamples}. 

\subsection{Connection to Type IIA}
\label{sec:IIAconn}
When considering type IIA string theory compactifications on $\mathbb{T}^6/(\mathbb{Z}_2 \times \mathbb{Z}_2)$ not all terms in \ref{eq:mtheoryW} are allowed \cite{Derendinger:2014wwa,Blaback:2018hdo,Villadoro:2007yq}. Namely, the terms proportional to the $c^i$ and the $d^k$ have to vanish in standard IIA orientifold compactifications. Additionally, $g_7$ gets replaced by the six-flux parameter $f_6$. In type IIA  we can give a physical interpretation to the various terms appearing in the superpotential:
\begin{itemize}
\item $a^i \frac{U_1 U_2 U_3}{U_i}$ are two-fluxes.
\item $U_i C^{ij} T_j$ and $S b^k U_k$ correspond to non-geometric fluxes.
\end{itemize}
There are more conditions that need to be satisfied, namely the tadpole constraints. Luckily, it is possible to include sources in order to fulfill them. The conditions are:
\begin{itemize}
\item $\sum_i a^i b^i = 0$ and $\sum_i a^i C^{ij} = 0$ that can be satisfied by $(O6/D6)$ sources.
\item $b^i C^{ij} + b^j C^{ii} = 0$ can be fulfilled by $(KK5/KKO5)$.
\item $C^{ij}C^{jk} + C^{ij} C^{jj} = 0$ gets relaxed by including $(KK5/KKO5)^\prime$.
\end{itemize}
The prime denotes a second set of $KK$-sources, wrapping a different cycle, see \cite{Blaback:2018hdo}.\\
In the various models we will discuss some of these conditions will be automatically satisfied while for others we will use them to further constrain some parameters. Then, no sources are necessary. 

\subsection{M-Theory Examples}
\label{sec:mtheoryexamples}
Here we will review explicit examples based on the potentials \eqref{eq:mtheorypots} that were discussed in detail in \cite{Cribiori:2019hrb}.
\paragraph{Including only single exponents in each direction} is the most straightforward thing we can try after exploring models with Kallosh-Linde type double exponents before. The superpotential we consider is:
\be 
W_1 = g_7 + b^k S U_k + C^{ij} U_i T_j + A_S \rme^{\rmi a_S S} + \sum_i \left( A_{T_i} \rme^{\rmi a_{T_i}T_i} + A_{U_i} \rme^{\rmi a_{U_i}U_i} \right)\,.
\ee
This model has 19 free parameters, only 8 of which will be required to solve the Minkowski equations $W = 0$ and $\partial_I W = 0$ ($I = \{S, T_i, U_i \}$) at the point $S_0$, $T_{i,0}$ and $U_{i,0}$. We choose to solve for the $A_I$ and $g_7$. The remaining parameters can be used to make sure all masses are positive. The uplift proceeds along the same lines as was discussed in the previous sections with the magnitude of the uplift given by $\mu$. Our choice of parameters can be found in table \ref{tab:3exppara} and the resulting masses are given in table \ref{tab:3expmass}.
\begin{table}[htb]
\center
\begin{tabular}{|c|c||c|c||c|c||c|c|}\hline
 $\,S_0\,$     & $1.0$   & $\,a_S\,$ & $\;1.0\,$ &$\,C^{11}\,$ & $\,0.11\,$ &  $\,C^{32}\,$ & $\,0.32\,$\\\hline
 $T_{1,0}$ & $\,1.1\,$ & $\,a_{T_1}\,$ & $\;1.1\,$ & $C^{12}$ & $0.12$ & $C^{33}$ & $0.33$ \\\hline
 $T_{2,0}$ & $1.2$ & $\,a_{T_2}\,$ & $\;1.1\,$ & $C^{13}$ & $0.13$ & $b^1$     & $0.55$\\\hline
 $T_{3,0}$ & $1.3$ & $\,a_{T_3}\,$ & $\;1.1\,$ & $C^{21}$ & $0.21$ & $b^2$     & $0.60$  \\\hline
 $U_{1,0}$ & $5.1$ & $\,a_{U_1}\,$ & $\;0.51\,$ & $C^{22}$ & $0.22$ & $b^3$     & $0.65$ \\\hline
 $U_{2,0}$ & $5.2$ & $\,a_{U_2}\,$ & $\;0.52\,$ & $C^{23}$ & $0.23$ & $\Delta g_7$ & $5 \cdot 10^{-3} $ \\\hline
 $U_{3,0}$ & $5.3$ & $\,a_{U_3}\,$ & $\;0.53\,$ & $C^{31}$ & $0.31$ & $\mu^4$ & $ 9 \cdot 10^{-9} $ \\\hline
\end{tabular}
\caption{Chosen parameters for the model with 3 single exponents.}
\label{tab:3exppara}
\end{table}
\begin{table}[htb]
\center
\begin{tabular}{|c|c|c|c|c|c|c|c|}\hline
     &$\,m_1^{\,2}\,$&$\,m_2^{\,2}\,$&$\,m_3^{\,2}\,$&$\,m_4^{\,2}\,$&$\,m_5^{\,2}\,$&$\,m_6^{\,2}\,$&$\,m_7^{\,2}\,$\\\hline
Mk & $\,0.41229\,$ & $\, 0.22090 \,$ & $\, 0.10343 \,$ & $\, 0.03087 \,$ & $\, 0.01977 \,$ & $\, 0.01275 \,$ & $\, 0.00676 \,$ \\\hline  
dS & $\, 0.41306 \,$ & $\, 0.22137 \,$ & $\, 0.10356 \,$ & $\, 0.03091 \,$ & $\, 0.01980 \,$ & $\, 0.01277 \,$ & $\,0.00677\,$ \\\hline  
\end{tabular}
\caption{The eigenvalues of the mass matrix for the model with 3 single exponents in Minkowski and de Sitter. The axions are omitted for brevity.}
\label{tab:3expmass}
\end{table}

In this model, the amount of free parameters allows us to fulfill the tadpole conditions without including sources. The conditions not automatically satisfied are:
\bea 
b^i C^{ij} + b^j C^{ii} = 0\,,\\
C^{ij} C^{jk} + C^{ik} C^{jj} = 0\,.
\eea
These conditions can, for example, be solved in terms of the $C^{ij}$ with $i \neq j$. If we keep the other parameters as before in table \ref{tab:3exppara} we obtain the masses in table \ref{tab:3expmasstad}. Satisfying the tadpole conditions without sources can be a great advantage for model building as one has to be very careful about the origins of the sources and their stability in some cases.
\begin{table}[htb]
\center
\begin{tabular}{|c|c|c|c|c|c|c|c|}\hline
     &$\,m_1^{\,2}\,$&$\,m_2^{\,2}\,$&$\,m_3^{\,2}\,$&$\,m_4^{\,2}\,$&$\,m_5^{\,2}\,$&$\,m_6^{\,2}\,$&$\,m_7^{\,2}\,$\\\hline
Mk &  $\, 0.09036 \,$  &  $\, 0.02693 \,$  &  $\, 0.01390 \,$  & $\, 0.00558 \,$ &  $\, 0.00388 \,$  &  $\, 0.00159 \,$  &  $\, 0.00063 \,$  \\\hline  
dS &  $\, 0.08982 \,$  &  $\, 0.02680 \,$  &  $\, 0.01383 \,$  & $\, 0.00555 \,$  &  $\, 0.00388 \,$  &  $\, 0.00158 \,$  &  $\, 0.00063 \,$  \\\hline   
\end{tabular}
\caption{The masses squared in the 3 exponent model where the tadpole conditions are satisfied.}
\label{tab:3expmasstad}
\end{table}

\FloatBarrier
\paragraph{Omitting the $\mathbf{S}$-exponent} can be done without including any new terms. The superpotential reads:
\be 
W_1 = g_7 + b^k S U_k + C^{ij} U_i T_j + \sum_i \left( A_{T_i} \rme^{\rmi a_{T_i}T_i} + A_{U_i} \rme^{\rmi a_{U_i}U_i} \right)\,,
\ee
and, because $A_S$ is absent, we solve for $b^1$ in addition to the usual parameters. Again, we use the parameters of table \ref{tab:3exppara}, except for those that are now determined by the equations $W = 0$ and $\partial_I W = 0$. The masses obtained this way are given in table \ref{tab:noSmass}.
\begin{table}[htb]
\center
\begin{tabular}{|c|c|c|c|c|c|c|c|}\hline
     &$\,m_1^{\,2}\,$&$\,m_2^{\,2}\,$&$\,m_3^{\,2}\,$&$\,m_4^{\,2}\,$&$\,m_5^{\,2}\,$&$\,m_6^{\,2}\,$&$\,m_7^{\,2}\,$\\\hline
Mk & $\, 0.40450 \,$ & $\, 0.21428 \,$ & $\, 0.10857 \,$ & $\, 0.02223 \,$ & $\, 0.01501 \,$ & $\, 0.00998 \,$ & $\, 0.00130 \,$ \\\hline  
dS & $\, 0.40513 \,$ & $\, 0.21465 \,$ & $\, 0.10870 \,$ & $\, 0.00223 \,$ & $\, 0.01503 \,$ & $\, 0.00999 \,$ & $\, 0.00130 \,$ \\\hline  
\end{tabular}
\caption{Eigenvalues of the mass matrix for the model without an exponent in the $S$-direction.}
\label{tab:noSmass}
\end{table}

\FloatBarrier
\paragraph{The 3 exponents in the $\mathbf{U_i}$ directions can be left out} by including the terms proportional to $a^i$ in \eqref{eq:mtheoryW}. The superpotential becomes:
\be
W = g_7 + S b^k U_k + U_i C^{ij} T_j + a^i \frac{U_1 U_2 U_3}{U_i} + A_S \rme^{\rmi a_S S} + \sum_i  A_{T_i} \rme^{\rmi a_{T_i}T_i}\,.
\ee
Instead of the $A_{U_i}$ we now solve for the parameters $a^i$. Keeping the remaining free parameters the same as in table \ref{tab:3exppara} we find the squared masses in table \ref{tab:noUmass}.
\begin{table}[htb]
\center
\begin{tabular}{|c|c|c|c|c|c|c|c|}\hline
     &$\,m_1^{\,2}\,$&$\,m_2^{\,2}\,$&$\,m_3^{\,2}\,$&$\,m_4^{\,2}\,$&$\,m_5^{\,2}\,$&$\,m_6^{\,2}\,$&$\,m_7^{\,2}\,$\\\hline
Mk & $\, 0.06600 \,$ & $\, 0.05485 \,$ & $\, 0.02910 \,$ & $\, 0.02028 \,$ & $\, 0.01588 \,$ & $\, 0.01061 \,$ & $\, 0.00066 \,$ \\\hline  
dS & $\, 0.06615 \,$ & $\, 0.05494 \,$ & $\, 0.02914 \,$ & $\, 0.02031 \,$ & $\, 0.01590 \,$ & $\, 0.01061 \,$ & $\, 0.00066 \,$  \\\hline  
\end{tabular}
\caption{Squared masses for the case without non-perturbative corrections in the $U$-directions.}
\label{tab:noUmass}
\end{table}

This model is in particular interesting as it still corresponds to a type IIA model. In \cite{Cribiori:2019drf,Cribiori:2019bfx} type IIA models were studied that had only 6-form flux as perturbative contributions but relied heavily on the inclusion of non-perturbative corrections in all directions. In particular, the origin of the exponents in the $U$-direction required some motivation as their origin is not fully understood.

\FloatBarrier
\paragraph{Building a model without $\mathbf{T}$ and $\mathbf{U}$ exponents} requires the inclusion of the terms proportional  to $c^i$. The superpotential becomes:
\be
W = g_7 + S b^k U_k + U_i C^{ij} T_j + a^i \frac{U_1 U_2 U_3}{U_i} + c^i \frac{T_1 T_2 T_3}{T_i} + A_S \rme^{\rmi a_S S}\,,
\ee
and we solve for $A_S$, the $a^i$ and the $c^i$ parameters. Note that this model can no longer be interpreted in terms of type IIA string theory, as discussed in section \ref{sec:IIAconn}. The parameters not set to zero or solved for are still taken from table \ref{tab:3exppara} and lead to the masses in table \ref{tab:noTUmass}.
\begin{table}[htb]
\center
\begin{tabular}{|c|c|c|c|c|c|c|c|}\hline
     &$\,m_1^{\,2}\,$&$\,m_2^{\,2}\,$&$\,m_3^{\,2}\,$&$\,m_4^{\,2}\,$&$\,m_5^{\,2}\,$&$\,m_6^{\,2}\,$&$\,m_7^{\,2}\,$\\\hline
Mk & $\, 0.06964 \,$ & $\, 0.06350 \,$ & $\, 0.02158 \,$ & $\, 0.00380 \,$ & $\, 0.00210 \,$ & $\, 0.00113 \,$ & $\, 0.00083 \,$ \\\hline  
dS & $\, 0.06948 \,$ & $\, 0.06315 \,$ & $\, 0.02152 \,$ & $\, 0.00380 \,$ & $\, 0.00208 \,$ & $\, 0.00113 \,$ & $\, 0.00082 \,$  \\\hline  
\end{tabular}
\caption{Eigenvalues of the mass matrix for the model without $T$ and $U$ exponents.}
\label{tab:noTUmass}
\end{table}

\FloatBarrier
\subsection{Type IIB Examples}
In type IIB models the superpotential has contributions from F-, H- and Q-flux \cite{Aldazabal:2006up,Dibitetto:2011gm,Blaback:2013ht}, which are all well-known and established. However, it has also been conjectured that, due to S-duality, P-fluxes should appear. These terms naturally appear in gauged supergravity in 4 dimensions \cite{Dibitetto:2011gm}. Keeping only terms even in the moduli we use the following superpotential:
\bea
\label{eq:noexpW}
W =\; &a_0 + a^i \frac{U_1 U_2 U_3}{U_i} + S \left( b^i U_i + b_3 U_1 U_2 U_3 \right) \\
&+ T_k \left( C^{ik}U_i - c^k U_1 U_2 U_3\right) - S T_k \left( d^k -D^{ik}\frac{U_1 U_2 U_3}{U_i}\right)\,.
\eea
The superpotential is made up of the following contributions:
\begin{itemize}
\item $a_0$ is a constant contribution, similar to $W_0$ in previous sections.
\item $a^i \frac{U_1 U_2 U_3}{U_i}$ comes from F-flux.
\item $S \left( b^i U_i + b_3 U_1 U_2 U_3 \right)$ gives H-flux contributions.
\item $T_k \left( C^{ik}U_i - c^k U_1 U_2 U_3\right)$ represents Q-flux.
\item $- S T_k \left( d^k -D^{ik}\frac{U_1 U_2 U_3}{U_i}\right)$ are the conjectured P-fluxes.
\end{itemize}
Using \eqref{eq:noexpW} we can build a model that does not rely at all on non-perturbative terms. First, we set the $D^{ik}=0$, as they are not required. Then, we find a Minkowski vacuum without flat directions by solving for $a_0$, $a^i$, $b_3$ and $c^k$ and using the parameters in table \ref{tab:noexppara}. Following the mass production procedure we then obtain a meta-stable de Sitter point with the downshift and uplift parameters:
\be 
\Delta a_0 = 5 \cdot 10^{-3}\qquad \text{and} \qquad \mu^4 = 9\cdot 10^{-9}\,.
\ee
The resulting masses are given in table \ref{tab:noexpmass}.
\begin{table}[htb]
\center
\begin{tabular}{|r|r||r|r||r|r||r|r||r|r|}\hline
$\,b^{1}$ & $\, 0.55 \,$  & $\,C^{11}$ & $ -0.11 \,$ & $\,C^{21}$ & $\, 0.21 \,$ & $\,C^{31}$ & $\, 0.31 \,$ & $\,d^{1}$ & $\, 5.1 \,$ \\\hline
$\,b^{2}$ & $\, 0.60 \,$ & $\,C^{12}$ & $\, 0.12 \,$ & $\,C^{22}$ & $ -0.22 \,$ & $\,C^{32}$ & $\, 0.32 \,$ & $\,d^{2}$ & $ -5.2 \,$ \\\hline
$\,b^{3}$ & $\, 0.65 \,$ & $\,C^{13}$ & $\, 0.13 \,$ & $\,C^{23}$ & $\, 0.23 \,$ & $\,C^{33}$ & $ -0.33 \,$ & $\,d^{3}$ & $\, 5.3 \,$ \\\hline
\end{tabular}
\caption{Parameter choices for the model without any exponents in IIB string theory/gauged supergravity.}
\label{tab:noexppara}
\end{table}
\begin{table}[htb]
\center
\begin{tabular}{|c|c|c|c|c|c|c|c|}\hline
     &$\,m_1^{\,2}\,$&$\,m_2^{\,2}\,$&$\,m_3^{\,2}\,$&$\,m_4^{\,2}\,$&$\,m_5^{\,2}\,$&$\,m_6^{\,2}\,$&$\,m_7^{\,2}\,$\\\hline
Mk & $\, 0.29074 \,$ & $\, 0.20712 \,$ & $\, 0.01075 \,$ & $\, 0.00383 \,$ & $\, 0.00287 \,$ & $\, 0.00057 \,$ & $\, 0.00016 \,$ \\\hline  
dS & $\, 0.29063 \,$ & $\, 0.20721 \,$ & $\, 0.01073 \,$ & $\, 0.00382 \,$ & $\, 0.00287 \,$ & $\, 0.00057 \,$ & $\, 0.00016 \,$  \\\hline  
\end{tabular}
\caption{Mass eigenvalues for the IIB model without non-perturbative corrections.}
\label{tab:noexpmass}
\end{table}

\FloatBarrier
\section{de Sitter Models from Anti-Branes - Interim Summary}
In this chapter we have investigated compactifications of string theory and M-theory that result in meta-stable de Sitter states where anti-branes were used in order to lift the vacuum energy to positive values. de Sitter spaces are of interest because the present day accelerated expansion of the universe matches very closely with the behavior of such a space with positive and constant, albeit very small, cosmological constant. The first model that can be argued to achieve a de Sitter space from string theory by uplifting with anti-branes is the KKLT scenario \cite{Kachru:2003aw,Kachru:2003sx}. This model is based in type IIB string theory and has an uplifting contribution from an anti-$D3$-brane. It remains one of the best studied de Sitter constructions to date and while there have been many criticisms raised, a lot of them have also been disproved or explained.\\
In \cite{Cribiori:2019bfx} and here in section \ref{sec:IIAuplift}, we translated this mechanism to type IIA string theory. There, the uplift is performed via the inclusion of an anti-$D6$-brane, which is the unique choice in type IIA for the manifolds we considered. The procedure relied on the inclusion of non-perturbative corrections in all moduli directions, some of which are non-standard. We motivated that they can still appear in the theory and that it is possible to use them to build meta-stable de Sitter states.\\
In section \ref{sec:massprod}, based on \cite{Kallosh:2019zgd,Cribiori:2019drf} we showed that there is a simple way to guarantee that a meta-stable de Sitter minimum can be obtained if one first constructs a Minkowski state without flat directions. We used a Kallosh-Linde type racetrack potential for all moduli directions. With this we will automatically not have any flat directions. Under certain conditions it is then straight forward to first go to anti-de Sitter space via a small downshift and subsequently to a de Sitter via an uplifting contribution from anti-$Dp$-branes. Going to AdS first allows to decouple the cosmological constant scale from the SUSY breaking scale when aiming to achieve a realistic value of $\Lambda = 10^{-120}$ in Planck units. We showed under what conditions it is possible to predict that the masses stay positive and thus also that the dS vacuum state is meta-stable. Several examples in type IIA and IIB were presented.\\
Models that do have perturbative contributions to the superpotential were explored in \cite{Cribiori:2019hrb} and section \ref{sec:mtheory}. These scenarios are motivated by M-theory constructions and can be matched with type IIA or IIB models in certain special cases. Using tree-level flux contributions to the superpotential it is possible to follow the mass production procedure of de Sitter vacua without including the double exponent contribution to the superpotential in all directions. In fact, in type IIA models were presented that include non-perturbative corrections only in some directions. Perhaps most interestingly, a model that does not need exponential terms in the $U$-directions and can still be matched to type IIA string theory was presented. Including atypical fluxes, conjectured via S-duality, in IIB one finds a gauged supergravity model that even allows the procedure to work without any non-perturbative contributions to the superpotential at all.\\
The extension of the KKLT scenario to type IIA and the development of the mass production process are interesting extensions of the string model builders toolkit. The inclusion of unusual contributions in some of the models can serve as an incentive to investigate the origin of these contributions further, especially with the growing interest in semi-realistic models that aim to describe future precision cosmology experiments.

\chapter{The de Sitter Swampland Conjecture}
\label{sec:swamplandconjectures}
\begin{figure}[htb]
     \centering
     \includegraphics[scale=0.7]{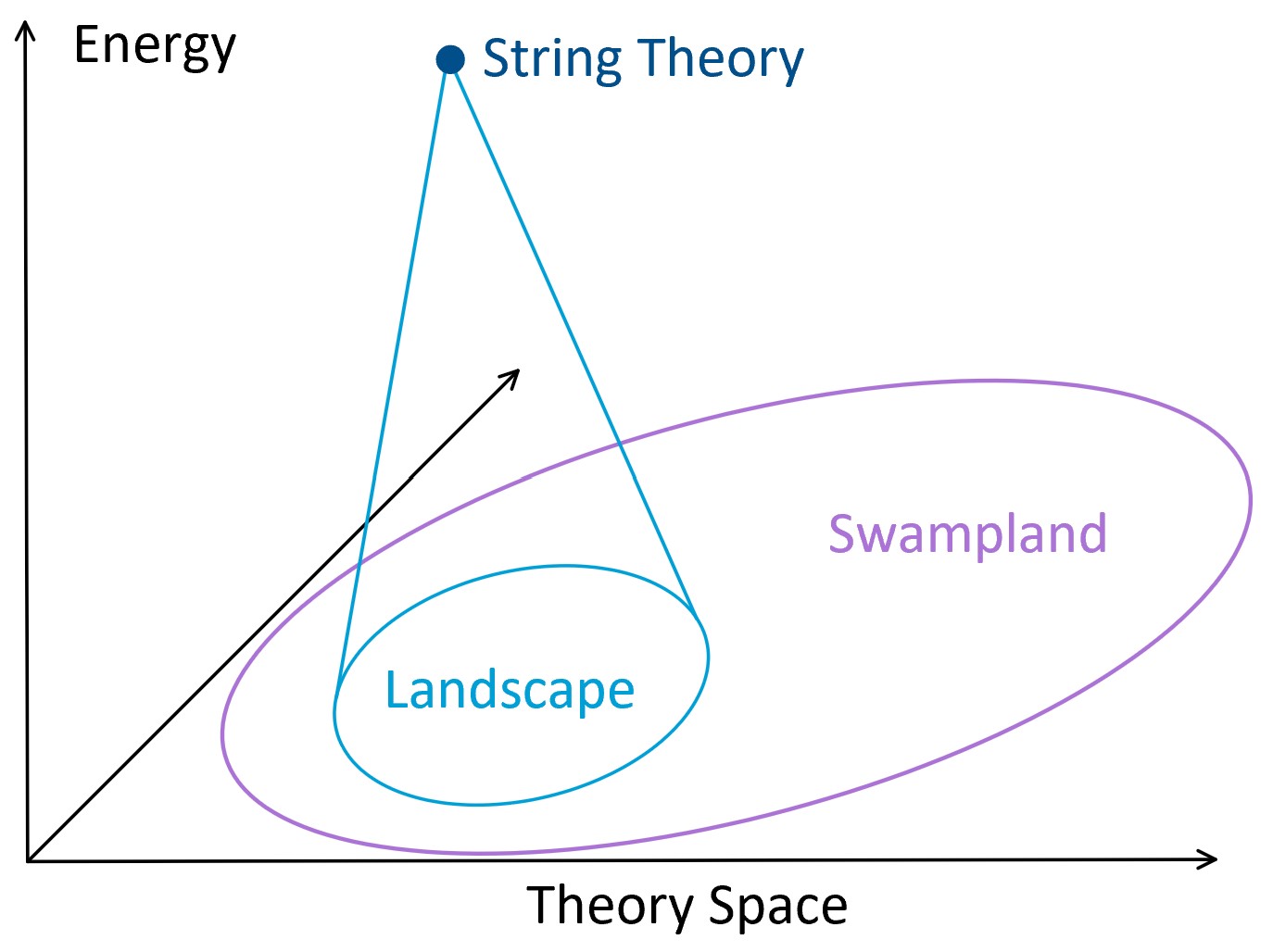}
     \caption{Schematic picture of the relation between the swampland, the landscape and string theory. Full string theory or quantum gravity arises at or around the Planck scale.}
     \label{fig:swamppic}
\end{figure}

Critical superstring theory lives naturally in $10$ spacetime dimensions. In nature or experiments we do not observe these extra dimensions and thus, in an attempt to describe the world we observe using string theory, we compactify down to $4$ extended dimensions and $6$ compact ones. The argument then goes such that at low energies $6$ dimensions curl up and become so small that we cannot observe them. If we go back to higher energies the compact directions will extend and become relevant again. Thus, a string compactification is a low-energy, effective theory that we usually describe in terms of supergravity.\\
The process of starting from $10d$ string theory and compactifying down to a realistic model is difficult and not every step is fully understood as of now. Not only would we need to start from full string theory but we need to choose a suitable compact manifold, extended objects like branes, that might be part of the model and fluxes on different cycles of these manifolds. This is incredibly difficult and thus we often build models directly in the low-energy theory and argue that they should be obtainable from string theory. This is achieved by certain requirements that we either obtain from full string theory or that are intuitive. One example is the requirement that the string coupling is weak. Only then are we able to neglect string loop corrections which would require control over string theory itself. The list of typical requirements goes on, with things like flux quantization, tadpole cancellation, large internal volume and so on. The idea behind swampland considerations is to take inspiration from these conditions to attempt to restrict possible models. Sometimes, conclusions drawn from the swampland are said to hold in any theory of quantum gravity. For some conjectures this is very reasonable. While for others the arguments are more suggestive. We will not go into the details of this discussion and will always assume the context of string theory.\\
In particular, the swampland program is concerned with whether the UV completion, in a sense the inverse of a compactification, of a given model is possible or not. This can be viewed as depicted in figure \ref{fig:swamppic}. There is a large amount of low-energy effective theories that can be seemingly consistent at low energies, however, only a subset, called the landscape, can be lifted to full quantum gravity. The rest of the theories lie within the swampland. For a good overview of the topic see the reviews \cite{Brennan:2017rbf,Palti:2019pca,vanBeest:2021lhn}. In this thesis we will focus on the de Sitter swampland conjecture that will be discussed in the following section. Based on \cite{Banlaki:2018ayh} we investigate the scaling limits of a certain class of type IIA compactifications for the possibility of finding stable de Sitter vacua. In section \ref{sec:dSextrema} we show how the original de Sitter swampland conjecture has a critical flaw \cite{Roupec:2018mbn} and in section \ref{sec:refconj} we present our own refined conjecture \cite{Andriot:2018mav}.
\section{de Sitter Space and the Swampland}
In order to describe the late time accelerated expansion of our universe it is necessary to have a positive vacuum energy. One way to describe such a situation is via a de Sitter vacuum. Models from string theory, that emulate such a situation, were studied in the previous chapter. Here, we consider that it might be futile to attempt to build such models. Previously, it was argued that there should be no obstruction to obtain dS space from string theory. One, very simple, motivation for this claim is that a rough estimate of the amount of possible, different models from string theory is given by the number of different Calabi-Yau manifolds, which can be estimated to be at the very least about $10^{500}$ \cite{Blumenhagen:2004xx}. Considering different setups, for example F-theory, the exponent can be even larger, by many orders of magnitude. Such a enormous numbers suggest that there should be many possible theories and certainly some that allow for a de Sitter vacuum with a cosmological constant that matches what we observe. Following the anthropic principle \cite{Weinberg:1987dv}, one would not be surprised if everything works out. Nevertheless, explicit constructions of de Sitter spaces remain very difficult and challenging in string theory setups. As a result, after a similar argument presented already in \cite{Danielsson:2018ztv}, the authors of \cite{Obied:2018sgi} proposed a conjecture that effectively rules out de Sitter vacua. For any scalar potential $V$ there should be a bound on its gradient of form:
\be 
|\nabla V| \geq \frac{c}{M_P} V\,,
\label{eq:dSconori}
\ee
with some $\mathcal{O}(1)$ number $c>0$. Since for dS $V>0$ and at a (meta-)stable minimum $|\nabla V| = 0$, we see that de Sitter vacua are not allowed. The original motivation for this conjecture were ``classical'' flux compactifications that do not rely on uplifting contributions from branes. In a sense these models are simpler than setups that produce de Sitter vacua via an anti-brane uplift. Later, in \cite{Ooguri:2018wrx} an argument, based on the entropy in de Sitter space was given in order to further motivate the conjecture.\\
The conjecture \eqref{eq:dSconori} immediately has some problems. First off, it also forbids unstable maxima with positive values for the potential. Such points, however, can be abundantly found, as we showed in \cite{Roupec:2018mbn}. We will review our findings in section \ref{sec:dSextrema}. Another problem appears when considering a coupling of the effective theory to the Standard Model \cite{Denef:2018etk}. It was shown there that the bound is violated by $50$ orders of magnitude when considering the Higgs potential as a scalar potential in string theory. While the number $c$ is not very well defined and one might even consider values ranging from $0.01$ to $100$ as ``$\mathcal{O}(1)$'', this is certainly problematic. All of these findings lead to a refinement of the conjecture, presented in \cite{Ooguri:2018wrx}, which takes the form of two conditions, combined with a logical ``or'':
\be
|\nabla V| \geq \frac{c}{M_P} V\quad \text{or} \quad \text{min}(\nabla_i \nabla_j V) \leq - \frac{c^\prime}{M_P^2} V\,.
\label{eq:dScon}
\ee
Here, both $c$ and $c^\prime$ are order one numbers. Any scalar potential then needs to satisfy either the first or the second condition. This disallows stable minima but does allow maxima. For the coupling to the Standard Model sector we find that:
\bea 
\frac{|\nabla V|}{V} &\sim \frac{10^{-55}}{M_P}\,,\\
\text{min}(\nabla_i \nabla_j V) &\sim -\frac{10^ {35}}{M_P^ 2}\,.
\eea 
We see that for the Higgs potential the second condition is satisfied. In \cite{Andriot:2018mav} we presented an alternative refinement and were able constrain the appearing parameters using data points from \cite{Roupec:2018mbn}. These results will be presented in section \ref{sec:refconj}. However, before we come to these direct investigations of the swampland conjecture, we want to have a look at scaling limits of classical type IIA compactifications and the general possibility of finding a stable de Sitter solution in such setups.

\section{Scaling Limits of de Sitter Vacua}
\label{sec:scaling}
In this section we review a peculiar feature of certain type IIA flux compactifications where a certain scaling limit exists that allows for controllable solutions of the low-energy effective theory. We investigate whether or not this limit does allow for consistent de Sitter solutions. This section is based on \cite{Banlaki:2018ayh} and similar research was done in \cite{Junghans:2018gdb}.
\subsection{Type II Flux Compactifications}
The specific setups we consider here are models that are compactifications of $10d$ type IIA supergravity on $SU(3)$ structure manifolds including RR- and NSNS-fluxes, orientifold planes, $Dp$-branes and potentially KK-monopoles and NS5-branes. After compactification, one obtains a $4d$, $\mathcal{N}=1$ supergravity model. A less studied alternative are type IIB models with $O5$/$O7$-planes \cite{Aharony:2010af,Lust:2009zb,Petrini:2013ika}.\\
One subset of these compactifications are massive type IIA models based on flat Calabi-Yau 3-folds that include $O6$-planes, NSNS-flux $H_3$ as well as RR-fluxes $F_0$, $F_2$ and $F_4$. The RR-fluxes are typically constrained by tadpole conditions but in the particular setup we are investigating the $F_4$-flux is unconstrained. The dimensional reduction of these theories was studied in \cite{Grimm:2004ua} and generically lead to AdS vacua \cite{Villadoro:2005cu,DeWolfe:2005uu,Camara:2005dc,Ihl:2006pp} and in a limit of large $F_4$-flux a separation between the AdS and KK-scale is obtained. In \cite{Hertzberg:2007ke} a numerical investigation of the scalar potential was performed in an attempt to find regions with positive values for the potential and very small slow-roll parameter:
\be 
\epsilon = \frac{1}{2} \frac{|\nabla V|^ 2}{V^ 2}\,,
\ee
corresponding to a flat region. For the models considered in \cite{Villadoro:2005cu,DeWolfe:2005uu,Ihl:2006pp} no such regions were found. This is relevant for our discussion here because the slow-roll parameter is related to the order one parameter of the initial de Sitter conjecture \eqref{eq:dSconori} in the following way:
\be 
\epsilon = \frac{c^ 2}{2}\,.
\ee
In fact it is possible to show the lack of these regions in an analytical way \cite{Hertzberg:2007wc}. For this we first define:
\bea 
\rho &= \left(vol_6\right)^ {\frac{1}{3}} \;\qquad \text{as the volume modulus and}\\
\tau &= \rme^ {-\phi} \sqrt{vol_6} \;\quad \text{as the dilaton.}
\eea 
With this we can schematically write the scalar potential of these models as:
\be 
V(\rho,\tau) = \frac{A_H}{\rho^3 \tau^2} + \sum_{p=0,2,4,5} \frac{A_p}{\rho^ {p-3} \tau^4} - \frac{A_{O6}}{\tau^3}\,,
\ee
with the $A$'s being positive functions of the flux quanta and possibly other moduli than $\rho$ and $\tau$. From this one can obtain the following condition:
\be 
-\rho \frac{\partial V}{\partial \rho} - 3 \tau \frac{\partial V}{\partial \tau} = 9 V + \sum_{p=0,2,4,5} \frac{A_p}{\rho^ {p-3} \tau^4} \geq 9V\,.
\label{eq:simpnogo}
\ee
This immediately forbids de Sitter critical points since for their existence the derivatives would vanish while the scalar potential is necessarily positive. For the models investigated in \cite{Hertzberg:2007wc} it is possible to find an explicit bound for the slow-roll parameter, and thus also $c$, as:
\be 
\epsilon \geq \frac{27}{13} \quad \Rightarrow \quad c \gtrsim 2\,.
\ee 
It is then natural to investigate broader classes of models that either relax some constraints or include more ingredients \cite{Caviezel:2008tf,Flauger:2008ad,Saltman:2004jh,Silverstein:2007ac,Haque:2008jz,Danielsson:2009ff,deCarlos:2009fq,deCarlos:2009qm,Caviezel:2009tu,Danielsson:2010bc,Dong:2010pm,Andriot:2010ju}. In order to avoid the above no-go theorem and similar ones presented, for example, in \cite{Flauger:2008ad,Silverstein:2007ac,Haque:2008jz,Danielsson:2009ff,Caviezel:2009tu,Caviezel:2008ik,Wrase:2010ew,Andriot:2016xvq,Andriot:2017jhf}\footnote{All of these satisfy the dS swampland conjecture with $c \gtrsim \mathcal{O}(1)$.}, the most straight forward, and also promising, generalization is to go towards non-flat manifolds. With the natural choice being $SU(3)$ structure manifolds. Furthermore, it is also possible to study type IIB models that can potentially evade all no-go theorems. The simplest example that can work is similar to the above Calabi-Yau case but now on an $SU(3)$ structure manifold with curvature. Additionally, we also allow for more general sources and let $A_{O6} \to A_{\text{sources}}$. The scalar potential then becomes:
\be 
V(\rho,\tau) = \frac{A_H}{\rho^3 \tau^2} + \sum_{p=0,2,4,5} \frac{A_p}{\rho^ {p-3} \tau^4} - \frac{A_{\text{sources}}}{\tau^3} + \frac{A_{R_6}}{\rho\tau}\,,
\label{eq:genscal}
\ee
where we have $A_{R_6}>0$ for spaces with negative curvature. $A_{\text{sources}}$ is composed of the contributions from $O6$-planes and (anti-) $D6$-branes and behaves like:
\be
A_{\text{sources}} \sim 2 N_{O6} - N_{D6} - N_{\overline{D6}}\,.
\ee
It turns out that for the non-flat case no simple no-go theorem in the spirit of \eqref{eq:simpnogo} exists. We would, however, like to stress that this does not mean that de Sitter extrema exist \cite{Obied:2018sgi} and, under certain conditions, there are still no-go theorems as we will review in the following. Even if no no-go theorem is known, other reasons for the lack of dS points can certainly be present.

\subsection{Maldacena-Nu\~nez like No-Go Theorem}
A well-known theorem by Maldacena-Nu\~nez \cite{Maldacena:2000mw} states that there are no de Sitter vacua in the absence of $O6$-planes. Here, we replicate this result for our model based on \eqref{eq:genscal}. First off, in our case we generalize to $A_{\text{sources}} \sim 2 N_{O6} - N_{D6} - N_{\overline{D6}} < 0$ as this is the more general form of $N_{O6}=0$ here. In order to arrive at a no-go theorem we calculate:
\be 
0= \partial_\tau V = -2 \frac{A_H}{\rho^ 3 \tau^3} - 4 \sum_{p=0,2,4,6} \frac{A_{F_p}}{\rho^ {p-3}\tau^ 5} - 3 \frac{|A_{\text{sources}}|}{\tau^ 4} - 2 \frac{A_{R_6}}{\rho \tau^3}\,,
\ee
at an extremum. In the case of manifolds with negative curvature ($A_{R_6} \propto -R_6 >0$) there exists no solution to the above equation since all terms are negative definite. If the manifold is positively curved we can solve the equation above and find:
\be 
\frac{A_{R_6}}{\rho \tau^ 2} = - \frac{A_H}{\rho^ 3 \tau^2} - 2 \sum_{p=0,2,4,6} \frac{A_{F_p}}{\rho^ {p-3}\tau^ 4} - \frac{3}{2} \frac{|A_{\text{sources}}|}{\tau^ 3}\,,
\ee
where we not only rearranged the terms but also multiplied the whole expression with $\tau$.\\
This expression can be plugged back into the scalar potential \eqref{eq:genscal} which yields that:
\be 
V = -\sum_{p=0,2,4,6} \frac{A_{F_p}}{\rho^ {p-3}\tau^ 4} - \frac{1}{2} \frac{|A_{\text{sources}}|}{\tau^ 3} < 0\,.
\ee
We thus find that if the amount of $D6$-branes plus the number of anti-$D6$-branes exceeds twice the amount of $O6$-planes we cannot obtain a de Sitter solution. While we can, in principle, add (anti-) $D6$-branes as much as we want the same does not hold for $O6$-planes. Their number is given by the number of fixed points under the orientifold projection which is a fixed property of the internal manifold.
\subsection[The large $F_4$-Flux Limit for Anti-de Sitter]{The large $\mathbb{F_4}$-Flux Limit for Anti-de Sitter}
The type IIA AdS solutions originally found in \cite{DeWolfe:2005uu} have the peculiar feature that the $F_4$ flux is not constrained by tadpole conditions and can become arbitrarily large. The authors were able to use this feature in order to obtain well-controlled, supersymmetric AdS solutions with scale separation, meaning that they have large internal volume and parametrically weak string coupling, see also \cite{Dasgupta:1999ss,Giddings:2001yu,Balasubramanian:2005zx,Gautason:2015tig,Westphal:2006tn,Farakos:2020phe} for more considerations about scale separation in string model constructions. In terms of the scaling behavior of $F_4$, which we are going to call $f_4$ \footnote{The $F_4$ flux is given as $a_{F_4} (f_4)^ 2$ with $a_{F_4}$ an order one parameter.}, the volume and string coupling respectively behave like:
\bea 
vol_6 &\propto (f_4)^ {\frac{3}{2}}\,,\\
\rme^ {-\phi} &\propto (f_4)^ {\frac{3}{4}}\,.
\eea
Additionally, the $4d$ Hubble constant is parametrically smaller than the KK-scale $1/R$ as $H\cdot R \propto (f_4)^ {-1/2}$. This means that the models are truly four dimensional.\\
We now attempt to use the scaling freedom of such models in order to circumvent or at least improve on the dS no-go theorems using our scalar  potential \eqref{eq:genscal}. First, let us start again with flat internal spaces, in particular Calabi-Yau manifolds. We then have $R_6=0 \Rightarrow A_{R_6} =0$ and if we take:
\bea 
A_{F_4} &= a_{F_4} (f_4)^ 2\,,\\
\rho &= (vol_6)^ {\frac{1}{3}} = \tilde{\rho} (f_4)^ {\frac{1}{2}}\,,\\
\tau &= \rme^ {-\phi} \sqrt{vol_6} = \tilde{\tau} (f_4)^ {\frac{3}{2}}\,,
\eea
we can write the scalar potential in the following way by making all scalings with $F_4$ explicit:
\be 
V = \frac{1}{(f_4)^ {\frac{9}{2}}} \left( \frac{A_H}{\tilde{\rho}^3 \tilde{\tau}^ 2} + \frac{A_{F_0}}{\tilde{\rho}^ {-3}\tilde{\tau}^ 4} + \frac{a_{F_4}}{\tilde{\rho}\tilde{\tau}^ 4} - \frac{A_{\text{sources}}}{\tilde{\tau}^ 3} \right) + \frac{1}{(f_4)^ {\frac{11}{2}}} \frac{A_{F_2}}{\tilde{\rho}^ {-1}\tilde{\tau}^ 4} + \frac{1}{(f_4)^ {\frac{15}{2}}} \frac{A_{F_6}}{\tilde{\rho}^3\tilde{\tau}^ 4}\,.
\ee
It is evident that the terms involving the $2$- and $6$-flux become irrelevant unless they are allowed to scale as well. Note that the $H$, $F_0$ and $F_2$ flux are constrained by the tadpole condition $dF_2 + F_0H = N_{O6}$ and thus cannot be arbitrarily large. Fortunately, $F_2$ and $F_6$ are not crucial for moduli stabilization \cite{DeWolfe:2005uu} as they only affect the values of the axions $B_2$ and $C_3$ at their minimum. This limit also allows for an order one number of $O6$-planes with small $H$- and $F_0$-flux as these three are related via the tadpole condition:
\be 
\sqrt{2} \int_{\Sigma_K} d \left(dC_1 + F_0 B\right) = \sqrt{2} \int_{\Sigma_K} F_0 H = (-2 N_{O6} + N_{D6} + N_{\overline{D6}})|_{\text{wrapping }\Sigma_K}\,,
\ee
which holds for all 3-cycles $\Sigma_K$ in integer 3-homology. The conclusion thus being that even for a small amount of $O6$-planes one can obtain an arbitrarily well controlled, supersymmetric AdS point. Such AdS solutions have dual CFT$_3$ theories with interesting features, some of which have been studied in \cite{Aharony:2008wz}.\\
If one moves away from flat CY-manifolds to internal spaces with non-vanishing curvature, the large $F_4$-flux limit becomes obstructed. This is because the curvature term is dominant in this limit:
\be 
V = \frac{1}{(f_4)^ {\frac{7}{2}}} \frac{A_{R_6}}{\tilde{\rho} \tilde{\tau}^ 2} +\frac{1}{(f_4)^ {\frac{9}{2}}} \left( \frac{A_H}{\tilde{\rho}^3 \tilde{\tau}^ 2} + \frac{A_{F_0}}{\tilde{\rho}^ {-3}\tilde{\tau}^ 4} + \frac{a_{F_4}}{\tilde{\rho}\tilde{\tau}^ 4} - \frac{A_{\text{sources}}}{\tilde{\tau}^ 3} \right) + \frac{1}{(f_4)^ {\frac{11}{2}}} \frac{A_{F_2}}{\tilde{\rho}^ {-1}\tilde{\tau}^ 4} + \frac{1}{(f_4)^ {\frac{15}{2}}} \frac{A_{F_6}}{\tilde{\rho}^3\tilde{\tau}^ 4}\,.
\ee
With such a potential it is not possible to achieve moduli stabilization at very large $F_4$ and hence no vacua can exist.\\
A particular concern that arises after considering the non-flat case is that even in Calabi-Yau compactifications the localization of the $O6$-planes might introduce an effective curvature via backreaction. In such a scenario, however, one first needs to identify the correct moduli as they will change as well due to the backreaction. Then it is necessary to fully understand the new, warped, effective field theory. Generically, it is expected that the effects of the $O$-plane backreaction should be negligible for weak coupling and large volume and thus the parametrically controlled AdS vacua we investigated here should be safe \cite{Acharya:2006ne,Saracco:2012wc,Gautason:2015tig,Junghans:2020acz}. 
\subsection[The large $F_4$-Flux Limit in de Sitter]{The large $\mathbf{F_4}$-Flux Limit in de Sitter}
\label{sec:largef4}
We would now like to study whether or not the the models we investigated in the last subsection for the case of AdS allow, in principle, de Sitter points. Numerous no-go theorems against dS solutions with vanishing mass parameter ($A_{F_0} = 0$) or positive internal curvature ($A_{R_6}\leq 0$) exist \cite{Wrase:2010ew,Flauger:2008ad,Junghans:2016uvg} but in general nothing forbids such points.\\
The case with vanishing mass parameter would be interesting because then one can lift the solutions to M-theory. Looking at our simplified scalar potential \eqref{eq:genscal} with $A_{F_0} = 0$ we find:
\be 
\left( - \rho \partial_\rho - \tau \partial_\tau \right) V = 5 \frac{A_{H}}{\rho^ 3 \tau^2} + \sum_{p=2,4,6} (p+1) \frac{A_{F_p}}{\rho^ {p-3} \tau} - 3 \frac{A_{\text{sources}}}{\tau^3} + 3 \frac{A_{R_6}}{\rho \tau^2} \geq 3 V\,,
\label{eq:masslessbound}
\ee 
so that again a stable de Sitter point cannot exist as for this $V>0$ and $\partial_\tau V = \partial_\rho V = 0$ would be required. One can go further by using the above condition to calculate a bound on the first slow-roll parameter. For this purpose, we need to identify the kinetic terms of $\rho$ and $\tau$ which are obtained from the dimensionally reduced and properly normalized Lagrangian \cite{Hertzberg:2007wc}:
\bea
\mathcal{L} &= \frac{1}{2} R - \frac{3}{4} \frac{\partial_\mu \rho \partial^ \mu \rho}{\rho^ 2} - \frac{\partial_\mu \tau \partial^ \mu \tau}{\tau^ 2} + \ldots\\
&= \frac{1}{2} R - \frac{1}{2} \partial_\mu \hat{\rho} \partial^ \mu \hat{\rho} - \frac{1}{2} \partial\mu \hat{\tau} \partial^ \mu \hat{\tau} + \ldots \,.
\eea
The $\ldots$ here stand for all other potential fields appearing in the action. Using this expression we find the first slow-roll parameter to be:
\be
\epsilon=\frac{1}{2} \frac{(\partial_{\hat{\rho}}\V)^ 2 + (\partial_{\hat{\tau}}\V)^ 2 + \ldots}{V^ 2} \geq \frac{1}{3} \left( \frac{\rho \partial_\rho V}{V}\right)^ 2 + \frac{1}{4} \left( \frac{\tau \partial_\tau V}{V} \right)^2\,.
\ee
Together with the condition we obtained for the case with $A_{F_0} = 0$ \eqref{eq:masslessbound} this yields a bound on $\epsilon$:
\be
\epsilon \geq \frac{1}{3} \left(3+\frac{\tau \partial_\tau V}{V} \right)^2 + \frac{1}{4} \left( \frac{\tau \partial_\tau V}{V} \right)^2\geq \frac{9}{7}\,,
\ee
where the numeric value can be obtained by minimizing the expression with respect to $\tau\cdot \partial_\tau V/V$.\\
A similar no-go arises if we require $A_{R_6} \leq 0$. Then one obtains:
\be
\left(-\rho \partial_\rho - 3 \tau \partial_\tau\right) V \geq 9V\,,
\ee
which leads to a numerical bound on the slow-roll parameter: $\epsilon \geq 27/13$ \cite{Hertzberg:2007wc}. We conclude that setups with vanishing mass parameter and/or negative internal curvature can never lead to stable de Sitter solutions.\\
While the large $F_4$ limit of the previous section is obstructed when we have non-zero curvature, there might still be other limits in the same vein that do allow for a controlled region with $\rho,\,\tau \gg 1$. The hope is then that such a solution would be likewise under good control. Thus far, we have seen that we require the following conditions:
\bea 
A_{\text{sources}} &> 0\,,\\
A_{F_0} &\neq 0\,,\\
A_{R_6} &>0\,.
\eea 
We now let the moduli scale in the following way:
\be 
\rho \propto \lambda^ {c_\rho} \quad \text{and} \quad \tau \propto \lambda^ {c_\tau}\,,
\ee
such that $\lambda \to \infty$ corresponds to a limit with large volume and small coupling. If we were to keep $A_{\text{sources}}$, $A_{F_0}$, and $A_{R_6}$ constant in the large $\lambda$ limit the corresponding terms would never scale the same as for that to happen one would need:
\be 
3 c_\tau = 4 c_\tau - 4 c_\rho = 2 c_\tau + c_\rho\,,
\ee
which is impossible to satisfy with just two free parameters, unless we allow the trivial solution $c_\rho = c_\tau = 0$. Then, however, the volume and dilaton do not scale as well and we gain nothing. Since, due to the no-goes discussed above, we require $A_{\text{sources}}$, $A_{F_0}$, and $A_{R_6}$ to be relevant, we cannot find a limit of parametrically large volume and weak coupling in this case.\\
The only way we can have a controlled limit is to allow scaling of other terms in the scalar potential \eqref{eq:genscal}. It is important to note that all terms in the scalar potential, except the source term, are quadratic in fluxes \cite{Herraez:2018vae} with the curvature being quadratic in the metric fluxes $\omega$. The final term in the scalar potential, $A_{\text{sources}}$ is linear in the number of $O6$-planes and (anti-) $D6$-branes:
\be 
A_{\text{sources}} \propto 2 N_{O6} - N_{D6} - N_{\overline{D6}}\,.
\ee
Nevertheless, the tadpole condition involving this source terms:
\be 
\sqrt{2} \int \left( \omega \cdot F_2 + F_0 H\right) = -2 N_{O6} + N_{D6} + N_{\overline{D6}}\,,
\ee
allows us to replace the $A_{\text{sources}}$ term by something that is quadratic in fluxes. More importantly though is that we notice from the above relation that the scaling of the sources has to be the same as $\sqrt{2} \int \left( \omega \cdot F_2 + F_0 H\right)$. We conclude that $-2 N_{O6} + N_{D6} + N_{\overline{D6}}$ is necessarily negative and, since the amount of $O6$-planes is fixed by the compactification manifold, $A_{\text{sources}}$ cannot be scaled to become arbitrarily large. This conversely means that $F_0$ and $\omega$ have to be of the same order and thus $A_{\text{sources}}$, $A_{F_0}$ should all be of the same order as well.\\
However, one loophole remains. Namely, we can have small $A_{\text{sources}}$ and large $\int \omega \cdot F_2$ and $\int F_0 H$ if they have approximately the same magnitude but opposite signs. The terms then almost cancel and allow a small contribution from the sources. In order to investigate this situation we allow the following additional scalings:
\be 
A_H \propto \lambda^ {c_H}, \quad A_{F_p} \propto \lambda^ {c_{F_p}},\quad A_{R_6} \propto \lambda^ {c_{R_6}}\,.
\ee
We still need $A_{\text{sources}}$, $A_{F_0}$, and $A_{R_6}$ to be of same order which boils down to the two conditions:
\be 
-3 c_\tau = c_{R_6} - c_\rho - 2 c_\tau \qquad \text{and} \qquad -3c_\tau = c_{F_0} + 3 c_\rho - 4 c_\tau\,.
\ee
This can easily be solved and the relevant terms scale the same for:
\be 
c_\tau = c_\rho - c_{R_6} \qquad \text{and} \qquad c_\rho = -\frac{1}{2} \left( c_{F_0} + c_{R_6} \right)\,.
\ee
Unfortunately, this does not facilitate our goal of having parametrically large volume and small coupling. The reason for this is that all fluxes are bounded from below and the second condition then implies that $c_\rho$ must be negative. Which, in turn, immediately leads to a small volume in the $\lambda \to \infty$ limit since $vol_6 = \rho^ 3 \propto \lambda^ {3 c_\rho} = \lambda^ {-3 |c_\rho|}\,$.\\
The conclusion of this investigation is that we are unable to establish parametrically controlled de Sitter solutions in the setup we discussed thus far. Nevertheless, there is still some hope left. If we can establish some hierarchies between the different contributions, represented by the $A$'s in \eqref{eq:genscal}, it could still be possible to arrive at a de Sitter point. In setups where only one $O6$-plane wraps each 3-cycle \cite{Danielsson:2011au} this seems, unfortunately, unlikely. Due to the limited amount of $O6$-planes, one has $F_0$ and $H$ fixed to a similarly small number as well. For example, let us consider a compactification on $T^ 6/(\mathbb{Z}_2 \times \mathbb{Z}_2)$ with an orientifold projection that flips the signs of the three circle directions. In this case one has $2^ 3 = 8$ $O6$-planes. However, due to the orbifold of $\mathbb{Z}_2 \times \mathbb{Z}_2$ plus an additional $\mathbb{Z}_2$ orientifolding leading us to an identification of $2^ 3 = 8$ fixed points. This leaves us with only one $O6$-plane in the end. Still, it is in principle possible to have complicated internal manifold with many fixed points under the orientifold projection and thus a larger number of $O$-planes. While we cannot present an explicit example here we will investigate the behavior of such models in general. To this end, we now allow the sources term to scale like $A_{\text{sources}} \propto \lambda^ {c_\text{s}}$, which means that we allow a large number $N_{O6}$ of $O6$-planes in our model. The terms $A_{\text{sources}}$, $A_{F_0}$, and $A_{R_6}$ then scale the same if we satisfy the conditions:
\be 
-3 c_\rho + 4 c_\tau -c_{F_0} = 3 c_\tau - c_{\text{s}} = c_\rho + 2 c_\tau -c_{R_6}\,.
\ee
Furthermore, if we keep in mind that the contributions from fluxes in the scalar potential are quadratic in flux quanta, we find from the tadpole condition that:
\be 
\frac{1}{2} \left( c_{R_6} + c_{F_2} \right) = \frac{1}{2} \left( c_{F_0} + c_H \right) = c_{\text{s}}\,.
\ee
With all of these conditions we can calculate a solution that allows for all required terms to stay relevant at the same time. This is achieved for:
\bea 
c_\rho &= \hspace{6pt} c_{\text{s}} - \frac{1}{2} \left( c_{F_0} + \;\, c_{R_6} \right)\,,\\
c_\tau &= 2 c_{\text{s}} - \frac{1}{2} \left(c_{F_0} + 3 c_{R_6} \right)\,,\\
c_{F_2} &= 2 c_{\text{s}} - c_{R_6}\,,\\
c_{H} &= 2 c_{\text{s}} - c_{F_0}\,.
\eea
Using these results we can now calculate $vol_6 = \rho^3$ and $\rme^ {-\phi}/\sqrt{vol_6}$ to behave like:
\bea 
vol_6 \propto \lambda^ {\frac{3}{2} \left[ 2 c_{\text{s}} - \left(c_{F_0} + c_{R_6}\right)\right]}\,,\\
\rme^ {-\phi} \propto \lambda^ {\frac{1}{4} \left[2c_{\text{s}}+c_{F_0} -3 c_{R_6}\right]}\,,
\eea
in the large $\lambda$ limit. For a parametrically controlled solution we need a large internal volume and small coupling. This can be obtained by requiring:
\be 
c_{\text{s}} > \frac{3}{2} c_{R_6} + \frac{1}{2} c_{F_0}\,.
\ee
The solution holds even for the case where $c_{F_0} = 0 = c_{R_6}$, which allows us to look at the interesting special case where only the number of $O6$-planes grows. Then we find:
\bea
vol_6 &\propto N^ 3_{O6}\,,\\
\rme^ {-\phi} &\propto \sqrt{N_{O6}}\,,
\eea
which tells us that controlled solutions can exists for a large number of orientifold planes. The question remaining is whether such compactifications, that allow for a large number of orientifold planes, can exist. In the case of F-theory and IIB compactifications on Calabi-Yau manifolds it is well understood what number of $O3$-planes one can expect, the same is not as clear for the case of compactifications on $SU(3)$-structure manifolds in type IIA. Constructing an explicit example with a large number of $O6$-planes remains an intriguing and challenging open problem.
\subsection{Additional Ingredients}
Thus far we have limited ourselves to the most typical ingredients that can appear in a scalar potential of a massive type IIA flux compactification. There are, however, additional contributions one might consider. The issue with including them, however, is that each new contribution comes with potential issues, like complicated backreaction effects and new degrees of freedom. Hence, one attempts to build the simplest model first and only includes more exotic contributions if absolutely necessary. Ingredients we will consider are $NS5$-branes, $NSO5$-planes, $KK$-monopoles and $KKO$-planes. Here, we will include them in the same manner as all other contributions to the scalar potential, reducing them to their scalings in $\rho$ and $\tau$ with all additional factors packed into a coefficient. The expressions are:
\bea 
V_{NS5} &= \frac{A_{NS5}}{\rho^ 2 \tau^ 2}\,, \qquad V_{NSO5} = - \frac{A_{NSO5}}{\rho^ 2 \tau^ 2}\,,\\
V_{KK} &= \frac{A_{KK}}{\rho \tau^ 2}\,, \qquad \; V_{KKO} = - \frac{A_{KKO}}{\rho \tau^ 2}\,,
\eea
where the coefficients are positive. For the complete expression of these terms we refer the reader to \cite{Silverstein:2007ac}.\\
For the $KK$-monopoles and $KKO$-planes the analysis is trivial since the corresponding terms scale exactly like the curvature does. This allows us to group all three terms together and define an effective curvature:
\be 
\tilde{A}_{R_6} = A_{R_6} + A_{KK} - A_{KKO}\,.
\ee
Again, de Sitter solutions can only exist for as long as this effective curvature term is positive: $\tilde{A}_{R_6} >0$. De Sitter solutions, found in \cite{Blaback:2018hdo}, including $KK$-sources then suffer from the same lack of a controlled limit as discussed before.\\
$NS5$-sources, on the other hand, do introduce a new scaling behavior into $V$. The complete scalar potential, again broken down to it's scaling behavior, is:
\be 
V(\rho,\tau) = \frac{A_H}{\rho^3 \tau^2} + \sum_{p=0,2,4,5} \frac{A_p}{\rho^ {p-3} \tau^4} - \frac{A_{\text{sources}}}{\tau^3} + \frac{\tilde{A}_{R_6}}{\rho\tau} + \frac{A_{NS5}}{\rho^ 2 \tau^ 2} - \frac{A_{NSO5}}{\rho^ 2 \tau^ 2}\,.
\label{eq:genscalplus}
\ee
We now proceed along similar lines as we did before and minimize the potential, first with respect to $\tau$:
\be
0 = \partial_\tau V = -2 \frac{A_H}{\rho^ 3 \tau^ 3} - 4 \sum_{p=0,2,4,5} \frac{A_p}{\rho^ {p-3} \tau^5} + 3 \frac{A_{\text{sources}}}{\tau^ 4} - 2 \frac{\tilde{A}_{R_6}}{\rho \tau^ 3} - 2 \frac{A_{NS5}}{\rho^ 2 \tau ^3} + 2 \frac{A_{NSO5}}{\rho^ 2 \tau ^3}\,,
\ee
which can be solved by:
\be 
\frac{A_H}{\rho^ 3 \tau^ 2} + \frac{\tilde{A}_{R_6}}{\rho \tau^ 2} + \frac{A_{NS5}}{\rho^ 2 \tau^ 2} - \frac{A_{NSO5}}{\rho^ 2 \tau^ 2} = -2 \sum_{p=0,2,4,5} \frac{A_p}{\rho^ {p-3} \tau^4} + \frac{3}{2} \frac{A_{\text{sources}}}{\tau^3}\,.
\ee
Using this in the scalar potential \eqref{eq:genscalplus} we find:
\be 
V(\rho,\tau) = - \sum_{p=0,2,4,5} \frac{A_p}{\rho^ {p-3} \tau^4} + \frac{1}{2} \frac{A_{\text{sources}}}{\tau^ 3}\,.
\ee
Again, we find that de Sitter solutions are only possible for $A_{\text{sources}} \propto 2 N_{O6} - N_{D6} -N_{\overline{D6}} >0$. This means that even the inclusion of a different kind of negative tension source does not relax the condition on the $O6$-planes. Furthermore, it is easy to check that we still require a massive compactification, if one only uses $NS5$-branes. In principle, however, it seems to be possible to have a dS solution with vanishing mass parameter $A_{F_0}$ in type IIA if we have $A_{NSO5} > A_{NS5}$. This then would allow for an M-theory uplift that is otherwise prohibited by the appearance of the Romans mass. The curvature term again has to satisfy $A_{R_6} \propto - R_6 \leq 0$ since otherwise the bound from above is obtained once more. For the massless case the relevant scalings now come from the curvature term, the sources and the $NSO5$ contribution, giving the relevant scalar potential:
\be 
V = \frac{A_{R_6}}{\rho \tau^ 2} - \frac{A_{\text{sources}}}{\tau^ 3} - \frac{A_{NSO5}}{\rho^ 2 \tau^ 2}\,,
\ee
with a fixed number of $O6$-planes and $NSO5$-planes. The scaling limit yields the following condition on the scale parameters:
\be 
c_{R_6} - c_\rho - 2 c_\tau = -3 c_\tau = -2 c_\rho -2 c_\tau\,.
\ee
This can be solved by:
\be 
c_\tau = 2 c_\rho \qquad \text{and} \qquad c_{R_6} = -c_\rho\,.
\ee
Letting $\lambda \to \infty$ and requiring $c_\rho >0$ implies that the curvature goes to zero:
\be 
A_{R_6} \propto - R_6 \propto \lambda^ {c_{R_6}} = \lambda^ {-c_\rho} \to 0\,.
\ee
However, the curvature is fixed for a given compactification manifold and thus cannot become parametrically small. We conclude that a de Sitter solution with a small, fixed amount of $NSO5$-planes is not possible.\\
Non-geometric Q- and R- fluxes are some more potential candidates for inclusion in type IIA compactifications. These can lead to stable de Sitter solutions \cite{deCarlos:2009fq,deCarlos:2009qm,Danielsson:2012by,Blaback:2013ht,Damian:2013dq,Blaback:2015zra}. The scaling behavior of these contributions goes like:
\be 
V_\text{Q} = \frac{A_{\text{Q}}}{\rho^{-1}\tau^ 2} \qquad \text{and} \qquad V_{\text{R}} = \frac{A_{\text{R}}}{\rho^ {-3} \tau^ 2}\,.
\ee
By duality there are also corresponding sources for these fluxes. The problem with these contributions is that, due to their non-geometric nature, it is not clear if and how $\alpha^ \prime$ corrections can be handled in their presence. Thus, we choose to omit a detailed discussion of their behavior.\\
In the case of a different class of compactification manifold, other than an $SU(3)$ structure manifold, it is possible to have non-trivial $1$- and $5$-cycles. This then allows for the corresponding $Dp$-branes and $Op$-planes to appear in the model. The scaling behavior of these sources goes like:
\be 
V_{Dp/\overline{Dp}} = \frac{A_{Dp/\overline{Dp}}}{\rho^ {\frac{6-p}{2}} \tau^ 3} \qquad \text{and} \qquad V_{O_p} = - \frac{A_{O_p}}{\rho^ {\frac{6-p}{2}}\tau^ 3}\,.
\ee
\subsection{Isotropic Compactification of massive Type IIA}
Up until this point our discussion has been fairly general. In this subsection, we want to give an explicit example of a compactification in order to complement the discussion. The isotropic compactification of massive type IIA on $S^ 3 \times S^ 3 /(\mathbb{Z}_2 \times \mathbb{Z}_2)$ is potentially the simplest model we could choose. It has been extensively studied in \cite{Roupec:2018mbn,Caviezel:2008tf,Flauger:2008ad,Danielsson:2009ff,Danielsson:2010bc,Danielsson:2012et,Junghans:2016uvg,Junghans:2016abx,Kallosh:2018nrk,Blaback:2018hdo,Villadoro:2005cu,Camara:2005dc,Aldazabal:2007sn,Blaback:2013fca} and admits dS critical points. This model might even be able to admit stable de Sitter points if one includes anti-$D6$-branes or $KK$-monopoles \cite{Kallosh:2018nrk,Blaback:2018hdo}. Often this particular setup is referred to as the STU model due to the fact that it has three complex moduli. This is because we identify the three 2-tori in the $T^ 6/(\mathbb{Z}_2 \times \mathbb{Z}_2)$ internal manifold. The inclusion of metric fluxes, that we set to one, effectively changes the internal manifold to $S^3\times S^ 3$. Furthermore, we include $H$-flux as well as $F_p$-fluxes with $p=0,\,2$ because the 4- and 6-flux can be set to zero here via a shift of the moduli. The real part of the Kähler modulus $T$ encodes the overall size of the compactification manifold. The dilaton is packaged together with the complex structure moduli into the two complex scalars $Z_1$ and $Z_2$. Relating these moduli back to what we had thus far in this section we find:
\be 
\rho = \rmre(T) \qquad \text{and} \qquad \tau^4 = \rmre(Z_1) \rmre(Z_2)^ 3\,.
\ee
In addition to the fluxes we also consider $\overline{D6}$-branes in the model that can wrap two different $3$-cycles \cite{Kallosh:2018nrk}, labeled $N_{\overline{D6},1}$ and $N_{\overline{D6},2}$. These anti-branes spontaneously break supersymmetry and we will use constrained multiplets in order to describe their contribution using only the Kähler- and superpotential as was discussed in chapter \ref{sec:antiD3}. The potentials we are going to use in this section are:
\bea 
K =& -\log\left[(T+\bar{T})^ 3 - \frac{1}{2} \frac{X \bar{X}}{ N_{\overline{D6},1}(Z_1 + \bar{Z}_1) + N_{\overline{D6},2}(Z_2 + \bar{Z}_2)}\right] \\
&- \log\left[2^ {-4} (Z_1 + \bar{Z}_1)(Z_2 + \bar{Z}_2)^ 3\right]\,,\\
W =& -3 f_2 T^ 2 + \rmi f_0 T^ 3 + (\rmi h - 3T) Z_1 - 3(\rmi h+T)Z_2 + X\,,
\eea
from which we can derive the scalar potential $V$ using the usual formula:
\be 
V = \rme^ K \left[D_I W K^ {I\bar{J}} \overline{D_J W} - 3 W \overline{W}\right]_{X\to 0\,,\bar{X} \to 0}\,,
\ee
where we have to set the nilpotent chiral fields $X$ and $\bar{X}$ to zero after performing the calculation since we are only interested in the purely bosonic scalar potential.\\
The resulting scalar potential has a number of re-scaling symmetries. One of those will precisely match the re-scaling to a large number of $O6$-planes, which was the only possibility for a stable de Sitter solution that we found in this section. Generally, we can scale the moduli any way we desire if the scalar potential does not change by more than an overall numeric factor. This does not change the eigenvalues of the mass matrix or the fact that critical points are present. Although, it in general changes the location of the critical points in moduli space. The most general re-scaling, satisfying these conditions, is $T \to aT$ and $f_0 \to b f_0$ which, in turn, fixes the required scaling of all other moduli:
\be 
T\to aT\,, \quad Z_1 \to a^ 2 b Z_1\,, \quad Z_2 \to a^ 2 b Z_2\,,\quad X \to a^ 3 b X\,,
\ee
where the same must hold for the complex conjugates as well. Since the real parts of the moduli have to be and remain positive we restrict to $a,\,b>0$. For the flux parameter and number of $O6$-planes this leads to:
\be 
f_0 \to b f_0\,,\quad f_2 \to abf_2\,,\quad h\to ah\,,\quad N_{\overline{D6},1} \to ab N_{\overline{D6},1}\,,\quad N_{\overline{D6},2} \to ab N_{\overline{D6},2}\,.
\ee
The effective re-scalings of the scalar- and superpotential that result from this are:
\be 
W \to a^ 3 b W \qquad \text{and} \qquad V \to a^ {-5}b^ {-2} V\,.
\ee
Evidently, this scaling freedom does have an effect on the physics of the model, by changing the cosmological constant. However, the existence of critical points is not effected. This allows us to set some fluxes to $1$ and thus fixing the symmetry. After finding a minimum one can once more employ the scaling symmetry to fix the values of the fluxes to appropriate numbers.\\
The model with standard ingredients, discussed in section \ref{sec:largef4}, has no scaling of $0$-flux and curvature. Relating this to the above discussion we have $b=0$. Then, we easily find that $a =N_{O6}$ and we can recover the familiar situation from the formulas we have in this section:
\bea 
vol_6 &\sim \rmre(T)^ 3 \propto N^ 3_{O6}\,,\\
\rme^ {-\phi} &\sim \frac{\sqrt[\leftroot{-2}\uproot{2} 4]{\rmre(Z_1)\rmre(Z_2)^ 3}}{\sqrt{vol_6}} \propto \sqrt{N_{O6}}\,.
\eea
\subsection{Explicit Solutions to the isotropic Compactification}
\begin{figure}[htb]
     \centering
     \includegraphics[scale=0.6]{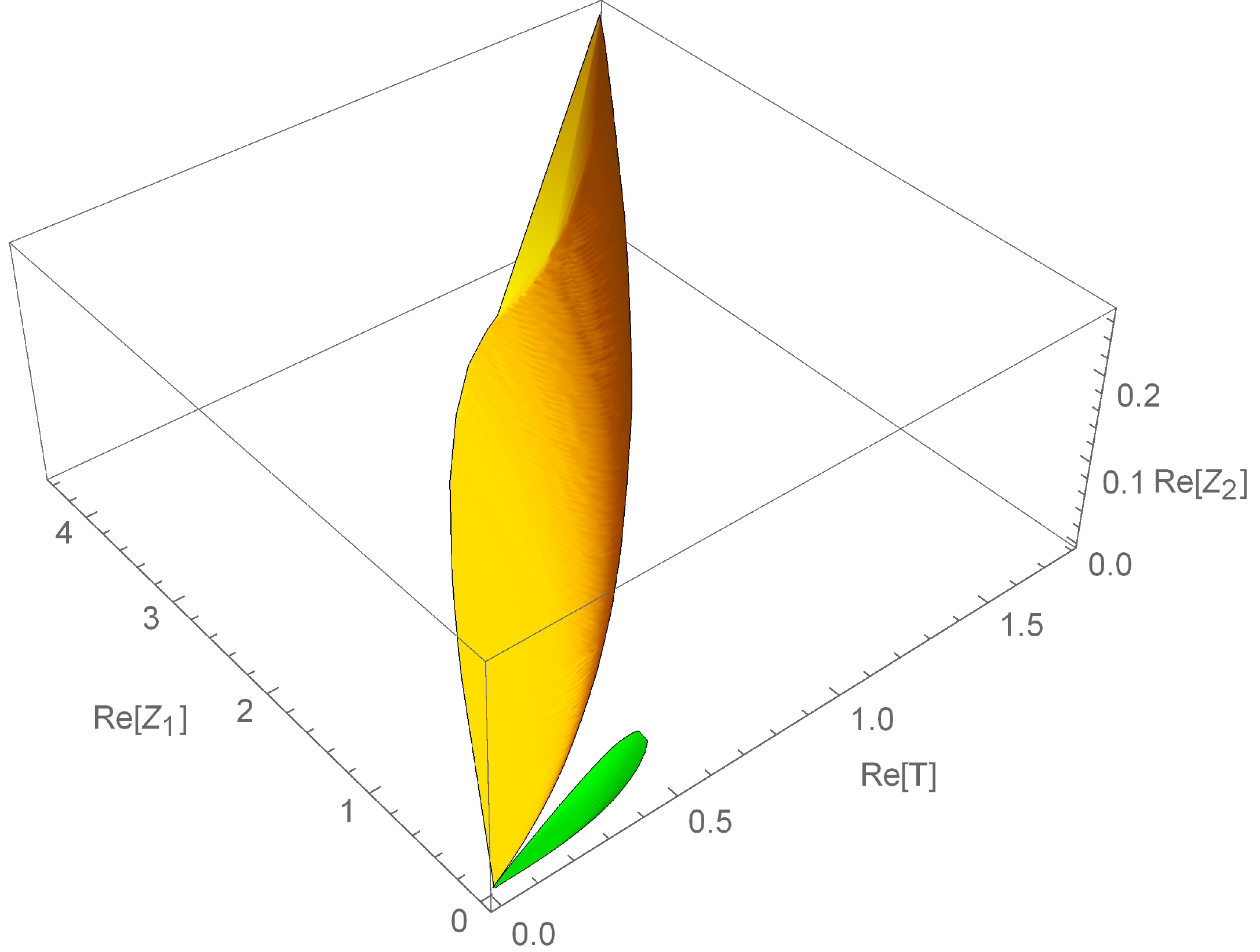}
     \caption{Two regions where stable de Sitter solutions are in principle possible. The larger region, in yellow, allows only for such solutions if the non-isotropic directions are fixed while the smaller, green, is tachyon free even after opening up the non-isotropic directions.}
     \label{fig:tworeg}
\end{figure}

Before we come to presenting explicit solutions to the rather simple model we are interested in, we need to mention that even here it can be very complicated to find a solution or even map out the part of parameter space where these solutions are possible. However, if we add anti-$D6$-branes, it becomes possible to find analytic solutions to the minimization problem. In order to do so, we first use the scaling symmetry of the scalar potential to fix $|f_0| = 1 = |f_2|$. The remaining parameters are the H-flux, represented by $h$ and the numbers of anti-branes: $N_{\overline{D6},1}$ and $N_{\overline{D6},2}$. It is then possible to solve the axionic derivatives, $\partial_{\rmim(T)} V = \partial_{\rmim(Z_1)} V = \partial_{\rmim(Z_2)} V= 0$ and so on, in terms of the corresponding axions, $\rmim(T)$, $\rmim(Z_1)$ and $\rmim(Z_2)$. The rest of the equations, $\partial_{\rmre(T)} V = \partial_{\rmre(Z_1)} V = \partial_{\rmre(Z_2)} V=0$ are more difficult to evaluate in terms of the moduli. Luckily, the remaining expressions with derivatives with respect to $Z_1$ and $Z_2$ are linear in the number of anti-branes. Likewise, $\partial_{\rmre(T)} V = 0$ is quadratic in the flux parameter $h$ and thus we can solve the expressions and obtain two sets of solutions. The solution then depends on the values of the moduli $\rmre(T)$, $\rmre(Z_1)$ and $\rmre(Z_2)$.\\
There are some additional criteria that need to be met in order to find a sensible solution. First, we are interested in solutions with $V_{\text{min}} >0$ and we require that we have positive numbers of anti-$D6$-branes. For stable extrema we also need to have all eigenvalues of the mass matrix positive. Disregarding quantization conditions, two regions emerge from our search that allow such points. They are depicted in figure \ref{fig:tworeg}. The smaller, green, region even allows for stable solutions if the non-isotropic directions are included, considering a total of $14$ real moduli.
In principle, we should now be able to use the scaling freedom of the model in order to change the values of the fluxes. Due to the fact that we have $N_{O6} = 1$ in this setup we are unfortunately restricted in doing so. In particular, we cannot re-scale to solution sufficiently in order to arrive at a trustworthy regime of $\rho = \rmre(T) \gg 1 $ and $\tau^ 4 \rmre(Z_1) \rmre(Z_2)^ 3 \gg 1$. Thus, we are not able to find de Sitter solutions that can be trusted in this simple model.\\
Since we have analytic control over the solutions we can check what the $\mathcal{O}(1)$ numbers, appearing in front of the scaling behavior, are in our solutions. We find that:
\bea 
vol_6 &= c_1 N^ 3_{O6} \qquad\quad \text{with} \qquad 8 \cdot 10^ {-8} \lesssim c_1 \lesssim 0.24\,,\\
\rme^ {-\phi} &= c_2 \sqrt{N_{O6}} \qquad \text{with} \qquad 4 \cdot 10^ {-5} \lesssim c_2 \lesssim 0.12\,.
\eea
These small parameters make it even harder to obtain a solution with a small number of $O6$-planes. Since we are restricted to one single $O6$-plane we are unable to use the current model to build a trustworthy de Sitter solution. The numbers above should give a reasonable estimate of the number of $O$-planes required in related models in order to find such dS points.
\subsection{Scaling Limits of de Sitter - Interim Summary}
In this part we have investigated simple type IIA flux compactifications regarding the possibility of constructing controlled de Sitter solutions. Inspired by the anti-de Sitter solutions of \cite{DeWolfe:2005uu}, where a large $F_4$-flux limit allows for solutions with large internal volume and small coupling, we investigate similar models for scaling behaviors that could lead to a similar behavior for dS. We find that the only limit is the one of a large number of $O6$-planes. Even considering more exotic contributions, like $KK$ and $NS$ sources does not improve the situation. The main problem is that the number of orientifold planes is limited by the geometry of the chosen compactification manifold. In the simple example we consider here this fixes the number of $O6$-planes to be $1$. While more complicated internal manifolds can certainly exist such models have not been explicitly investigated thus far and are likely hard, if not practically impossible, to find, given the tools currently available. We furthermore looked into a concrete example and find that our theoretical conclusions are practically realized. Due to the good control of the model we are able to give an estimate of the order of $O6$-planes one would require and find that we would at least require one or two orders of magnitude more orientifold planes than currently available models admit in order to arrive at stable de Sitter solutions. Of course, conclusions drawn from one class of models cannot be transferred one-to-one to others but we believe that the rough estimate still holds.

\section{de Sitter Extrema}
\label{sec:dSextrema}
In this section we will show that it is relatively easy to find de Sitter extrema from classical type II flux compactifications. This is exactly the setup that the conjecture \eqref{eq:dSconori} should hold in. These points violate the original conjecture but the refined version \eqref{eq:dScon} still holds due to the condition on the Hessian. The contents discussed in the following are based on \cite{Roupec:2018mbn}.\\
We are going to find numerical de Sitter points by looking explicitly at the scalar potentials of the models described in the previous section, as was first done in \cite{Caviezel:2008tf,Flauger:2008ad} and later extended upon by us in \cite{Roupec:2018mbn}. Note that the mere existence of critical de Sitter points is still not necessarily an objection to the original dS swampland conjecture. If the solutions found numerically cannot satisfy certain requirements in order to be consistent with string theory they should not be considered. These include tadpole cancellation, flux quantization and the requirement that string loop corrections should be negligible.

\subsection{Potential Issues}
\label{sec:stringrequ}
Before we come to presenting our results we discuss some potential problems and required consistency conditions that string compactifications should satisfy.
\paragraph{First, we need to identify the relevant moduli.} For $SU(3)$ compactifications it is problematic to correctly isolate the lightest fields in the theory \cite{Kashani-Poor:2006ofe,Kashani-Poor:2007nby}. This essentially disallows us from studying the low-energy effective theory. Luckily, a consistent truncation exists \cite{Cassani:2009ck} for compactifications on group manifolds. One can then expand in left-invariant forms and the $4d$ theory will lift consistently to $10d$. The orientifold projection, which we include in the models we investigate here, should not change this expansion. It is still not guaranteed that the scalars are the lightest fields, or even below the KK-scale \cite{Andriot:2018wzk} but, since we found a solution in $4d$ that lifts to solutions in $10d$, this is no issue. Furthermore, giving more evidence for the validity of this procedure, some of the $4d$ dS extrema can be explicitly lifted to $10d$ \cite{Danielsson:2010bc}.
\paragraph{Another concern are integrated equations of motion of intersecting sources.} This can be an issue since it is not possible to solve the equations of motion for intersecting sources. Due to this, one usually restricts to smearing the sources. A priori this should not be allowed as, for example, $O$-planes are localized objects and thus cannot be smeared. For initial discussions on this topic see \cite{Douglas:2010rt,Blaback:2010sj,Blaback:2011nz,Blaback:2011pn,Saracco:2012wc,McOrist:2012yc}. Rather recently, however, it has been argued that this is no issue as the smearing should give the effective result in the low-energy theory. First order localizations of this problem were performed in \cite{Baines:2020dmu,Junghans:2020acz,Marchesano:2020qvg,Cribiori:2021djm} and it seems that the smearing is an appropriate approximation. Still, in the case of only parallel sources the solution can be obtained by including simple warp factors and no no-go theorems exist when restricting to such setups \cite{Andriot:2016xvq}. However, it is not clear if geometries exist that allow for such situations. Our setups here all include intersecting sources such that we rely on the recent investigations regarding first-order localization and the smearing argument. 
\paragraph{Due to our orbifold compactification, blow-up modes} could pose a potential problem for our models. The compactifications we consider are based on orbifolds of group manifolds, or so-called twisted tori. Only considering standard abelian orbifolds, we are limited to $\mathbb{Z}_2 \times \mathbb{Z}_2$ orbifolds as for all other cases no-go theorems, forbidding de Sitter critical points, exist \cite{Flauger:2008ad}. For non-abelian orbifolds one can find de Sitter critical points for $\Delta(12)$ which interestingly contains $\mathbb{Z}_2 \times \mathbb{Z}_2$ as a subgroup, as discussed in \cite{Danielsson:2011au}. Note however, that this study only contains a limited amount of such orbifolds. On such compactifications, singularities can appear due to the orbifolding. For some examples the impact of such divergences has been studied \cite{DeWolfe:2005uu,Ihl:2006pp} and it was found that one can remove the singularities with blow-ups. These then come with additional moduli but also allow additional fluxes on the newly created cycles. It is possible, at least in these examples, to stabilize the new moduli at a smaller scale. We are going to assume that similar mechanisms will work out in the examples we consider here. Unfortunately, the new modes from the blow-ups will not be left invariant forms and thus we no longer have a consistent truncation if these modes become relevant.
\paragraph{The mass parameter $\mathbf{m_0 = F_0}$ in type IIA} plays an important role in compactifications \cite{Romans:1985tz}. If present, it prevents the uplift to M-theory and it is not clear if perturbative string theory can describe the situation and one might not be able to define orientifold planes. These problems are not limited to solutions with positive vacuum energy but even supersymmetric AdS vacua \cite{Banks:2006hg}. On the other hand, massive type IIA models cannot be strongly coupled in regions with weak spacetime curvature \cite{Aharony:2010af}, which is an interesting feature certainly worth investigating.\\
Fortunately, the mass parameter is not crucial in order to obtain dS extrema. This can be seen by formally T-dualizing the type IIA theory with curvature and $O6$-planes which yields a type IIB theory with $O5$ and $O7$-planes compactified on $SU(2)$ structure manifolds \cite{Caviezel:2009tu}. There, the dual of the mass parameter $F_0$ is $F_1$ which can be present for manifolds with non-trivial 1- and 5-cycles. Explicit compactifications are known that have large volume and small coupling and even one explicit dS point was found using such setups. In type IIB all know solutions require intersecting sources, this time $O5$- and $O7$-planes, which means one relies once again on a smeared approximation.
\paragraph{Finally, let us mention flux quantization} as a condition that comes directly from string theory and restricts the low-energy effective theory. When one searches numerically for solutions this condition is usually not imposed from the beginning and we will follow this approach in our analysis as well. The solutions that are obtained often are valid for a range of fluxes but not always.\\
For the NSNS-flux $H_3$ the condition boils down to the requirement that the integral of the flux over all cycles in integer holonomy has to return an integer. For the RR-fluxes the flux quantization condition changes in the presence of $H_3$. The correct framework to investigate this issue would be $H_3$-twisted K-theory \cite{Moore:1999gb,Minasian:1997mm}. Luckily, for simply connected $6d$ manifolds this theory is isomorphic to $H_3$-twisted cohomology \cite{Collinucci:2006ug}. Thus far, this has only be done for $SU(2)\times SU(2) = S^ 3 \times S^ 3$ \cite{Danielsson:2011au}.
For our case of $SU(3)$ structure manifolds, where we expand everything in left-invariant forms, it seems that the fluxes do not have to be in twisted cohomology and thus not quantized. Nevertheless, the fluxes appear in the tadpole conditions and have to cancel quantized charges. Then, they still have to be quantized in that sense.
\subsection{Critical de Sitter Points}
We now come to review de Sitter extrema that were found via numerical models in type IIA compactifications on $SU(3)$ structure manifolds including curvature, fluxes and $O6$-planes. These models are similar to compactifications on $T^ 6$. The scalar potential that should allow for the dS points comes from $F-$ and $D-$terms \cite{Grimm:2004ua,Robbins:2007yv} in supergravity. A detailed review of these constructions, tailored for our purpose here, can be found in \cite{Danielsson:2011au}.\\
In $4d$ we have a scalar potential that depends on 7 moduli at most. Among them are the axio-dilaton and the geometric scalars that appear when expanding the 3-form $\Omega$ and the Kähler form $J$. Appearing in $V$ we also have parameters that come from the NSNS- and RR-fluxes and, furthermore, geometric fluxes from the $SU(3)$ structure. The latter ones incorporate the curvature of the manifold. One can attempt to find critical points of this scalar potential using the moduli and parameters and some further conditions such as the positivity of the potential in order to find dS points. The earliest results were obtained in \cite{Caviezel:2008tf,Flauger:2008ad}, where 3 dS Extrema were found. In \cite{Danielsson:2012et} a study covering different internal manifolds found over $100$ de Sitter extrema. Then later in \cite{Roupec:2018mbn} we were able to extend this number by obtaining over $300$ more such points. In the following we will discuss all of these points and their importance for the initial de Sitter swampland conjecture.\\
One immediate issue with the numerical method of finding these points, however, is that all quantities, including the fluxes, are allowed to vary continuously. The conclusions of this would be that flux quantization or the tadpole conditions, cannot be satisfied. Fortunately, due to scaling freedoms in the models, this is not a fatal flaw. It is possible to re-scale moduli and fluxes while staying at a critical point and thus the flux quantization and tadpole conditions can be satisfied. In the particularly simple example of compactifications on $S^ 3 \times S^ 3/(\mathbb{Z}_2 \times \mathbb{Z}_2)$ the parameter space has been mapped out and one finds that it is possible to set all but one of the fluxes equal to $1$. Effectively, the scalar potential then depends on one parameter and it is possible to attempt to find critical points. Indeed, dS critical points have been found \cite{Danielsson:2010bc} for a certain range of the parameter. Furthermore, in \cite{Danielsson:2011au} the flux quantization conditions have been worked out. Unfortunately, this leads to a small volume and strong coupling, which means that $\alpha^ \prime$ and string loop corrections are significant and one cannot trust these critical points.\\
We now move on to the different setups that were considered in \cite{Danielsson:2012et} and, following the initial de Sitter swampland conjecture, by us in \cite{Roupec:2018mbn}. The numerical search was implemented using Mathematica and returned critical points with values for the moduli and fluxes around $1$. In natural units, this typically leads to a volume of order $\mathcal{O}(10)-\mathcal{O}(100)$ and an inverse string coupling of order $\mathcal{O}(10)$. These values can be considered to be acceptable, at least for our principal considerations here. The tadpole conditions are more troublesome with contributions of either sign of order $\mathcal{O}(10)$. For positive tadpoles one can, in principle, add an arbitrary number of $D6$-branes. On the other hand, for negative tadpoles, it would be necessary to add $O6$-planes. It, however, is not possible to add a large number of those as their number is fixed by the compactification to be of order one. A feature that cannot be changed and was already discussed in the previous section. In fact, the negative contributions to the tadpole from the $O6$-planes play an important role in avoiding the Maldacena-N\'{u}\~{n}ez no-go theorem \cite{Maldacena:2000mw}. Hence, we conclude that proper dS critical points should have one or more negative tadpole of $\mathcal{O}(1)$. Still, the points obtained in the numerical search are not doomed, as there is a scaling symmetry, universal to all investigated examples, that can be used to adjust the tadpole. Importantly, this re-scaling does not change any physical results that we are interested in here. In practice, we have to scale the Kähler moduli by a constant and the complex structure moduli by its inverse. Requiring that the scalar potential changes at most by an overall multiplicative factor, which has no implications on the critical points, fixes the scaling behavior of the fluxes. The relevant physical quantities, like for example, the slow-roll parameter $\epsilon$, do not change at all. The geometric fluxes, depending on the model, stay invariant while the mass parameter and the tadpole change in a non-trivial manner. The models we consider have $4$ different, left-invariant $3$-forms that each come with their own tadpole condition. Using the re-scaling one can change the largest, negative tadpole such that it can be satisfied by a single $O6$-plane. The rest of the tadpoles then can be accounted either by a single $O6$ again or some number of $D6$-branes. After the re-scaling one has to check the internal volume and string coupling in order to see which models still have acceptable values for these two quantities. Here we found that there seems to be a correlation between the volume and the mass parameter $F_0$. All remaining models have:
\be 
F_0^2 \cdot vol_6 \sim \mathcal{O}(1) \; - \; \mathcal{O}(100)\,.
\ee
Thus, for large internal volume one would need particularly small $F_0$ flux. This, however, cannot be since the quantization condition for $F_0$ requires it to be an integer. While $F_0=1$ and a volume of order $100$ can be acceptable one might not be quite certain if it is safe to neglect $\alpha^ \prime$ and string loop corrections. This problem deserves further attention in order to reach a satisfying conclusion. As it stands we have found models with $vol_6 \sim \mathcal{O}(100)$ and $\rme^ {-\phi} \sim \mathcal{O}(10)$, which are definitely in tension with the original de Sitter swampland condition.\\
We would like to point out that the conclusion that no particularly large volume seems to be possible might be a limitation of the numerical implementation of the search. In the AdS vacua, based on CY-compactifications, found in \cite{DeWolfe:2005uu}, it is analytically known that solutions with large volume exist in a limit with large $F_4$ flux, as discussed in \ref{sec:scaling}. Even though the same statements are not known for compactifications with curvature, examples with flux limits do exist, again, for AdS. In dS no flux limit has been identified thus far. Nevertheless, we also searched for anti-de Sitter points in our setups using the same numerical methods as for de Sitter. The solutions obtained in this way showed the same property that $F_0^ 2 \cdot vol_6$ is of order $1$ to $100$. Due to the analytical existence of solutions with large volume this hints that there might be a limitation of the methods employed at work, rather than a physical obstacle.

\subsection{de Sitter Extrema - Interim Summary}
In this section we discussed models where it is feasible to find de Sitter points that do not rely on an uplifting procedure and are in that sense classical. Such setups are the main target of the initial de Sitter swampland conjecture of \cite{Obied:2018sgi}. We discussed several obstacles and potential resolutions that these constructions face. Conditions that need to be met include intersecting $O$-planes and the flux quantization and tadpole conditions. For the first of these, recent progress \cite{Junghans:2020acz,Marchesano:2020qvg,Cribiori:2021djm} suggests that a solution is possible and that intersecting sources pose no problem. The second issue needs to be addressed in explicit examples. The numerical solutions from \cite{Danielsson:2012et,Roupec:2018mbn} are only on the verge of being acceptable. With satisfied flux quantization and tadpole conditions one is left with a volume that is at best around $100$ in natural units. At this level it is not quite clear that $\alpha^ \prime$ and string loop corrections are negligible. Still, we argue that the solutions that were obtained are at least in tension with the de Sitter swampland conjecture and a refinement of it had to be the natural conclusion.

\section{A refined dS Conjecture}
\label{sec:refconj}
Due to the criticism from both, classical dS constructions \cite{Andriot:2018ept,Roupec:2018mbn} and Standard Model scalar potentials \cite{Denef:2018etk,Murayama:2018lie,Choi:2018rze,Hamaguchi:2018vtv}, the initial de Sitter swampland conjecture \cite{Obied:2018sgi}, presented in \eqref{eq:dSconori}, needed to be improved \cite{Andriot:2018wzk,Garg:2018reu}. In \cite{Ooguri:2018wrx} the refined conjecture in \eqref{eq:dScon} was presented which allows de Sitter extrema that are unstable and is no longer in tension with the Standard Model. In \cite{Andriot:2018mav} we presented an alternative that has the advantage of condensing both conditions into one inequality and we support the parameters in the condition by the de Sitter extrema that were discussed in the previous section.
\subsection{Single Expression Conjecture}
When discussing the de Sitter conjecture we have an effective $4d$ theory in mind with scalars $\phi^ i$ that couple to gravity. The action for such a theory can be written as:
\be 
S = \int d^4x \sqrt{g_4} \left(\frac{M_P^ 2}{2} R_4 - \frac{1}{2} K_{ij} \partial_\mu \phi^ i \partial^ \mu \phi^ j - V(\phi^ i)\right)\,,
\label{eq:leeact}
\ee
where $R_4$ is the $4d$ Ricci scalar and the scalar potential $V(\phi^ i)$ determines the dynamics of the scalars $\phi^ i$. In this section we will include the explicit factors of $M_P$ when necessary due to a point that will be raised further down below. In order to be precise, let us (re-) define some quantities we will be using in the following. The gradient of the scalar potential is:
\be 
|\nabla V| = \sqrt{K^{ij}\partial_i V \partial_j V}\,,
\ee
and for the condition on the smallest value of the Hessian we use:
\be 
\text{min}(\nabla^i \nabla_j V) = \text{smallest eigenvalue of }(\nabla^i \nabla_j V)\,.
\ee
For a positive value of the potential we define the first and second slow-roll parameters\footnote{Strictly speaking the slow-roll parameters we use here are the ones only valid for single field slow-roll inflation, often denoted $\epsilon_V$ and $\eta_V$.} to be:
\bea 
\epsilon &= \frac{M_P^ 2}{2} \frac{|\nabla V|^2}{V^2}\,,\\
\eta &= M_P^ 2 \, \frac{\text{min}(\nabla^i \nabla_j V)}{V}\,.
\eea

\begin{figure}[htp]
     \centering
     \includegraphics[width=0.9 \textwidth]{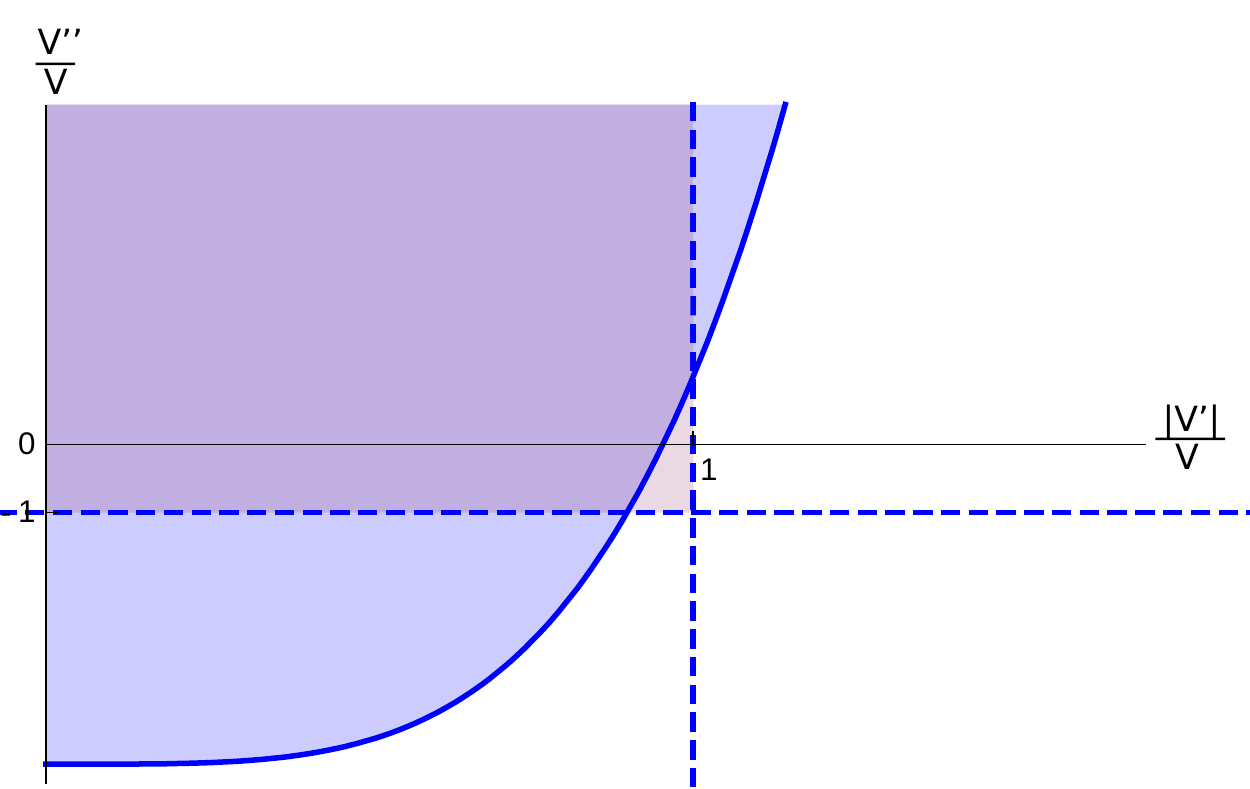}
     \caption{The excluded regions of parameter space for both refined conjectures, \eqref{eq:dScon} and \eqref{eq:ourdScon}. The axis are $(\sqrt{2\epsilon},\eta)$, denoted in the graphs as $(|V^ \prime|/V,V^ {\prime \prime}/V)$. The upper left quadrant of the dashed lines is excluded for the refined conjecture of \cite{Ooguri:2018wrx} and everything up and left of the curved solid line is excluded for our refined conjecture. The differences between the two conjectures are the approximately triangular shaped region near the cross (shaded pink) which is allowed by our conjecture and the regions right of and below the dashed lines up to the solid curved line (shaded light blue). Those are excluded by our conjecture but allowed by the conjecture \eqref{eq:dScon}. The selected value of $a = 1/5.7$ in this picture gives a good overview of the different regions. The choices $c=c^ \prime = 1$ puts the dashed lines at $-1$ and $1$. For $q$ we chose $4$.}
     \label{fig:conjreg}
\end{figure}

\begin{figure}[htp]
     \centering
     \includegraphics[width=0.45 \textwidth]{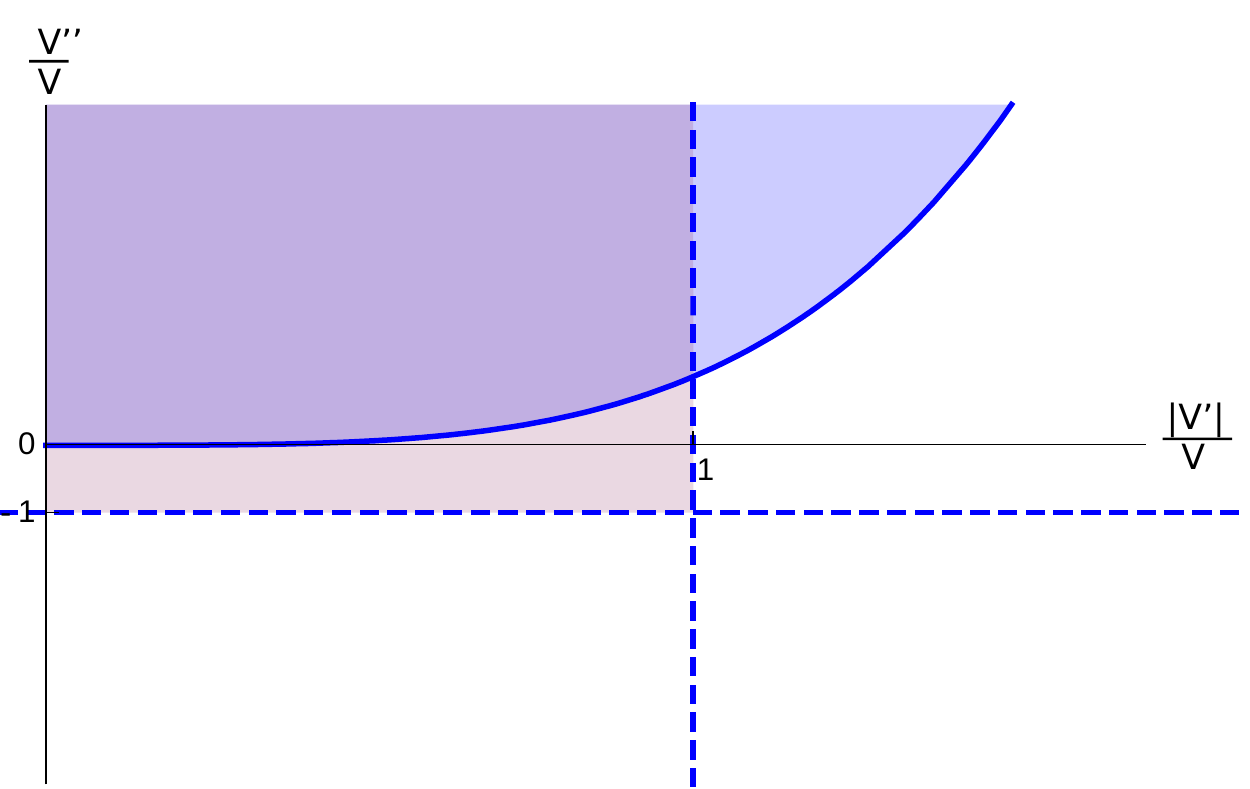}
     \includegraphics[width=0.45 \textwidth]{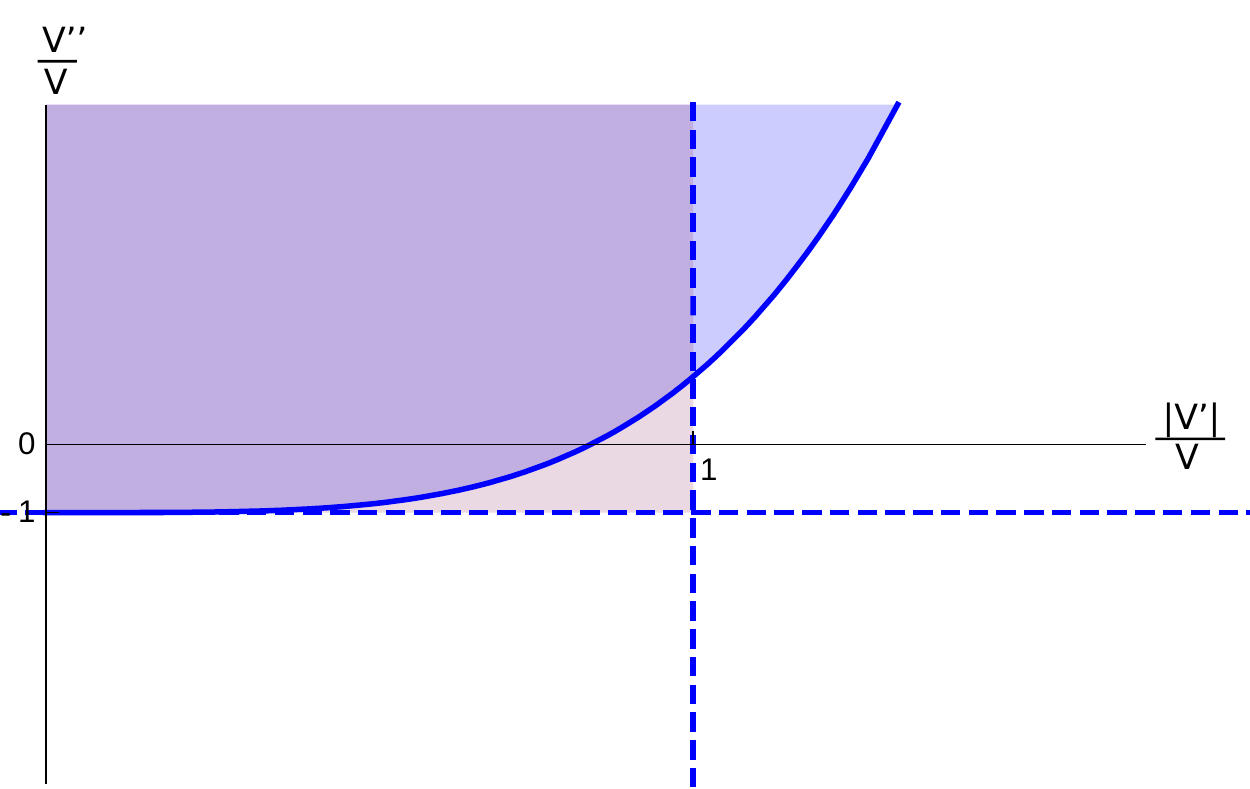}
     \caption{For different values of the parameter $a$ the differences in allowed/excluded regions change as some regions merge or disappear compared to \ref{fig:conjreg}. On the left we have $a = 1$. Here the region below the curved line is only excluded by the conjecture \eqref{eq:dScon}. On the right we have $a=1/2$. There the conjectures agree for very small $\epsilon$. Again, the other parameters are $c=c^ \prime = 1$ and $q=4$.}
     \label{fig:conjreg2}
\end{figure}

Finally, for constant fields, the mass matrix is:
\be
M^i_j= \nabla^i \nabla_j V\,,
\ee
and the eigenvalues are the masses of the scalars. If negative eigenvalues are present we know that a tachyon is present and the extremum will be unstable. Likewise, a zero value of an eigenvalue means that we have a flat direction that might be problematic as well. The refined de Sitter conjecture \eqref{eq:dScon}, in terms of the quantities defined here and including factors of $M_P$, is:
\be 
|\nabla V| \geq \frac{c}{M_P} V \quad \text{or} \quad \text{min}(\nabla^i \nabla_j V) \leq - \frac{c^ \prime}{M_P^ 2} V\,,
\label{eq:dSconMP}
\ee
with our two order one parameters $c$ and $c^ \prime$. In terms of the slow-roll parameters, the refined de Sitter conjecture \eqref{eq:dScon} reads:
\be 
\sqrt{2 \epsilon \,} \geq c \quad \text{or} \quad \eta \leq - c^\prime\,,
\ee
for as long as the value of the potential at the extremum is positive. Disregarding, for now, the motivation given in \cite{Ooguri:2018wrx}, the second part of the above expression is easily motivated if one wants to allow unstable dS extrema. The logical ``or'' is a strange sight in a physical expression and in \cite{Andriot:2018wzk} an attempt was made to propose a singular expression that incorporates both quantities, $\epsilon$ and $\eta$, that reproduces the same key features, namely disallowing stable de Sitter points while allowing unstable, positive extrema. The problem of this proposal was that it was not formulated in a covariant manner. We not only rectified this issue in \cite{Andriot:2018mav}, motivate the parameter range of the parameters in the conjecture with data from \cite{Roupec:2018mbn} but also discussed the non-trivial limit of $M_P \to \infty$ of the expression. This (formal) limit is a key point in the swampland program as it corresponds to the decoupling of gravity from the field theory. For the conjectures this means that they have to be trivially satisfied in the decoupling limit as they should be valid in a regime where we have a consistent, effective theory of quantum gravity. For the refined conjecture \eqref{eq:dSconMP} this is trivially the case, in no small part due to the ``or'' appearing therein.\\
We, on the other hand, propose a single expression as an alternative to the refined de Sitter conjecture that gives a simultaneous condition on $\epsilon$ and $\eta$. Formulated as a conjecture:\\
\emph{Any consistent, low-energy, effective quantum gravity theory, that can be written as in \eqref{eq:leeact}, should satisfy the following condition whenever $V>0$:}
\be 
\left(M_P \frac{|\nabla V|}{V}\right)^ q - a M_P^2 \frac{\text{min}(\nabla^i \nabla_j V)}{V} \geq b\,,
\label{eq:ourdScon}
\ee
\emph{where the parameters are free but need to satisfy:}
\be 
a,\, b > 0,\qquad a+b = 1, \qquad \text{and}\qquad q>2\,.
\ee
Written in terms of the slow-roll parameters this reads:
\be 
(2\epsilon)^ {\frac{q}{2}} - a \eta \geq b\,.
\label{eq:ourdSconslowroll}
\ee
We will motivate this conjecture, consisting of the inequality \eqref{eq:ourdScon} and the bounds on the parameters further down below.\\
Before we come to that, let us mention that the decoupling limit is trivially satisfied by first dividing by $M_P^q$. Then one recovers $(|\nabla V|/V)^q \geq 0$ which is always satisfied. When going to an extremum one finds, due to the positivity of the parameters in the conjecture, that $(\text{min}(\nabla^i \nabla_j V)/V)|_{\text{ext}}<0$ which means that only unstable de Sitter points are allowed because we have at least one tachyon. Our conjecture thus has the same basic features as the refined dS conjecture of \cite{Ooguri:2018wrx} but there are differences in the allowed values of $\epsilon$ and $\eta$. Both conjectures disallow small values of $\epsilon$ with positive $\eta$ but differ in certain regions of the $(\epsilon,\eta)$ parameter space. It, however, is not the case that one conjecture is strictly stronger than the other. This differences, depicted in figures \ref{fig:conjreg} and \ref{fig:conjreg2}, mean that the conjectures draw a different border of the swampland in the landscape and potentially allow or disallow certain models. For example, as we will discuss later on, single field slow-roll inflation is not impossible in our conjecture.
\paragraph{We now move on to motivate our conjecture } \eqref{eq:ourdScon} by an argument based on the weak coupling regime similar to what was done in \cite{Ooguri:2018wrx}. The idea is that when going to large distances in field space, corresponding to weak coupling, one is in a controlled regime of string theory. This is usually the case when the string coupling is small, the dilaton is large and negative, $\alpha^ \prime$ corrections are negligible and the internal volume is large. When going to large distances in field space one also has to consider the swampland distance conjecture \cite{Baume:2016psm} for the case of a non-zero scalar potential \cite{Klaewer:2016kiy,Ooguri:2006in}. In short, the distance conjecture proposes that infinite towers of light modes appear when going to large distances in moduli space. Since more degrees of freedom enter the effective theory one can relate this to entropy. De Sitter space has a finite amount of entropy given by the Gibbons-Hawking entropy \cite{Gibbons:1977mu}. In the weak coupling limit, we assume that the Bousso bound \cite{Bousso:1999xy} is saturated. In order to apply these tools for our purpose in a universe that undergoes accelerated expansion we have to make sure that we remain within a semi-classical description. Concretely, this boils down to the fact that the second slow-roll parameter $\eta$ has to be below $-1$, which corresponds to the second condition in \eqref{eq:dSconMP}. The goal of the following derivation is to motivate the conjecture \eqref{eq:ourdScon} and the parameter ranges of $a$, $b$, and $q$.\\
Let us start the motivation of our conjecture by considering a single, canonically normalized scalar field $\phi$ with scalar potential $V$ in the low-energy effective theory. As is argued in \cite{Ooguri:2018wrx}, the scalar potential takes the form:
\be 
V(\phi)=\left(n(\phi) \rme^ {d\phi}\right)^ {-k}\,,
\label{eq:expdec}
\ee
where both parameters $d$ and $k$ are positive. The function $n(\phi)$ counts the number of towers that become light in the asymptotic regime. This means that $n(\phi)>0$ and $\partial_\phi n =: n^ \prime > 0$ as the number of contributing towers grows when going to large distances in field space. Using this we can write:
\be 
\frac{V^ \prime}{V} = -k \left(d+ \frac{n^ \prime}{n}\right)\,,
\ee
using $n^ \prime >0$ this leads to the first condition of \eqref{eq:dSconMP}:
\be 
\frac{|V^ \prime|}{V} \geq k\cdot d =: c\,,
\ee
where we now defined the familiar order one parameter $c$. Raising this to the power of $q$ we find:
\be 
\left(\frac{|V^ \prime|}{V}\right)^ q = k^ q \left(d+\frac{n^ \prime}{n}\right)^ q \geq k^ q \left(d^q + q \frac{n^ \prime}{n} d^ {q-1}\right)\,,
\label{eq:partcon1}
\ee
which holds since all terms in the power expansion are positive. This is true not only for integers $q>2$ but also more generally for real valued $q$ since:
\be 
(x+y)^ q -x^q = \int_x^ {x+y} dt \, q t^ {q-1} \geq \int_x^ {x+y} dt\, q x^ {q-1} = q y x^ {q-1}\,.
\ee
In order to arrive at our conjecture \eqref{eq:ourdScon} we also use the second derivative of the scalar potential to compute:
\be 
\frac{V^{\prime\prime}}{V} = k^ 2 d^ 2 + \frac{n^ \prime}{n} k \left( \frac{k+1}{n} + 2 kd\right) - k\frac{n^ {\prime\prime}}{n}\,.
\ee
We then assume $n\geq 1$ as in the asymptotic regime the number of towers that become light grows and it is therefore reasonable to start with at least one tower. Furthermore, we also consider $n^{\prime\prime} \geq 0$ as the number of towers that become relevant should continue to grow the farther one goes in field space. Writing these conditions as:
\be 
n\geq 1 \qquad \text{and} \qquad \frac{n^ {\prime\prime}}{n} \geq 0\,,
\ee we deduce from the above expression that:
\be 
\frac{V^ {\prime\prime}}{V} \leq k^ 2 d^ 2+ \frac{n^ \prime}{n} k (k+1+2kd)\,.
\label{eq:partcon2}
\ee
Combining \eqref{eq:partcon1} and \eqref{eq:partcon2} we find:
\be
\left(\frac{|V^\prime|}{V}\right)^ q -a \frac{V^ {\prime\prime}}{V} \geq k^ q d^ q - a k^ 2 d^ 2 + \frac{n^ \prime}{n} \left[q k^ q d^ {q-1} - a k (k+1+2kd)\right]\,,
\ee
where we assumed $n^ \prime/n$ to be constant and introduced the parameters $a$ and $b$ as:
\bea
a &= \frac{q(kd)^ {q-1}}{k+1+2kd}\,,\\
b &= (kd)^q \frac{k+1+kd(2-q)}{k+1+2kd}\,.
\label{eq:paradef}
\eea
With this we obtain our conjecture that was already given previously without motivation in \eqref{eq:ourdScon}:
\be 
\left(\frac{|V^ \prime|}{V}\right)^ q - a \frac{V^ {\prime\prime}}{V}\geq b\,.
\ee
If we require $c$ to be of order one, as was already done in \cite{Obied:2018sgi}, we find that since $c = kd \sim 1$:
\bea
a &= \frac{q}{k+3}\,,\\
b &= 1- \frac{q}{k+3}\,,\\
a+b &= 1\,.
\eea
For consistency we also have to require:
\be 
q < k+3\,,
\ee
because $a$ and $b$ have to be positive (see equation \eqref{eq:paradef}). This certainly can hold for $q>2$ where we will allow real valued $q$ as well.

\subsection{Constraining Parameters with known Examples}
\begin{figure}[htb]
     \centering
     \includegraphics[trim=0 210 0 170,clip,width=0.95\textwidth]{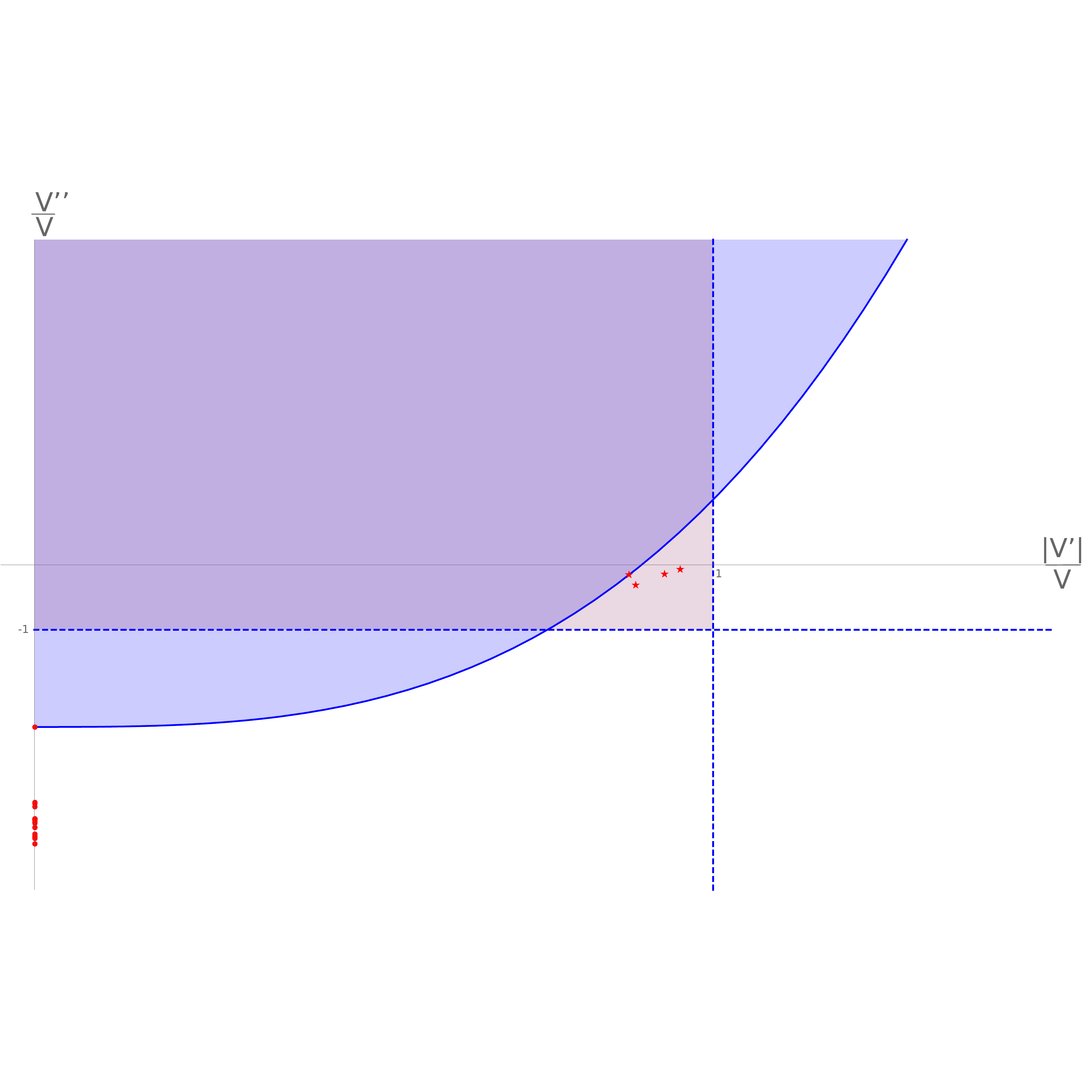}
     \caption{In this figure we present the way how known solutions, listed in table \ref{tab:dSpoints}, constrain the parameters of the conjecture \eqref{eq:ourdScon}. The red dots are the de Sitter extrema while the red stars give the saddle points. We adjusted the parameters such that the conjecture agrees with the known solutions which is in agreement with the general expected values allowed for these parameters. For the lines symbolizing the typical de Sitter conjecture $c=c^ \prime = 1$ was used again.}
     \label{fig:constconj}
\end{figure}

Here, we will use solutions, like the ones discussed in the previous section, in order to constrain the parameters $a$, $b$ and $q$ appearing in our conjecture \eqref{eq:ourdScon}. We use points obtained from classical solutions as they have less intrinsic problems as constructions relying on anti-$Dp$-brane uplifts. Still, we have the same requirements as discussed in section \ref{sec:stringrequ}. For a discussion about the construction and search for such classical de Sitter points look there or in \cite{Caviezel:2008tf,Flauger:2008ad,Danielsson:2012et}. We use two types of points, classical unstable de Sitter points that are extrema, found in \cite{Danielsson:2012et} and later extended by us in \cite{Roupec:2018mbn} and non-extrema with $V>0$, obtained in \cite{Blaback:2013fca}. The latter were obtained from similar setups as our unstable dS points and they face the same issues as well. We select several points from both data sets that have acceptable values for the string coupling and internal volume in such a way that they limit the parameters of the conjecture the most. This basically means that they allow as many models in the landscape as possible from the data we have at our disposal. One interesting thing to note is that the curve, given by the conjecture \eqref{eq:ourdScon}, always runs through $(\sqrt{2\epsilon},\eta) = (1,1)$, meaning that points with $\sqrt{2\epsilon} \geq 1$ and $\eta \leq 1$ do not constrain the parameters. The solutions we selected are given in table \ref{tab:dSpoints} and we illustrate the situation in figure \ref{fig:constconj}. We note that these points also conform with the refined de Sitter conjecture \eqref{eq:dScon}, for reasonable values of $c$ and $c^ \prime$. The deviation from $1$ is typically done as there is no reason that forces them to strictly be equal to that value.

\begin{table}[H]
     \begin{center}
     \begin{tabular}{|c|c|c|c||c|c|}\hline
     $\sqrt{2 \epsilon_V}$ & $\eta_V$ &$\sqrt{2 \epsilon_V}$ & $\eta_V$&$\sqrt{2 \epsilon_V}$ & $\eta_V$\\\hline
     $3.738\cdot 10^{-10}$& 	$-2.495$ &$1.728\cdot10^{-6}$&	$-3.927$&$ 0.950$  & $-0.077$\\
     $3.711\cdot10^{-7}$&	$-3.663$ &$1.894\cdot10^{-6}$&	$-3.964$&$ 0.927$  & $-0.151$\\
     $1.479\cdot10^{-6}$&	$-3.673$ &$2.378\cdot10^{-6}$&	$-3.973$&$ 0.875$  & $-0.162$\\
     $7.733\cdot10^{-6}$&	$-3.727$ &$1.129\cdot10^{-6}$&	$-4.049$&$ 0.885$  & $-0.318$\\
     $4.487\cdot10^{-6}$&	$-3.904$ &$2.859\cdot10^{-6}$&	$-4.049$&&\\
     $2.712\cdot10^{-6}$&	$-3.914$ &$6.255\cdot10^{-6}$&	$-4.145$&&\\
     $8.251\cdot10^{-7}$&	$-3.915$ &$1.998\cdot10^{-6}$&	$-4.186$&&\\
     $1.105\cdot10^{-6}$&	$-3.915$ &$4.634\cdot10^{-6}$&	$-4.187$&&\\
     $9.719\cdot10^{-7}$& 	$-3.918$ &$4.289 \cdot 10^ {-5}$&	$-4.211$&&\\
     $1.248\cdot10^{-6}$&	$-3.922$ &$2.562 \cdot 10^ {-5}$&	$-4.297$&&\\\hline
     \end{tabular} \caption{20 data points of de Sitter extrema from \cite{Danielsson:2012et,Roupec:2018mbn} in the left and middle column and 4 points of classical non-extrema from \cite{Blaback:2013fca} in the rightmost column.} 
     \label{tab:dSpoints}
     \end{center}
\end{table}

\newpage

The data points constrain the parameters in \eqref{eq:ourdScon} in the following way:
\begin{itemize}
     \item The points that have $\epsilon \sim 0$ give:
     \be 
     a \geq \frac{1}{1-\eta} \quad \text{which leads to} \quad a\geq 0.286\,.
     \ee
     \item $b$ is then given by $a+b = 1$ to $b \leq 0.714$.
     \item With these values the point $(\epsilon,\eta) = (0.875,-0.162)$, extracted from table \ref{tab:dSpoints}, provides an upper bound of $q \leq 3.03$.
\end{itemize}
Note, however, that the bound for $q$ is a conservative value, where we assumed that the bound of $a$ is saturated. If we choose a higher value of $a$ then $q$ can likewise be a higher number.\\
As a final remark we want to mention some more points that were obtained in the literature. They are not more constraining than what is presented here but are generally of the same order which gives some weight to our discussion.\\
In similar type IIA setups as discussed here the authors of \cite{Flauger:2008ad} mention de Sitter points with $\eta \simeq -3.7$ and $\eta \simeq -2.5$.\\
In type IIB solutions found in \cite{Caviezel:2009tu} obtained a value of $\eta \simeq -3.1$. This is a nice hint at the universality of these bounds. \\
Lastly, there is an interesting analytical upper bound on $\eta$ that was derived in \cite{Junghans:2016abx} for specific constructions in $4d$, $\mathcal{N}=1$ supergravities that are parametrically close to Minkowski solutions. The bound is $\eta \leq -4/3$ which would be a bit more constraining than our explicit examples. It is, unfortunately, not clear to us if this bound can be reached via an explicit and consistent string construction.

\subsection{A Note on cosmological Implications}
\begin{figure}[htb]
     \centering
     \includegraphics[scale=0.9]{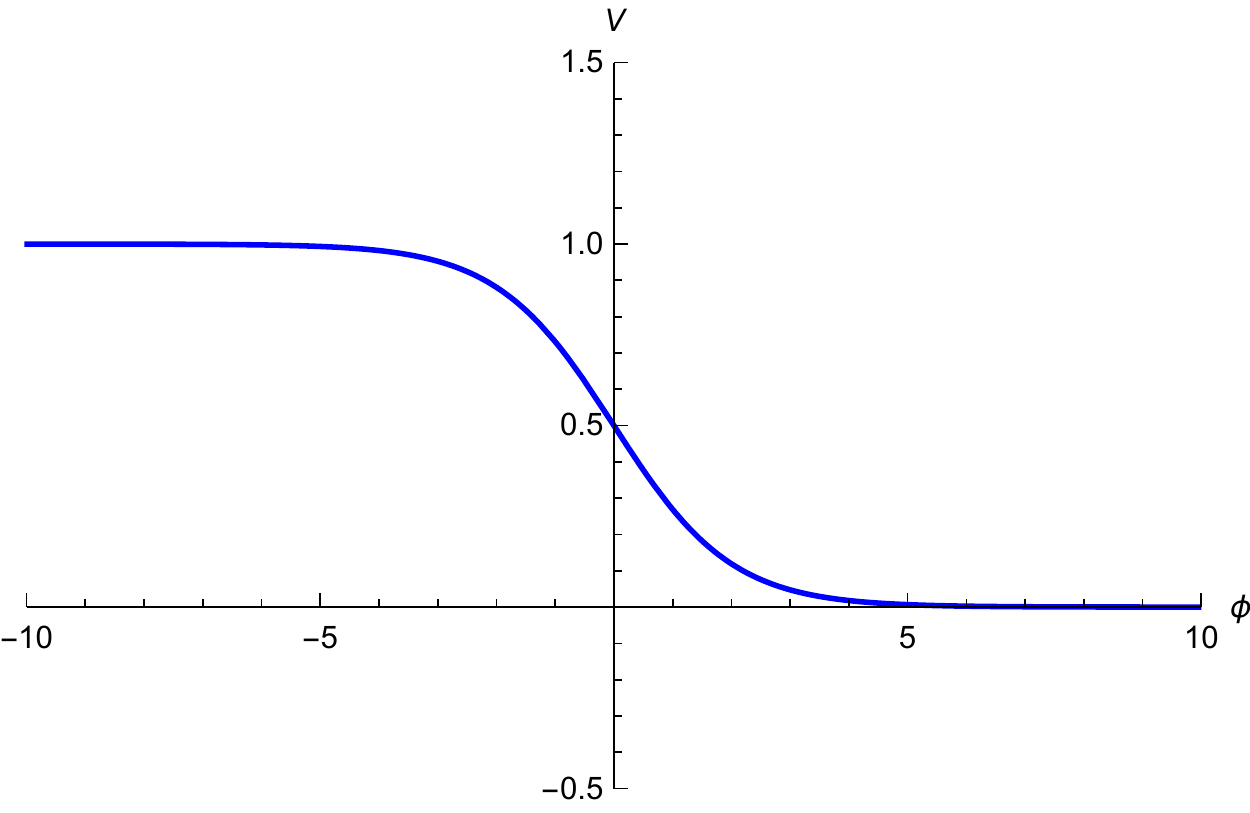}
     \caption{The improved quintessence potential from equation \eqref{eq:improvquint} with $V_0=1$ and $\lambda= 1/2$.}
     \label{fig:quintpot1}
\end{figure}

Since the refined de Sitter conjecture gives constraints on the two slow-roll parameters $\epsilon$ and $\eta$, as evident from \eqref{eq:ourdSconslowroll}, it is natural to consider potential implications for the theory of inflation. In inflationary models one does not sit at an extremum but instead on a flat slope. Nevertheless, the conjecture should hold in general, not only for specific points. Hence, we can restrict the space of allowed inflationary theories if we take the conditions of \eqref{eq:ourdSconslowroll} serious.\\
First off, our conjecture favors models with $\eta \leq 0$ which implies that single field models that have a positive scalar potential should have a concave form of the potential. This is even more interesting as such potentials have recently been found to be favored by experimental data as well \cite{Planck:2018jri}.\\
In particular, single field inflation models are only allowed by our conjecture if they have parameters that satisfy $a \simeq 1$ and $b \ll 1$, which is possible. A setup that has $\epsilon \leq - \eta \simeq b \ll 1 $ is favored by current observations \cite{Planck:2018jri} and satisfies our conjecture for all $q$. For $q\simeq 2$ even convex potentials are allowed and thus our conjecture \eqref{eq:ourdScon} does allow single field slow-roll models as opposed to the original refined conjecture \eqref{eq:dScon} of \cite{Ooguri:2018wrx} where more complicated setups, in order to safe inflationary models, were required \cite{Achucarro:2018vey,Kehagias:2018uem}.\\
As an alternative to inflation models, works on the de Sitter swampland conjecture, often advertise so-called quintessence \cite{Agrawal:2018own}. Such models have a potential with a very flat runaway direction. The simplest example that is often used for this potential is:
\be 
V(\phi) = V_0 \rme^ {-\lambda \phi}\,,
\label{eq:quint}
\ee
where $\lambda >0$ is some parameter. Note that this potential is convex, which is not favored by our conjecture. The current experimental bound on the parameter in \eqref{eq:quint} is $\lambda \leq 0.6$ \cite{Agrawal:2018own} which, as we will show, is not in agreement with our conjecture. Our condition in equation \eqref{eq:ourdScon}, with the quintessence potential in \eqref{eq:quint}, leads to:
\be 
\lambda^ q - a \lambda^ 2 - b \geq 0\,,
\ee
which is saturated for $\lambda = 1$, independent of the parameters $a$, $b$ and $q$ and only values of $\lambda \geq 1$ are allowed. Since experimentally $\lambda \leq 0.6$ such simple quintessence models are thus excluded by our conjecture. Quintessence models in general, however, are not completely excluded. If one still wants $\rme^ {-\lambda \phi}$ with $\lambda \geq 0$ exponentially, either due to the Dine-Seiberg argument \cite{Dine:1985he}, because of the upper bound obtained in \cite{Hebecker:2018vxz} or the exponential decay discussed around equation \eqref{eq:expdec} \cite{Ooguri:2018wrx}, one can make the proposal for the scalar potential:
\be 
V(\phi) = \frac{V_0}{2} \left(1- \tanh (\lambda \phi)\right) = V_0\frac{\rme^ {-\lambda \phi}}{\rme^ {\lambda \phi}+ \rme^ {-\lambda \phi}}\,.
\label{eq:improvquint}
\ee
Here, importantly, $V_0 >0$ and $\lambda >0$. The form of the potential is depicted in figure \ref{fig:quintpot1}.
This potential has a very flat slope for negative values of $\phi$, below about $\phi = -5$. There, the potential is concave and quintessence seems possible with good slow-roll parameters. We can check the first and second slow-roll parameters by computing:
\bea 
\frac{|V^ \prime|}{V} &= 2 \lambda \frac{\rme^ {\lambda \phi}}{\rme^ {\lambda \phi} + \rme^ {-\lambda \phi}}\,,\\
\frac{V^{\prime \prime}}{V} &= 4 \lambda^ 2 \rme^ {\lambda \phi} \frac{\rme^ {\lambda \phi}-\rme^ {-\lambda \phi}}{(\rme^ {\lambda \phi} + \rme^ {-\lambda \phi})^2}\,.
\eea
In the limit where we let $\phi \to -\infty$, we get:
\bea 
\frac{|V^ \prime|}{V} &\sim 2 \lambda \rme^ {-2\lambda |\phi|}\,,\\
\frac{V^ {\prime \prime}}{V} &\sim 4 \lambda^ 2 \rme^{-2 \lambda |\phi|}\,,
\eea
which means that the slow-roll parameters $\epsilon$ and $\eta$ are exponentially suppressed in this limit. The ratio of the pressure and energy density of dark energy $w$ is given to be \cite{Agrawal:2018own}:
\be 
w+1 = \frac{2}{3} \epsilon\,,
\ee
in this regime and since $\epsilon$ basically vanishes we obtain $w=-1$ to very high precision, as is required by observations.\\
What is left is to verify that this potential also works with our refined conjecture \eqref{eq:ourdScon}, for which we have to find suitable and allowed parameters $b$ (or $a$) and $q$. As the condition is saturated for $\rme^ {-\phi}$ we can fix $\lambda = 1/2$ such that we get back the critical expression from $V \sim \rme^ {-2\lambda \phi}$ to which \eqref{eq:improvquint} converges for $\phi \to +\infty$. For the limit of $\phi \to - \infty$, the $\eta$ part of the conjecture dominates and we can use this to fix:
\be 
b = \frac{1}{2} \left( 4 \lambda^ 2 \rme^ {2\lambda \phi_0}\right) = \frac{1}{2} \rme^ {-10}\,,
\ee
where we have used $\phi_0=-10$ as a somewhat arbitrary cutoff related to the time in the past when quintessence started.
\begin{figure}[htb]
     \centering
     \includegraphics[scale=0.9]{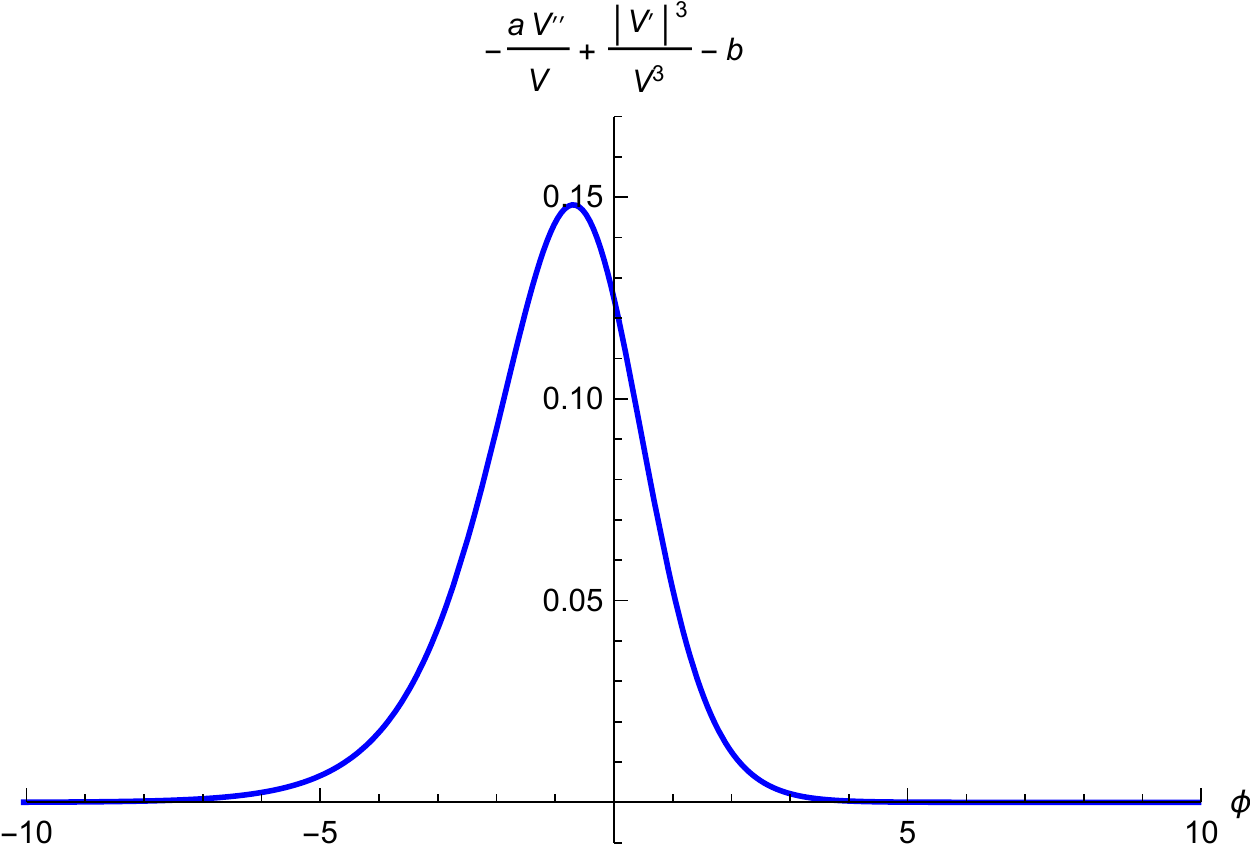}
     \caption{The conjecture \eqref{eq:ourdScon} checked for the scalar potential \eqref{eq:improvquint} with $\lambda= 1/2$, $b=\rme^ {-10}/2$ and $q=3$.}
     \label{fig:quintpot2}
\end{figure}
The scalar potential is concave from $-\infty$ until the origin and thus there is no issue satisfying the conjecture for that region. However, it needs to hold for any part of the range of $\phi$ and starting from $\phi = 0$ the potential is convex. Due to $\epsilon < 1$ we cannot have $q$ too big. As it turns out $q=3$ is acceptable. The situation with these parameters is shown in figure \ref{fig:quintpot2}.\\
As a word of caution we would like to mention that the potential we used for quintessence here \eqref{eq:improvquint} is purpose built in order to illustrate how our refined de Sitter conjecture \eqref{eq:ourdScon} can accommodate quintessence. Whether such a model can realistically arise in string cosmology needs to be investigated.\\
Finally we note that both, slow-roll inflation and quintessence models, compatible with our conjecture need to have parameters $a \simeq 1$, $b\ll1$. 

\subsection{Refined de Sitter Conjecture - Interim Summary}
In this subsection we discussed an alternative to the refined de Sitter conjecture \eqref{eq:dScon} of \cite{Ooguri:2018wrx}. Our condition \eqref{eq:ourdScon}, first presented in \cite{Andriot:2018mav}, has the advantage of giving a single expression that incorporates the same basic features as the original refined conjecture. Both conjectures allow positive extrema and forbid stable de Sitter minima. Nevertheless, the conjectures differ in certain regions and allow different kinds of models. For example, our condition does allow certain single field slow-roll inflation models but gives restrictions on quintessence. The constructions that are favored by our conjecture also are currently prime candidates according to cosmological observations \cite{Planck:2018jri}. Interestingly, cosmological models allowed by our conjecture seem to prefer $a\simeq 1$ and $b \ll 1$. We also attempted to motivate our parameters $a$, $b$ and $q$ by known classical de Sitter constructions from \cite{Danielsson:2012et,Roupec:2018mbn}. This lead to a lower bound $a \geq 0.286$ and an upper bound of $q\leq 3.03$ if one saturates the bound on $a$.

\chapter{Summary and Outlook}
In the present thesis we investigated several aspects about constructing de Sitter spaces in low-energy effective theories (supergravities) originating from string theory \cite{Roupec:2018mbn,Banlaki:2018ayh,Andriot:2018mav,Cribiori:2019hod,Cribiori:2019bfx,Cribiori:2019drf,Cribiori:2019hrb,Cribiori:2020bgt}.\\
In chapter \ref{sec:antiD3} we focused on $Dp$-branes and their description in $4d$, $\mathcal{N}=1$ supergravity. For this we utilized constrained multiplets \cite{Rocek:1978nb,Lindstrom:1979kq,Samuel:1982uh,Komargodski:2009rz}, mainly the nilpotent chiral field $X$. After reviewing non-linear supergravity \cite{Farakos:2013ih,Dudas:2015eha,Bergshoeff:2015tra,Hasegawa:2015bza,Ferrara:2015gta,DallAgata:2016syy,Ferrara:2016een} in general we proceeded to give the correct description of supersymmetry breaking branes. In the case of intersecting $D6$-branes, we cleaned up a long-standing misconception about the correct description of the situation in $4d$ supergravity. Previously, it was claimed that this situation can be described using the standard $D$-term of supergravity \cite{Villadoro:2006ia,Blumenhagen:2002wn,Kachru:1999vj,Cvetic:2001nr}. We showed, however, that this cannot be the case, as a setup with linear and non-linear supersymmetry necessarily leads to a non-linear description. In section \ref{sec:D3barKKLTmain}, we used the same basic concepts in order to give the complete description of the uplifting anti-$D3$-brane at the bottom of a warped throat in the KKL(MM)T model \cite{Kachru:2003aw,Kachru:2003sx}. We showed \cite{GarciadelMoral:2017vnz,Cribiori:2019hod} how all terms can be included using constrained superfields and used the so-called new $D$-term \cite{Cribiori:2017laj,Cribiori:2018dlc,Cribiori:2018hxv} in order to incorporate the sign flip that is present for the Chern-Simons part of the action. Using the new D-term we wrote this behavior in terms of a holomorphic gauge kinetic function where, ordinarily, it would seem that we require an anti-holomorphic one, which would clash with supersymmetry. In the end, we were able to write down the complete action of the the $\overline{D3}$-brane at the bottom of a warped throat in the KKLT setup, including all world-volume fields.\\
The two applications that were discussed here highlight the power and utility of the formulation of non-linear supergravity using constrained multiplets. It stands to reason that the formalism can be applied to many more situations. Some possibilities would be the description of other non-supersymmetric objects, such as $NS$-sources. In principle, the methods discussed in section \ref{sec:nonsusybranessec}, are applicable to any source, if one knows the correct scalar potential contribution. Nevertheless, working out the details requires great care. A different kind of extension would be to include all fields for a given situation. We showed this only for the anti-$D3$-brane in the KKLT background. For other situations this has not been done thus far but would be equally interesting. The case of intersecting $D6$-branes in type IIA would be a worthwhile first example due to its relevance for both cosmological and Standard Model like scenarios \cite{Grana:2005jc,Douglas:2006es,Blumenhagen:2006ci,Blumenhagen:2005mu}.\\
Chapter \ref{sec:kkltconstr} introduced new ways to construct de Sitter vacua, along the lines of the familiar KKLT construction \cite{Cribiori:2019bfx,Kallosh:2019zgd,Cribiori:2019drf,Cribiori:2019hrb}. All of these model rely on an uplifting anti-brane and thus are usually not considered as classical constructions in the sense of the de Sitter swampland conjecture. We extended the KKLT formalism to type IIA in section \ref{sec:IIAuplift} where we found that the construction relies on non-perturbative corrections in all moduli directions. We claimed that these indeed can appear \cite{Palti:2008mg} for all moduli. For the direction of the volume modulus we argued that such corrections are motivated by M-theory triality \cite{Hull:1994ys,Schwarz:1996bh,Acharya:2007rc}. Explicitly, these corrections can be sourced by worldsheet instantons \cite{Kachru:2000ih,Blumenhagen:2009qh}. With this, it is relatively easy to construct models similar to the original KKLT proposal. Using the non-perturbative corrections and extending the setup to the Kallosh-Linde racetrack double exponent \cite{Kallosh:2004yh}, we introduced the mass production mechanism of de Sitter vacua in section \ref{sec:massprod}. By first constructing a Minkowski progenitor solution one can guarantee that, at least for the racetrack superpotential, a stable de Sitter space can be constructed without any tachyonic directions, if the Minkowski space has no flat directions. An additional nice feature of this procedure is that one performs a downshift from Minkowski to anti-de Sitter before introducing the uplift to de Sitter. This means that the uplift does not have to be incredibly small in order to match the cosmological constant and disentangles the SUSY breaking scale from the scale of the cosmological constant. We tested the mass production procedure for various models with different internal geometry and confirmed that it is easy to produce stable de Sitter minima in each case, both in type IIA and IIB. Extending the range of this construction tool even further, we investigated models based on M-theory compactifications in section \ref{sec:mtheory}. The models are based on compactifications on the generalized, twisted $7$-torus \cite{DallAgata:2005zlf,Duff:2010vy,Derendinger:2014wwa,Ferrara:2016fwe} and we found that in certain cases it is possible to forego non-perturbative corrections in some moduli directions. This is an interesting feature since such corrections are traditionally tedious to deal with. Including conjectured but not yet confirmed $S$-fluxes it is even possible to find a model that has no non-perturbative corrections at all. If confirmed, this would be a KKLT like construction that does not require such corrections in order to find a stable AdS point. We also considered the relation of the M-theory inspired models to type II compactifications and found that they generally can be translated to either type IIA or IIB if one excludes particularly exotic contributions.\\
Having a way to consistently and easily build de Sitter models is a great opportunity for string phenomenology. The methods we presented here are accessible and allow for a wide variety of parameters. That means one can attempt to construct models that match desired physical observations, like the current expansion of our universe or the measured parameters from inflation. If it is possible to find a sufficiently long  and flat slope for a period of inflation in our models here, it would certainly be a great success. Other than that, there are many open questions regarding the details of our setup and its relation to full string theory. Investigating whether or not the conjectured $S$-fluxes can explicitly appear in type IIB string theory and its compactifications would be a worthwhile effort.\\
Finally, in chapter \ref{sec:swamplandconjectures} we discussed matters regarding the swampland program \cite{Brennan:2017rbf,Palti:2019pca,vanBeest:2021lhn}, focusing on the conjectures that restrict de Sitter constructions \cite{Danielsson:2018ztv,Obied:2018sgi,Ooguri:2018wrx,Andriot:2018mav}. In section \ref{sec:scaling} we reviewed simple, classical type IIA compactifications regarding the principle possibility for controlled de Sitter vacua. Inspired by the AdS solutions of \cite{DeWolfe:2005uu}, we looked at different scaling limits where we might obtain parametrically large volume and small coupling. The result of this investigation is that, even considering some more non-standard ingredients \cite{Caviezel:2008tf,Flauger:2008ad,Saltman:2004jh,Silverstein:2007ac,Haque:2008jz,Danielsson:2009ff,deCarlos:2009fq,deCarlos:2009qm,Caviezel:2009tu,Danielsson:2010bc,Dong:2010pm,Andriot:2010ju}, like $KK$- and $NSO$-sources, the only possible limit that allows for controlled vacua is the one of a large amount of $O6$-planes. This is particularly troublesome since the number of orientifold planes in a compactification is fixed by the geometry of the internal manifold. In typical, known examples this number is restricted to be of order one. We confirmed these findings with an explicit example and found that the proportionality constants between the number of $O6$-planes and the behavior of the volume of the compactification manifold and the coupling do not favor such a limit as well. While these constants were found in a particular setup, they still should give an idea of what kinds of manifolds we need to look for regarding the number of $O6$-planes if we hope to find de Sitter solutions.\\
In section \ref{sec:dSextrema} we gave several counter examples to the original de Sitter conjecture \cite{Roupec:2018mbn} that also disallowed unstable de Sitter extrema. We performed a numerical search in type IIA compactifications and found many new such points, in addition to highlighting known ones from \cite{Caviezel:2008tf,Flauger:2008ad}, that explicitly violate the original conjecture that was formulated only giving a constraint on the first derivative on the scalar potential. Several features of the considered models were reviewed and we showed that they can be in controlled regions. Together with other criticisms this lead to a necessary refinement of the conjecture \cite{Ooguri:2018wrx} where a second condition on the Hessian of the mass matrix was added, thus allowing for unstable dS extrema.\\
Our own refined de Sitter swampland conjecture \cite{Andriot:2018mav} was presented in section \ref{sec:refconj}. Motivated by a simple effective action of a scalar field we combined conditions on the first and second derivative of the scalar potential into one inequality. This, alternative, refined de Sitter conjecture differs from the one given in \cite{Ooguri:2018wrx} in physical ways but paints the same basic picture, stable de Sitter points are prohibited. The differences manifest in cosmological applications. Our conjecture allows for slow-roll inflation and favors concave potentials, a feature that aligns well with current cosmological observations \cite{Planck:2018jri}. Interestingly, the most simple quintessence models are not allowed by our version of the conjecture and a more complicated setup is required there as well.\\
The question if de Sitter vacua from string theory are principally impossible deserves certainly more attention. The way to proceed is twofold. The conjectures are motivated by known examples but are not mathematically proven from string theory. Thus, if one finds an explicit solution that is in a controlled regime, satisfies quantization conditions and so on, the de Sitter swampland conjectures would need to be abandoned. The two ways are thus either proofing the conjecture, or at least, supplying more evidence or constructing an explicit solution. Both ways are worth pursuing and in the process it is clear that we will learn many things about the merits and limitations of string theory. The end result is not yet clear.\vspace{12pt}\\
Unraveling the true nature of our universe is an ongoing and daunting task. However, it is full of excitement and wonder. With the onset of precision cosmology experiments we can hope to come closer to a true and fundamental understanding of nature. The author hopes that his work has made a contribution to these efforts, however small it may be.

\newpage
\section*{Acknowledgments}
During the course of my doctoral studies I have been funded by an FWF grant with the number P 30265, a DOC grant of the Austrian Academy of Sciences and via the Doctoral College Particles  and Interactions with the project number W1252-N27. My extended stay at Stanford University was made possible by an Austrian Marshall Plan Fellowship.\\
I would like to offer my sincerest thanks to my advisor Timm Wrase, for all his help and patience in all things related to my doctorate, both scientific and organizational. His explanations taught me more than any book could ever have.\\
Many thanks to Niccol\`o Cribiori for many explanations, discussions and encouragement. Many more thanks as well to Andreas Banlaki for stimulating discussions and to Abhiram Kidambi for forcing me to consider topics beyond my comfort.\\
Special thanks to Renata Kallosh and Andrei Linde for the warm welcome at the Stanford Institute for Theoretical Physics. The discussions with you opened my eyes in many regards.\\
Furthermore I would like to thank David Andriot 
and Antoine Van Proeyen for the fruitful collaborations and kind help offered and Anton Rebhan as well as all the people at the Institute  for Theoretical Physics at TU Wien for all their continued support.\\
Personal thanks go to all my friends and family for their continuous support, especially to Lukas Lüftinger for being a pillar of support when needed and to Matthias Röcklinger for additional encouragement.\\
To my wife, Anneliese, I cannot offer sufficient thanks for supporting me over so many years but I will try my best.\\
Finally, to my parents, I am thankful for encouraging me to pursue my dreams and supporting me each step of the way.

\appendix
\chapter{Some more Details}

\section[Modular Properties of the anti-$D3$-brane Action]{Modular Properties of the anti-$\mathbf{D3}$-brane Action}
\label{app:modinv}
\subsection{Modular Invariance of the Goldstino and the Fermions}
The anti-$D3$-brane has to be invariant under moduli transformations. For the Goldstino and fermion sector the transformations under $SL(2,\mathbb{Z})$ are:
\bea 
\tau \quad &\to \quad \frac{a\tau + b}{c \tau + d}\,,\\
G_3 \quad &\to \quad \frac{G_3}{c\tau + d}\,,
\eea 
where the $a$, $b$, $c$ and $d$ are integer numbers and $G_3$ is the background 3-flux. For invariance, the fermions need to transform as well, in the following way:
\bea 
P_L \lambda \quad &\to \quad \rme^{-\rmi \delta} P_L \lambda\,,\\
P_L \chi^i \quad &\to \quad \rme^{-\rmi \delta} P_L \chi^i\,,
\eea 
with the phase:
\be 
\rme^{-2\rmi\delta} = \left( \frac{c\bar{\tau}+d}{c \tau + d} \right)^{\frac{1}{2}}\,.
\ee
To compensate for the transformation of $G_3$, the nilpotent multiplet $X$, likewise, has to change in the following way:
\be 
X \quad \to \quad \frac{X}{c\tau + d}\,,
\ee
in order such that the superpotential remains constant. This, in turn, implies that the coupling of $X$ to the axio-dilaton:
\be 
\frac{X \bar{X}}{\tau-\bar{\tau}}\,,
\ee
has to be invariant. This fixes the coupling of the two because if the superpotential changes in the following way:
\be 
W \quad \to \quad \frac{W}{c\tau + d}\,,
\ee
then we have to change the Kähler potential by a Kähler transformation:
\be 
K \quad \to \quad K + \log | c \tau + d |^2\,,
\ee
which in turn ensures that the supergravity theory is modular invariant. Since we can only transform the first term in the Kähler potential \eqref{eq:D3barKKfinal} the remaining ones all have to be modular invariant individually which fixes the remaining couplings. With an analogous argument we find that:
\be 
\delta_{i\jb} \frac{Y^i \bar{Y}^\jb}{\tau-\bar{\tau}}\,,
\ee
is modular invariant if the multiplets $Y^i$ transform exactly like the $X$:
\be 
Y^i \quad \to \quad \frac{Y^i}{c\tau +d}\,.
\ee
The remaining terms of the Kähler and superpotential then already have the correct behavior under modular transformations.
\subsection{Self-Duality of the Vector}
We still have to discuss the properties of the $U(1)$ vector. When coupled to the axio-dilaton its gauge group will be enhanced to $SL(2,\mathbb{Z})$ \cite{Gaillard:1981rj}, which is just the modular group we are discussing here. As opposed to the other contributions, discussed in the previous subsection, the action of the vector will not be invariant. The behavior was calculated in \cite{Gaillard:1981rj} and it was found that a specific symmetry is present for the $U(1)$ sector, called self-duality. It exchanges the electric with the magnetic field strength and the coupling goes to its inverse. We need to verify that this holds for our description in terms of multiplets. For this we let:
\bea 
P_L\Lambda_\alpha \quad &\leftrightarrow \quad P_L \Lambda_{\alpha}^D\,,\\
\hat{f}_{\overline{D3}} \quad &\leftrightarrow \quad \left( \hat{f}_{\overline{D3}} \right)^{-1}\,,
\eea
with $P_L \Lambda_\alpha^D$ the dual of $P_L \Lambda_\alpha$. Self-duality is an on-shell property which allows us to consider any action that reduces the 
\be 
S_V = -\frac{1}{4} \left[\hat{f}_{\overline{D3}} \bar{\Lambda}P_L\Lambda\right]_F\,,
\label{eq:veconshell}
\ee
after imposing the on-shell conditions. For the discussion here, it is useful to not consider constrained multiplets (in particular $P_L \Lambda_\alpha$) but rather impose the constraint explicitly in form of a Lagrange multiplier:
\be
\tilde{S}_V = -\frac{1}{4} \left[\hat{f}_{\overline{D3}} \bar{\Lambda} P_L \Lambda\right]_F + \frac{1}{2} \left[\bar{\Phi}P_L\Lambda X\right]_F + \left[\frac{\rmi}{2}\bar{\Lambda}^DP_L\Lambda\right]_F\,.
\label{eq:lagmult}
\ee
Here $P_L \Lambda_\alpha$ is a chiral multiplet, explicitly not constrained under \eqref{eq:fermconst}, and $P_L \Phi_\alpha$ is the chiral Lagrange multiplier multiplet that reinstates the constraint. We now go on and write the dual chiral multiplet $P_L \Lambda_\alpha^D$ using $\Sigma$ and an operator $\mathcal{D}_\alpha$, that can be thought of as a superspace derivative \cite{Kugo:1983mv}:
\be 
\mathcal{D}_\alpha:\quad (w,c) \quad \to \quad \left(w + \frac{1}{2},c-\frac{3}{2}\right)\,.
\ee
Using this, we write:
\be
P_L\Lambda_\alpha^D = \Sigma\left(\mathcal{D}_\alpha U\right)\,,
\ee
where $U$ is a vector multiplet with vanishing weights: $(0,0)$. This allows us to compute:
\bea 
\left[\rmi \bar{\Lambda}_D P_L \Lambda \right]_F &= \; \left[\rmi \Sigma\left(\mathcal{D}^\alpha U P_L \Lambda_\alpha\right)\right]_F\\
&= + \left[\rmi \mathcal{D}^\alpha U P_L \Lambda_\alpha - \rmi \mathcal{D}^\alpha U P_R \Lambda_\alpha \right]_D\\
&= - \left[\rmi U \mathcal{D}^\alpha P_L \Lambda_\alpha - \rmi U \mathcal{D}^\alpha P_R \Lambda_\alpha \right]_D\,,
\eea
where boundary terms were neglected. If we vary this $F$-term action with respect to $U$ we find:
\be 
\delta U: \quad \mathcal{D}^\alpha P_L \Lambda_\alpha = \mathcal{D}^\alpha P_R \Lambda_\alpha\,.
\ee
This is just the supersymmetric version of the Bianchi identity which in turn suggests that $P_L \Lambda_\alpha$ is the field strength of a vector multiplet via: $V:\, P_L \Lambda_\alpha = \Sigma (\mathcal{D}_\alpha V)$. Plugging these results into the Lagrange multiplier action \eqref{eq:lagmult} gives back the correct on-shell expression of equation \eqref{eq:veconshell}. If we, however, instead vary the other fields:
\bea 
\delta P_L \Phi_\alpha: \quad P_L \Lambda_\alpha &= 0\,,\\
\delta P_L \hat{f}_{\overline{D3}} P_L \Lambda_\alpha &= \rmi P_L \Lambda_\alpha^D - P_L \Phi_\alpha X\,,
\eea 
and use that $X$ is nilpotent, we find, by multiplying the second line with $X$ that:
\be 
P_L \Lambda_\alpha^D  X = 0\,.
\ee
Thus, we have removed the fermion from the dual multiplet $P_L \Lambda_\alpha^D$. Using these conditions for the action \eqref{eq:lagmult}, we find the following on-shell action:
\be 
\tilde{S}_V = -\frac{1}{4} \left[\hat{f}^{-1}_{\overline{D3}} \bar{\Lambda}^D P_L \Lambda^D\right]_F\,,
\ee
and so we have shown the self dual property of the $U(1)$ sector. As a last remark we notice that, due to the nilpotency of $X$, the gauge kinetic function satisfies:
\be 
\left( \hat{f}_{\overline{D3}} (\bar{f})\right)^{-1} = \hat{f}_{\overline{D3}}(\bar{f}^{-1})\,,
\ee
which can be proven by using the definition of the generalized gauge kinetic function \eqref{eq:gengaugekin}. For our choice of $f(\tau) = -\rmi \tau$, this sends $\tau \to -1/\tau$. Furthermore, $\tau \to \tau + 1$ is a trivial symmetry of the action and thus we have the complete $SL(2,\mathbb{Z})$ group realized.

\newpage
\makeatletter
\renewenvironment{thebibliography}[1]
     {\chapter*{\bibname}
      \@mkboth{\MakeUppercase\bibname}{\MakeUppercase\bibname}%
      \list{\@biblabel{\@arabic\c@enumiv}}%
           {\settowidth\labelwidth{\@biblabel{#1}}%
            \leftmargin\labelwidth
            \advance\leftmargin\labelsep
            \@openbib@code
            \usecounter{enumiv}%
            \let\p@enumiv\@empty
            \renewcommand\theenumiv{\@arabic\c@enumiv}}%
      \sloppy
      \clubpenalty4000
      \@clubpenalty \clubpenalty
      \widowpenalty4000%
      \sfcode`\.\@m}
     {\def\@noitemerr
       {\@latex@warning{Empty `thebibliography' environment}}%
      \endlist}
\makeatother
\clearpage
\addcontentsline{toc}{chapter}{Bibliography}
\bibliographystyle{JHEP}
\singlespace
\bibliography{refs}

\providecommand{\href}[2]{#2}\begingroup\raggedright\begin{thebibliography}{100}

\bibitem{Roupec:2018mbn}
C.~Roupec and T.~Wrase, \emph{{de Sitter Extrema and the Swampland}},
  \href{https://doi.org/10.1002/prop.201800082}{\emph{Fortsch. Phys.}
  {\bfseries 67} (2019) 1800082}
  [\href{https://arxiv.org/abs/1807.09538}{{\ttfamily 1807.09538}}].

\bibitem{Banlaki:2018ayh}
A.~Banlaki, A.~Chowdhury, C.~Roupec and T.~Wrase, \emph{{Scaling limits of dS
  vacua and the swampland}},
  \href{https://doi.org/10.1007/JHEP03(2019)065}{\emph{JHEP} {\bfseries 03}
  (2019) 065} [\href{https://arxiv.org/abs/1811.07880}{{\ttfamily
  1811.07880}}].

\bibitem{Andriot:2018mav}
D.~Andriot and C.~Roupec, \emph{{Further refining the de Sitter swampland
  conjecture}}, \href{https://doi.org/10.1002/prop.201800105}{\emph{Fortsch.
  Phys.} {\bfseries 67} (2019) 1800105}
  [\href{https://arxiv.org/abs/1811.08889}{{\ttfamily 1811.08889}}].

\bibitem{Cribiori:2019hod}
N.~Cribiori, C.~Roupec, T.~Wrase and Y.~Yamada, \emph{{Supersymmetric
  anti-D3-brane action in the Kachru-Kallosh-Linde-Trivedi setup}},
  \href{https://doi.org/10.1103/PhysRevD.100.066001}{\emph{Phys. Rev.}
  {\bfseries D100} (2019) 066001}
  [\href{https://arxiv.org/abs/1906.07727}{{\ttfamily 1906.07727}}].

\bibitem{Cribiori:2019bfx}
N.~Cribiori, R.~Kallosh, C.~Roupec and T.~Wrase, \emph{{Uplifting
  Anti-D6-brane}}, \href{https://doi.org/10.1007/JHEP12(2019)171}{\emph{JHEP}
  {\bfseries 12} (2019) 171}
  [\href{https://arxiv.org/abs/1909.08629}{{\ttfamily 1909.08629}}].

\bibitem{Cribiori:2019drf}
N.~Cribiori, R.~Kallosh, A.~Linde and C.~Roupec, \emph{{Mass Production of IIA
  and IIB dS Vacua}},
  \href{https://doi.org/10.1007/JHEP02(2020)063}{\emph{JHEP} {\bfseries 02}
  (2020) 063} [\href{https://arxiv.org/abs/1912.00027}{{\ttfamily
  1912.00027}}].

\bibitem{Cribiori:2019hrb}
N.~Cribiori, R.~Kallosh, A.~Linde and C.~Roupec, \emph{{de Sitter Minima from M
  theory and String theory}},
  \href{https://doi.org/10.1103/PhysRevD.101.046018}{\emph{Phys. Rev. D}
  {\bfseries 101} (2020) 046018}
  [\href{https://arxiv.org/abs/1912.02791}{{\ttfamily 1912.02791}}].

\bibitem{Cribiori:2020bgt}
N.~Cribiori, C.~Roupec, M.~Tournoy, A.~Van~Proeyen and T.~Wrase,
  \emph{{Non-supersymmetric branes}},
  \href{https://doi.org/10.1007/JHEP07(2020)189}{\emph{JHEP} {\bfseries 07}
  (2020) 189} [\href{https://arxiv.org/abs/2004.13110}{{\ttfamily
  2004.13110}}].

\bibitem{Planck:2018jri}
{\scshape Planck} collaboration, Y.~Akrami et~al., \emph{{Planck 2018 results.
  X. Constraints on inflation}},
  \href{https://doi.org/10.1051/0004-6361/201833887}{\emph{Astron. Astrophys.}
  {\bfseries 641} (2020) A10}
  [\href{https://arxiv.org/abs/1807.06211}{{\ttfamily 1807.06211}}].

\bibitem{Brennan:2017rbf}
T.~D. Brennan, F.~Carta and C.~Vafa, \emph{{The String Landscape, the
  Swampland, and the Missing Corner}},
  \href{https://doi.org/10.22323/1.305.0015}{\emph{PoS} {\bfseries TASI2017}
  (2017) 015} [\href{https://arxiv.org/abs/1711.00864}{{\ttfamily
  1711.00864}}].

\bibitem{Palti:2019pca}
E.~Palti, \emph{{The Swampland: Introduction and Review}},
  \href{https://doi.org/10.1002/prop.201900037}{\emph{Fortsch. Phys.}
  {\bfseries 67} (2019) 1900037}
  [\href{https://arxiv.org/abs/1903.06239}{{\ttfamily 1903.06239}}].

\bibitem{vanBeest:2021lhn}
M.~van Beest, J.~Calder\'on-Infante, D.~Mirfendereski and I.~Valenzuela,
  \emph{{Lectures on the Swampland Program in String Compactifications}},
  \href{https://arxiv.org/abs/2102.01111}{{\ttfamily 2102.01111}}.

\bibitem{Obied:2018sgi}
G.~Obied, H.~Ooguri, L.~Spodyneiko and C.~Vafa, \emph{{De Sitter Space and the
  Swampland}},  \href{https://arxiv.org/abs/1806.08362}{{\ttfamily
  1806.08362}}.

\bibitem{Kachru:2003aw}
S.~Kachru, R.~Kallosh, A.~D. Linde and S.~P. Trivedi, \emph{{De Sitter vacua in
  string theory}},
  \href{https://doi.org/10.1103/PhysRevD.68.046005}{\emph{Phys. Rev.}
  {\bfseries D68} (2003) 046005}
  [\href{https://arxiv.org/abs/hep-th/0301240}{{\ttfamily hep-th/0301240}}].

\bibitem{Rocek:1978nb}
M.~Rocek, \emph{{Linearizing the Volkov-Akulov Model}},
  \href{https://doi.org/10.1103/PhysRevLett.41.451}{\emph{Phys. Rev. Lett.}
  {\bfseries 41} (1978) 451}.

\bibitem{Lindstrom:1979kq}
U.~Lindstrom and M.~Rocek, \emph{{CONSTRAINED LOCAL SUPERFIELDS}},
  \href{https://doi.org/10.1103/PhysRevD.19.2300}{\emph{Phys. Rev. D}
  {\bfseries 19} (1979) 2300}.

\bibitem{Samuel:1982uh}
S.~Samuel and J.~Wess, \emph{{A Superfield Formulation of the Nonlinear
  Realization of Supersymmetry and Its Coupling to Supergravity}},
  \href{https://doi.org/10.1016/0550-3213(83)90622-3}{\emph{Nucl. Phys. B}
  {\bfseries 221} (1983) 153}.

\bibitem{Komargodski:2009rz}
Z.~Komargodski and N.~Seiberg, \emph{{From Linear SUSY to Constrained
  Superfields}},
  \href{https://doi.org/10.1088/1126-6708/2009/09/066}{\emph{JHEP} {\bfseries
  09} (2009) 066} [\href{https://arxiv.org/abs/0907.2441}{{\ttfamily
  0907.2441}}].

\bibitem{DallAgata:2016syy}
G.~Dall'Agata, E.~Dudas and F.~Farakos, \emph{{On the origin of constrained
  superfields}}, \href{https://doi.org/10.1007/JHEP05(2016)041}{\emph{JHEP}
  {\bfseries 05} (2016) 041}
  [\href{https://arxiv.org/abs/1603.03416}{{\ttfamily 1603.03416}}].

\bibitem{Ferrara:2016een}
S.~Ferrara, R.~Kallosh, A.~Van~Proeyen and T.~Wrase, \emph{{Linear Versus
  Non-linear Supersymmetry, in General}},
  \href{https://doi.org/10.1007/JHEP04(2016)065}{\emph{JHEP} {\bfseries 04}
  (2016) 065} [\href{https://arxiv.org/abs/1603.02653}{{\ttfamily
  1603.02653}}].

\bibitem{Sugimoto:1999tx}
S.~Sugimoto, \emph{{Anomaly cancellations in type I D-9 - anti-D-9 system and
  the USp(32) string theory}},
  \href{https://doi.org/10.1143/PTP.102.685}{\emph{Prog. Theor. Phys.}
  {\bfseries 102} (1999) 685}
  [\href{https://arxiv.org/abs/hep-th/9905159}{{\ttfamily hep-th/9905159}}].

\bibitem{Antoniadis:1999xk}
I.~Antoniadis, E.~Dudas and A.~Sagnotti, \emph{{Brane supersymmetry breaking}},
  \href{https://doi.org/10.1016/S0370-2693(99)01023-0}{\emph{Phys. Lett. B}
  {\bfseries 464} (1999) 38}
  [\href{https://arxiv.org/abs/hep-th/9908023}{{\ttfamily hep-th/9908023}}].

\bibitem{Angelantonj:1999jh}
C.~Angelantonj, \emph{{Comments on open string orbifolds with a nonvanishing
  B(ab)}}, \href{https://doi.org/10.1016/S0550-3213(99)00662-8}{\emph{Nucl.
  Phys. B} {\bfseries 566} (2000) 126}
  [\href{https://arxiv.org/abs/hep-th/9908064}{{\ttfamily hep-th/9908064}}].

\bibitem{Aldazabal:1999jr}
G.~Aldazabal and A.~M. Uranga, \emph{{Tachyon free nonsupersymmetric type IIB
  orientifolds via Brane - anti-brane systems}},
  \href{https://doi.org/10.1088/1126-6708/1999/10/024}{\emph{JHEP} {\bfseries
  10} (1999) 024} [\href{https://arxiv.org/abs/hep-th/9908072}{{\ttfamily
  hep-th/9908072}}].

\bibitem{Angelantonj:1999ms}
C.~Angelantonj, I.~Antoniadis, G.~D'Appollonio, E.~Dudas and A.~Sagnotti,
  \emph{{Type I vacua with brane supersymmetry breaking}},
  \href{https://doi.org/10.1016/S0550-3213(00)00052-3}{\emph{Nucl. Phys. B}
  {\bfseries 572} (2000) 36}
  [\href{https://arxiv.org/abs/hep-th/9911081}{{\ttfamily hep-th/9911081}}].

\bibitem{Dudas:2000nv}
E.~Dudas and J.~Mourad, \emph{{Consistent gravitino couplings in
  nonsupersymmetric strings}},
  \href{https://doi.org/10.1016/S0370-2693(01)00777-8}{\emph{Phys. Lett. B}
  {\bfseries 514} (2001) 173}
  [\href{https://arxiv.org/abs/hep-th/0012071}{{\ttfamily hep-th/0012071}}].

\bibitem{Pradisi:2001yv}
G.~Pradisi and F.~Riccioni, \emph{{Geometric couplings and brane supersymmetry
  breaking}}, \href{https://doi.org/10.1016/S0550-3213(01)00441-2}{\emph{Nucl.
  Phys. B} {\bfseries 615} (2001) 33}
  [\href{https://arxiv.org/abs/hep-th/0107090}{{\ttfamily hep-th/0107090}}].

\bibitem{Villadoro:2006ia}
G.~Villadoro and F.~Zwirner, \emph{{D terms from D-branes, gauge invariance and
  moduli stabilization in flux compactifications}},
  \href{https://doi.org/10.1088/1126-6708/2006/03/087}{\emph{JHEP} {\bfseries
  03} (2006) 087} [\href{https://arxiv.org/abs/hep-th/0602120}{{\ttfamily
  hep-th/0602120}}].

\bibitem{Blumenhagen:2002wn}
R.~Blumenhagen, V.~Braun, B.~Kors and D.~Lust, \emph{{Orientifolds of K3 and
  Calabi-Yau manifolds with intersecting D-branes}},
  \href{https://doi.org/10.1088/1126-6708/2002/07/026}{\emph{JHEP} {\bfseries
  07} (2002) 026} [\href{https://arxiv.org/abs/hep-th/0206038}{{\ttfamily
  hep-th/0206038}}].

\bibitem{Kachru:1999vj}
S.~Kachru and J.~McGreevy, \emph{{Supersymmetric three cycles and supersymmetry
  breaking}}, \href{https://doi.org/10.1103/PhysRevD.61.026001}{\emph{Phys.
  Rev. D} {\bfseries 61} (2000) 026001}
  [\href{https://arxiv.org/abs/hep-th/9908135}{{\ttfamily hep-th/9908135}}].

\bibitem{Cvetic:2001nr}
M.~Cvetic, G.~Shiu and A.~M. Uranga, \emph{{Chiral four-dimensional N=1
  supersymmetric type 2A orientifolds from intersecting D6 branes}},
  \href{https://doi.org/10.1016/S0550-3213(01)00427-8}{\emph{Nucl. Phys. B}
  {\bfseries 615} (2001) 3}
  [\href{https://arxiv.org/abs/hep-th/0107166}{{\ttfamily hep-th/0107166}}].

\bibitem{GarciadelMoral:2017vnz}
M.~P. Garcia~del Moral, S.~Parameswaran, N.~Quiroz and I.~Zavala,
  \emph{{Anti-D3 branes and moduli in non-linear supergravity}},
  \href{https://doi.org/10.1007/JHEP10(2017)185}{\emph{JHEP} {\bfseries 10}
  (2017) 185} [\href{https://arxiv.org/abs/1707.07059}{{\ttfamily
  1707.07059}}].

\bibitem{Kachru:2003sx}
S.~Kachru, R.~Kallosh, A.~D. Linde, J.~M. Maldacena, L.~P. McAllister and S.~P.
  Trivedi, \emph{{Towards inflation in string theory}},
  \href{https://doi.org/10.1088/1475-7516/2003/10/013}{\emph{JCAP} {\bfseries
  0310} (2003) 013} [\href{https://arxiv.org/abs/hep-th/0308055}{{\ttfamily
  hep-th/0308055}}].

\bibitem{Kallosh:2019zgd}
R.~Kallosh and A.~Linde, \emph{{Mass Production of Type IIA dS Vacua}},
  \href{https://doi.org/10.1007/JHEP01(2020)169}{\emph{JHEP} {\bfseries 01}
  (2020) 169} [\href{https://arxiv.org/abs/1910.08217}{{\ttfamily
  1910.08217}}].

\bibitem{Kallosh:2004yh}
R.~Kallosh and A.~D. Linde, \emph{{Landscape, the scale of SUSY breaking, and
  inflation}}, \href{https://doi.org/10.1088/1126-6708/2004/12/004}{\emph{JHEP}
  {\bfseries 12} (2004) 004}
  [\href{https://arxiv.org/abs/hep-th/0411011}{{\ttfamily hep-th/0411011}}].

\bibitem{DallAgata:2005zlf}
G.~Dall'Agata and N.~Prezas, \emph{{Scherk-Schwarz reduction of M-theory on
  G2-manifolds with fluxes}},
  \href{https://doi.org/10.1088/1126-6708/2005/10/103}{\emph{JHEP} {\bfseries
  10} (2005) 103} [\href{https://arxiv.org/abs/hep-th/0509052}{{\ttfamily
  hep-th/0509052}}].

\bibitem{Duff:2010vy}
M.~J. Duff and S.~Ferrara, \emph{{Four curious supergravities}},
  \href{https://doi.org/10.1103/PhysRevD.83.046007}{\emph{Phys. Rev. D}
  {\bfseries 83} (2011) 046007}
  [\href{https://arxiv.org/abs/1010.3173}{{\ttfamily 1010.3173}}].

\bibitem{Derendinger:2014wwa}
J.-P. Derendinger and A.~Guarino, \emph{{A second look at gauged supergravities
  from fluxes in M-theory}},
  \href{https://doi.org/10.1007/JHEP09(2014)162}{\emph{JHEP} {\bfseries 09}
  (2014) 162} [\href{https://arxiv.org/abs/1406.6930}{{\ttfamily 1406.6930}}].

\bibitem{Ferrara:2016fwe}
S.~Ferrara and R.~Kallosh, \emph{{Seven-disk manifold, $\alpha$-attractors, and
  $B$ modes}}, \href{https://doi.org/10.1103/PhysRevD.94.126015}{\emph{Phys.
  Rev.} {\bfseries D94} (2016) 126015}
  [\href{https://arxiv.org/abs/1610.04163}{{\ttfamily 1610.04163}}].

\bibitem{Ooguri:2018wrx}
H.~Ooguri, E.~Palti, G.~Shiu and C.~Vafa, \emph{{Distance and de Sitter
  Conjectures on the Swampland}},
  \href{https://doi.org/10.1016/j.physletb.2018.11.018}{\emph{Phys. Lett. B}
  {\bfseries 788} (2019) 180}
  [\href{https://arxiv.org/abs/1810.05506}{{\ttfamily 1810.05506}}].

\bibitem{Danielsson:2018ztv}
U.~H. Danielsson and T.~Van~Riet, \emph{{What if string theory has no de Sitter
  vacua?}}, \href{https://doi.org/10.1142/S0218271818300070}{\emph{Int. J. Mod.
  Phys. D} {\bfseries 27} (2018) 1830007}
  [\href{https://arxiv.org/abs/1804.01120}{{\ttfamily 1804.01120}}].

\bibitem{DeWolfe:2005uu}
O.~DeWolfe, A.~Giryavets, S.~Kachru and W.~Taylor, \emph{{Type IIA moduli
  stabilization}},
  \href{https://doi.org/10.1088/1126-6708/2005/07/066}{\emph{JHEP} {\bfseries
  07} (2005) 066} [\href{https://arxiv.org/abs/hep-th/0505160}{{\ttfamily
  hep-th/0505160}}].

\bibitem{Caviezel:2008tf}
C.~Caviezel, P.~Koerber, S.~Kors, D.~Lust, T.~Wrase and M.~Zagermann, \emph{{On
  the Cosmology of Type IIA Compactifications on SU(3)-structure Manifolds}},
  \href{https://doi.org/10.1088/1126-6708/2009/04/010}{\emph{JHEP} {\bfseries
  04} (2009) 010} [\href{https://arxiv.org/abs/0812.3551}{{\ttfamily
  0812.3551}}].

\bibitem{Flauger:2008ad}
R.~Flauger, S.~Paban, D.~Robbins and T.~Wrase, \emph{{Searching for slow-roll
  moduli inflation in massive type IIA supergravity with metric fluxes}},
  \href{https://doi.org/10.1103/PhysRevD.79.086011}{\emph{Phys. Rev.}
  {\bfseries D79} (2009) 086011}
  [\href{https://arxiv.org/abs/0812.3886}{{\ttfamily 0812.3886}}].

\bibitem{Denef:2018etk}
F.~Denef, A.~Hebecker and T.~Wrase, \emph{{de Sitter swampland conjecture and
  the Higgs potential}},
  \href{https://doi.org/10.1103/PhysRevD.98.086004}{\emph{Phys. Rev. D}
  {\bfseries 98} (2018) 086004}
  [\href{https://arxiv.org/abs/1807.06581}{{\ttfamily 1807.06581}}].

\bibitem{Volkov:1972jx}
D.~V. Volkov and V.~P. Akulov, \emph{{Possible universal neutrino
  interaction}}, {\emph{JETP Lett.} {\bfseries 16} (1972) 438}.

\bibitem{Kallosh:2016aep}
R.~Kallosh, B.~Vercnocke and T.~Wrase, \emph{{String Theory Origin of
  Constrained Multiplets}},
  \href{https://doi.org/10.1007/JHEP09(2016)063}{\emph{JHEP} {\bfseries 09}
  (2016) 063} [\href{https://arxiv.org/abs/1606.09245}{{\ttfamily
  1606.09245}}].

\bibitem{Vercnocke:2016fbt}
B.~Vercnocke and T.~Wrase, \emph{{Constrained superfields from an anti-D3-brane
  in KKLT}}, \href{https://doi.org/10.1007/JHEP08(2016)132}{\emph{JHEP}
  {\bfseries 08} (2016) 132}
  [\href{https://arxiv.org/abs/1605.03961}{{\ttfamily 1605.03961}}].

\bibitem{Kallosh:2018nrk}
R.~Kallosh and T.~Wrase, \emph{{dS Supergravity from 10d}},
  \href{https://doi.org/10.1002/prop.201800071}{\emph{Fortsch. Phys.}
  {\bfseries 2018} (2018) 1800071}
  [\href{https://arxiv.org/abs/1808.09427}{{\ttfamily 1808.09427}}].

\bibitem{Kallosh:2019apq}
R.~Kallosh and Y.~Yamada, \emph{{Simple sinflaton-less $\alpha$-attractors}},
  \href{https://doi.org/10.1007/JHEP03(2019)139}{\emph{JHEP} {\bfseries 03}
  (2019) 139} [\href{https://arxiv.org/abs/1901.09046}{{\ttfamily
  1901.09046}}].

\bibitem{Cribiori:2020zoh}
N.~Cribiori, \emph{{Constructing the Supersymmetric anti-D3-brane action in
  KKLT}}, \href{https://doi.org/10.22323/1.376.0137}{\emph{PoS} {\bfseries
  CORFU2019} (2020) 137} [\href{https://arxiv.org/abs/2003.09937}{{\ttfamily
  2003.09937}}].

\bibitem{Bergshoeff:2015jxa}
E.~A. Bergshoeff, K.~Dasgupta, R.~Kallosh, A.~Van~Proeyen and T.~Wrase,
  \emph{{$ \overline{\mathrm{D}3} $ and dS}},
  \href{https://doi.org/10.1007/JHEP05(2015)058}{\emph{JHEP} {\bfseries 05}
  (2015) 058} [\href{https://arxiv.org/abs/1502.07627}{{\ttfamily
  1502.07627}}].

\bibitem{Bandos:2016xyu}
I.~Bandos, M.~Heller, S.~M. Kuzenko, L.~Martucci and D.~Sorokin, \emph{{The
  Goldstino brane, the constrained superfields and matter in $ \mathcal{N}=1 $
  supergravity}}, \href{https://doi.org/10.1007/JHEP11(2016)109}{\emph{JHEP}
  {\bfseries 11} (2016) 109}
  [\href{https://arxiv.org/abs/1608.05908}{{\ttfamily 1608.05908}}].

\bibitem{DallAgata:2015zxp}
G.~Dall'Agata and F.~Farakos, \emph{{Constrained superfields in Supergravity}},
  \href{https://doi.org/10.1007/JHEP02(2016)101}{\emph{JHEP} {\bfseries 02}
  (2016) 101} [\href{https://arxiv.org/abs/1512.02158}{{\ttfamily
  1512.02158}}].

\bibitem{Cribiori:2018hxv}
N.~Cribiori, \emph{{Non-linear realisations in global and local
  supersymmetry}}, Ph.D. thesis, Padua U., 2018.
\newblock \href{https://arxiv.org/abs/1901.02097}{{\ttfamily 1901.02097}}.

\bibitem{Freedman:2012zz}
D.~Z. Freedman and A.~Van~Proeyen, \emph{{Supergravity}}. Cambridge Univ.
  Press, Cambridge, UK, 2012.

\bibitem{Kuzenko:2011tj}
S.~M. Kuzenko and S.~J. Tyler, \emph{{On the Goldstino actions and their
  symmetries}}, \href{https://doi.org/10.1007/JHEP05(2011)055}{\emph{JHEP}
  {\bfseries 05} (2011) 055} [\href{https://arxiv.org/abs/1102.3043}{{\ttfamily
  1102.3043}}].

\bibitem{Farakos:2013ih}
F.~Farakos and A.~Kehagias, \emph{{Decoupling Limits of sGoldstino Modes in
  Global and Local Supersymmetry}},
  \href{https://doi.org/10.1016/j.physletb.2013.06.001}{\emph{Phys. Lett. B}
  {\bfseries 724} (2013) 322}
  [\href{https://arxiv.org/abs/1302.0866}{{\ttfamily 1302.0866}}].

\bibitem{Dudas:2015eha}
E.~Dudas, S.~Ferrara, A.~Kehagias and A.~Sagnotti, \emph{{Properties of
  Nilpotent Supergravity}},
  \href{https://doi.org/10.1007/JHEP09(2015)217}{\emph{JHEP} {\bfseries 09}
  (2015) 217} [\href{https://arxiv.org/abs/1507.07842}{{\ttfamily
  1507.07842}}].

\bibitem{Bergshoeff:2015tra}
E.~A. Bergshoeff, D.~Z. Freedman, R.~Kallosh and A.~Van~Proeyen, \emph{{Pure de
  Sitter Supergravity}}, \href{https://doi.org/10.1103/PhysRevD.93.069901,
  10.1103/PhysRevD.92.085040}{\emph{Phys. Rev.} {\bfseries D92} (2015) 085040}
  [\href{https://arxiv.org/abs/1507.08264}{{\ttfamily 1507.08264}}].

\bibitem{Hasegawa:2015bza}
F.~Hasegawa and Y.~Yamada, \emph{{Component action of nilpotent multiplet
  coupled to matter in 4 dimensional $ \mathcal{N}=1 $ supergravity}},
  \href{https://doi.org/10.1007/JHEP10(2015)106}{\emph{JHEP} {\bfseries 10}
  (2015) 106} [\href{https://arxiv.org/abs/1507.08619}{{\ttfamily
  1507.08619}}].

\bibitem{Ferrara:2015gta}
S.~Ferrara, M.~Porrati and A.~Sagnotti, \emph{{Scale invariant
  Volkov\textendash{}Akulov supergravity}},
  \href{https://doi.org/10.1016/j.physletb.2015.08.066}{\emph{Phys. Lett. B}
  {\bfseries 749} (2015) 589}
  [\href{https://arxiv.org/abs/1508.02939}{{\ttfamily 1508.02939}}].

\bibitem{Brignole:1997pe}
A.~Brignole, F.~Feruglio and F.~Zwirner, \emph{{On the effective interactions
  of a light gravitino with matter fermions}},
  \href{https://doi.org/10.1088/1126-6708/1997/11/001}{\emph{JHEP} {\bfseries
  11} (1997) 001} [\href{https://arxiv.org/abs/hep-th/9709111}{{\ttfamily
  hep-th/9709111}}].

\bibitem{DallAgata:2015pdd}
G.~Dall'Agata, S.~Ferrara and F.~Zwirner, \emph{{Minimal scalar-less
  matter-coupled supergravity}},
  \href{https://doi.org/10.1016/j.physletb.2015.11.066}{\emph{Phys. Lett. B}
  {\bfseries 752} (2016) 263}
  [\href{https://arxiv.org/abs/1509.06345}{{\ttfamily 1509.06345}}].

\bibitem{Cribiori:2017laj}
N.~Cribiori, F.~Farakos, M.~Tournoy and A.~van Proeyen, \emph{{Fayet-Iliopoulos
  terms in supergravity without gauged R-symmetry}},
  \href{https://doi.org/10.1007/JHEP04(2018)032}{\emph{JHEP} {\bfseries 04}
  (2018) 032} [\href{https://arxiv.org/abs/1712.08601}{{\ttfamily
  1712.08601}}].

\bibitem{Cribiori:2018dlc}
N.~Cribiori, F.~Farakos and M.~Tournoy, \emph{{Supersymmetric Born-Infeld
  actions and new Fayet-Iliopoulos terms}},
  \href{https://doi.org/10.1007/JHEP03(2019)050}{\emph{JHEP} {\bfseries 03}
  (2019) 050} [\href{https://arxiv.org/abs/1811.08424}{{\ttfamily
  1811.08424}}].

\bibitem{Kuzenko:2018jlz}
S.~M. Kuzenko, \emph{{Taking a vector supermultiplet apart: Alternative
  Fayet\textendash{}Iliopoulos-type terms}},
  \href{https://doi.org/10.1016/j.physletb.2018.04.051}{\emph{Phys. Lett. B}
  {\bfseries 781} (2018) 723}
  [\href{https://arxiv.org/abs/1801.04794}{{\ttfamily 1801.04794}}].

\bibitem{Antoniadis:2018cpq}
I.~Antoniadis, A.~Chatrabhuti, H.~Isono and R.~Knoops,
  \emph{{Fayet\textendash{}Iliopoulos terms in supergravity and D-term
  inflation}}, \href{https://doi.org/10.1140/epjc/s10052-018-5861-6}{\emph{Eur.
  Phys. J. C} {\bfseries 78} (2018) 366}
  [\href{https://arxiv.org/abs/1803.03817}{{\ttfamily 1803.03817}}].

\bibitem{Fayet:1974jb}
P.~Fayet and J.~Iliopoulos, \emph{{Spontaneously Broken Supergauge Symmetries
  and Goldstone Spinors}},
  \href{https://doi.org/10.1016/0370-2693(74)90310-4}{\emph{Phys. Lett. B}
  {\bfseries 51} (1974) 461}.

\bibitem{Freedman:1976uk}
D.~Z. Freedman, \emph{{Supergravity with Axial Gauge Invariance}},
  \href{https://doi.org/10.1103/PhysRevD.15.1173}{\emph{Phys. Rev. D}
  {\bfseries 15} (1977) 1173}.

\bibitem{Antoniadis:2018oeh}
I.~Antoniadis, A.~Chatrabhuti, H.~Isono and R.~Knoops, \emph{{The cosmological
  constant in Supergravity}},
  \href{https://doi.org/10.1140/epjc/s10052-018-6175-4}{\emph{Eur. Phys. J. C}
  {\bfseries 78} (2018) 718}
  [\href{https://arxiv.org/abs/1805.00852}{{\ttfamily 1805.00852}}].

\bibitem{Mourad:2017rrl}
J.~Mourad and A.~Sagnotti, \emph{{An Update on Brane Supersymmetry Breaking}},
  \href{https://arxiv.org/abs/1711.11494}{{\ttfamily 1711.11494}}.

\bibitem{Lust:2008zd}
D.~Lust, F.~Marchesano, L.~Martucci and D.~Tsimpis, \emph{{Generalized
  non-supersymmetric flux vacua}},
  \href{https://doi.org/10.1088/1126-6708/2008/11/021}{\emph{JHEP} {\bfseries
  11} (2008) 021} [\href{https://arxiv.org/abs/0807.4540}{{\ttfamily
  0807.4540}}].

\bibitem{Polchinski:1998rr}
J.~Polchinski, \emph{{String theory. Vol. 2: Superstring theory and beyond}},
  Cambridge Monographs on Mathematical Physics. Cambridge University Press, 12,
  2007,
  \href{https://doi.org/10.1017/CBO9780511618123}{10.1017/CBO9780511618123}.

\bibitem{Cederwall:1996pv}
M.~Cederwall, A.~von Gussich, B.~E.~W. Nilsson and A.~Westerberg, \emph{{The
  Dirichlet super three-brane in ten-dimensional type IIB supergravity}},
  \href{https://doi.org/10.1016/S0550-3213(97)00071-0}{\emph{Nucl. Phys. B}
  {\bfseries 490} (1997) 163}
  [\href{https://arxiv.org/abs/hep-th/9610148}{{\ttfamily hep-th/9610148}}].

\bibitem{Aganagic:1996pe}
M.~Aganagic, C.~Popescu and J.~H. Schwarz, \emph{{D-brane actions with local
  kappa symmetry}},
  \href{https://doi.org/10.1016/S0370-2693(96)01643-7}{\emph{Phys. Lett. B}
  {\bfseries 393} (1997) 311}
  [\href{https://arxiv.org/abs/hep-th/9610249}{{\ttfamily hep-th/9610249}}].

\bibitem{Cederwall:1996ri}
M.~Cederwall, A.~von Gussich, B.~E.~W. Nilsson, P.~Sundell and A.~Westerberg,
  \emph{{The Dirichlet super p-branes in ten-dimensional type IIA and IIB
  supergravity}},
  \href{https://doi.org/10.1016/S0550-3213(97)00075-8}{\emph{Nucl. Phys. B}
  {\bfseries 490} (1997) 179}
  [\href{https://arxiv.org/abs/hep-th/9611159}{{\ttfamily hep-th/9611159}}].

\bibitem{Bergshoeff:1996tu}
E.~Bergshoeff and P.~K. Townsend, \emph{{Super D-branes}},
  \href{https://doi.org/10.1016/S0550-3213(97)00072-2}{\emph{Nucl. Phys. B}
  {\bfseries 490} (1997) 145}
  [\href{https://arxiv.org/abs/hep-th/9611173}{{\ttfamily hep-th/9611173}}].

\bibitem{Aganagic:1996nn}
M.~Aganagic, C.~Popescu and J.~H. Schwarz, \emph{{Gauge invariant and gauge
  fixed D-brane actions}},
  \href{https://doi.org/10.1016/S0550-3213(97)00180-6}{\emph{Nucl. Phys. B}
  {\bfseries 495} (1997) 99}
  [\href{https://arxiv.org/abs/hep-th/9612080}{{\ttfamily hep-th/9612080}}].

\bibitem{Johnson:2003glb}
C.~V. Johnson, \emph{{D-branes}}, Cambridge Monographs on Mathematical Physics.
  Cambridge University Press, 2005,
  \href{https://doi.org/10.1017/CBO9780511606540}{10.1017/CBO9780511606540}.

\bibitem{Koerber:2010bx}
P.~Koerber, \emph{{Lectures on Generalized Complex Geometry for Physicists}},
  \href{https://doi.org/10.1002/prop.201000083}{\emph{Fortsch. Phys.}
  {\bfseries 59} (2011) 169} [\href{https://arxiv.org/abs/1006.1536}{{\ttfamily
  1006.1536}}].

\bibitem{Antoniadis:2014oya}
I.~Antoniadis, E.~Dudas, S.~Ferrara and A.~Sagnotti, \emph{{The
  Volkov\textendash{}Akulov\textendash{}Starobinsky supergravity}},
  \href{https://doi.org/10.1016/j.physletb.2014.04.015}{\emph{Phys. Lett. B}
  {\bfseries 733} (2014) 32} [\href{https://arxiv.org/abs/1403.3269}{{\ttfamily
  1403.3269}}].

\bibitem{Kallosh:2015sea}
R.~Kallosh, \emph{{Matter-coupled de Sitter Supergravity}},
  \href{https://doi.org/10.1134/S0040577916050068}{\emph{Theor. Math. Phys.}
  {\bfseries 187} (2016) 695}
  [\href{https://arxiv.org/abs/1509.02136}{{\ttfamily 1509.02136}}].

\bibitem{Schillo:2015ssx}
M.~Schillo, E.~van~der Woerd and T.~Wrase, \emph{{The general de Sitter
  supergravity component action}},
  \href{https://doi.org/10.1002/prop201500074}{\emph{Fortsch. Phys.} {\bfseries
  64} (2016) 292} [\href{https://arxiv.org/abs/1511.01542}{{\ttfamily
  1511.01542}}].

\bibitem{Volkov:1973ix}
D.~V. Volkov and V.~P. Akulov, \emph{{Is the Neutrino a Goldstone Particle?}},
  \href{https://doi.org/10.1016/0370-2693(73)90490-5}{\emph{Phys. Lett. B}
  {\bfseries 46} (1973) 109}.

\bibitem{Casalbuoni:1988sx}
R.~Casalbuoni, S.~De~Curtis, D.~Dominici, F.~Feruglio and R.~Gatto, \emph{{WHEN
  DOES SUPERGRAVITY BECOME STRONG?}},
  \href{https://doi.org/10.1016/0370-2693(89)91123-4}{\emph{Phys. Lett. B}
  {\bfseries 216} (1989) 325}.

\bibitem{Kallosh:2000ve}
R.~Kallosh, L.~Kofman, A.~D. Linde and A.~Van~Proeyen, \emph{{Superconformal
  symmetry, supergravity and cosmology}},
  \href{https://doi.org/10.1088/0264-9381/17/20/308}{\emph{Class. Quant. Grav.}
  {\bfseries 17} (2000) 4269}
  [\href{https://arxiv.org/abs/hep-th/0006179}{{\ttfamily hep-th/0006179}}].

\bibitem{DallAgata:2014qsj}
G.~Dall'Agata and F.~Zwirner, \emph{{On sgoldstino-less supergravity models of
  inflation}}, \href{https://doi.org/10.1007/JHEP12(2014)172}{\emph{JHEP}
  {\bfseries 12} (2014) 172} [\href{https://arxiv.org/abs/1411.2605}{{\ttfamily
  1411.2605}}].

\bibitem{Ferrara:2015tyn}
S.~Ferrara, R.~Kallosh and J.~Thaler, \emph{{Cosmology with orthogonal
  nilpotent superfields}},
  \href{https://doi.org/10.1103/PhysRevD.93.043516}{\emph{Phys. Rev.}
  {\bfseries D93} (2016) 043516}
  [\href{https://arxiv.org/abs/1512.00545}{{\ttfamily 1512.00545}}].

\bibitem{Carrasco:2015iij}
J.~J.~M. Carrasco, R.~Kallosh and A.~Linde, \emph{{Minimal supergravity
  inflation}}, \href{https://doi.org/10.1103/PhysRevD.93.061301}{\emph{Phys.
  Rev.} {\bfseries D93} (2016) 061301}
  [\href{https://arxiv.org/abs/1512.00546}{{\ttfamily 1512.00546}}].

\bibitem{Klebanov:2000hb}
I.~R. Klebanov and M.~J. Strassler, \emph{{Supergravity and a confining gauge
  theory: Duality cascades and chi SB resolution of naked singularities}},
  \href{https://doi.org/10.1088/1126-6708/2000/08/052}{\emph{JHEP} {\bfseries
  08} (2000) 052} [\href{https://arxiv.org/abs/hep-th/0007191}{{\ttfamily
  hep-th/0007191}}].

\bibitem{Kachru:2002gs}
S.~Kachru, J.~Pearson and H.~L. Verlinde, \emph{{Brane / flux annihilation and
  the string dual of a nonsupersymmetric field theory}},
  \href{https://doi.org/10.1088/1126-6708/2002/06/021}{\emph{JHEP} {\bfseries
  06} (2002) 021} [\href{https://arxiv.org/abs/hep-th/0112197}{{\ttfamily
  hep-th/0112197}}].

\bibitem{Aalsma:2018pll}
L.~Aalsma, M.~Tournoy, J.~P. Van Der~Schaar and B.~Vercnocke,
  \emph{{Supersymmetric embedding of antibrane polarization}},
  \href{https://doi.org/10.1103/PhysRevD.98.086019}{\emph{Phys. Rev.}
  {\bfseries D98} (2018) 086019}
  [\href{https://arxiv.org/abs/1807.03303}{{\ttfamily 1807.03303}}].

\bibitem{Berkooz:1996km}
M.~Berkooz, M.~R. Douglas and R.~G. Leigh, \emph{{Branes intersecting at
  angles}}, \href{https://doi.org/10.1016/S0550-3213(96)00452-X}{\emph{Nucl.
  Phys. B} {\bfseries 480} (1996) 265}
  [\href{https://arxiv.org/abs/hep-th/9606139}{{\ttfamily hep-th/9606139}}].

\bibitem{Anastasopoulos:2011hj}
P.~Anastasopoulos, M.~Bianchi and R.~Richter, \emph{{Light stringy states}},
  \href{https://doi.org/10.1007/JHEP03(2012)068}{\emph{JHEP} {\bfseries 03}
  (2012) 068} [\href{https://arxiv.org/abs/1110.5424}{{\ttfamily 1110.5424}}].

\bibitem{Anastasopoulos:2016cmg}
P.~Anastasopoulos and M.~Bianchi, \emph{{Revisiting light stringy states in
  view of the 750 GeV diphoton excess}},
  \href{https://doi.org/10.1016/j.nuclphysb.2016.08.033}{\emph{Nucl. Phys. B}
  {\bfseries 911} (2016) 928}
  [\href{https://arxiv.org/abs/1601.07584}{{\ttfamily 1601.07584}}].

\bibitem{Grana:2005jc}
M.~Grana, \emph{{Flux compactifications in string theory: A Comprehensive
  review}}, \href{https://doi.org/10.1016/j.physrep.2005.10.008}{\emph{Phys.
  Rept.} {\bfseries 423} (2006) 91}
  [\href{https://arxiv.org/abs/hep-th/0509003}{{\ttfamily hep-th/0509003}}].

\bibitem{Douglas:2006es}
M.~R. Douglas and S.~Kachru, \emph{{Flux compactification}},
  \href{https://doi.org/10.1103/RevModPhys.79.733}{\emph{Rev. Mod. Phys.}
  {\bfseries 79} (2007) 733}
  [\href{https://arxiv.org/abs/hep-th/0610102}{{\ttfamily hep-th/0610102}}].

\bibitem{Blumenhagen:2006ci}
R.~Blumenhagen, B.~Kors, D.~Lust and S.~Stieberger, \emph{{Four-dimensional
  String Compactifications with D-Branes, Orientifolds and Fluxes}},
  \href{https://doi.org/10.1016/j.physrep.2007.04.003}{\emph{Phys. Rept.}
  {\bfseries 445} (2007) 1}
  [\href{https://arxiv.org/abs/hep-th/0610327}{{\ttfamily hep-th/0610327}}].

\bibitem{Blumenhagen:2005mu}
R.~Blumenhagen, M.~Cvetic, P.~Langacker and G.~Shiu, \emph{{Toward realistic
  intersecting D-brane models}},
  \href{https://doi.org/10.1146/annurev.nucl.55.090704.151541}{\emph{Ann. Rev.
  Nucl. Part. Sci.} {\bfseries 55} (2005) 71}
  [\href{https://arxiv.org/abs/hep-th/0502005}{{\ttfamily hep-th/0502005}}].

\bibitem{Aalsma:2017ulu}
L.~Aalsma, J.~P. van~der Schaar and B.~Vercnocke, \emph{{Constrained
  superfields on metastable anti-D3-branes}},
  \href{https://doi.org/10.1007/JHEP05(2017)089}{\emph{JHEP} {\bfseries 05}
  (2017) 089} [\href{https://arxiv.org/abs/1703.05771}{{\ttfamily
  1703.05771}}].

\bibitem{Martucci:2005ht}
L.~Martucci and P.~Smyth, \emph{{Supersymmetric D-branes and calibrations on
  general N=1 backgrounds}},
  \href{https://doi.org/10.1088/1126-6708/2005/11/048}{\emph{JHEP} {\bfseries
  11} (2005) 048} [\href{https://arxiv.org/abs/hep-th/0507099}{{\ttfamily
  hep-th/0507099}}].

\bibitem{Martucci:2006ij}
L.~Martucci, \emph{{D-branes on general N=1 backgrounds: Superpotentials and
  D-terms}}, \href{https://doi.org/10.1088/1126-6708/2006/06/033}{\emph{JHEP}
  {\bfseries 06} (2006) 033}
  [\href{https://arxiv.org/abs/hep-th/0602129}{{\ttfamily hep-th/0602129}}].

\bibitem{Martucci:2011dn}
L.~Martucci, \emph{{Electrified branes}},
  \href{https://doi.org/10.1007/JHEP02(2012)097}{\emph{JHEP} {\bfseries 02}
  (2012) 097} [\href{https://arxiv.org/abs/1110.0627}{{\ttfamily 1110.0627}}].

\bibitem{Kallosh:2014wsa}
R.~Kallosh and T.~Wrase, \emph{{Emergence of Spontaneously Broken Supersymmetry
  on an Anti-D3-Brane in KKLT dS Vacua}},
  \href{https://doi.org/10.1007/JHEP12(2014)117}{\emph{JHEP} {\bfseries 12}
  (2014) 117} [\href{https://arxiv.org/abs/1411.1121}{{\ttfamily 1411.1121}}].

\bibitem{Grimm:2004ua}
T.~W. Grimm and J.~Louis, \emph{{The Effective action of type IIA Calabi-Yau
  orientifolds}},
  \href{https://doi.org/10.1016/j.nuclphysb.2005.04.007}{\emph{Nucl. Phys.}
  {\bfseries B718} (2005) 153}
  [\href{https://arxiv.org/abs/hep-th/0412277}{{\ttfamily hep-th/0412277}}].

\bibitem{Ferrara:2014kva}
S.~Ferrara, R.~Kallosh and A.~Linde, \emph{{Cosmology with Nilpotent
  Superfields}}, \href{https://doi.org/10.1007/JHEP10(2014)143}{\emph{JHEP}
  {\bfseries 10} (2014) 143} [\href{https://arxiv.org/abs/1408.4096}{{\ttfamily
  1408.4096}}].

\bibitem{Kallosh:2015nia}
R.~Kallosh, F.~Quevedo and A.~M. Uranga, \emph{{String Theory Realizations of
  the Nilpotent Goldstino}},
  \href{https://doi.org/10.1007/JHEP12(2015)039}{\emph{JHEP} {\bfseries 12}
  (2015) 039} [\href{https://arxiv.org/abs/1507.07556}{{\ttfamily
  1507.07556}}].

\bibitem{Garcia-Etxebarria:2015lif}
I.~n. Garc\'\i{}a-Etxebarria, F.~Quevedo and R.~Valandro, \emph{{Global String
  Embeddings for the Nilpotent Goldstino}},
  \href{https://doi.org/10.1007/JHEP02(2016)148}{\emph{JHEP} {\bfseries 02}
  (2016) 148} [\href{https://arxiv.org/abs/1512.06926}{{\ttfamily
  1512.06926}}].

\bibitem{Grana:2002tu}
M.~Grana, \emph{{D3-brane action in a supergravity background: The Fermionic
  story}}, \href{https://doi.org/10.1103/PhysRevD.66.045014}{\emph{Phys. Rev.
  D} {\bfseries 66} (2002) 045014}
  [\href{https://arxiv.org/abs/hep-th/0202118}{{\ttfamily hep-th/0202118}}].

\bibitem{Grana:2003ek}
M.~Grana, T.~W. Grimm, H.~Jockers and J.~Louis, \emph{{Soft supersymmetry
  breaking in Calabi-Yau orientifolds with D-branes and fluxes}},
  \href{https://doi.org/10.1016/j.nuclphysb.2004.04.021}{\emph{Nucl. Phys. B}
  {\bfseries 690} (2004) 21}
  [\href{https://arxiv.org/abs/hep-th/0312232}{{\ttfamily hep-th/0312232}}].

\bibitem{Marolf:2003ye}
D.~Marolf, L.~Martucci and P.~J. Silva, \emph{{Fermions, T duality and
  effective actions for D-branes in bosonic backgrounds}},
  \href{https://doi.org/10.1088/1126-6708/2003/04/051}{\emph{JHEP} {\bfseries
  04} (2003) 051} [\href{https://arxiv.org/abs/hep-th/0303209}{{\ttfamily
  hep-th/0303209}}].

\bibitem{Tripathy:2005hv}
P.~K. Tripathy and S.~P. Trivedi, \emph{{D3 brane action and fermion zero modes
  in presence of background flux}},
  \href{https://doi.org/10.1088/1126-6708/2005/06/066}{\emph{JHEP} {\bfseries
  06} (2005) 066} [\href{https://arxiv.org/abs/hep-th/0503072}{{\ttfamily
  hep-th/0503072}}].

\bibitem{Martucci:2005rb}
L.~Martucci, J.~Rosseel, D.~Van~den Bleeken and A.~Van~Proeyen, \emph{{Dirac
  actions for D-branes on backgrounds with fluxes}},
  \href{https://doi.org/10.1088/0264-9381/22/13/014}{\emph{Class. Quant. Grav.}
  {\bfseries 22} (2005) 2745}
  [\href{https://arxiv.org/abs/hep-th/0504041}{{\ttfamily hep-th/0504041}}].

\bibitem{Bergshoeff:2013pia}
E.~Bergshoeff, F.~Coomans, R.~Kallosh, C.~S. Shahbazi and A.~Van~Proeyen,
  \emph{{Dirac-Born-Infeld-Volkov-Akulov and Deformation of Supersymmetry}},
  \href{https://doi.org/10.1007/JHEP08(2013)100}{\emph{JHEP} {\bfseries 08}
  (2013) 100} [\href{https://arxiv.org/abs/1303.5662}{{\ttfamily 1303.5662}}].

\bibitem{Dasgupta:2016prs}
K.~Dasgupta, M.~Emelin and E.~McDonough, \emph{{Fermions on the antibrane:
  Higher order interactions and spontaneously broken supersymmetry}},
  \href{https://doi.org/10.1103/PhysRevD.95.026003}{\emph{Phys. Rev.}
  {\bfseries D95} (2017) 026003}
  [\href{https://arxiv.org/abs/1601.03409}{{\ttfamily 1601.03409}}].

\bibitem{Giddings:2001yu}
S.~B. Giddings, S.~Kachru and J.~Polchinski, \emph{{Hierarchies from fluxes in
  string compactifications}},
  \href{https://doi.org/10.1103/PhysRevD.66.106006}{\emph{Phys. Rev. D}
  {\bfseries 66} (2002) 106006}
  [\href{https://arxiv.org/abs/hep-th/0105097}{{\ttfamily hep-th/0105097}}].

\bibitem{Frey:2008xw}
A.~R. Frey, G.~Torroba, B.~Underwood and M.~R. Douglas, \emph{{The Universal
  Kahler Modulus in Warped Compactifications}},
  \href{https://doi.org/10.1088/1126-6708/2009/01/036}{\emph{JHEP} {\bfseries
  01} (2009) 036} [\href{https://arxiv.org/abs/0810.5768}{{\ttfamily
  0810.5768}}].

\bibitem{deAlwis:2016cty}
S.~P. de~Alwis, \emph{{Constraints on Dbar Uplifts}},
  \href{https://doi.org/10.1007/JHEP11(2016)045}{\emph{JHEP} {\bfseries 11}
  (2016) 045} [\href{https://arxiv.org/abs/1605.06456}{{\ttfamily
  1605.06456}}].

\bibitem{McGuirk:2012sb}
P.~McGuirk, G.~Shiu and F.~Ye, \emph{{Soft branes in supersymmetry-breaking
  backgrounds}}, \href{https://doi.org/10.1007/JHEP07(2012)188}{\emph{JHEP}
  {\bfseries 07} (2012) 188} [\href{https://arxiv.org/abs/1206.0754}{{\ttfamily
  1206.0754}}].

\bibitem{DeWolfe:2002nn}
O.~DeWolfe and S.~B. Giddings, \emph{{Scales and hierarchies in warped
  compactifications and brane worlds}},
  \href{https://doi.org/10.1103/PhysRevD.67.066008}{\emph{Phys. Rev. D}
  {\bfseries 67} (2003) 066008}
  [\href{https://arxiv.org/abs/hep-th/0208123}{{\ttfamily hep-th/0208123}}].

\bibitem{Baumann:2007ah}
D.~Baumann, A.~Dymarsky, I.~R. Klebanov and L.~McAllister, \emph{{Towards an
  Explicit Model of D-brane Inflation}},
  \href{https://doi.org/10.1088/1475-7516/2008/01/024}{\emph{JCAP} {\bfseries
  0801} (2008) 024} [\href{https://arxiv.org/abs/0706.0360}{{\ttfamily
  0706.0360}}].

\bibitem{Moritz:2017xto}
J.~Moritz, A.~Retolaza and A.~Westphal, \emph{{Toward de Sitter space from ten
  dimensions}}, \href{https://doi.org/10.1103/PhysRevD.97.046010}{\emph{Phys.
  Rev. D} {\bfseries 97} (2018) 046010}
  [\href{https://arxiv.org/abs/1707.08678}{{\ttfamily 1707.08678}}].

\bibitem{Moritz:2018sui}
J.~Moritz and T.~Van~Riet, \emph{{Racing through the swampland: de Sitter
  uplift vs weak gravity}},
  \href{https://doi.org/10.1007/JHEP09(2018)099}{\emph{JHEP} {\bfseries 09}
  (2018) 099} [\href{https://arxiv.org/abs/1805.00944}{{\ttfamily
  1805.00944}}].

\bibitem{Kallosh:2018wme}
R.~Kallosh, A.~Linde, E.~McDonough and M.~Scalisi, \emph{{de Sitter Vacua with
  a Nilpotent Superfield}},
  \href{https://doi.org/10.1002/prop.201800068}{\emph{Fortsch. Phys.}
  {\bfseries 67} (2019) 1800068}
  [\href{https://arxiv.org/abs/1808.09428}{{\ttfamily 1808.09428}}].

\bibitem{Moritz:2018ani}
J.~Moritz, A.~Retolaza and A.~Westphal, \emph{{On uplifts by warped
  anti-D3-branes}},
  \href{https://doi.org/10.1002/prop.201800098}{\emph{Fortsch. Phys.}
  {\bfseries 67} (2019) 1800098}
  [\href{https://arxiv.org/abs/1809.06618}{{\ttfamily 1809.06618}}].

\bibitem{Kallosh:2018psh}
R.~Kallosh, A.~Linde, E.~McDonough and M.~Scalisi, \emph{{4D models of de
  Sitter uplift}},
  \href{https://doi.org/10.1103/PhysRevD.99.046006}{\emph{Phys. Rev. D}
  {\bfseries 99} (2019) 046006}
  [\href{https://arxiv.org/abs/1809.09018}{{\ttfamily 1809.09018}}].

\bibitem{Gautason:2018gln}
F.~Gautason, V.~Van~Hemelryck and T.~Van~Riet, \emph{{The Tension between 10D
  Supergravity and dS Uplifts}},
  \href{https://doi.org/10.1002/prop.201800091}{\emph{Fortsch. Phys.}
  {\bfseries 67} (2019) 1800091}
  [\href{https://arxiv.org/abs/1810.08518}{{\ttfamily 1810.08518}}].

\bibitem{Hamada:2018qef}
Y.~Hamada, A.~Hebecker, G.~Shiu and P.~Soler, \emph{{On brane gaugino
  condensates in 10d}},
  \href{https://doi.org/10.1007/JHEP04(2019)008}{\emph{JHEP} {\bfseries 04}
  (2019) 008} [\href{https://arxiv.org/abs/1812.06097}{{\ttfamily
  1812.06097}}].

\bibitem{Kallosh:2019axr}
R.~Kallosh, A.~Linde, E.~McDonough and M.~Scalisi, \emph{{dS Vacua and the
  Swampland}}, \href{https://doi.org/10.1007/JHEP03(2019)134}{\emph{JHEP}
  {\bfseries 03} (2019) 134}
  [\href{https://arxiv.org/abs/1901.02022}{{\ttfamily 1901.02022}}].

\bibitem{Kallosh:2019oxv}
R.~Kallosh, \emph{{Gaugino Condensation and Geometry of the Perfect Square}},
  \href{https://doi.org/10.1103/PhysRevD.99.066003}{\emph{Phys. Rev. D}
  {\bfseries 99} (2019) 066003}
  [\href{https://arxiv.org/abs/1901.02023}{{\ttfamily 1901.02023}}].

\bibitem{Hamada:2019ack}
Y.~Hamada, A.~Hebecker, G.~Shiu and P.~Soler, \emph{{Understanding KKLT from a
  10d perspective}}, \href{https://doi.org/10.1007/JHEP06(2019)019}{\emph{JHEP}
  {\bfseries 06} (2019) 019}
  [\href{https://arxiv.org/abs/1902.01410}{{\ttfamily 1902.01410}}].

\bibitem{Carta:2019rhx}
F.~Carta, J.~Moritz and A.~Westphal, \emph{{Gaugino condensation and small
  uplifts in KKLT}}, \href{https://doi.org/10.1007/JHEP08(2019)141}{\emph{JHEP}
  {\bfseries 08} (2019) 141}
  [\href{https://arxiv.org/abs/1902.01412}{{\ttfamily 1902.01412}}].

\bibitem{Gautason:2019jwq}
F.~F. Gautason, V.~Van~Hemelryck, T.~Van~Riet and G.~Venken, \emph{{A 10d view
  on the KKLT AdS vacuum and uplifting}},
  \href{https://doi.org/10.1007/JHEP06(2020)074}{\emph{JHEP} {\bfseries 06}
  (2020) 074} [\href{https://arxiv.org/abs/1902.01415}{{\ttfamily
  1902.01415}}].

\bibitem{Kachru:2019dvo}
S.~Kachru, M.~Kim, L.~Mcallister and M.~Zimet, \emph{{de Sitter Vacua from Ten
  Dimensions}},  \href{https://arxiv.org/abs/1908.04788}{{\ttfamily
  1908.04788}}.

\bibitem{Gukov:1999ya}
S.~Gukov, C.~Vafa and E.~Witten, \emph{{CFT's from Calabi-Yau four folds}},
  \href{https://doi.org/10.1016/S0550-3213(00)00373-4}{\emph{Nucl. Phys. B}
  {\bfseries 584} (2000) 69}
  [\href{https://arxiv.org/abs/hep-th/9906070}{{\ttfamily hep-th/9906070}}].

\bibitem{Baumann:2010sx}
D.~Baumann, A.~Dymarsky, S.~Kachru, I.~R. Klebanov and L.~McAllister,
  \emph{{D3-brane Potentials from Fluxes in AdS/CFT}},
  \href{https://doi.org/10.1007/JHEP06(2010)072}{\emph{JHEP} {\bfseries 06}
  (2010) 072} [\href{https://arxiv.org/abs/1001.5028}{{\ttfamily 1001.5028}}].

\bibitem{Dymarsky:2010mf}
A.~Dymarsky and L.~Martucci, \emph{{D-brane non-perturbative effects and
  geometric deformations}},
  \href{https://doi.org/10.1007/JHEP04(2011)061}{\emph{JHEP} {\bfseries 04}
  (2011) 061} [\href{https://arxiv.org/abs/1012.4018}{{\ttfamily 1012.4018}}].

\bibitem{Bergshoeff:2005yp}
E.~Bergshoeff, R.~Kallosh, A.-K. Kashani-Poor, D.~Sorokin and A.~Tomasiello,
  \emph{{An Index for the Dirac operator on D3 branes with background fluxes}},
  \href{https://doi.org/10.1088/1126-6708/2005/10/102}{\emph{JHEP} {\bfseries
  10} (2005) 102} [\href{https://arxiv.org/abs/hep-th/0507069}{{\ttfamily
  hep-th/0507069}}].

\bibitem{Parameswaran:2020ukp}
S.~Parameswaran and F.~Tonioni, \emph{{Non-supersymmetric String Models from
  Anti-D3-/D7-branes in Strongly Warped Throats}},
  \href{https://arxiv.org/abs/2007.11333}{{\ttfamily 2007.11333}}.

\bibitem{Balasubramanian:2005zx}
V.~Balasubramanian, P.~Berglund, J.~P. Conlon and F.~Quevedo,
  \emph{{Systematics of moduli stabilisation in Calabi-Yau flux
  compactifications}},
  \href{https://doi.org/10.1088/1126-6708/2005/03/007}{\emph{JHEP} {\bfseries
  03} (2005) 007} [\href{https://arxiv.org/abs/hep-th/0502058}{{\ttfamily
  hep-th/0502058}}].

\bibitem{Conlon:2005ki}
J.~P. Conlon, F.~Quevedo and K.~Suruliz, \emph{{Large-volume flux
  compactifications: Moduli spectrum and D3/D7 soft supersymmetry breaking}},
  \href{https://doi.org/10.1088/1126-6708/2005/08/007}{\emph{JHEP} {\bfseries
  08} (2005) 007} [\href{https://arxiv.org/abs/hep-th/0505076}{{\ttfamily
  hep-th/0505076}}].

\bibitem{Andriot:2019wrs}
D.~Andriot, \emph{{Open problems on classical de Sitter solutions}},
  \href{https://doi.org/10.1002/prop.201900026}{\emph{Fortsch. Phys.}
  {\bfseries 67} (2019) 1900026}
  [\href{https://arxiv.org/abs/1902.10093}{{\ttfamily 1902.10093}}].

\bibitem{Bena:2019mte}
I.~Bena, M.~Gra\~na, N.~Kovensky and A.~Retolaza, \emph{{K\"ahler moduli
  stabilization from ten dimensions}},
  \href{https://doi.org/10.1007/JHEP10(2019)200}{\emph{JHEP} {\bfseries 10}
  (2019) 200} [\href{https://arxiv.org/abs/1908.01785}{{\ttfamily
  1908.01785}}].

\bibitem{Koerber:2007xk}
P.~Koerber and L.~Martucci, \emph{{From ten to four and back again: How to
  generalize the geometry}},
  \href{https://doi.org/10.1088/1126-6708/2007/08/059}{\emph{JHEP} {\bfseries
  08} (2007) 059} [\href{https://arxiv.org/abs/0707.1038}{{\ttfamily
  0707.1038}}].

\bibitem{Armas:2018rsy}
J.~Armas, N.~Nguyen, V.~Niarchos, N.~A. Obers and T.~Van~Riet,
  \emph{{Meta-stable non-extremal anti-branes}},
  \href{https://doi.org/10.1103/PhysRevLett.122.181601}{\emph{Phys. Rev. Lett.}
  {\bfseries 122} (2019) 181601}
  [\href{https://arxiv.org/abs/1812.01067}{{\ttfamily 1812.01067}}].

\bibitem{Blaback:2012nf}
J.~Blaback, U.~H. Danielsson and T.~Van~Riet, \emph{{Resolving anti-brane
  singularities through time-dependence}},
  \href{https://doi.org/10.1007/JHEP02(2013)061}{\emph{JHEP} {\bfseries 02}
  (2013) 061} [\href{https://arxiv.org/abs/1202.1132}{{\ttfamily 1202.1132}}].

\bibitem{Bena:2009xk}
I.~Bena, M.~Grana and N.~Halmagyi, \emph{{On the Existence of Meta-stable Vacua
  in Klebanov-Strassler}},
  \href{https://doi.org/10.1007/JHEP09(2010)087}{\emph{JHEP} {\bfseries 09}
  (2010) 087} [\href{https://arxiv.org/abs/0912.3519}{{\ttfamily 0912.3519}}].

\bibitem{Bena:2012tx}
I.~Bena, D.~Junghans, S.~Kuperstein, T.~Van~Riet, T.~Wrase and M.~Zagermann,
  \emph{{Persistent anti-brane singularities}},
  \href{https://doi.org/10.1007/JHEP10(2012)078}{\emph{JHEP} {\bfseries 10}
  (2012) 078} [\href{https://arxiv.org/abs/1205.1798}{{\ttfamily 1205.1798}}].

\bibitem{McGuirk:2009xx}
P.~McGuirk, G.~Shiu and Y.~Sumitomo, \emph{{Non-supersymmetric infrared
  perturbations to the warped deformed conifold}},
  \href{https://doi.org/10.1016/j.nuclphysb.2010.09.008}{\emph{Nucl. Phys. B}
  {\bfseries 842} (2011) 383}
  [\href{https://arxiv.org/abs/0910.4581}{{\ttfamily 0910.4581}}].

\bibitem{Dymarsky:2011pm}
A.~Dymarsky, \emph{{On gravity dual of a metastable vacuum in
  Klebanov-Strassler theory}},
  \href{https://doi.org/10.1007/JHEP05(2011)053}{\emph{JHEP} {\bfseries 05}
  (2011) 053} [\href{https://arxiv.org/abs/1102.1734}{{\ttfamily 1102.1734}}].

\bibitem{Polchinski:2000uf}
J.~Polchinski and M.~J. Strassler, \emph{{The String dual of a confining
  four-dimensional gauge theory}},
  \href{https://arxiv.org/abs/hep-th/0003136}{{\ttfamily hep-th/0003136}}.

\bibitem{Blaback:2019ucp}
J.~Bl\r{a}b\"ack, F.~F. Gautason, A.~Ruip\'erez and T.~Van~Riet,
  \emph{{Anti-brane singularities as red herrings}},
  \href{https://doi.org/10.1007/JHEP12(2019)125}{\emph{JHEP} {\bfseries 12}
  (2019) 125} [\href{https://arxiv.org/abs/1907.05295}{{\ttfamily
  1907.05295}}].

\bibitem{Danielsson:2016cit}
U.~H. Danielsson, F.~F. Gautason and T.~Van~Riet, \emph{{Unstoppable brane-flux
  decay of $ \overline{\mathrm{D}6} $ branes}},
  \href{https://doi.org/10.1007/JHEP03(2017)141}{\emph{JHEP} {\bfseries 03}
  (2017) 141} [\href{https://arxiv.org/abs/1609.06529}{{\ttfamily
  1609.06529}}].

\bibitem{Bena:2013hr}
I.~Bena, J.~Blaback, U.~H. Danielsson and T.~Van~Riet, \emph{{Antibranes cannot
  become black}}, \href{https://doi.org/10.1103/PhysRevD.87.104023}{\emph{Phys.
  Rev. D} {\bfseries 87} (2013) 104023}
  [\href{https://arxiv.org/abs/1301.7071}{{\ttfamily 1301.7071}}].

\bibitem{Blaback:2014tfa}
J.~Bl\r{a}b\"ack, U.~H. Danielsson, D.~Junghans, T.~Van~Riet and S.~C. Vargas,
  \emph{{Localised anti-branes in non-compact throats at zero and finite $T$}},
  \href{https://doi.org/10.1007/JHEP02(2015)018}{\emph{JHEP} {\bfseries 02}
  (2015) 018} [\href{https://arxiv.org/abs/1409.0534}{{\ttfamily 1409.0534}}].

\bibitem{Polchinski:2015bea}
J.~Polchinski, \emph{{Brane/antibrane dynamics and KKLT stability}},
  \href{https://arxiv.org/abs/1509.05710}{{\ttfamily 1509.05710}}.

\bibitem{Cohen-Maldonado:2015ssa}
D.~Cohen-Maldonado, J.~Diaz, T.~van Riet and B.~Vercnocke, \emph{{Observations
  on fluxes near anti-branes}},
  \href{https://doi.org/10.1007/JHEP01(2016)126}{\emph{JHEP} {\bfseries 01}
  (2016) 126} [\href{https://arxiv.org/abs/1507.01022}{{\ttfamily
  1507.01022}}].

\bibitem{Gautason:2015tla}
F.~F. Gautason, B.~Truijen and T.~Van~Riet, \emph{{The many faces of brane-flux
  annihilation}}, \href{https://doi.org/10.1007/JHEP10(2015)152}{\emph{JHEP}
  {\bfseries 10} (2015) 152}
  [\href{https://arxiv.org/abs/1505.00159}{{\ttfamily 1505.00159}}].

\bibitem{Kuperstein:2014zda}
S.~Kuperstein, B.~Truijen and T.~Van~Riet, \emph{{Non-SUSY fractional branes}},
  \href{https://doi.org/10.1007/JHEP03(2015)161}{\emph{JHEP} {\bfseries 03}
  (2015) 161} [\href{https://arxiv.org/abs/1411.3358}{{\ttfamily 1411.3358}}].

\bibitem{Danielsson:2014yga}
U.~H. Danielsson and T.~Van~Riet, \emph{{Fatal attraction: more on decaying
  anti-branes}}, \href{https://doi.org/10.1007/JHEP03(2015)087}{\emph{JHEP}
  {\bfseries 03} (2015) 087} [\href{https://arxiv.org/abs/1410.8476}{{\ttfamily
  1410.8476}}].

\bibitem{Gao:2020xqh}
X.~Gao, A.~Hebecker and D.~Junghans, \emph{{Control issues of KKLT}},
  \href{https://doi.org/10.1002/prop.202000089}{\emph{Fortsch. Phys.}
  {\bfseries 68} (2020) 2000089}
  [\href{https://arxiv.org/abs/2009.03914}{{\ttfamily 2009.03914}}].

\bibitem{Dibitetto:2011gm}
G.~Dibitetto, A.~Guarino and D.~Roest, \emph{{Charting the landscape of N=4
  flux compactifications}},
  \href{https://doi.org/10.1007/JHEP03(2011)137}{\emph{JHEP} {\bfseries 03}
  (2011) 137} [\href{https://arxiv.org/abs/1102.0239}{{\ttfamily 1102.0239}}].

\bibitem{Danielsson:2013rza}
U.~Danielsson and G.~Dibitetto, \emph{{An alternative to anti-branes and
  O-planes?}}, \href{https://doi.org/10.1007/JHEP05(2014)013}{\emph{JHEP}
  {\bfseries 05} (2014) 013} [\href{https://arxiv.org/abs/1312.5331}{{\ttfamily
  1312.5331}}].

\bibitem{Retolaza:2015sta}
A.~Retolaza, A.~M. Uranga and A.~Westphal, \emph{{Bifid Throats for Axion
  Monodromy Inflation}},
  \href{https://doi.org/10.1007/JHEP07(2015)099}{\emph{JHEP} {\bfseries 07}
  (2015) 099} [\href{https://arxiv.org/abs/1504.02103}{{\ttfamily
  1504.02103}}].

\bibitem{Palti:2008mg}
E.~Palti, G.~Tasinato and J.~Ward, \emph{{WEAKLY-coupled IIA Flux
  Compactifications}},
  \href{https://doi.org/10.1088/1126-6708/2008/06/084}{\emph{JHEP} {\bfseries
  06} (2008) 084} [\href{https://arxiv.org/abs/0804.1248}{{\ttfamily
  0804.1248}}].

\bibitem{Hull:1994ys}
C.~M. Hull and P.~K. Townsend, \emph{{Unity of superstring dualities}},
  \href{https://doi.org/10.1016/0550-3213(94)00559-W}{\emph{Nucl. Phys.}
  {\bfseries B438} (1995) 109}
  [\href{https://arxiv.org/abs/hep-th/9410167}{{\ttfamily hep-th/9410167}}].

\bibitem{Schwarz:1996bh}
J.~H. Schwarz, \emph{{Lectures on superstring and M theory dualities: Given at
  ICTP Spring School and at TASI Summer School}},
  \href{https://doi.org/10.1016/S0920-5632(97)00070-4}{\emph{Nucl. Phys. Proc.
  Suppl.} {\bfseries 55B} (1997) 1}
  [\href{https://arxiv.org/abs/hep-th/9607201}{{\ttfamily hep-th/9607201}}].

\bibitem{Acharya:2007rc}
B.~S. Acharya, K.~Bobkov, G.~L. Kane, P.~Kumar and J.~Shao, \emph{{Explaining
  the Electroweak Scale and Stabilizing Moduli in M Theory}},
  \href{https://doi.org/10.1103/PhysRevD.76.126010}{\emph{Phys. Rev. D}
  {\bfseries 76} (2007) 126010}
  [\href{https://arxiv.org/abs/hep-th/0701034}{{\ttfamily hep-th/0701034}}].

\bibitem{Behrndt:1996hu}
K.~Behrndt, R.~Kallosh, J.~Rahmfeld, M.~Shmakova and W.~K. Wong, \emph{{STU
  black holes and string triality}},
  \href{https://doi.org/10.1103/PhysRevD.54.6293}{\emph{Phys. Rev.} {\bfseries
  D54} (1996) 6293} [\href{https://arxiv.org/abs/hep-th/9608059}{{\ttfamily
  hep-th/9608059}}].

\bibitem{Kachru:2000ih}
S.~Kachru, S.~H. Katz, A.~E. Lawrence and J.~McGreevy, \emph{{Open string
  instantons and superpotentials}},
  \href{https://doi.org/10.1103/PhysRevD.62.026001}{\emph{Phys. Rev.}
  {\bfseries D62} (2000) 026001}
  [\href{https://arxiv.org/abs/hep-th/9912151}{{\ttfamily hep-th/9912151}}].

\bibitem{Blumenhagen:2009qh}
R.~Blumenhagen, M.~Cvetic, S.~Kachru and T.~Weigand, \emph{{D-Brane Instantons
  in Type II Orientifolds}},
  \href{https://doi.org/10.1146/annurev.nucl.010909.083113}{\emph{Ann. Rev.
  Nucl. Part. Sci.} {\bfseries 59} (2009) 269}
  [\href{https://arxiv.org/abs/0902.3251}{{\ttfamily 0902.3251}}].

\bibitem{BlancoPillado:2005fn}
J.~J. Blanco-Pillado, R.~Kallosh and A.~D. Linde, \emph{{Supersymmetry and
  stability of flux vacua}},
  \href{https://doi.org/10.1088/1126-6708/2006/05/053}{\emph{JHEP} {\bfseries
  05} (2006) 053} [\href{https://arxiv.org/abs/hep-th/0511042}{{\ttfamily
  hep-th/0511042}}].

\bibitem{Kallosh:2011qk}
R.~Kallosh, A.~Linde, K.~A. Olive and T.~Rube, \emph{{Chaotic inflation and
  supersymmetry breaking}},
  \href{https://doi.org/10.1103/PhysRevD.84.083519}{\emph{Phys. Rev.}
  {\bfseries D84} (2011) 083519}
  [\href{https://arxiv.org/abs/1106.6025}{{\ttfamily 1106.6025}}].

\bibitem{Denef:2004ze}
F.~Denef and M.~R. Douglas, \emph{{Distributions of flux vacua}},
  \href{https://doi.org/10.1088/1126-6708/2004/05/072}{\emph{JHEP} {\bfseries
  05} (2004) 072} [\href{https://arxiv.org/abs/hep-th/0404116}{{\ttfamily
  hep-th/0404116}}].

\bibitem{Kallosh:2014oja}
R.~Kallosh, A.~Linde, B.~Vercnocke and T.~Wrase, \emph{{Analytic Classes of
  Metastable de Sitter Vacua}},
  \href{https://doi.org/10.1007/JHEP10(2014)011}{\emph{JHEP} {\bfseries 10}
  (2014) 011} [\href{https://arxiv.org/abs/1406.4866}{{\ttfamily 1406.4866}}].

\bibitem{Linde:2011ja}
A.~Linde, Y.~Mambrini and K.~A. Olive, \emph{{Supersymmetry Breaking due to
  Moduli Stabilization in String Theory}},
  \href{https://doi.org/10.1103/PhysRevD.85.066005}{\emph{Phys. Rev.}
  {\bfseries D85} (2012) 066005}
  [\href{https://arxiv.org/abs/1111.1465}{{\ttfamily 1111.1465}}].

\bibitem{Derendinger:2004jn}
J.-P. Derendinger, C.~Kounnas, P.~M. Petropoulos and F.~Zwirner,
  \emph{{Superpotentials in IIA compactifications with general fluxes}},
  \href{https://doi.org/10.1016/j.nuclphysb.2005.02.038}{\emph{Nucl. Phys.}
  {\bfseries B715} (2005) 211}
  [\href{https://arxiv.org/abs/hep-th/0411276}{{\ttfamily hep-th/0411276}}].

\bibitem{Villadoro:2005cu}
G.~Villadoro and F.~Zwirner, \emph{{N=1 effective potential from dual type-IIA
  D6/O6 orientifolds with general fluxes}},
  \href{https://doi.org/10.1088/1126-6708/2005/06/047}{\emph{JHEP} {\bfseries
  06} (2005) 047} [\href{https://arxiv.org/abs/hep-th/0503169}{{\ttfamily
  hep-th/0503169}}].

\bibitem{Kallosh:2017ced}
R.~Kallosh, A.~Linde, T.~Wrase and Y.~Yamada, \emph{{Maximal Supersymmetry and
  B-Mode Targets}}, \href{https://doi.org/10.1007/JHEP04(2017)144}{\emph{JHEP}
  {\bfseries 04} (2017) 144}
  [\href{https://arxiv.org/abs/1704.04829}{{\ttfamily 1704.04829}}].

\bibitem{Cicoli:2008va}
M.~Cicoli, J.~P. Conlon and F.~Quevedo, \emph{{General Analysis of LARGE Volume
  Scenarios with String Loop Moduli Stabilisation}},
  \href{https://doi.org/10.1088/1126-6708/2008/10/105}{\emph{JHEP} {\bfseries
  10} (2008) 105} [\href{https://arxiv.org/abs/0805.1029}{{\ttfamily
  0805.1029}}].

\bibitem{Cicoli:2008gp}
M.~Cicoli, C.~P. Burgess and F.~Quevedo, \emph{{Fibre Inflation: Observable
  Gravity Waves from IIB String Compactifications}},
  \href{https://doi.org/10.1088/1475-7516/2009/03/013}{\emph{JCAP} {\bfseries
  0903} (2009) 013} [\href{https://arxiv.org/abs/0808.0691}{{\ttfamily
  0808.0691}}].

\bibitem{Burgess:2016owb}
C.~P. Burgess, M.~Cicoli, S.~de~Alwis and F.~Quevedo, \emph{{Robust Inflation
  from Fibrous Strings}},
  \href{https://doi.org/10.1088/1475-7516/2016/05/032}{\emph{JCAP} {\bfseries
  1605} (2016) 032} [\href{https://arxiv.org/abs/1603.06789}{{\ttfamily
  1603.06789}}].

\bibitem{Kallosh:2017wku}
R.~Kallosh, A.~Linde, D.~Roest, A.~Westphal and Y.~Yamada, \emph{{Fibre
  Inflation and $\alpha$-attractors}},
  \href{https://doi.org/10.1007/JHEP02(2018)117}{\emph{JHEP} {\bfseries 02}
  (2018) 117} [\href{https://arxiv.org/abs/1707.05830}{{\ttfamily
  1707.05830}}].

\bibitem{Bobkov:2010rf}
K.~Bobkov, V.~Braun, P.~Kumar and S.~Raby, \emph{{Stabilizing All Kahler Moduli
  in Type IIB Orientifolds}},
  \href{https://doi.org/10.1007/JHEP12(2010)056}{\emph{JHEP} {\bfseries 12}
  (2010) 056} [\href{https://arxiv.org/abs/1003.1982}{{\ttfamily 1003.1982}}].

\bibitem{Grimm:2004uq}
T.~W. Grimm and J.~Louis, \emph{{The Effective action of N = 1 Calabi-Yau
  orientifolds}},
  \href{https://doi.org/10.1016/j.nuclphysb.2004.08.005}{\emph{Nucl. Phys.}
  {\bfseries B699} (2004) 387}
  [\href{https://arxiv.org/abs/hep-th/0403067}{{\ttfamily hep-th/0403067}}].

\bibitem{Denef:2004dm}
F.~Denef, M.~R. Douglas and B.~Florea, \emph{{Building a better racetrack}},
  \href{https://doi.org/10.1088/1126-6708/2004/06/034}{\emph{JHEP} {\bfseries
  06} (2004) 034} [\href{https://arxiv.org/abs/hep-th/0404257}{{\ttfamily
  hep-th/0404257}}].

\bibitem{Linde:2020mdk}
A.~Linde, \emph{{KKLT without AdS}},
  \href{https://doi.org/10.1007/JHEP05(2020)076}{\emph{JHEP} {\bfseries 05}
  (2020) 076} [\href{https://arxiv.org/abs/2002.01500}{{\ttfamily
  2002.01500}}].

\bibitem{Scherk:1978ta}
J.~Scherk and J.~H. Schwarz, \emph{{Spontaneous Breaking of Supersymmetry
  Through Dimensional Reduction}},
  \href{https://doi.org/10.1016/0370-2693(79)90425-8}{\emph{Phys. Lett. B}
  {\bfseries 82} (1979) 60}.

\bibitem{Scherk:1979zr}
J.~Scherk and J.~H. Schwarz, \emph{{How to Get Masses from Extra Dimensions}},
  \href{https://doi.org/10.1016/0550-3213(79)90592-3}{\emph{Nucl. Phys. B}
  {\bfseries 153} (1979) 61}.

\bibitem{Mueller-Hoissen:1987cwl}
F.~Mueller-Hoissen and R.~Stuckl, \emph{{Coset Spaces and Ten-dimensional
  Unified Theories}},
  \href{https://doi.org/10.1088/0264-9381/5/1/011}{\emph{Class. Quant. Grav.}
  {\bfseries 5} (1988) 27}.

\bibitem{Grana:2013ila}
M.~Gra\~na, R.~Minasian, H.~Triendl and T.~Van~Riet, \emph{{Quantization
  problem in Scherk-Schwarz compactifications}},
  \href{https://doi.org/10.1103/PhysRevD.88.085018}{\emph{Phys. Rev. D}
  {\bfseries 88} (2013) 085018}
  [\href{https://arxiv.org/abs/1305.0785}{{\ttfamily 1305.0785}}].

\bibitem{Bergshoeff:2001pv}
E.~Bergshoeff, R.~Kallosh, T.~Ortin, D.~Roest and A.~Van~Proeyen, \emph{{New
  formulations of D = 10 supersymmetry and D8 - O8 domain walls}},
  \href{https://doi.org/10.1088/0264-9381/18/17/303}{\emph{Class. Quant. Grav.}
  {\bfseries 18} (2001) 3359}
  [\href{https://arxiv.org/abs/hep-th/0103233}{{\ttfamily hep-th/0103233}}].

\bibitem{Blaback:2018hdo}
J.~Bl\r{a}b\"ack, U.~Danielsson and G.~Dibitetto, \emph{{A new light on the
  darkest corner of the landscape}},
  \href{https://arxiv.org/abs/1810.11365}{{\ttfamily 1810.11365}}.

\bibitem{Villadoro:2007yq}
G.~Villadoro and F.~Zwirner, \emph{{Beyond Twisted Tori}},
  \href{https://doi.org/10.1016/j.physletb.2007.07.002}{\emph{Phys. Lett. B}
  {\bfseries 652} (2007) 118}
  [\href{https://arxiv.org/abs/0706.3049}{{\ttfamily 0706.3049}}].

\bibitem{Aldazabal:2006up}
G.~Aldazabal, P.~G. Camara, A.~Font and L.~E. Ibanez, \emph{{More dual fluxes
  and moduli fixing}},
  \href{https://doi.org/10.1088/1126-6708/2006/05/070}{\emph{JHEP} {\bfseries
  05} (2006) 070} [\href{https://arxiv.org/abs/hep-th/0602089}{{\ttfamily
  hep-th/0602089}}].

\bibitem{Harvey:1999as}
J.~A. Harvey and G.~W. Moore, \emph{{Superpotentials and membrane instantons}},
   \href{https://arxiv.org/abs/hep-th/9907026}{{\ttfamily hep-th/9907026}}.

\bibitem{Blaback:2013ht}
J.~Blaback, U.~Danielsson and G.~Dibitetto, \emph{{Fully stable dS vacua from
  generalised fluxes}},
  \href{https://doi.org/10.1007/JHEP08(2013)054}{\emph{JHEP} {\bfseries 08}
  (2013) 054} [\href{https://arxiv.org/abs/1301.7073}{{\ttfamily 1301.7073}}].

\bibitem{Blumenhagen:2004xx}
R.~Blumenhagen, F.~Gmeiner, G.~Honecker, D.~Lust and T.~Weigand, \emph{{The
  Statistics of supersymmetric D-brane models}},
  \href{https://doi.org/10.1016/j.nuclphysb.2005.02.005}{\emph{Nucl. Phys. B}
  {\bfseries 713} (2005) 83}
  [\href{https://arxiv.org/abs/hep-th/0411173}{{\ttfamily hep-th/0411173}}].

\bibitem{Weinberg:1987dv}
S.~Weinberg, \emph{{Anthropic Bound on the Cosmological Constant}},
  \href{https://doi.org/10.1103/PhysRevLett.59.2607}{\emph{Phys. Rev. Lett.}
  {\bfseries 59} (1987) 2607}.

\bibitem{Junghans:2018gdb}
D.~Junghans, \emph{{Weakly Coupled de Sitter Vacua with Fluxes and the
  Swampland}}, \href{https://doi.org/10.1007/JHEP03(2019)150}{\emph{JHEP}
  {\bfseries 03} (2019) 150}
  [\href{https://arxiv.org/abs/1811.06990}{{\ttfamily 1811.06990}}].

\bibitem{Aharony:2010af}
O.~Aharony, D.~Jafferis, A.~Tomasiello and A.~Zaffaroni, \emph{{Massive type
  IIA string theory cannot be strongly coupled}},
  \href{https://doi.org/10.1007/JHEP11(2010)047}{\emph{JHEP} {\bfseries 11}
  (2010) 047} [\href{https://arxiv.org/abs/1007.2451}{{\ttfamily 1007.2451}}].

\bibitem{Lust:2009zb}
D.~Lust and D.~Tsimpis, \emph{{Classes of AdS(4) type IIA/IIB compactifications
  with SU(3) x SU(3) structure}},
  \href{https://doi.org/10.1088/1126-6708/2009/04/111}{\emph{JHEP} {\bfseries
  04} (2009) 111} [\href{https://arxiv.org/abs/0901.4474}{{\ttfamily
  0901.4474}}].

\bibitem{Petrini:2013ika}
M.~Petrini, G.~Solard and T.~Van~Riet, \emph{{AdS vacua with scale separation
  from IIB supergravity}},
  \href{https://doi.org/10.1007/JHEP11(2013)010}{\emph{JHEP} {\bfseries 11}
  (2013) 010} [\href{https://arxiv.org/abs/1308.1265}{{\ttfamily 1308.1265}}].

\bibitem{Camara:2005dc}
P.~G. Camara, A.~Font and L.~E. Ibanez, \emph{{Fluxes, moduli fixing and
  MSSM-like vacua in a simple IIA orientifold}},
  \href{https://doi.org/10.1088/1126-6708/2005/09/013}{\emph{JHEP} {\bfseries
  09} (2005) 013} [\href{https://arxiv.org/abs/hep-th/0506066}{{\ttfamily
  hep-th/0506066}}].

\bibitem{Ihl:2006pp}
M.~Ihl and T.~Wrase, \emph{{Towards a Realistic Type IIA T**6/Z(4) Orientifold
  Model with Background Fluxes. Part 1. Moduli Stabilization}},
  \href{https://doi.org/10.1088/1126-6708/2006/07/027}{\emph{JHEP} {\bfseries
  07} (2006) 027} [\href{https://arxiv.org/abs/hep-th/0604087}{{\ttfamily
  hep-th/0604087}}].

\bibitem{Hertzberg:2007ke}
M.~P. Hertzberg, M.~Tegmark, S.~Kachru, J.~Shelton and O.~Ozcan,
  \emph{{Searching for Inflation in Simple String Theory Models: An
  Astrophysical Perspective}},
  \href{https://doi.org/10.1103/PhysRevD.76.103521}{\emph{Phys. Rev. D}
  {\bfseries 76} (2007) 103521}
  [\href{https://arxiv.org/abs/0709.0002}{{\ttfamily 0709.0002}}].

\bibitem{Hertzberg:2007wc}
M.~P. Hertzberg, S.~Kachru, W.~Taylor and M.~Tegmark, \emph{{Inflationary
  Constraints on Type IIA String Theory}},
  \href{https://doi.org/10.1088/1126-6708/2007/12/095}{\emph{JHEP} {\bfseries
  12} (2007) 095} [\href{https://arxiv.org/abs/0711.2512}{{\ttfamily
  0711.2512}}].

\bibitem{Saltman:2004jh}
A.~Saltman and E.~Silverstein, \emph{{A New handle on de Sitter
  compactifications}},
  \href{https://doi.org/10.1088/1126-6708/2006/01/139}{\emph{JHEP} {\bfseries
  01} (2006) 139} [\href{https://arxiv.org/abs/hep-th/0411271}{{\ttfamily
  hep-th/0411271}}].

\bibitem{Silverstein:2007ac}
E.~Silverstein, \emph{{Simple de Sitter Solutions}},
  \href{https://doi.org/10.1103/PhysRevD.77.106006}{\emph{Phys. Rev. D}
  {\bfseries 77} (2008) 106006}
  [\href{https://arxiv.org/abs/0712.1196}{{\ttfamily 0712.1196}}].

\bibitem{Haque:2008jz}
S.~S. Haque, G.~Shiu, B.~Underwood and T.~Van~Riet, \emph{{Minimal simple de
  Sitter solutions}},
  \href{https://doi.org/10.1103/PhysRevD.79.086005}{\emph{Phys. Rev.}
  {\bfseries D79} (2009) 086005}
  [\href{https://arxiv.org/abs/0810.5328}{{\ttfamily 0810.5328}}].

\bibitem{Danielsson:2009ff}
U.~H. Danielsson, S.~S. Haque, G.~Shiu and T.~Van~Riet, \emph{{Towards
  Classical de Sitter Solutions in String Theory}},
  \href{https://doi.org/10.1088/1126-6708/2009/09/114}{\emph{JHEP} {\bfseries
  09} (2009) 114} [\href{https://arxiv.org/abs/0907.2041}{{\ttfamily
  0907.2041}}].

\bibitem{deCarlos:2009fq}
B.~de~Carlos, A.~Guarino and J.~M. Moreno, \emph{{Flux moduli stabilisation,
  Supergravity algebras and no-go theorems}},
  \href{https://doi.org/10.1007/JHEP01(2010)012}{\emph{JHEP} {\bfseries 01}
  (2010) 012} [\href{https://arxiv.org/abs/0907.5580}{{\ttfamily 0907.5580}}].

\bibitem{deCarlos:2009qm}
B.~de~Carlos, A.~Guarino and J.~M. Moreno, \emph{{Complete classification of
  Minkowski vacua in generalised flux models}},
  \href{https://doi.org/10.1007/JHEP02(2010)076}{\emph{JHEP} {\bfseries 02}
  (2010) 076} [\href{https://arxiv.org/abs/0911.2876}{{\ttfamily 0911.2876}}].

\bibitem{Caviezel:2009tu}
C.~Caviezel, T.~Wrase and M.~Zagermann, \emph{{Moduli Stabilization and
  Cosmology of Type IIB on SU(2)-Structure Orientifolds}},
  \href{https://doi.org/10.1007/JHEP04(2010)011}{\emph{JHEP} {\bfseries 04}
  (2010) 011} [\href{https://arxiv.org/abs/0912.3287}{{\ttfamily 0912.3287}}].

\bibitem{Danielsson:2010bc}
U.~H. Danielsson, P.~Koerber and T.~Van~Riet, \emph{{Universal de Sitter
  solutions at tree-level}},
  \href{https://doi.org/10.1007/JHEP05(2010)090}{\emph{JHEP} {\bfseries 05}
  (2010) 090} [\href{https://arxiv.org/abs/1003.3590}{{\ttfamily 1003.3590}}].

\bibitem{Dong:2010pm}
X.~Dong, B.~Horn, E.~Silverstein and G.~Torroba, \emph{{Micromanaging de Sitter
  holography}},
  \href{https://doi.org/10.1088/0264-9381/27/24/245020}{\emph{Class. Quant.
  Grav.} {\bfseries 27} (2010) 245020}
  [\href{https://arxiv.org/abs/1005.5403}{{\ttfamily 1005.5403}}].

\bibitem{Andriot:2010ju}
D.~Andriot, E.~Goi, R.~Minasian and M.~Petrini, \emph{{Supersymmetry breaking
  branes on solvmanifolds and de Sitter vacua in string theory}},
  \href{https://doi.org/10.1007/JHEP05(2011)028}{\emph{JHEP} {\bfseries 05}
  (2011) 028} [\href{https://arxiv.org/abs/1003.3774}{{\ttfamily 1003.3774}}].

\bibitem{Caviezel:2008ik}
C.~Caviezel, P.~Koerber, S.~Kors, D.~Lust, D.~Tsimpis and M.~Zagermann,
  \emph{{The Effective theory of type IIA AdS(4) compactifications on
  nilmanifolds and cosets}},
  \href{https://doi.org/10.1088/0264-9381/26/2/025014}{\emph{Class. Quant.
  Grav.} {\bfseries 26} (2009) 025014}
  [\href{https://arxiv.org/abs/0806.3458}{{\ttfamily 0806.3458}}].

\bibitem{Wrase:2010ew}
T.~Wrase and M.~Zagermann, \emph{{On Classical de Sitter Vacua in String
  Theory}}, \href{https://doi.org/10.1002/prop.201000053}{\emph{Fortsch. Phys.}
  {\bfseries 58} (2010) 906} [\href{https://arxiv.org/abs/1003.0029}{{\ttfamily
  1003.0029}}].

\bibitem{Andriot:2016xvq}
D.~Andriot and J.~Blaback, \emph{{Refining the boundaries of the classical de
  Sitter landscape}}, \href{https://doi.org/10.1007/JHEP03(2017)102,
  10.1007/JHEP03(2018)083}{\emph{JHEP} {\bfseries 03} (2017) 102}
  [\href{https://arxiv.org/abs/1609.00385}{{\ttfamily 1609.00385}}].

\bibitem{Andriot:2017jhf}
D.~Andriot, \emph{{On classical de Sitter and Minkowski solutions with
  intersecting branes}},
  \href{https://doi.org/10.1007/JHEP03(2018)054}{\emph{JHEP} {\bfseries 03}
  (2018) 054} [\href{https://arxiv.org/abs/1710.08886}{{\ttfamily
  1710.08886}}].

\bibitem{Maldacena:2000mw}
J.~M. Maldacena and C.~Nunez, \emph{{Supergravity description of field theories
  on curved manifolds and a no go theorem}},
  \href{https://doi.org/10.1142/S0217751X01003937}{\emph{Int. J. Mod. Phys. A}
  {\bfseries 16} (2001) 822}
  [\href{https://arxiv.org/abs/hep-th/0007018}{{\ttfamily hep-th/0007018}}].

\bibitem{Dasgupta:1999ss}
K.~Dasgupta, G.~Rajesh and S.~Sethi, \emph{{M theory, orientifolds and G -
  flux}}, \href{https://doi.org/10.1088/1126-6708/1999/08/023}{\emph{JHEP}
  {\bfseries 08} (1999) 023}
  [\href{https://arxiv.org/abs/hep-th/9908088}{{\ttfamily hep-th/9908088}}].

\bibitem{Gautason:2015tig}
F.~Gautason, M.~Schillo, T.~Van~Riet and M.~Williams, \emph{{Remarks on scale
  separation in flux vacua}},
  \href{https://doi.org/10.1007/JHEP03(2016)061}{\emph{JHEP} {\bfseries 03}
  (2016) 061} [\href{https://arxiv.org/abs/1512.00457}{{\ttfamily
  1512.00457}}].

\bibitem{Westphal:2006tn}
A.~Westphal, \emph{{de Sitter string vacua from Kahler uplifting}},
  \href{https://doi.org/10.1088/1126-6708/2007/03/102}{\emph{JHEP} {\bfseries
  03} (2007) 102} [\href{https://arxiv.org/abs/hep-th/0611332}{{\ttfamily
  hep-th/0611332}}].

\bibitem{Farakos:2020phe}
F.~Farakos, G.~Tringas and T.~Van~Riet, \emph{{No-scale and scale-separated
  flux vacua from IIA on G2 orientifolds}},
  \href{https://doi.org/10.1140/epjc/s10052-020-8247-5}{\emph{Eur. Phys. J. C}
  {\bfseries 80} (2020) 659}
  [\href{https://arxiv.org/abs/2005.05246}{{\ttfamily 2005.05246}}].

\bibitem{Aharony:2008wz}
O.~Aharony, Y.~E. Antebi and M.~Berkooz, \emph{{On the Conformal Field Theory
  Duals of type IIA AdS(4) Flux Compactifications}},
  \href{https://doi.org/10.1088/1126-6708/2008/02/093}{\emph{JHEP} {\bfseries
  02} (2008) 093} [\href{https://arxiv.org/abs/0801.3326}{{\ttfamily
  0801.3326}}].

\bibitem{Acharya:2006ne}
B.~S. Acharya, F.~Benini and R.~Valandro, \emph{{Fixing moduli in exact type
  IIA flux vacua}},
  \href{https://doi.org/10.1088/1126-6708/2007/02/018}{\emph{JHEP} {\bfseries
  02} (2007) 018} [\href{https://arxiv.org/abs/hep-th/0607223}{{\ttfamily
  hep-th/0607223}}].

\bibitem{Saracco:2012wc}
F.~Saracco and A.~Tomasiello, \emph{{Localized O6-plane solutions with Romans
  mass}}, \href{https://doi.org/10.1007/JHEP07(2012)077}{\emph{JHEP} {\bfseries
  07} (2012) 077} [\href{https://arxiv.org/abs/1201.5378}{{\ttfamily
  1201.5378}}].

\bibitem{Junghans:2020acz}
D.~Junghans, \emph{{O-Plane Backreaction and Scale Separation in Type IIA Flux
  Vacua}}, \href{https://doi.org/10.1002/prop.202000040}{\emph{Fortsch. Phys.}
  {\bfseries 68} (2020) 2000040}
  [\href{https://arxiv.org/abs/2003.06274}{{\ttfamily 2003.06274}}].

\bibitem{Junghans:2016uvg}
D.~Junghans, \emph{{Tachyons in Classical de Sitter Vacua}},
  \href{https://doi.org/10.1007/JHEP06(2016)132}{\emph{JHEP} {\bfseries 06}
  (2016) 132} [\href{https://arxiv.org/abs/1603.08939}{{\ttfamily
  1603.08939}}].

\bibitem{Herraez:2018vae}
A.~Herraez, L.~E. Ibanez, F.~Marchesano and G.~Zoccarato, \emph{{The Type IIA
  Flux Potential, 4-forms and Freed-Witten anomalies}},
  \href{https://doi.org/10.1007/JHEP09(2018)018}{\emph{JHEP} {\bfseries 09}
  (2018) 018} [\href{https://arxiv.org/abs/1802.05771}{{\ttfamily
  1802.05771}}].

\bibitem{Danielsson:2011au}
U.~H. Danielsson, S.~S. Haque, P.~Koerber, G.~Shiu, T.~Van~Riet and T.~Wrase,
  \emph{{De Sitter hunting in a classical landscape}},
  \href{https://doi.org/10.1002/prop.201100047}{\emph{Fortsch. Phys.}
  {\bfseries 59} (2011) 897} [\href{https://arxiv.org/abs/1103.4858}{{\ttfamily
  1103.4858}}].

\bibitem{Danielsson:2012by}
U.~Danielsson and G.~Dibitetto, \emph{{On the distribution of stable de Sitter
  vacua}}, \href{https://doi.org/10.1007/JHEP03(2013)018}{\emph{JHEP}
  {\bfseries 03} (2013) 018} [\href{https://arxiv.org/abs/1212.4984}{{\ttfamily
  1212.4984}}].

\bibitem{Damian:2013dq}
C.~Damian, L.~R. Diaz-Barron, O.~Loaiza-Brito and M.~Sabido, \emph{{Slow-Roll
  Inflation in Non-geometric Flux Compactification}},
  \href{https://doi.org/10.1007/JHEP06(2013)109}{\emph{JHEP} {\bfseries 06}
  (2013) 109} [\href{https://arxiv.org/abs/1302.0529}{{\ttfamily 1302.0529}}].

\bibitem{Blaback:2015zra}
J.~Bl\r{a}b\"ack, U.~H. Danielsson, G.~Dibitetto and S.~C. Vargas,
  \emph{{Universal dS vacua in STU-models}},
  \href{https://doi.org/10.1007/JHEP10(2015)069}{\emph{JHEP} {\bfseries 10}
  (2015) 069} [\href{https://arxiv.org/abs/1505.04283}{{\ttfamily
  1505.04283}}].

\bibitem{Danielsson:2012et}
U.~H. Danielsson, G.~Shiu, T.~Van~Riet and T.~Wrase, \emph{{A note on obstinate
  tachyons in classical dS solutions}},
  \href{https://doi.org/10.1007/JHEP03(2013)138}{\emph{JHEP} {\bfseries 03}
  (2013) 138} [\href{https://arxiv.org/abs/1212.5178}{{\ttfamily 1212.5178}}].

\bibitem{Junghans:2016abx}
D.~Junghans and M.~Zagermann, \emph{{A Universal Tachyon in Nearly No-scale de
  Sitter Compactifications}},
  \href{https://doi.org/10.1007/JHEP07(2018)078}{\emph{JHEP} {\bfseries 07}
  (2018) 078} [\href{https://arxiv.org/abs/1612.06847}{{\ttfamily
  1612.06847}}].

\bibitem{Aldazabal:2007sn}
G.~Aldazabal and A.~Font, \emph{{A Second look at N=1 supersymmetric AdS(4)
  vacua of type IIA supergravity}},
  \href{https://doi.org/10.1088/1126-6708/2008/02/086}{\emph{JHEP} {\bfseries
  02} (2008) 086} [\href{https://arxiv.org/abs/0712.1021}{{\ttfamily
  0712.1021}}].

\bibitem{Blaback:2013fca}
J.~Bl\r{a}b\"ack, U.~Danielsson and G.~Dibitetto, \emph{{Accelerated Universes
  from type IIA Compactifications}},
  \href{https://doi.org/10.1088/1475-7516/2014/03/003}{\emph{JCAP} {\bfseries
  03} (2014) 003} [\href{https://arxiv.org/abs/1310.8300}{{\ttfamily
  1310.8300}}].

\bibitem{Kashani-Poor:2006ofe}
A.-K. Kashani-Poor and R.~Minasian, \emph{{Towards reduction of type II
  theories on SU(3) structure manifolds}},
  \href{https://doi.org/10.1088/1126-6708/2007/03/109}{\emph{JHEP} {\bfseries
  03} (2007) 109} [\href{https://arxiv.org/abs/hep-th/0611106}{{\ttfamily
  hep-th/0611106}}].

\bibitem{Kashani-Poor:2007nby}
A.-K. Kashani-Poor, \emph{{Nearly Kaehler Reduction}},
  \href{https://doi.org/10.1088/1126-6708/2007/11/026}{\emph{JHEP} {\bfseries
  11} (2007) 026} [\href{https://arxiv.org/abs/0709.4482}{{\ttfamily
  0709.4482}}].

\bibitem{Cassani:2009ck}
D.~Cassani and A.-K. Kashani-Poor, \emph{{Exploiting N=2 in consistent coset
  reductions of type IIA}},
  \href{https://doi.org/10.1016/j.nuclphysb.2009.03.011}{\emph{Nucl. Phys. B}
  {\bfseries 817} (2009) 25} [\href{https://arxiv.org/abs/0901.4251}{{\ttfamily
  0901.4251}}].

\bibitem{Andriot:2018wzk}
D.~Andriot, \emph{{On the de Sitter swampland criterion}},
  \href{https://doi.org/10.1016/j.physletb.2018.09.022}{\emph{Phys. Lett. B}
  {\bfseries 785} (2018) 570}
  [\href{https://arxiv.org/abs/1806.10999}{{\ttfamily 1806.10999}}].

\bibitem{Douglas:2010rt}
M.~R. Douglas and R.~Kallosh, \emph{{Compactification on negatively curved
  manifolds}}, \href{https://doi.org/10.1007/JHEP06(2010)004}{\emph{JHEP}
  {\bfseries 06} (2010) 004} [\href{https://arxiv.org/abs/1001.4008}{{\ttfamily
  1001.4008}}].

\bibitem{Blaback:2010sj}
J.~Blaback, U.~H. Danielsson, D.~Junghans, T.~Van~Riet, T.~Wrase and
  M.~Zagermann, \emph{{Smeared versus localised sources in flux
  compactifications}},
  \href{https://doi.org/10.1007/JHEP12(2010)043}{\emph{JHEP} {\bfseries 12}
  (2010) 043} [\href{https://arxiv.org/abs/1009.1877}{{\ttfamily 1009.1877}}].

\bibitem{Blaback:2011nz}
J.~Blaback, U.~H. Danielsson, D.~Junghans, T.~Van~Riet, T.~Wrase and
  M.~Zagermann, \emph{{The problematic backreaction of SUSY-breaking branes}},
  \href{https://doi.org/10.1007/JHEP08(2011)105}{\emph{JHEP} {\bfseries 08}
  (2011) 105} [\href{https://arxiv.org/abs/1105.4879}{{\ttfamily 1105.4879}}].

\bibitem{Blaback:2011pn}
J.~Blaback, U.~H. Danielsson, D.~Junghans, T.~Van~Riet, T.~Wrase and
  M.~Zagermann, \emph{{(Anti-)Brane backreaction beyond perturbation theory}},
  \href{https://doi.org/10.1007/JHEP02(2012)025}{\emph{JHEP} {\bfseries 02}
  (2012) 025} [\href{https://arxiv.org/abs/1111.2605}{{\ttfamily 1111.2605}}].

\bibitem{McOrist:2012yc}
J.~McOrist and S.~Sethi, \emph{{M-theory and Type IIA Flux Compactifications}},
  \href{https://doi.org/10.1007/JHEP12(2012)122}{\emph{JHEP} {\bfseries 12}
  (2012) 122} [\href{https://arxiv.org/abs/1208.0261}{{\ttfamily 1208.0261}}].

\bibitem{Baines:2020dmu}
S.~Baines and T.~Van~Riet, \emph{{Smearing orientifolds in flux
  compactifications can be OK}},
  \href{https://doi.org/10.1088/1361-6382/aba8e0}{\emph{Class. Quant. Grav.}
  {\bfseries 37} (2020) 195015}
  [\href{https://arxiv.org/abs/2005.09501}{{\ttfamily 2005.09501}}].

\bibitem{Marchesano:2020qvg}
F.~Marchesano, E.~Palti, J.~Quirant and A.~Tomasiello, \emph{{On supersymmetric
  AdS$_{4}$ orientifold vacua}},
  \href{https://doi.org/10.1007/JHEP08(2020)087}{\emph{JHEP} {\bfseries 08}
  (2020) 087} [\href{https://arxiv.org/abs/2003.13578}{{\ttfamily
  2003.13578}}].

\bibitem{Cribiori:2021djm}
N.~Cribiori, D.~Junghans, V.~Van~Hemelryck, T.~Van~Riet and T.~Wrase,
  \emph{{Scale-separated AdS$_4$ vacua of IIA orientifolds and M-theory}},
  \href{https://arxiv.org/abs/2107.00019}{{\ttfamily 2107.00019}}.

\bibitem{Romans:1985tz}
L.~J. Romans, \emph{{Massive N=2a Supergravity in Ten-Dimensions}},
  \href{https://doi.org/10.1016/0370-2693(86)90375-8}{\emph{Phys. Lett. B}
  {\bfseries 169} (1986) 374}.

\bibitem{Banks:2006hg}
T.~Banks and K.~van~den Broek, \emph{{Massive IIA flux compactifications and
  U-dualities}},
  \href{https://doi.org/10.1088/1126-6708/2007/03/068}{\emph{JHEP} {\bfseries
  03} (2007) 068} [\href{https://arxiv.org/abs/hep-th/0611185}{{\ttfamily
  hep-th/0611185}}].

\bibitem{Moore:1999gb}
G.~W. Moore and E.~Witten, \emph{{Selfduality, Ramond-Ramond fields, and K
  theory}}, \href{https://doi.org/10.1088/1126-6708/2000/05/032}{\emph{JHEP}
  {\bfseries 05} (2000) 032}
  [\href{https://arxiv.org/abs/hep-th/9912279}{{\ttfamily hep-th/9912279}}].

\bibitem{Minasian:1997mm}
R.~Minasian and G.~W. Moore, \emph{{K theory and Ramond-Ramond charge}},
  \href{https://doi.org/10.1088/1126-6708/1997/11/002}{\emph{JHEP} {\bfseries
  11} (1997) 002} [\href{https://arxiv.org/abs/hep-th/9710230}{{\ttfamily
  hep-th/9710230}}].

\bibitem{Collinucci:2006ug}
A.~Collinucci and J.~Evslin, \emph{{Twisted Homology}},
  \href{https://doi.org/10.1088/1126-6708/2007/03/058}{\emph{JHEP} {\bfseries
  03} (2007) 058} [\href{https://arxiv.org/abs/hep-th/0611218}{{\ttfamily
  hep-th/0611218}}].

\bibitem{Robbins:2007yv}
D.~Robbins and T.~Wrase, \emph{{D-terms from generalized NS-NS fluxes in type
  II}}, \href{https://doi.org/10.1088/1126-6708/2007/12/058}{\emph{JHEP}
  {\bfseries 12} (2007) 058} [\href{https://arxiv.org/abs/0709.2186}{{\ttfamily
  0709.2186}}].

\bibitem{Andriot:2018ept}
D.~Andriot, \emph{{New constraints on classical de Sitter: flirting with the
  swampland}}, \href{https://doi.org/10.1002/prop.201800103}{\emph{Fortsch.
  Phys.} {\bfseries 67} (2019) 1800103}
  [\href{https://arxiv.org/abs/1807.09698}{{\ttfamily 1807.09698}}].

\bibitem{Murayama:2018lie}
H.~Murayama, M.~Yamazaki and T.~T. Yanagida, \emph{{Do We Live in the
  Swampland?}}, \href{https://doi.org/10.1007/JHEP12(2018)032}{\emph{JHEP}
  {\bfseries 12} (2018) 032}
  [\href{https://arxiv.org/abs/1809.00478}{{\ttfamily 1809.00478}}].

\bibitem{Choi:2018rze}
K.~Choi, D.~Chway and C.~S. Shin, \emph{{The dS swampland conjecture with the
  electroweak symmetry and QCD chiral symmetry breaking}},
  \href{https://doi.org/10.1007/JHEP11(2018)142}{\emph{JHEP} {\bfseries 11}
  (2018) 142} [\href{https://arxiv.org/abs/1809.01475}{{\ttfamily
  1809.01475}}].

\bibitem{Hamaguchi:2018vtv}
K.~Hamaguchi, M.~Ibe and T.~Moroi, \emph{{The swampland conjecture and the
  Higgs expectation value}},
  \href{https://doi.org/10.1007/JHEP12(2018)023}{\emph{JHEP} {\bfseries 12}
  (2018) 023} [\href{https://arxiv.org/abs/1810.02095}{{\ttfamily
  1810.02095}}].

\bibitem{Garg:2018reu}
S.~K. Garg and C.~Krishnan, \emph{{Bounds on Slow Roll and the de Sitter
  Swampland}}, \href{https://doi.org/10.1007/JHEP11(2019)075}{\emph{JHEP}
  {\bfseries 11} (2019) 075}
  [\href{https://arxiv.org/abs/1807.05193}{{\ttfamily 1807.05193}}].

\bibitem{Baume:2016psm}
F.~Baume and E.~Palti, \emph{{Backreacted Axion Field Ranges in String
  Theory}}, \href{https://doi.org/10.1007/JHEP08(2016)043}{\emph{JHEP}
  {\bfseries 08} (2016) 043}
  [\href{https://arxiv.org/abs/1602.06517}{{\ttfamily 1602.06517}}].

\bibitem{Klaewer:2016kiy}
D.~Klaewer and E.~Palti, \emph{{Super-Planckian Spatial Field Variations and
  Quantum Gravity}}, \href{https://doi.org/10.1007/JHEP01(2017)088}{\emph{JHEP}
  {\bfseries 01} (2017) 088}
  [\href{https://arxiv.org/abs/1610.00010}{{\ttfamily 1610.00010}}].

\bibitem{Ooguri:2006in}
H.~Ooguri and C.~Vafa, \emph{{On the Geometry of the String Landscape and the
  Swampland}},
  \href{https://doi.org/10.1016/j.nuclphysb.2006.10.033}{\emph{Nucl. Phys. B}
  {\bfseries 766} (2007) 21}
  [\href{https://arxiv.org/abs/hep-th/0605264}{{\ttfamily hep-th/0605264}}].

\bibitem{Gibbons:1977mu}
G.~W. Gibbons and S.~W. Hawking, \emph{{Cosmological Event Horizons,
  Thermodynamics, and Particle Creation}},
  \href{https://doi.org/10.1103/PhysRevD.15.2738}{\emph{Phys. Rev. D}
  {\bfseries 15} (1977) 2738}.

\bibitem{Bousso:1999xy}
R.~Bousso, \emph{{A Covariant entropy conjecture}},
  \href{https://doi.org/10.1088/1126-6708/1999/07/004}{\emph{JHEP} {\bfseries
  07} (1999) 004} [\href{https://arxiv.org/abs/hep-th/9905177}{{\ttfamily
  hep-th/9905177}}].

\bibitem{Achucarro:2018vey}
A.~Ach\'ucarro and G.~A. Palma, \emph{{The string swampland constraints require
  multi-field inflation}},
  \href{https://doi.org/10.1088/1475-7516/2019/02/041}{\emph{JCAP} {\bfseries
  02} (2019) 041} [\href{https://arxiv.org/abs/1807.04390}{{\ttfamily
  1807.04390}}].

\bibitem{Kehagias:2018uem}
A.~Kehagias and A.~Riotto, \emph{{A note on Inflation and the Swampland}},
  \href{https://doi.org/10.1002/prop.201800052}{\emph{Fortsch. Phys.}
  {\bfseries 66} (2018) 1800052}
  [\href{https://arxiv.org/abs/1807.05445}{{\ttfamily 1807.05445}}].

\bibitem{Agrawal:2018own}
P.~Agrawal, G.~Obied, P.~J. Steinhardt and C.~Vafa, \emph{{On the Cosmological
  Implications of the String Swampland}},
  \href{https://doi.org/10.1016/j.physletb.2018.07.040}{\emph{Phys. Lett. B}
  {\bfseries 784} (2018) 271}
  [\href{https://arxiv.org/abs/1806.09718}{{\ttfamily 1806.09718}}].

\bibitem{Dine:1985he}
M.~Dine and N.~Seiberg, \emph{{Is the Superstring Weakly Coupled?}},
  \href{https://doi.org/10.1016/0370-2693(85)90927-X}{\emph{Phys. Lett. B}
  {\bfseries 162} (1985) 299}.

\bibitem{Hebecker:2018vxz}
A.~Hebecker and T.~Wrase, \emph{{The Asymptotic dS Swampland Conjecture - a
  Simplified Derivation and a Potential Loophole}},
  \href{https://doi.org/10.1002/prop.201800097}{\emph{Fortsch. Phys.}
  {\bfseries 67} (2019) 1800097}
  [\href{https://arxiv.org/abs/1810.08182}{{\ttfamily 1810.08182}}].

\bibitem{Gaillard:1981rj}
M.~K. Gaillard and B.~Zumino, \emph{{Duality Rotations for Interacting
  Fields}}, \href{https://doi.org/10.1016/0550-3213(81)90527-7}{\emph{Nucl.
  Phys. B} {\bfseries 193} (1981) 221}.

\bibitem{Kugo:1983mv}
T.~Kugo and S.~Uehara, \emph{{$N=1$ Superconformal Tensor Calculus: Multiplets
  With External Lorentz Indices and Spinor Derivative Operators}},
  \href{https://doi.org/10.1143/PTP.73.235}{\emph{Prog. Theor. Phys.}
  {\bfseries 73} (1985) 235}.

\end{thebibliography}\endgroup

\end{document}